# Introduction to Univalent Foundations of Mathematics with Agda

Martín Hötzel Escardó, School of Computer Science, University of Birmingham, UK.

**Abstract.** We introduce Voevodsky's univalent foundations and univalent mathematics, and explain how to develop them with the computer system Agda, which is based on Martin-Löf type theory. Agda allows us to write mathematical definitions, constructions, theorems and proofs, for example in number theory, analysis, group theory, topology, category theory or programming language theory, checking them for logical and mathematical correctness.

Agda is a constructive mathematical system by default, which amounts to saying that it can also be considered as a programming language for manipulating mathematical objects. But we can assume the axiom of choice or the principle of excluded middle for pieces of mathematics that require them, at the cost of losing the implicit programming-language character of the system. For a fully constructive development of univalent mathematics in Agda, we would need to use its new cubical flavour, and we hope these notes provide a base for researchers interested in learning cubical type theory and cubical Agda as the next step.

Compared to most expositions of the subject, we work with explicit universe levels.

We also fully discuss and emphasize that non-constructive classical axioms can be assumed consistently in univalent mathematics.

**About this document.** This is a set of so-called literate Agda files, with the formal, verified, mathematical development within *code* environments, and the usual mathematical discussion outside them. Most of this file is not Agda code, and is in markdown format, and the html web page is generated automatically from it using Agda and other tools. Github issues or pull requests by students to fix typos or mistakes and clarify ambiguities are welcome. There is also a pdf version with internal links to sections and Agda definitions, which is automatically generated from the html version. And a pdf version is also available at the arxiv, updated from time to time.

These notes were originally developed for the Midlands Graduate School 2019. They will evolve for a while.

How to cite this document.

## Introduction

A univalent type theory is the underlying formal system for a foundation of univalent mathematics as conceived by Voevodsky.

In the same way as there isn't just one set theory (we have e.g. ZFC and NBG among others), there isn't just one univalent type theory (we have e.g. the underlying type theory used in UniMath, HoTT-book type theory, and cubical type theory, among others, and more are expected to come in the foreseeable future before the foundations of univalent mathematics stabilize).

The salient differences between univalent mathematics and traditional, set-based mathematics may be shocking at first sight:

1. The kinds of objects we take as basic.

    - Certain things called types, or higher groupoids, rather than sets, are the primitive objects.
    - Sets, also called 0-groupoids, are particular kinds of types.
    - So we have more general objects as a starting point.

- E.g. the type $\mathbb{N}$ of natural numbers is a set, and this is a theorem, not a definition.
- E.g. the type of monoids is not a set, but instead a $1$-groupoid, automatically.
- E.g. the type of categories is a $2$-groupoid, again automatically.

2. The treatment of logic.

    - Mathematical statements are interpreted as types rather than truth values.
    - Truth values are particular kinds of types, called $-1$-groupoids, with at most one element.
    - Logical operations are particular cases of mathematical operations on types.
    - The mathematics comes first, with logic as a derived concept.
    - E.g. when we say "and", we are taking the cartesian product of two types, which may or may not be truth values.

3. The treatment of equality.

    - The value of an equality $x = y$ is a type, called the *identity type*, which is not necessarily a truth value.
    - It collects the ways in which the mathematical objects $x$ and $y$ are identified.
    - E.g. it is a truth value for elements of $\mathbb{N}$, as there is at most one way for two natural numbers to be equal.
    - E.g. for the type of monoids, it is a set, amounting to the type of monoid isomorphisms, automatically.
    - E.g. for the type of categories, it is a $1$-groupoid, amounting to the type of equivalences of categories, again automatically.

The important word in the above description of univalent foundations is *automatic*. For example, we don't *define* equality of monoids to be isomorphism. Instead, we define the collection of monoids as the large type of small types that are sets, equipped with a binary multiplication operation and a unit satisfying associativity of multiplication and neutrality of the unit in the usual way, and then we *prove* that the native notion of equality that comes with univalent type theory (inherited from Martin-Löf type theory) happens to coincide with the notion of monoid isomorphism. Largeness and smallness are taken as relative concepts, with type *universes* incorporated in the theory to account for the size distinction.

In particular, properties of monoids are automatically invariant under isomorphism, properties of categories are automatically invariant under equivalence, and so on.

Voevodsky's way to achieve this is to start with a Martin-Löf type theory (MLTT), including identity types and type universes, and postulate a single axiom, named *univalence*. This axiom stipulates a canonical bijection between *type equivalences* (in a suitable sense defined by Voevodsky in type theory) and *type identifications* (in the original sense of Martin-Löf's identity type). Voevodsky's notion of type equivalence, formulated in MLTT, is a refinement of the notion of isomorphism, which works uniformly for all higher groupoids, i.e. types.

In particular, Voevodsky didn't design a new type theory, but instead gave an axiom for an existing type theory (or any of a family of possible type theories, to be more precise).

The main *technical* contributions in type theory by Voevodsky are:

4. The definition of type levels in MLTT, classifying them as n-groupoids including the possibility $n=\infty$.
5. The (simple and elegant) definition of type equivalence that works uniformly for all type levels in MLTT.
6. The formulation of the univalence axiom in MLTT.

Univalent mathematics begins within MLTT with (4) and (5) before we postulate univalence. In fact, as the reader will see, we will do a fair amount of univalent mathematics before we formulate or assume the univalence axiom.

All of (4)-(6) crucially rely on Martin-Löf's identity type. Initially, Voevodsky thought that a new concept would be needed in the type theory to achieve (4)-(6) and hence (1) and (3) above. But he eventually discovered that Martin-Löf's identity type is precisely what he needed.

It may be considered somewhat miraculous that the addition of the univalence axiom alone to MLTT can achieve (1) and (3). Martin-Löf type theory was designed to achieve (2), and, regarding (1), types were imagined/conceived as sets (and even named *sets* in some of the original expositions by Martin-Löf), and, regarding (3), the identity type was imagined/conceived as having at most one element, even if MLTT cannot prove or disprove this statement, as was eventually shown by Hofmann and Streicher with their groupoid model of types in the early 1990's.

Another important aspect of univalent mathematics is the presence of explicit mechanisms for distinguishing

7. property (e.g. an unspecified thing exists),
8. data or structure (e.g. a designated thing exists or is given),

which are common place in current mathematical practice (e.g. cartesian closedness of a category is a property in some sense (up to isomorphism), whereas monoidal closedness is given structure).

In summary, univalent mathematics is characterized by (1)-(8) and not by the univalence axiom alone. In fact, half of these notes begin *without* the univalence axiom.

Lastly, univalent type theories don't assume the axiom of choice or the principle of excluded middle, and so in some sense they are constructive by default. But we emphasize that these two axioms are consistent and hence can be safely used as assumptions. However, virtually all theorems of univalent mathematics, e.g. in UniMath, have been proved without assuming them, with natural mathematical arguments. The formulations of these principles in univalent mathematics differ from their traditional formulations in MLTT, and hence we sometimes refer to them as the *univalent* principle of excluded middle and the *univalent* axiom of choice.

In these notes we will explore the above ideas, using Agda to write MLTT definitions, constructions, theorems and proofs, with univalence as an explicit assumption each time it is needed. We will have a further assumption, the existence of certain subsingleton (or propositional, or truth-value) truncations in order to be able to deal with the distinction between property and data, and in particular with the distinction between designated and unspecified existence (for example to be able to define the notions of image of a function and of surjective function).

We will not assume univalence and truncation globally, so that the students can see clearly when they are or are not needed. In fact, the foundational definitions, constructions, theorems and proofs of univalent mathematics don't require univalence or propositional truncation, and so can be developed in a version of the original Martin-Löf type theories, and this is what happens in these notes, and what Voevodsky did in his brilliant original development in the computer system Coq. Our use of Agda, rather than Coq, is a personal matter of taste only, and the students are encouraged to learn Coq, too.

### Homotopy type theory

Univalent type theory is often called *homotopy type theory*. Here we are following Voevodsky, who coined the phrases *univalent foundations* and *univalent mathematics*. We regard the terminology *homotopy type theory* as probably more appropriate for referring to the *synthetic* development of homotopy theory within univalent mathematics, for which we refer the reader to the HoTT book.

However, the terminology *homotopy type theory* is also used as a synonym for univalent type theory, not only because univalent type theory has a model in homotopy types (as defined in homotopy theory), but also because, without considering models, types do behave like homotopy types, automatically. We will not discuss how to do homotopy theory using univalent type theory in these notes. We refer the reader to the HoTT book as a starting point.

A common compromise is to refer to the subject as HoTT/UF.

### General references

- Papers by Martin-Löf.
- Homotopy type theory website references.
- HoTT book.
- `ncatlab` references.

In particular, it is recommended to read the concluding notes for each chapter in the HoTT book for discussion of original sources. Moreover, the whole HoTT book is a recommended complementary reading for this course.

And, after the reader has gained enough experience:

- Voevodsky's original foundations of univalent mathematics in Coq.
- UniMath project in Coq.
- Agda UniMath library.

- Coq HoTT library.
- Agda HoTT library.

Regarding the computer language Agda, we recommend the following as starting points:

- Agda wiki.
- Dependent types at work by Ana Bove and Peter Dybjer.
- Agda reference manual.
- Agda further references.
- Cubical Agda blog post by Anders Mörtberg.
- Cubical Agda documentation.

Regarding the genesis of the subject:

- A very short note on homotopy λ-calculus.
- Notes on homotopy λ-calculus.

Voevodsky says that he was influenced by Makkai's thinking:

- FOLDS.
- The theory of abstract sets based on first-order logic with dependent types.

An important foundational reference, by Steve Awodey and Michael A. Warren, is

- Homotopy theoretic models of identity types.

Additional expository material:

- An introduction to univalent foundations for mathematicians, a paper at the Bulletin of the AMS by Dan Grayson.
- Voevodsky's Memorial talk by Dan Grayson.
- Univalent foundations - an introduction by Benedikt Ahrens.
- Introduction to Homotopy Type Theory by Egbert Rijke.
- A course on homotopy (type) theory by Andrej Bauer and Jaka Smrekar.
- 15-819 Homotopy Type Theory by Bob Harper.
- Homotopy type theory: the logic of space by Mike Shulman.
- Logic in univalent type theory by Martin Escardo.

More references as clickable links are given in the course of the notes.

We also have an Agda development of univalent foundations which is applied to work on injective types, compact (or searchable) types, compact ordinals and more.

## Choice of material

This is intended as an introductory graduate course. We include what we regard as the essence of univalent foundations and univalent mathematics, but we are certainly omitting important material that is needed to do univalent mathematics in practice, and the readers who wish to practice univalent mathematics should consult the above references.

# MLTT in Agda

## What is Agda?

There are two views:

1. Agda is a dependently-typed functional programming language.

2. Agda is a language for defining mathematical notions (e.g. group or topological space), formulating constructions to be performed (e.g. a type of real numbers, a group structure on the integers, a topology on the reals), formulating theorems (e.g. a certain construction is indeed a group structure, there are infinitely many primes), and proving theorems (e.g. the infinitude of the collection of primes with Euclid's argument).

This doesn't mean that Agda has two sets of features, one for (1) and the other for (2). The same set of features account simultaneously for (1) and (2). Programs are mathematical constructions that happen not to use non-constructive principles such as excluded middle.

In these notes we study a minimal univalent type theory and do mathematics with it using a minimal subset of the computer language Agda as a vehicle.

Agda allows one to construct proofs interactively, but we will not discuss how to do this in these notes. Agda is not an automatic theorem prover. We have to come up with our own proofs, which Agda checks for correctness. We do get some form of interactive help to input our proofs and render them as formal objects.

## A spartan Martin-Löf type theory (MLTT)

Before embarking into a full definition of our Martin-Löf type theory in Agda, we summarize the particular Martin-Löf type theory that we will consider, by naming the concepts that we will include. We will have:

- An empty type `𝟘`.

- A one-element type `𝟙`.

- A type of `ℕ` natural numbers.

- Type formers `+` (binary sum), `Π` (product), `Σ` (sum), `Id` (identity type).

- Universes (types of types), ranged over by $\mathcal{U}, \mathcal{V}, \mathcal{W}$.

This is enough to do number theory, analysis, group theory, topology, category theory and more.

spartan /ˈspɑːt(ə)n/ adjective:

> showing or characterized by austerity or a lack of comfort or luxury.

We will also be rather spartan with the subset of Agda that we choose to discuss. Many things we do here can be written in more concise ways using more advanced features. Here we introduce a minimal subset of Agda where everything in our spartan MLTT can be expressed.

## Getting started with Agda

We don't use any Agda library. For pedagogical purposes, we start from scratch, and here are our first two lines of code:

```
{-# OPTIONS --without-K --exact-split --safe --auto-inline #-}

module HoTT-UF-Agda where
```

- The option `--without-K` disables Streicher's `K` axiom, which we don't want for univalent mathematics.

- The option `--exact-split` makes Agda to only accept definitions with the equality sign “`=`” that behave like so-called *judgmental* or *definitional* equalities.

- The option `--safe` disables features that may make Agda inconsistent, such as `--type-in-type`, postulates and more.

- Every Agda file is a module. These lecture notes are a set of Agda files, which are converted to html by Agda after it successfully checks the mathematical development for correctness.

The Agda code in these notes has syntax highlighting and links (in the html and pdf versions), so that we can navigate to the definition of a name or symbol by clicking at it.

## Type universes

A universe $\mathcal{U}$ is a type of types.

1. One use of universes is to define families of types indexed by a type `X` as functions `X → 𝓤`.

2. Such a function is sometimes seen as a property of elements of `X`.

3. Another use of universes, as we shall see, is to define types of mathematical structures, such as monoids, groups, topological spaces, categories etc.

Sometimes we need more than one universe. For example, the type of groups in a universe lives in a bigger universe, and given a category in one universe, its presheaf category also lives in a larger universe.

We will work with a tower of type universes

```
𝓤₀, 𝓤₁, 𝓤₂, 𝓤₃, ...
```

In practice, the first one, two or three universes suffice, but it will be easier to formulate definitions, constructions and theorems in full generality, to avoid making universe choices before knowing how they are going to be applied.

These are actually universe names (also called levels, not to be confused with hlevels). We reference the universes themselves by a deliberately almost-invisible superscript dot. For example, we will have

```
𝟙 : 𝓤₀ ̇
```

where `𝟙` is the one-point type to be defined shortly. We will sometimes omit this superscript in our discussions, but are forced to write it in Agda code. We have that the universe $\mathcal{U}_0$ is a type in the universe $\mathcal{U}_1$, which is a type in the universe $\mathcal{U}_2$ and so on.

```
𝓤₀ ̇   : 𝓤₁ ̇

𝓤₁ ̇   : 𝓤₂ ̇

𝓤₂ ̇   : 𝓤₃ ̇

 ⋮
```

The assumption that `𝓤₀ : 𝓤₀` or that any universe is in itself or a smaller universe gives rise to a contradiction, similar to Russell’s Paradox.

Given a universe $\mathcal{U}$, we denote by

```
𝓤 ⁺
```

its successor universe. For example, if $\mathcal{U}$ is $\mathcal{U}_0$ then $\mathcal{U}^+$ is $\mathcal{U}_1$. According to the above discussion, we have

```
𝓤 ̇ : 𝓤 ⁺ ̇
```

The least upper bound of two universes $\mathcal{U}$ and $\mathcal{V}$ is written

$\mathcal{U} \sqcup \mathcal{V}$.

For example, if $\mathcal{U}$ is $\mathcal{U}_0$ and $\mathcal{V}$ is $\mathcal{U}_1$, then $\mathcal{U} \sqcup \mathcal{V}$ is $\mathcal{U}_1$.

We now bring our notation for universes by importing our Agda file `Universes`. The Agda keyword `open` asks to make all definitions in the file `Universe` visible in our file here.

```
open import Universes public
```

The keyword `public` makes the contents of the file `Universes` available to importers of our module `HoTT-UF-Agda`.

We will refer to universes by letters $\mathcal{U}, \mathcal{V}, \mathcal{W}, \mathcal{T}$:

```
variable
 𝓤 𝓥 𝓦 𝓣 : Universe
```

In some type theories, the universes are cumulative "on the nose", in the sense that from `X : 𝓤` we derive that `X : 𝓤 ⊔ 𝓥`. We will instead have an embedding $\mathcal{U} \to \mathcal{U} \sqcup \mathcal{V}$ of universes into larger universes.

## The one-element type 𝟙

We place it in the first universe, and we name its unique element "⋆". We use the `data` declaration in Agda to introduce it:

```
data 𝟙 : 𝓤₀ ̇  where
 ⋆ : 𝟙
```

It is important that the point `⋆` lives in the type `𝟙` and in no other type. There isn't dual citizenship in our type theory. When we create a type, we also create freshly new elements for it, in this case "⋆". (However, Agda has a limited form of overloading, which allows us to sometimes use the same name for different things.)

Next we want to give a mechanism to prove that all points of the type `𝟙` satisfy a given property `A`.

1. The property is a function `A : 𝟙 → 𝓤` for some universe $\mathcal{U}$.

2. The type `A(x)`, which we will write simply `A x`, doesn't need to be a truth value. It can be any type. We will see examples shortly.

3. In MLTT, mathematical statements are types, such as

   `Π A : 𝟙 → 𝓤, A ⋆ → Π x : 𝟙, A x.`

4. We read this in natural language as "for any given property `A` of elements of the type `𝟙`, if `A ⋆` holds, then it follows that `A x` holds for all `x : 𝟙`".

5. In Agda, the above type is written as

   `(A : 𝟙 → 𝓤 ̇ ) → A ⋆ → (x : 𝟙) → A x.`

   This is the type of functions with three arguments `A : 𝟙 → 𝓤 ̇` and `a : A ⋆` and `x : 𝟙`, with values in the type `A x`.

6. A proof of a mathematical statement rendered as a type is a construction of an element of the type. In our example, we have to construct a function, which we will name `𝟙-induction`.

We do this as follows in Agda, where we first declare the type of the function `𝟙-induction` with ":" and then define the function by an equation:

```
𝟙-induction : (A : 𝟙 → 𝓤 ̇ ) → A ⋆ → (x : 𝟙) → A x
𝟙-induction A a ⋆ = a
```

The universe $\mathcal{U}$ is arbitrary, and Agda knows $\mathcal{U}$ is a universe variable because we said so above.

Notice that we supply `A` and `a` as arbitrary arguments, but instead of an arbitrary `x : 𝟙` we have written "`⋆`". Agda accepts this because it knows from the definition of `𝟙` that "`⋆`" is the only element of the type `𝟙`. This mechanism is called *pattern matching*.

A particular case of `𝟙-induction` occurs when the family `A` is constant with value `B`, which can be written variously as

```
A x = B
```

or

```
A = λ (x : 𝟙) → B,
```

or

```
A = λ x → B
```

if we want Agda to figure out the type of `x` by itself, or

```
A = λ _ → B
```

if we don't want to name the argument of `A` because it is not used. In usual mathematical practice, such a lambda expression is often written

`x ↦ B` (`x` is mapped to `B`)

so that the above amount to `A = (x ↦ B)`.

Given a type `B` and a point `b : B`, we construct the function `𝟙 → B` that maps any given `x : 𝟙` to `b`.

```
𝟙-recursion : (B : 𝓤 ̇ ) → B → (𝟙 → B)
𝟙-recursion B b x = 𝟙-induction (λ _ → B) b x
```

The above expression `B → (𝟙 → B)` can be written as `B → 𝟙 → B`, omitting the brackets, as the function-type formation symbol `→` is taken to be right associative.

Not all types have to be seen as mathematical statements (for example the type `ℕ` of natural numbers defined below). But the above definition has a dual interpretation as a mathematical function, and as the statement "`B` implies (*true* implies `B`)" where `𝟙` is the type encoding the truth value *true*.

The unique function to `𝟙` will be named `!𝟙`. We define two versions to illustrate implicit arguments, which correspond in mathematics to "subscripts that are omitted when the reader can safely infer them", as for example for the identity function of a set or the identity arrow of an object of a category.

```
!𝟙' : (X : 𝓤 ̇ ) → X → 𝟙
!𝟙' X x = ⋆

!𝟙 : {X : 𝓤 ̇ } → X → 𝟙
!𝟙 x = ⋆
```

This means that when we write

```
!𝟙 x
```

we have to recover the (uniquely determined) missing type `X` with `x : X` "from the context". When Agda can't figure it out, we need to supply it and write

```
!𝟙 {𝓤} {X} x.
```

This is because `𝓤` is also an implicit argument (all things declared with the Agda keyword *variable* as above are implicit arguments). There are other, non-positional, ways to indicate `X` without having to indicate `𝓤` too. Occasionally, people define variants of a function with different choices of "implicitness", as above.

### The empty type 𝟘

It is defined like 𝟙, except that no elements are listed for it:

```
data 𝟘 : 𝓤₀ ̇  where
```

That's the complete definition. This has a dual interpretation, mathematically as the empty set (we can actually prove that this type is a set, once we know the definition of set), and logically as the truth value *false*. To prove that a property of elements of the empty type holds for all elements of the empty type, we have to do nothing.

```
𝟘-induction : (A : 𝟘 → 𝓤 ̇ ) → (x : 𝟘) → A x
𝟘-induction A ()
```

When we write the pattern `()`, Agda checks if there is any case we missed. If there is none, our definition is accepted. The expression `()` corresponds to the mathematical phrase vacuously true. The unique function from 𝟘 to any type is a particular case of `𝟘-induction`.

```
𝟘-recursion : (A : 𝓤 ̇ ) → 𝟘 → A
𝟘-recursion A a = 𝟘-induction (λ _ → A) a
```

We will use the following categorical notation for `𝟘-recursion`:

```
!𝟘 : (A : 𝓤 ̇ ) → 𝟘 → A
!𝟘 = 𝟘-recursion
```

We give the two names `is-empty` and `¬` to the same function now:

```
is-empty : 𝓤 ̇ → 𝓤 ̇
is-empty X = X → 𝟘

¬ : 𝓤 ̇ → 𝓤 ̇
¬ X = X → 𝟘
```

This says that a type is empty precisely when we have a function to the empty type. Assuming univalence, once we have defined the identity type former `_≡_`, we will be able to prove that `(is-empty X) ≡ (X ≃ 𝟘)`, where `X ≃ 𝟘` is the type of bijections, or equivalences, from `X` to 𝟘. We will also be able to prove things like `(2 + 2 ≡ 5) ≡ 𝟘` and `(2 + 2 ≡ 4) ≡ 𝟙`.

This is for *numbers*. If we define *types* `𝟚 = 𝟙 + 𝟙` and `𝟜 = 𝟚 + 𝟚` with two and four elements respectively, where we are anticipating the definition of `_+_` for types, then we will instead have that `𝟚 + 𝟚 ≡ 𝟜` is a type with `4!` elements, which is the number of permutations of a set with four elements, rather than a truth value 𝟘 or 𝟙, as a consequence of the univalence axiom. That is, we will have `(𝟚 + 𝟚 ≡ 𝟜) ≃ (𝟜 + 𝟜 + 𝟜 + 𝟜 + 𝟜 + 𝟜)`, so that the type identity `𝟚 + 𝟚 ≡ 𝟜` holds in many more ways than the numerical equation `2 + 2 ≡ 4`.

The above is possible only because universes are genuine types and hence their elements (that is, types) have identity types themselves, so that writing `X ≡ Y` for types `X` and `Y` (inhabiting the same universe) is allowed.

When we view 𝟘 as *false*, we can read the definition of the *negation* `¬ X` as saying that "`X` implies *false*". With univalence we will be able to show that "(*false* → *true*) ≡ *true*", which amounts to `(𝟘 → 𝟙) ≡ 𝟙`, which in turn says that there is precisely one function `𝟘 → 𝟙`, namely the vacuous function.

### The type ℕ of natural numbers

The definition is similar but not quite the same as the one via Peano Axioms.

We stipulate an element `zero : ℕ` and a successor function `succ : ℕ → ℕ`, and then define induction. Once we have defined the identity type former `_≡_`, we will *prove* the other peano axioms.

```
data ℕ : 𝓤₀ ̇  where
 zero : ℕ
 succ : ℕ → ℕ
```

In general, declarations with `data` are inductive definitions. To write the number `5`, we have to write

```
succ (succ (succ (succ (succ zero))))
```

We can use the following Agda *built-in* to be able to just write `5` as a shorthand:

```
{-# BUILTIN NATURAL ℕ #-}
```

Apart from this notational effect, the above declaration doesn't play any role in the Agda development of these lecture notes.

In the following, the type family `A` can be seen as playing the role of a property of elements of `ℕ`, except that it doesn't need to be necessarily subsingleton valued. When it is, the *type* of the function gives the familiar principle of mathematical induction for natural numbers, whereas, in general, its definition says how to compute with induction.

```
ℕ-induction : (A : ℕ → 𝓤 ̇ )
            → A 0
            → ((n : ℕ) → A n → A (succ n))
            → (n : ℕ) → A n

ℕ-induction A a₀ f = h
 where
  h : (n : ℕ) → A n
  h 0        = a₀
  h (succ n) = f n (h n)
```

The definition of `ℕ-induction` is itself by induction. It says that given a point

```
a₀ : A 0
```

and a function

```
f : (n : ℕ) → A n → A (succ n),
```

in order to calculate an element of `A n` for a given `n : ℕ`, we just calculate `h n`, that is,

```
f n (f (n-1) (f (n-2) (... (f 0 a₀)...))).
```

So the principle of induction is a construction that given a *base case* `a₀ : A 0`, an *induction step* `f : (n : ℕ) → A n → A (succ n)` and a number `n`, says how to get an element of the type `A n` by primitive recursion.

Notice also that `ℕ-induction` is the dependently typed version of primitive recursion, where the non-dependently typed version is

```
ℕ-recursion : (X : 𝓤 ̇ )
            → X
            → (ℕ → X → X)
            → ℕ → X

ℕ-recursion X = ℕ-induction (λ _ → X)
```

The following special case occurs often (and is related to the fact that `ℕ` is the initial algebra of the functor `𝟙 + (-)`):

```
ℕ-iteration : (X : 𝓤 ̇ )
            → X
            → (X → X)
            → ℕ → X

ℕ-iteration X x f = ℕ-recursion X x (λ _ x → f x)
```

Agda checks that any recursive definition as above is well founded, with recursive invocations with structurally smaller arguments only. If it isn't, the definition is not accepted.

In official Martin-Löf type theories, we have to use the `ℕ-induction` functional to define everything else with the natural numbers. But Agda allows us to define functions by structural recursion, like we defined `ℕ-induction`.

We now define addition and multiplication for the sake of illustration. We first do it in Peano style. We will create a local `module` so that the definitions are not globally visible, as we want to have the symbols `+` and `×` free for type operations of MLTT to be defined soon. The things in the module are indented and are visible outside the module only if we `open` the module or if we write them as e.g. `Arithmetic._+_` in the following example.

```
module Arithmetic where

  _+_  _×_ : ℕ → ℕ → ℕ

  x + 0      = x
  x + succ y = succ (x + y)

  x × 0      = 0
  x × succ y = x + x × y

  infixl 20 _+_
  infixl 21 _×_
```

The above "fixity" declarations allow us to indicate the precedences (multiplication has higher precedence than addition) and their associativity (here we take left-associativity as the convention, so that e.g. `x+y+z` parses as `(x+y)+z`).

Equivalent definitions use `ℕ-induction` on the second argument `y`, via `ℕ-iteration`:

```
module Arithmetic' where

  _+_  _×_ : ℕ → ℕ → ℕ

  x + y = h y
   where
    h : ℕ → ℕ
    h = ℕ-iteration ℕ x succ

  x × y = h y
   where
    h : ℕ → ℕ
    h = ℕ-iteration ℕ 0 (x +_)

  infixl 20 _+_
  infixl 21 _×_
```

Here the expression "`x +_`" stands for the function `ℕ → ℕ` that adds `x` to its argument. So to multiply `x` by `y`, we apply `y` times the function "`x +_`" to `0`.

As another example, we define the less-than-or-equal relation by nested induction, on the first argument and then the second, but we use pattern matching for the sake of readability.

*Exercise*. Write it using `ℕ-induction`, recursion or iteration, as appropriate.

```
module ℕ-order where

  _≤_ _≥_ : ℕ → ℕ → 𝓤₀ ̇
  0      ≤ y      = 𝟙
  succ x ≤ 0      = 𝟘
  succ x ≤ succ y = x ≤ y

  x ≥ y = y ≤ x

  infix 10 _≤_
  infix 10 _≥_
```

*Exercise*. After learning `Σ` and `_≡_` explained below, prove that

> `x ≤ y` if and only if `Σ z ꞉ ℕ , x + z ≡ y`.

Later, after learning univalence prove that in this case this implies

> `(x ≤ y) ≡ Σ z ꞉ ℕ , x + z ≡ y`.

That bi-implication can be turned into equality only holds for types that are subsingletons (and this is called propositional extensionality).

If we are doing applied mathematics and want to actually compute, we can define a type for binary notation for the sake of efficiency, and of course people have done that. Here we are not concerned with efficiency but only with understanding how to codify mathematics in (univalent) type theory and in Agda.

Table of contents ⇑

## The binary sum type constructor `_+_`

We now define the disjoint sum of two types `X` and `Y`. The elements of the type

```
X + Y
```

are stipulated to be of the forms

`inl x` and `inr y`

with `x : X` and `y : Y`. If `X : 𝓤` and `Y : 𝓥`, we stipulate that `X + Y : 𝓤 ⊔ 𝓥`, where

```
𝓤 ⊔ 𝓥
```

is the least upper bound of the two universes `𝓤` and `𝓥`. In Agda we can define this as follows.

```
data _+_ {𝓤 𝓥} (X : 𝓤 ̇ ) (Y : 𝓥 ̇ ) : 𝓤 ⊔ 𝓥 ̇  where
 inl : X → X + Y
 inr : Y → X + Y
```

To prove that a property `A` of the sum holds for all `z : X + Y`, it is enough to prove that `A (inl x)` holds for all `x : X` and that `A (inr y)` holds for all `y : Y`. This amounts to definition by cases:

```
+-induction : {X : 𝓤 ̇ } {Y : 𝓥 ̇ } (A : X + Y → 𝓦 ̇ )
            → ((x : X) → A (inl x))
            → ((y : Y) → A (inr y))
            → (z : X + Y) → A z

+-induction A f g (inl x) = f x
+-induction A f g (inr y) = g y

+-recursion : {X : 𝓤 ̇ } {Y : 𝓥 ̇ } {A : 𝓦 ̇ } → (X → A) → (Y → A) → X + Y → A
+-recursion {𝓤} {𝓥} {𝓦} {X} {Y} {A} = +-induction (λ _ → A)
```

When the types `A` and `B` are understood as mathematical statements, the type `A + B` is understood as the statement "`A` or `B`", because to prove "`A` or `B`" we have to prove one of `A` and `B`. When `A` and `B` are simultaneously possible, we have two proofs, but sometimes we want to deliberately ignore which one we get, when we want to get a truth value rather than a possibly more general type, and in this case we use the truncation `∥ A + B ∥`.

But also `_+_` is used to construct mathematical objects. For example, we can define a two-point type:

```
𝟚 : 𝓤₀ ̇
𝟚 = 𝟙 + 𝟙
```

We can name the left and right points as follows, using patterns, so that they can be used in pattern matching:

```
pattern ₀ = inl ⋆
pattern ₁ = inr ⋆
```

We can define induction on `𝟚` directly by pattern matching:

```
𝟚-induction : (A : 𝟚 → 𝓤 ̇ ) → A ₀ → A ₁ → (n : 𝟚) → A n
𝟚-induction A a₀ a₁ ₀ = a₀
𝟚-induction A a₀ a₁ ₁ = a₁
```

Or we can prove it by induction on `_+_` and `𝟙`:

```
𝟚-induction' : (A : 𝟚 → 𝓤 ̇ ) → A ₀ → A ₁ → (n : 𝟚) → A n
𝟚-induction' A a₀ a₁ = +-induction A
                        (𝟙-induction (λ (x : 𝟙) → A (inl x)) a₀)
                        (𝟙-induction (λ (y : 𝟙) → A (inr y)) a₁)
```

## Σ types

Given universes 𝓤 and 𝓥, a type

```
X : 𝓤
```

and a type family

```
Y : X → 𝓥,
```

we want to construct its sum, which is a type whose elements are of the form

```
(x , y)
```

with `x : X` and `y : Y x`. This sum type will live in the least upper bound

```
𝓤 ⊔ 𝓥
```

of the universes 𝓤 and 𝓥. We will write this sum

```
Σ Y,
```

with `X`, as well as the universes, implicit. Often Agda, and people, can figure out what the unwritten type `X` is, from the definition of `Y`. But sometimes there may be either lack of enough information, or of enough concentration power by people, or of sufficiently powerful inference algorithms in the implementation of Agda. In such cases we can write

```
Σ λ(x : X) → Y x,
```

because `Y = λ (x : X) → Y x` by a so-called η-rule. However, we will define syntax to be able to write this in more familiar form as

```
Σ x ꞉ X , Y x.
```

In MLTT we can write this in other ways, for example with the indexing `x : X` written as a subscript of `Σ` or under it.

Or it may be that the name `Y` is not defined, and we work with a nameless family defined on the fly, as in the exercise proposed above:

```
Σ z ꞉ ℕ , x + z ≡ y,
```

where `Y z = (x + z ≡ y)` in this case, and where we haven't defined the identity type former `_≡_` yet.

We can construct the `Σ` type former as follows in Agda:

```
record Σ {𝓤 𝓥} {X : 𝓤 ̇ } (Y : X → 𝓥 ̇ ) : 𝓤 ⊔ 𝓥 ̇  where
  constructor
   _,_
  field
   x : X
   y : Y x
```

This says we are defining a binary operator `_,_` to construct the elements of this type as `x , y`. The brackets are not needed, but we will often write them to get the more familiar notation `(x , y)`. The definition says that an element of `Σ Y` has two `fields`, giving the types for them.

*Remark.* Beginners may safely ignore this remark: Normally people will call these two fields `x` and `y` something like `pr₁` and `pr₂`, or `fst` and `snd`, for first and second projection, rather than `x` and `y`, and then do `open Σ public` and have the projections available as functions automatically. But we will deliberately not do that, and instead define the projections ourselves, because this is confusing for beginners, no matter how mathematically or computationally versed they may be, in particular because it will not be immediately clear that the projections have the following types.

```
pr₁ : {X : 𝓤 ̇ } {Y : X → 𝓥 ̇ } → Σ Y → X
pr₁ (x , y) = x

pr₂ : {X : 𝓤 ̇ } {Y : X → 𝓥 ̇ } → (z : Σ Y) → Y (pr₁ z)
pr₂ (x , y) = y
```

We now introduce syntax to be able to write `Σ x ꞉ X , Y` instead of `Σ λ(x : X) → Y`. For this purpose, we first define a version of `Σ` making the index type explicit.

```
-Σ : {𝓤 𝓥 : Universe} (X : 𝓤 ̇ ) (Y : X → 𝓥 ̇ ) → 𝓤 ⊔ 𝓥 ̇
-Σ X Y = Σ Y

syntax -Σ X (λ x → y) = Σ x ꞉ X , y
```

For some reason, Agda has this kind of definition backwards: the *definiendum* and the *definiens* are swapped with respect to the normal convention of writing what is defined on the left-hand side of the equality sign.

Notice also that "꞉" in the above syntax definition is not the same as ":", even though they may look the same. For the above notation `Σ x ꞉ A , b`, the symbol "꞉" has to be typed "\:4" in the emacs Agda mode.

To prove that `A z` holds for all `z : Σ Y`, for a given property `A`, we just prove that we have `A (x , y)` for all `x : X` and `y : Y x`. This is called `Σ` induction or `Σ` elimination, or `uncurry`, after Haskell Curry.

```
Σ-induction : {X : 𝓤 ̇ } {Y : X → 𝓥 ̇ } {A : Σ Y → 𝓦 ̇ }
            → ((x : X) (y : Y x) → A (x , y))
            → ((x , y) : Σ Y) → A (x , y)

Σ-induction g (x , y) = g x y
```

This function has an inverse:

```
curry : {X : 𝓤 ̇ } {Y : X → 𝓥 ̇ } {A : Σ Y → 𝓦 ̇ }
      → (((x , y) : Σ Y) → A (x , y))
      → ((x : X) (y : Y x) → A (x , y))

curry f x y = f (x , y)
```

An important particular case of the sum type is the binary cartesian product, when the type family doesn't depend on the indexing type:

```
_×_ : 𝓤 ̇ → 𝓥 ̇ → 𝓤 ⊔ 𝓥 ̇
X × Y = Σ x ꞉ X , Y
```

We have seen by way of examples that the function type symbol `→`
represents logical implication, and that a dependent function type `(x
X) → A x represents a universal quantification. We have the following uses of Σ`.

1. The binary cartesian product represents conjunction "and". If the types `A` and `B` stand for mathematical statements, then the mathematical statement

   `A` and `B`

   is codified as

   `A × B`,

   because to establish "`A` and `B`", we have to provide a pair `(a , b)` of proofs `a : A` and `b : B`.

   So notice that in type theory proofs are mathematical objects, rather than meta-mathematical entities like in set theory. They are just elements of types.

2. The more general type

   `Σ x ꞉ X , A x`,

   if the type `X` stands for a mathematical object and `A` stands for a mathematical statement, represents *designated* existence

   > there is a designated `x ꞉ X` with `A x`.

   To establish this, we have to provide a specific element `x ꞉ X` and a proof `a ꞉ A x`, together in a pair `(x , a)`.

3. Later we will discuss *unspecified* existence

   `∃ x ꞉ X , A x`,

   which will be obtained by a sort of quotient of `Σ x ꞉ X , A x`, written

   `∥ Σ x ꞉ X , A x ∥`,

   that identifies all the elements of the type `Σ x ꞉ X , A x` in a single equivalence class, called its subsingleton (or truth value or propositional) truncation.

4. Another reading of

   `Σ x ꞉ X , A x`

   is as

   > the type of `x ꞉ X` with `A x`,

   similar to set-theoretical notation

   `{ x ∈ X | A x }`.

   But we have to be careful because if there is more than one element in the type `A x`, then `x` will occur more than once in this type. More precisely, for `a₀ a₁ ꞉ A x` we have inhabitants `(x , a₀)` and `(x , a₁)` of the type `Σ x ꞉ X , A x`.

   In such situations, if we don't want that, we have to either ensure that the type `A x` has at most one element for every `x ꞉ X`, or instead consider the truncated type `∥ A x ∥` and write

   `Σ x ꞉ X , ∥ A x ∥`.

   An example is the image of a function `f : X → Y`, which will be defined to be

   `Σ y ꞉ Y , ∥ Σ x ꞉ X , f x ≡ y ∥`.

   This is the type of `y ꞉ Y` for which there is an unspecified `x ꞉ X` with `f x ≡ y`.

   (For constructively minded readers, we emphasize that this doesn't erase the witness `x ꞉ X`.)

## Π types

`Π` types are builtin with a different notation in Agda, as discussed above, but we can introduce the notation `Π` for them, similar to that for `Σ`:

```
Π : {X : 𝓤 ̇ } (A : X → 𝓥 ̇ ) → 𝓤 ⊔ 𝓥 ̇
Π {𝓤} {𝓥} {X} A = (x : X) → A x

-Π : {𝓤 𝓥 : Universe} (X : 𝓤 ̇ ) (Y : X → 𝓥 ̇ ) → 𝓤 ⊔ 𝓥 ̇
-Π X Y = Π Y
```

```
syntax -Π A (λ x → b) = Π x ꞉ A , b
```

Notice that the function type `X → Y` is the particular case of the `Π` type when the family `A` is constant with value `Y`.

We take the opportunity to define the identity function (in two versions with different implicit arguments) and function composition:

```
id : {X : 𝓤 ̇ } → X → X
id x = x

𝑖𝑑 : (X : 𝓤 ̇ ) → X → X
𝑖𝑑 X = id
```

Usually the type of function composition `_∘_` is given as simply

```
(Y → Z) → (X → Y) → (X → Z).
```

With dependent functions, we can give it a more general type:

```
_∘_ : {X : 𝓤 ̇ } {Y : 𝓥 ̇ } {Z : Y → 𝓦 ̇ }
    → ((y : Y) → Z y)
    → (f : X → Y)
    → (x : X) → Z (f x)

g ∘ f = λ x → g (f x)
```

Notice that this type for the composition function can be read as a mathematical statement: If `Z y` holds for all `y : Y`, then for any given `f : X → Y` we have that `Z (f x)` holds for all `x : X`. And the non-dependent one is just the transitivity of implication.

The following functions are sometimes useful when we are using implicit arguments, in order to recover them explicitly without having to list them as given arguments:

```
domain : {X : 𝓤 ̇ } {Y : 𝓥 ̇ } → (X → Y) → 𝓤 ̇
domain {𝓤} {𝓥} {X} {Y} f = X

codomain : {X : 𝓤 ̇ } {Y : 𝓥 ̇ } → (X → Y) → 𝓥 ̇
codomain {𝓤} {𝓥} {X} {Y} f = Y

type-of : {X : 𝓤 ̇ } → X → 𝓤 ̇
type-of {𝓤} {X} x = X
```

## The identity type former `Id`, also written `_=_`

We now introduce the central type constructor of MLTT from the point of view of univalent mathematics. In Agda we can define Martin-Löf's identity type as follows:

```
data Id {𝓤} (X : 𝓤 ̇ ) : X → X → 𝓤 ̇  where
 refl : (x : X) → Id X x x
```

Intuitively, the above definition would say that the only element of the type `Id X x x` is something called `refl x` (for reflexivity). But, as we shall see in a moment, this intuition turns out to be incorrect.

Notice a crucial difference with the previous definitions using `data` or induction: In the previous cases, we defined *types*, namely `𝟘`, `𝟙`, `ℕ`, or a *type depending on type parameters*, namely `_+_`, with `𝓤` and `𝓥` fixed,

```
_+_ : 𝓤 ̇ → 𝓥 ̇ → 𝓤 ⊔ 𝓥 ̇
```

But here we are defining a *type family* indexed by the *elements* of a given type, rather than a new type from old types. Given a type `X` in a universe `𝓤`, we define a *function*

```
Id X : X → X → 𝓤
```

by some mysterious sort of induction. It is this that prevents us from being able to prove that the only element of the type `Id X x x` would be `refl x`, or that the type `Id X x y` would have at most one element no matter what `y : X` is.

There is however, one interesting, and crucial, thing we can prove, namely that for any fixed element `x : X`, the type

```
Σ y : Y , Id X x y
```

is always a singleton.

We will use the following alternative notation for the identity type former `Id`, where the symbol "_" in the right-hand side of the definition indicates that we ask Agda to infer which type we are talking about (which is `X`, but this name is not available in the scope of the *defining equation* of the type former _＝_):

```
_＝_ : {X : 𝓤 ̇ } → X → X → 𝓤 ̇
x ＝ y = Id _ x y
```

**Notational remark**. The symbol "＝" for the identity type is full width equal sign because the normal Ascii symbol "=" is reserved in Agda. To set-up emacs to type our symbol for the identity type, please follow these instructions.

Another intuition for this type family `_＝_ : X → X → 𝓤` is that it gives the least reflexive relation on the type `X`, as suggested by Martin-Löf's induction principle `J` discussed below.

Whereas we can make the intuition that `x ＝ x` has precisely one element good by *postulating* a certain `K` axiom due to Thomas Streicher, which comes with Agda by default but we have disabled above, we cannot *prove* that `refl x` is the only element of `x ＝ x` for an arbitrary type `X`. This non-provability result was established by Hofmann and Streicher, by giving a model of type theory in which types are interpreted as `1`-groupoids. This is in spirit similar to the non-provability of the parallel postulate in Euclidean geometry, which also considers models, which in turn are interesting in their own right.

However, for the elements of *some* types, such as the type `ℕ` of natural numbers, it is possible to prove that any identity type `x ＝ y` has at most one element. Such types are called sets in univalent mathematics.

If instead of the axiom `K` we adopt Voevodsky's univalence axiom, we get specific examples of objects `x` and `y` such that the type `x ＝ y` has multiple elements, *within* the type theory. It follows that the identity type `x ＝ y` is fairly under-specified in general, in that we can't prove or disprove that it has at most one element.

There are two opposing ways to resolve the ambiguity or under-specification of the identity types: (1) We can consider the `K` axiom, which postulates that all types are sets, or (2) we can consider the univalence axiom, arriving at univalent mathematics, which gives rise to types that are more general than sets, the `n`-groupoids and `∞`-groupoids. In fact, the univalence axiom will say, in particular, that for some types `X` and elements `x y : X`, the identity type `x ＝ y` does have more than one element.

A possible way to understand the element `refl x` of the type `x ＝ x` is as the "generic identification" between the point `x` and itself, but which is by no means necessarily the *only* identitification in univalent foundations. It is generic in the sense that to explain what happens with all identifications `p : x ＝ y` between any two points `x` and `y` of a type `X`, it suffices to explain what happens with the identification `refl x : x ＝ x` for all points `x : X`. This is what the induction principle for identity given by Martin-Löf says, which he called `J` (we could have called it `＝-induction`, but we prefer to honour MLTT tradition):

```
𝕁 : (X : 𝓤 ̇ ) (A : (x y : X) → x ＝ y → 𝓥 ̇ )
  → ((x : X) → A x x (refl x))
  → (x y : X) (p : x ＝ y) → A x y p

𝕁 X A f x x (refl x) = f x
```

This is related to the Yoneda Lemma in category theory, for readers familiar with the subject, which says that certain natural transformations are *uniquely determined* by their *action on the identity arrows*, even if they are *defined for all arrows*. Similarly here, `𝕁` is uniquely determined by its action on reflexive identifications, but is *defined for all identifications between any two points*, not just reflexivities.

In summary, Martin-Löf's identity type is given by the data

- `Id`,

- refl,
- 𝕁,
- the above equation for 𝕁.

However, we will not always use this induction principle, because we can instead work with the instances we need by pattern matching on refl (which is just what we did to define the principle itself) and there is a theorem by Jesper Cockx that says that with the Agda option without-K, pattern matching on refl can define/prove precisely what 𝕁 can.

*Exercise*. Define

```
ℍ : {X : 𝓤 ̇ } (x : X) (B : (y : X) → x ≡ y → 𝓥 ̇ )
  → B x (refl x)
  → (y : X) (p : x ≡ y) → B y p

ℍ x B b x (refl x) = b
```

Then we can define 𝕁 from ℍ as follows:

```
𝕁' : (X : 𝓤 ̇ ) (A : (x y : X) → x ≡ y → 𝓥 ̇ )
   → ((x : X) → A x x (refl x))
   → (x y : X) (p : x ≡ y) → A x y p

𝕁' X A f x = ℍ x (A x) (f x)

𝕁s-agreement : (X : 𝓤 ̇ ) (A : (x y : X) → x ≡ y → 𝓥 ̇ )
               (f : (x : X) → A x x (refl x)) (x y : X) (p : x ≡ y)
             → 𝕁 X A f x y p ≡ 𝕁' X A f x y p

𝕁s-agreement X A f x x (refl x) = refl (f x)
```

Similarly define ℍ' from 𝕁 without using pattern matching on refl and show that it coincides with ℍ (possibly using pattern matching on refl). This is harder.

With this, we have concluded the rendering of our spartan MLTT in Agda notation. Before embarking on the development of univalent mathematics within our spartan MLTT, we pause to discuss some basic examples of mathematics in Martin-Löf type theory.

## Basic constructions with the identity type

*Transport along an identification.*

```
transport : {X : 𝓤 ̇ } (A : X → 𝓥 ̇ ) {x y : X}
          → x ≡ y → A x → A y

transport A (refl x) = id (A x)
```

We can equivalently define transport using 𝕁 as follows:

```
transport𝕁 : {X : 𝓤 ̇ } (A : X → 𝓥 ̇ ) {x y : X}
           → x ≡ y → A x → A y

transport𝕁 {𝓤} {𝓥} {X} A {x} {y} = 𝕁 X (λ x y _ → A x → A y) (λ x → id (A x)) x y
```

In the same way ℕ-recursion can be seen as the non-dependent special case of ℕ-induction, the following transport function can be seen as the non-dependent special case of the ≡-induction principle ℍ with some of the arguments permuted and made implicit:

```
nondep-ℍ : {X : 𝓤 ̇ } (x : X) (A : X → 𝓥 ̇ )
         → A x → (y : X) → x ≡ y → A y
nondep-ℍ x A = ℍ x (λ y _ → A y)
```

```
transportℍ : {X : 𝓤 ̇ } (A : X → 𝓥 ̇ ) {x y : X}
           → x ＝ y → A x → A y
transportℍ A {x} {y} p a = nondep-ℍ x A a y p
```

All the above transports coincide:

```
transports-agreement : {X : 𝓤 ̇ } (A : X → 𝓥 ̇ ) {x y : X} (p : x ＝ y)
                     → (transportℍ A p ＝ transport A p)
                     × (transport𝕁 A p ＝ transport A p)

transports-agreement A (refl x) = refl (transport A (refl x)) ,
                                  refl (transport A (refl x))
```

The following is for use when we want to recover implicit arguments without mentioning them.

```
lhs : {X : 𝓤 ̇ } {x y : X} → x ＝ y → X
lhs {𝓤} {X} {x} {y} p = x

rhs : {X : 𝓤 ̇ } {x y : X} → x ＝ y → X
rhs {𝓤} {X} {x} {y} p = y
```

*Composition of identifications*. Given two identifications `p : x ＝ y` and `q : y ＝ z`, we can compose them to get an identification `p ∙ q : x ＝ z`. This can also be seen as transitivity of equality. Because the type of composition doesn't mention `p` and `q`, we can use the non-dependent version of ＝-induction.

```
_∙_ : {X : 𝓤 ̇ } {x y z : X} → x ＝ y → y ＝ z → x ＝ z
p ∙ q = transport (lhs p ＝_) q p
```

Here we are considering the family `A t = (x ＝ t)`, and using the identification `q : y ＝ z` to transport `A y` to `A z`, that is `x ＝ y` to `x ＝ z`.

*Exercise*. Can you define an alternative version that uses `p` to transport. Do the two versions give equal results?

When writing `p ∙ q`, we lose information on the lhs and the rhs of the identifications `p : x ＝ y` and `q : y ＝ z`, which makes some definitions hard to read. We now introduce notation to be able to write e.g.

```
    x ＝⟨ p ⟩

    y ＝⟨ q ⟩

    z ∎
```

as a synonym of the expression `p ∙ q` with some of the implicit arguments of `_∙_` made explicit. We have one ternary mixfix operator `_＝⟨_⟩_` and one unary `postfix` operator `_∎`.

```
_＝⟨_⟩_ : {X : 𝓤 ̇ } (x : X) {y z : X} → x ＝ y → y ＝ z → x ＝ z
x ＝⟨ p ⟩ q = p ∙ q

_∎ : {X : 𝓤 ̇ } (x : X) → x ＝ x
x ∎ = refl x
```

*Inversion of identifications*. Given an identification, we get an identification in the opposite direction:

```
_⁻¹ : {X : 𝓤 ̇ } → {x y : X} → x ＝ y → y ＝ x
p ⁻¹ = transport (_＝ lhs p) p (refl (lhs p))
```

We can define an alternative of identification composition with this:

```
_∙'_ : {X : 𝓤 ̇ } {x y z : X} → x ＝ y → y ＝ z → x ＝ z
p ∙' q = transport (_＝ rhs q) (p ⁻¹) q
```

This agrees with the previous one:

```
∙agreement : {X : 𝓤 ̇ } {x y z : X} (p : x ≡ y) (q : y ≡ z)
           → p ∙' q ≡ p ∙ q

∙agreement (refl x) (refl x) = refl (refl x)
```

But `refl y` is a definitional neutral element for one of them on the right and for the other one on the left,

- `p ∙ refl y = p`,
- `refl y ∙' q = q`,

which can be checked as follows

```
rdnel : {X : 𝓤 ̇ } {x y : X} (p : x ≡ y)
      → p ∙ refl y ≡ p

rdnel p = refl p

rdner : {X : 𝓤 ̇ } {y z : X} (q : y ≡ z)
      → refl y  ∙' q ≡ q

rdner q = refl q
```

*Exercise*. The identification `refl y` is neutral on both sides of each of the two operations `_∙_` and `_∙'_`, although not definitionally. This has to be proved by induction on identifications, as in `∙-agreement`.

*Application of a function to an identification*. Given an identification `p : x ≡ x'` we get an identification `ap f p : f x ≡ f x'` for any `f : X → Y`:

```
ap : {X : 𝓤 ̇ } {Y : 𝓥 ̇ } (f : X → Y) {x x' : X} → x ≡ x' → f x ≡ f x'
ap f {x} {x'} p = transport (λ - → f x ≡ f -) p (refl (f x))
```

Here the symbol "-", which is not to be confused with the symbol "_", is a variable. We will adopt the convention in these notes of using this variable name "-" to make clear which part of an expression we are replacing with `transport`.

Notice that we have so far used the recursion principle `transport` only. To reason about `transport`, `_∙_`, `_⁻¹` and `ap`, we will need to use the full induction principle `𝕁` (or equivalently pattern matching on `refl`).

*Pointwise equality of functions*. We will work with pointwise equality of functions, defined as follows, which, using univalence, will be equivalent to equality of functions.

```
_∼_ : {X : 𝓤 ̇ } {A : X → 𝓥 ̇ } → Π A → Π A → 𝓤 ⊔ 𝓥 ̇
f ∼ g = ∀ x → f x ≡ g x
```

The symbol `∀` is a built-in notation for `Π` . We could equivalently write the *definiens* as

```
(x : _) → f x ≡ g x,
```

or, with our `domain` notation

```
(x : domain f) → f x ≡ g x.
```

## Reasoning with negation

We first introduce notation for double and triple negation to avoid the use of brackets.

```
¬¬ ¬¬¬ : 𝓤 ̇ → 𝓤 ̇
¬¬  A = ¬(¬ A)
¬¬¬ A = ¬(¬¬ A)
```

To prove that `A → ¬¬ A`, that is, `A → ((A → 𝟘) → 𝟘)`, we start with a hypothetical element `a : A` and a hypothetical function `u : A → 𝟘` and the goal is to get an element of `𝟘`. All we need to do is to apply the function `u` to `a`. This gives double-negation introduction:

```
dni : (A : 𝓤 ̇ ) → A → ¬¬ A
dni A a u = u a
```

Mathematically, this says that if we have a point of `A` (we say that `A` is pointed) then `A` is nonempty. There is no general procedure to implement the converse, that is, from a function `(A → 𝟘) → 𝟘` to get a point of `A`. For truth values `A`, we can assume this as an axiom if we wish, because it is equivalent to the principle excluded middle. For arbitrary types `A`, this would be a form of global choice for type theory. However, global choice is inconsistent with univalence [HoTT book, Theorem 3.2.2], because there is no way to choose an element of every non-empty type in a way that is invariant under automorphisms. But the axiom of choice *is* consistent with univalent type theory, as stated in the introduction.

In the proof of the following, we are given hypothetical functions `f : A → B` and `v : B → 𝟘`, and a hypothetical element `a : A`, and our goal is to get an element of `𝟘`. But this is easy, because `f a : B` and hence `v (f a) : 𝟘`.

```
contrapositive : {A : 𝓤 ̇ } {B : 𝓥 ̇ } → (A → B) → (¬ B → ¬ A)
contrapositive f v a = v (f a)
```

We have given a logical name to this function. Mathematically, this says that if we have a function `A → B` and `B` is empty, then `A` must be empty, too. The proof is by assuming that `A` would have an inhabitant `a`, to get a contradiction, namely that `B` would have an inhabitant, too, showing that there can't be any such inhabitant `a` of `A` after all. See Bauer for a general discussion.

And from this we get that three negations imply one:

```
tno : (A : 𝓤 ̇ ) → ¬¬¬ A → ¬ A
tno A = contrapositive (dni A)
```

Hence, using `dni` once again, we get that `¬¬¬ A` if and only if `¬ A`. It is entertaining to see how Brouwer formulated and proved this fact in his Cambridge Lectures on Intuitionism:

> Theorem. Absurdity of absurdity of absurdity is equivalent to absurdity.
>
> Proof. *Firstly*, since implication of the assertion *y* by the assertion *x* implies implication of absurdity of *x* by absurdity of *y*, the implication of *absurdity of absurdity* by *truth* (which is an established fact) implies the implication of *absurdity of truth*, that is to say of *absurdity*, by *absurdity of absurdity of absurdity*. *Secondly*, since truth of an assertion implies absurdity of its absurdity, in particular truth of absurdity implies absurdity of absurdity of absurdity.

If we define *logical equivalence* by

```
_⇔_ : 𝓤 ̇ → 𝓥 ̇ → 𝓤 ⊔ 𝓥 ̇
X ⇔ Y = (X → Y) × (Y → X)

lr-implication : {X : 𝓤 ̇ } {Y : 𝓥 ̇ } → (X ⇔ Y) → (X → Y)
lr-implication = pr₁

rl-implication : {X : 𝓤 ̇ } {Y : 𝓥 ̇ } → (X ⇔ Y) → (Y → X)
rl-implication = pr₂
```

then we can render Brouwer's argument in Agda as follows, where the "established fact" is `dni`:

```
absurdity³-is-absurdity : {A : 𝓤 ̇ } → ¬¬¬ A ⇔ ¬ A
absurdity³-is-absurdity {𝓤} {A} = firstly , secondly
 where
  firstly : ¬¬¬ A → ¬ A
  firstly = contrapositive (dni A)

  secondly : ¬ A → ¬¬¬ A
  secondly = dni (¬ A)
```

But of course Brouwer, as is well known, was averse to formalism, and hence wouldn't approve of such a sacrilege.

We now define a symbol for the negation of equality.

```
_≢_ : {X : 𝓤 ̇ } → X → X → 𝓤 ̇
x ≢ y = ¬(x ≡ y)
```

In the following proof, we have `u : x ≡ y → 𝟘` and need to define a function `y ≡ x → 𝟘`. So all we need to do is to compose the function that inverts identifications with `u`:

```
≢-sym : {X : 𝓤 ̇ } {x y : X} → x ≢ y → y ≢ x
≢-sym {𝓤} {X} {x} {y} u = λ (p : y ≡ x) → u (p ⁻¹)
```

To show that the type `𝟙` is not equal to the type `𝟘`, we use that `transport id` gives `𝟙 ≡ 𝟘 → id 𝟙 → id 𝟘` where `id` is the identity function of the universe $\mathcal{U}_0$. More generally, we have the following conversion of type identifications into functions:

```
Id→Fun : {X Y : 𝓤 ̇ } → X ≡ Y → X → Y
Id→Fun {𝓤} = transport (id (𝓤 ̇ ))
```

Here the identity function is that of the universe $\mathcal{U}$ where the types `X` and `Y` live. An equivalent definition is the following, where this time the identity function is that of the type `X`:

```
Id→Fun' : {X Y : 𝓤 ̇ } → X ≡ Y → X → Y
Id→Fun' (refl X) = id X

Id→Funs-agree : {X Y : 𝓤 ̇ } (p : X ≡ Y)
              → Id→Fun p ≡ Id→Fun' p

Id→Funs-agree (refl X) = refl (id X)
```

So if we have a hypothetical identification `p : 𝟙 ≡ 𝟘`, then we get a function `𝟙 → 𝟘`. We apply this function to `⋆ : 𝟙` to conclude the proof.

```
𝟙-is-not-𝟘 : 𝟙 ≢ 𝟘
𝟙-is-not-𝟘 p = Id→Fun p ⋆
```

To show that the elements `₁` and `₀` of the two-point type `𝟚` are not equal, we reduce to the above case. We start with a hypothetical identification `p : ₁ ≡ ₀`.

```
₁-is-not-₀ : ₁ ≢ ₀
₁-is-not-₀ p = 𝟙-is-not-𝟘 q
 where
  f : 𝟚 → 𝓤₀ ̇
  f ₀ = 𝟘
  f ₁ = 𝟙

  q : 𝟙 ≡ 𝟘
  q = ap f p
```

*Remark.* Agda allows us to use a pattern `()` to get the following quick proof. However, this method of proof doesn't belong to the realm of MLTT. Hence we will use the pattern `()` only in the above definition of `𝟘-induction` and nowhere else in these notes.

```
₁-is-not-₀[not-an-MLTT-proof] : ¬(₁ ≡ ₀)
₁-is-not-₀[not-an-MLTT-proof] ()
```

Perhaps the following is sufficiently self-explanatory given the above:

```
decidable : 𝓤 ̇ → 𝓤 ̇
decidable A = A + ¬ A

has-decidable-equality : 𝓤 ̇ → 𝓤 ̇
has-decidable-equality X = (x y : X) → decidable (x ≡ y)

𝟚-has-decidable-equality : has-decidable-equality 𝟚
𝟚-has-decidable-equality ₀ ₀ = inl (refl ₀)
𝟚-has-decidable-equality ₀ ₁ = inr (≢-sym ₁-is-not-₀)
```

```
𝟚-has-decidable-equality ₁ ₀ = inr ₁-is-not-₀
𝟚-has-decidable-equality ₁ ₁ = inl (refl ₁)
```

So we consider four cases. In the first and the last, we have equal things and so we give an answer in the left-hand side of the sum. In the middle two, we give an answer in the right-hand side, where we need functions `₀ ≡ ₁ → 𝟘` and `₁ ≡ ₀ → 𝟘`, which we can take to be `≢-sym ₁-is-not-₀` and `₁-is-not-₀` respectively.

The following is more interesting. We consider the two possible cases for `n`. The first one assumes a hypothetical function `f : ₀ ≡ ₀ → 𝟘`, from which we get `f (refl ₀) : 𝟘`, and then, using `!𝟘`, we get an element of any type we like, which we choose to be `₀ ≡ ₁`, and we are done. Of course, we will never be able to use the function `not-zero-is-one` with such outrageous arguments. The other case `n = ₁` doesn't need to use the hypothesis `f : ₁ ≡ ₀ → 𝟘`, because the desired conclusion holds right away, as it is `₁ ≡ ₁`, which is proved by `refl ₁`. But notice that there is nothing wrong with the hypothesis `f : ₁ ≡ ₀ → 𝟘`. For example, we can use `not-zero-is-one` taking `n` to be `₀` and `f` to be `₁-is-not-₀`, so that the hypotheses can be fulfilled in the second equation.

```
not-zero-is-one : (n : 𝟚) → n ≢ ₀ → n ≡ ₁
not-zero-is-one ₀ f = !𝟘 (₀ ≡ ₁) (f (refl ₀))
not-zero-is-one ₁ f = refl ₁
```

The following generalizes `₁-is-not-₀`, both in its statement and its proof (so we could have formulated it first and then used it to deduce `₁-is-not-₀`):

```
inl-inr-disjoint-images : {X : 𝓤 ̇ } {Y : 𝓥 ̇ } {x : X} {y : Y} → inl x ≢ inr y
inl-inr-disjoint-images {𝓤} {𝓥} {X} {Y} p = 𝟙-is-not-𝟘 q
 where
  f : X + Y → 𝓤₀ ̇
  f (inl x) = 𝟙
  f (inr y) = 𝟘

  q : 𝟙 ≡ 𝟘
  q = ap f p
```

If `P or Q` holds and `Q` fails, then `P` holds:

```
right-fails-gives-left-holds : {P : 𝓤 ̇ } {Q : 𝓥 ̇ } → P + Q → ¬ Q → P
right-fails-gives-left-holds (inl p) u = p
right-fails-gives-left-holds (inr q) u = !𝟘 _ (u q)
```

## Example: formulation of the twin-prime conjecture

We illustrate the above constructs of MLTT to formulate this conjecture.

```
module twin-primes where

 open Arithmetic renaming (_×_ to _*_ ; _+_ to _∔_)
 open ℕ-order

 is-prime : ℕ → 𝓤₀ ̇
 is-prime n = (n ≥ 2) × ((x y : ℕ) → x * y ≡ n → (x ≡ 1) + (x ≡ n))

 twin-prime-conjecture : 𝓤₀ ̇
 twin-prime-conjecture = (n : ℕ) → Σ p ꞉ ℕ , (p ≥ n)
                                           × is-prime p
                                           × is-prime (p ∔ 2)
```

Thus, not only can we write down definitions, constructions, theorems and proofs, but also conjectures. They are just definitions of types. Likewise, the univalence axiom, to be formulated in due course, is a type.

## Remaining Peano axioms and basic arithmetic

We first prove the remaining Peano axioms.

```
positive-not-zero : (x : ℕ) → succ x ≢ 0
positive-not-zero x p = 𝟙-is-not-𝟘 (g p)
 where
  f : ℕ → 𝓤₀ ̇
  f 0        = 𝟘
  f (succ x) = 𝟙

  g : succ x ≡ 0 → 𝟙 ≡ 𝟘
  g = ap f
```

To show that the successor function is left cancellable, we can use the following predecessor function.

```
pred : ℕ → ℕ
pred 0        = 0
pred (succ n) = n

succ-lc : {x y : ℕ} → succ x ≡ succ y → x ≡ y
succ-lc = ap pred
```

With this we have proved all the Peano axioms.

Without assuming the principle of excluded middle, we can prove that `ℕ` has decidable equality:

```
ℕ-has-decidable-equality : has-decidable-equality ℕ
ℕ-has-decidable-equality 0 0               = inl (refl 0)
ℕ-has-decidable-equality 0 (succ y)        = inr (≢-sym (positive-not-zero y))
ℕ-has-decidable-equality (succ x) 0        = inr (positive-not-zero x)
ℕ-has-decidable-equality (succ x) (succ y) = f (ℕ-has-decidable-equality x y)
 where
  f : decidable (x ≡ y) → decidable (succ x ≡ succ y)
  f (inl p) = inl (ap succ p)
  f (inr k) = inr (λ (s : succ x ≡ succ y) → k (succ-lc s))
```

*Exercise*. Students should do this kind of thing at least once in their academic life: rewrite the above proof of the decidability of equality of `ℕ` to use the `ℕ-induction` principle instead of pattern matching and recursion, to understand by themselves that this can be done.

We now move to basic arithmetic, and we use a module for that.

```
module basic-arithmetic-and-order where

  open ℕ-order public
  open Arithmetic renaming (_+_ to _∔_) hiding (_×_)
```

We can show that addition is associative as follows, by induction on `z`, where `IH` stands for "induction hypothesis":

```
  +-assoc : (x y z : ℕ) → (x ∔ y) ∔ z ≡ x ∔ (y ∔ z)

  +-assoc x y 0        = (x ∔ y) ∔ 0 ≡⟨ refl _ ⟩
                          x ∔ (y ∔ 0) ∎

  +-assoc x y (succ z) = (x ∔ y) ∔ succ z   ≡⟨ refl _     ⟩
                          succ ((x ∔ y) ∔ z) ≡⟨ ap succ IH ⟩
                          succ (x ∔ (y ∔ z)) ≡⟨ refl _     ⟩
                          x ∔ (y ∔ succ z)   ∎
   where
    IH : (x ∔ y) ∔ z ≡ x ∔ (y ∔ z)
    IH = +-assoc x y z
```

Notice that the proofs `refl _` should be read as "by definition" or "by construction". They are not necessary, because Agda knows the definitions and silently expands them when necessary, but we are writing them here for the sake of clarity. Here is the version with the silent expansion of definitions, for the sake of illustration (the author of these notes can write, but not read it the absence of the above verbose version):

```
+-assoc' : (x y z : ℕ) → (x ∔ y) ∔ z ≡ x ∔ (y ∔ z)
+-assoc' x y zero     = refl _
+-assoc' x y (succ z) = ap succ (+-assoc' x y z)
```

We defined addition by induction on the second argument. Next we show that the base case and induction step of a definition by induction on the first argument hold (but of course not definitionally). We do this by induction on the second argument.

```
+-base-on-first : (x : ℕ) → 0 ∔ x ≡ x

+-base-on-first 0        = refl 0

+-base-on-first (succ x) = 0 ∔ succ x   ≡⟨ refl _      ⟩
                           succ (0 ∔ x) ≡⟨ ap succ IH ⟩
                           succ x       ∎
 where
  IH : 0 ∔ x ≡ x
  IH = +-base-on-first x

+-step-on-first : (x y : ℕ) → succ x ∔ y ≡ succ (x ∔ y)

+-step-on-first x zero     = refl (succ x)

+-step-on-first x (succ y) = succ x ∔ succ y   ≡⟨ refl _      ⟩
                             succ (succ x ∔ y) ≡⟨ ap succ IH ⟩
                             succ (x ∔ succ y) ∎
 where
  IH : succ x ∔ y ≡ succ (x ∔ y)
  IH = +-step-on-first x y
```

Using this, the commutativity of addition can be proved by induction on the first argument.

```
+-comm : (x y : ℕ) → x ∔ y ≡ y ∔ x

+-comm 0 y = 0 ∔ y ≡⟨ +-base-on-first y ⟩
             y     ≡⟨ refl _ ⟩
             y ∔ 0 ∎

+-comm (succ x) y = succ x ∔ y  ≡⟨ +-step-on-first x y ⟩
                    succ(x ∔ y) ≡⟨ ap succ IH          ⟩
                    succ(y ∔ x) ≡⟨ refl _              ⟩
                    y ∔ succ x  ∎
  where
   IH : x ∔ y ≡ y ∔ x
   IH = +-comm x y
```

We now show that addition is cancellable in its left argument, by induction on the left argument:

```
+-lc : (x y z : ℕ) → x ∔ y ≡ x ∔ z → y ≡ z

+-lc 0        y z p = y     ≡⟨ (+-base-on-first y)⁻¹ ⟩
                      0 ∔ y ≡⟨ p                     ⟩
                      0 ∔ z ≡⟨ +-base-on-first z     ⟩
                      z     ∎

+-lc (succ x) y z p = IH
 where
  q = succ (x ∔ y) ≡⟨ (+-step-on-first x y)⁻¹ ⟩
      succ x ∔ y   ≡⟨ p                       ⟩
      succ x ∔ z   ≡⟨ +-step-on-first x z     ⟩
      succ (x ∔ z) ∎
```

```
  IH : y ≡ z
  IH = +-lc x y z (succ-lc q)
```

Now we solve part of an exercise given above, namely that `(x ≤ y) ⇔ Σ z ꞉ ℕ , x + z ≡ y`.

First we name the alternative definition of `≤`:

```
_≼_ : ℕ → ℕ → 𝓤₀ ̇
x ≼ y = Σ z ꞉ ℕ , x ∔ z ≡ y
```

Next we show that the two relations `≤` and `≼` imply each other.

In both cases, we proceed by induction on both arguments.

```
≤-gives-≼ : (x y : ℕ) → x ≤ y → x ≼ y
≤-gives-≼ 0 0               l = 0 , refl 0
≤-gives-≼ 0 (succ y)        l = succ y , ∔-base-on-first (succ y)
≤-gives-≼ (succ x) 0        l = !𝟘 (succ x ≼ zero) l
≤-gives-≼ (succ x) (succ y) l = γ
 where
  IH : x ≼ y
  IH = ≤-gives-≼ x y l

  z : ℕ
  z = pr₁ IH

  p : x ∔ z ≡ y
  p = pr₂ IH

  γ : succ x ≼ succ y
  γ = z , (succ x ∔ z   ≡⟨ ∔-step-on-first x z ⟩
           succ (x ∔ z) ≡⟨ ap succ p          ⟩
           succ y       ∎)

≼-gives-≤ : (x y : ℕ) → x ≼ y → x ≤ y

≼-gives-≤ 0 0               (z , p) = ⋆

≼-gives-≤ 0 (succ y)        (z , p) = ⋆

≼-gives-≤ (succ x) 0        (z , p) = positive-not-zero (x ∔ z) q
 where
  q = succ (x ∔ z) ≡⟨ (∔-step-on-first x z)⁻¹ ⟩
      succ x ∔ z   ≡⟨ p                       ⟩
      zero         ∎

≼-gives-≤ (succ x) (succ y) (z , p) = IH
 where
  q = succ (x ∔ z) ≡⟨ (∔-step-on-first x z)⁻¹ ⟩
      succ x ∔ z   ≡⟨ p                       ⟩
      succ y       ∎

  IH : x ≤ y
  IH = ≼-gives-≤ x y (z , succ-lc q)
```

Later we will show that `(x ≤ y) ≡ Σ z ꞉ ℕ , x + z ≡ y`, using univalence.

We now develop some generally useful material regarding the order `≤` on natural numbers. First, it is reflexive, transitive and antisymmetric:

```
≤-refl : (n : ℕ) → n ≤ n
≤-refl zero     = ⋆
≤-refl (succ n) = ≤-refl n

≤-trans : (l m n : ℕ) → l ≤ m → m ≤ n → l ≤ n
```

```
≤-trans zero     m        n        p q = ⋆
≤-trans (succ l) zero     n        p q = !𝟘 (succ l ≤ n) p
≤-trans (succ l) (succ m) zero     p q = q
≤-trans (succ l) (succ m) (succ n) p q = ≤-trans l m n p q

≤-anti : (m n : ℕ) → m ≤ n → n ≤ m → m ≡ n
≤-anti zero     zero     p q = refl zero
≤-anti zero     (succ n) p q = !𝟘 (zero ≡ succ n) q
≤-anti (succ m) zero     p q = !𝟘 (succ m ≡ zero) p
≤-anti (succ m) (succ n) p q = ap succ (≤-anti m n p q)

≤-succ : (n : ℕ) → n ≤ succ n
≤-succ zero     = ⋆
≤-succ (succ n) = ≤-succ n

zero-minimal : (n : ℕ) → zero ≤ n
zero-minimal n = ⋆

unique-minimal : (n : ℕ) → n ≤ zero → n ≡ zero
unique-minimal zero     p = refl zero
unique-minimal (succ n) p = !𝟘 (succ n ≡ zero) p

≤-split : (m n : ℕ) → m ≤ succ n → (m ≤ n) + (m ≡ succ n)
≤-split zero     n        l = inl l
≤-split (succ m) zero     l = inr (ap succ (unique-minimal m l))
≤-split (succ m) (succ n) l = +-recursion inl (inr ∘ ap succ) (≤-split m n l)

_<_ : ℕ → ℕ → 𝓤₀ ̇
x < y = succ x ≤ y

infix 10 _<_

not-<-gives-≥ : (m n : ℕ) → ¬(n < m) → m ≤ n
not-<-gives-≥ zero     n        u = zero-minimal n
not-<-gives-≥ (succ m) zero     u = dni (zero < succ m) (zero-minimal m) u
not-<-gives-≥ (succ m) (succ n) u = not-<-gives-≥ m n u

bounded-∀-next : (A : ℕ → 𝓤 ̇ ) (k : ℕ)
               → A k
               → ((n : ℕ) → n < k → A n)
               → (n : ℕ) → n < succ k → A n
bounded-∀-next A k a φ n l = +-recursion f g s
 where
  s : (n < k) + (succ n ≡ succ k)
  s = ≤-split (succ n) k l

  f : n < k → A n
  f = φ n

  g : succ n ≡ succ k → A n
  g p = transport A ((succ-lc p)⁻¹) a
```

The type of roots of a function:

```
root : (ℕ → ℕ) → 𝓤₀ ̇
root f = Σ n ꞉ ℕ , f n ≡ 0

_has-no-root<_ : (ℕ → ℕ) → ℕ → 𝓤₀ ̇
f has-no-root< k = (n : ℕ) → n < k → f n ≢ 0

is-minimal-root : (ℕ → ℕ) → ℕ → 𝓤₀ ̇
is-minimal-root f m = (f m ≡ 0) × (f has-no-root< m)

at-most-one-minimal-root : (f : ℕ → ℕ) (m n : ℕ)
                         → is-minimal-root f m → is-minimal-root f n → m ≡ n

at-most-one-minimal-root f m n (p , φ) (q , ψ) = c m n a b
```

```
  where
   a : ¬(m < n)
   a u = ψ m u p

   b : ¬(n < m)
   b v = φ n v q

   c : (m n : ℕ) → ¬(m < n) → ¬(n < m) → m ≡ n
   c m n u v = ≤-anti m n (not-<-gives-≥ m n v) (not-<-gives-≥ n m u)
```

The type of minimal roots of a function:

```
minimal-root : (ℕ → ℕ) → 𝓤₀ ̇
minimal-root f = Σ m ꞉ ℕ , is-minimal-root f m

minimal-root-is-root : ∀ f → minimal-root f → root f
minimal-root-is-root f (m , p , _) = m , p

bounded-ℕ-search : ∀ k f → (minimal-root f) + (f has-no-root< k)
bounded-ℕ-search zero f = inr (λ n → !𝟘 (f n ≢ 0))
bounded-ℕ-search (succ k) f = +-recursion φ γ (bounded-ℕ-search k f)
 where
  A : ℕ → (ℕ → ℕ) → 𝓤₀ ̇
  A k f = (minimal-root f) + (f has-no-root< k)

  φ : minimal-root f → A (succ k) f
  φ = inl

  γ : f has-no-root< k → A (succ k) f
  γ u = +-recursion γ₀ γ₁ (ℕ-has-decidable-equality (f k) 0)
   where
    γ₀ : f k ≡ 0 → A (succ k) f
    γ₀ p = inl (k , p , u)

    γ₁ : f k ≢ 0 → A (succ k) f
    γ₁ v = inr (bounded-∀-next (λ n → f n ≢ 0) k v u)
```

Given any root, we can find a minimal root.

```
root-gives-minimal-root : ∀ f → root f → minimal-root f
root-gives-minimal-root f (n , p) = γ
 where
  g : ¬(f has-no-root< (succ n))
  g φ = φ n (≤-refl n) p

  γ : minimal-root f
  γ = right-fails-gives-left-holds (bounded-ℕ-search (succ n) f) g
```

# Univalent Mathematics in Agda

## Our univalent type theory

- A spartan MLTT as above.
- Univalence axiom as below.
- Subsingleton (or truth-value or propositional) truncations as below.

But, as discussed above, rather than postulating univalence and truncation, we will use them as explicit assumptions each time they are needed.

We emphasize that there are univalent type theories in which univalence and existence of truncations are theorems, for example cubical type theory, which has a version available in Agda, called cubical Agda.

## Singletons, subsingletons and sets

### Singleton (or contractible) types

Voevodsky defined a notion of *contractible type*, which we refer to here as *singleton type*. We say that a type is a singleton if there is a designated `c : X` that is identified with each `x : X`.

```
is-singleton : 𝓤 ̇ → 𝓤 ̇
is-singleton X = Σ c ꞉ X , ((x : X) → c ≡ x)
```

Such an element `c` is sometimes referred to as a *center of contraction* of `X`, in connection with homotopy theory.

```
is-center : (X : 𝓤 ̇ ) → X → 𝓤 ̇
is-center X c = (x : X) → c ≡ x
```

The canonical singleton type is `𝟙`:

```
𝟙-is-singleton : is-singleton 𝟙
𝟙-is-singleton = ⋆ , 𝟙-induction (λ x → ⋆ ≡ x) (refl ⋆)
```

Once we have defined the notion of type equivalence, we will have that a type is a singleton if and only if it is equivalent to `𝟙`.

When working with singleton types, it will be convenient to have distinguished names for the two projections:

```
center : (X : 𝓤 ̇ ) → is-singleton X → X
center X (c , φ) = c

centrality : (X : 𝓤 ̇ ) (i : is-singleton X) (x : X) → center X i ≡ x
centrality X (c , φ) = φ
```

### Subsingletons (or propositions or truth values)

A type is a subsingleton if it has at most one element, that is, any two of its elements are equal, or identified.

```
is-subsingleton : 𝓤 ̇ → 𝓤 ̇
is-subsingleton X = (x y : X) → x ≡ y

𝟘-is-subsingleton : is-subsingleton 𝟘
𝟘-is-subsingleton x y = !𝟘 (x ≡ y) x

singletons-are-subsingletons : (X : 𝓤 ̇ ) → is-singleton X → is-subsingleton X
singletons-are-subsingletons X (c , φ) x y = x ≡⟨ (φ x)⁻¹ ⟩
                                             c ≡⟨ φ y      ⟩
                                             y ∎

𝟙-is-subsingleton : is-subsingleton 𝟙
𝟙-is-subsingleton = singletons-are-subsingletons 𝟙 𝟙-is-singleton

pointed-subsingletons-are-singletons : (X : 𝓤 ̇ )
                                     → X → is-subsingleton X → is-singleton X

pointed-subsingletons-are-singletons X x s = (x , s x)

singleton-iff-pointed-and-subsingleton : {X : 𝓤 ̇ }
                                       → is-singleton X ⇔ (X × is-subsingleton X)

singleton-iff-pointed-and-subsingleton {𝓤} {X} = (a , b)
 where
  a : is-singleton X → X × is-subsingleton X
  a s = center X s , singletons-are-subsingletons X s

  b : X × is-subsingleton X → is-singleton X
  b (x , t) = pointed-subsingletons-are-singletons X x t
```

The terminology stands for *subtype of a singleton* and is justified by the fact that a type `X` is a subsingleton according to the above definition if and only if the map `X → 𝟙` is an embedding, if and only if `X` is embedded into a singleton.

Under univalent excluded middle, the empty type `𝟘` and the singleton type `𝟙` are the only subsingletons, up to equivalence, or up to identity if we further assume univalence.

Subsingletons are also called propositions or truth values:

```
is-prop is-truth-value : 𝓤 ̇ → 𝓤 ̇
is-prop        = is-subsingleton
is-truth-value = is-subsingleton
```

The terminology `is-subsingleton` is more mathematical and avoids the clash with the slogan propositions as types, which is based on the interpretation of mathematical statements as arbitrary types, rather than just subsingletons. In these notes we prefer the terminology *subsingleton*, but we occasionally use the terminologies *proposition* and *truth value*. When we wish to emphasize this notion of proposition adopted in univalent mathematics, as opposed to "propositions as (arbitrary) types", we may say *univalent proposition*.

In some formal systems, for example set theory based on first-order logic, one can tell that something is a proposition by looking at its syntactical form, e.g. "there are infinitely many prime numbers" is a proposition, by construction, in such a theory.

In univalent mathematics, however, propositions are mathematical, rather than meta-mathematical, objects, and the fact that a type turns out to be a proposition requires mathematical proof. Moreover, such a proof may be subtle and not immediate and require significant preparation. An example is the proof that the univalence axiom is a proposition, which relies on the fact that univalence implies function extensionality, which in turn implies that the statement that a map is an equivalence is a proposition. Another non-trivial example, which again relies on univalence or at least function extensionality, is the proof that the statement that a type `X` is a proposition is itself a proposition, which can be written as `is-prop (is-prop X)`.

Singletons and subsingletons are also called -2-groupoids and -1-groupoids respectively.

### Sets (or 0-groupoids)

A type is defined to be a set if there is at most one way for any two of its elements to be equal:

```
is-set : 𝓤 ̇ → 𝓤 ̇
is-set X = (x y : X) → is-subsingleton (x ＝ y)
```

At this point, with the definition of these notions, we are entering the realm of univalent mathematics, but not yet needing the univalence axiom.

## Univalent excluded middle

As mentioned above, under excluded middle, the only two subsingletons, up to equivalence, are `𝟘` and `𝟙`. In fact, excluded middle in univalent mathematics says precisely that.

```
EM EM' : ∀ 𝓤 → 𝓤 ⁺ ̇
EM  𝓤 = (X : 𝓤 ̇ ) → is-subsingleton X → X + ¬ X
EM' 𝓤 = (X : 𝓤 ̇ ) → is-subsingleton X → is-singleton X + is-empty X
```

Notice that the above two definitions don't assert excluded middle, but instead say what excluded middle is (like when we said what the twin-prime conjecture is), in two logically equivalent ways:

```
EM-gives-EM' : EM 𝓤 → EM' 𝓤
EM-gives-EM' em X s = γ (em X s)
 where
  γ : X + ¬ X → is-singleton X + is-empty X
  γ (inl x) = inl (pointed-subsingletons-are-singletons X x s)
  γ (inr x) = inr x
```

```
EM'-gives-EM : EM' 𝓤 → EM 𝓤
EM'-gives-EM em' X s = γ (em' X s)
 where
  γ : is-singleton X + is-empty X → X + ¬ X
  γ (inl i) = inl (center X i)
  γ (inr x) = inr x
```

We will not assume or deny excluded middle, which is an independent statement in our spartan univalent type theory - it can't be proved or disproved, just as the parallel postulate in Euclidean geometry can't be proved or disproved. We will deliberately keep it undecided, adopting a neutral approach to the constructive vs. non-constructive dichotomy. We will however prove a couple of consequences of excluded middle in discussions of foundational issues such as size and existence of subsingleton truncations. We will also prove that excluded middle is a consequence of the axiom of choice.

It should be emphasized that the potential failure of excluded middle doesn't say that there may be mysterious subsingletons that fail to be singletons and also fail to be empty. No such things occur in mathematical nature:

```
no-unicorns : ¬(Σ X ꞉ 𝓤 ̇ , is-subsingleton X × ¬(is-singleton X) × ¬(is-empty X))
no-unicorns (X , i , f , g) = c
 where
  e : is-empty X
  e x = f (pointed-subsingletons-are-singletons X x i)

  c : 𝟘
  c = g e
```

Given this, what does the potential failure of excluded middle *mean*? That there is no general way, provided by our spartan univalent type theory, to *determine which of the two cases* `is-singleton X` and `is-empty X` applies for a given subsingleton `X`. This kind of phenomenon should be familiar even in classical, non-constructive mathematics: although we are entitled to believe that the Goldbach conjecture either holds or fails, we still don't know which one is the case, despite efforts by the sharpest mathematical minds. A hypothetical element of the type `EM` would, in particular, be able to solve the Goldbach conjecture. There is nothing wrong or contradictory with assuming the existence of such a magic blackbox (EM stands for "there `E`xists such a `M`agic box"). There is only loss of generality and of the implicit algorithmic character of our spartan base type theory, both of which most mathematicians will be perfectly happy to live with.

In these notes we don't advocate any particular philosophy for or against excluded middle and other non-constructive principles. We confine ourselves to discussing mathematical facts. Axioms that can be assumed consistently but reduce generality and/or break the implicit computational character of our base type theory are discussed at various parts of these lecture notes, and are summarized at the end.

*Exercise*. We also have that it is impossible for `is-singleton X + is-empty X` to fail for a given subsingleton `X`, which amounts to saying that

```
¬¬(is-singleton X + is-empty X)
```

always holds.

Occasionaly one hears people asserting that this says that the double negation of excluded middle holds. But this is incorrect. The above statement, when written in full, is

```
(X ꞉ 𝓤 ̇ ) → is-subsingleton X → ¬¬(is-singleton X + is-empty X).
```

This is a theorem, which is quite different from the double negation of excluded middle, which is not a theorem and is

```
¬¬((X ꞉ 𝓤 ̇ ) → is-subsingleton X → is-singleton X + is-empty X).
```

Just as excluded middle, this is an independent statement.

*Exercise*. Continued from the previous exercise. Also for any type `R` replacing the empty type, there is a function `((X + (X → R)) → R) → R`, so that the kind of phenomenon illustrated in the previous exercise has little to do with the emptiness of the empty type.

## The types of magmas and monoids

A magma is a *set* equipped with a binary operation subject to no laws [Bourbaki]. We can define the type of magmas in a universe 𝓤 as follows:

```
module magmas where

 Magma : (𝓤 : Universe) → 𝓤 ⁺ ̇
 Magma 𝓤 = Σ X ꞉ 𝓤 ̇ , is-set X × (X → X → X)
```

The type `Magma 𝓤` collects all magmas in a universe 𝓤, and lives in the successor universe 𝓤 ⁺. Thus, this doesn't define what a magma is as a property. It defines the type of magmas. A magma is an element of this type, that is, a triple `(X , i , _·_)` with `X : 𝓤` and `i : is-set X` and `_·_ : X → X → X`.

Given a magma `M = (X , i , _·_)` we denote by `⟨ M ⟩` its underlying set `X` and by `magma-operation M` its multiplication `_·_`:

```
 ⟨_⟩ : Magma 𝓤 → 𝓤 ̇
 ⟨ X , i , _·_ ⟩ = X

 magma-is-set : (M : Magma 𝓤) → is-set ⟨ M ⟩
 magma-is-set (X , i , _·_) = i

 magma-operation : (M : Magma 𝓤) → ⟨ M ⟩ → ⟨ M ⟩ → ⟨ M ⟩
 magma-operation (X , i , _·_) = _·_
```

The following syntax declaration allows us to write `x ·⟨ M ⟩ y` as an abbreviation of `magma-operation M x y`:

```
 syntax magma-operation M x y = x ·⟨ M ⟩ y
```

The point is that this time we need such a mechanism because in order to be able to mention arbitrary `x` and `y`, we first need to know their types, which is `⟨ M ⟩` and hence `M` has to occur before `x` and `y` in the definition of `magma-operation`. The syntax declaration circumvents this.

A function of the underlying sets of two magmas is a called a homomorphism when it commutes with the magma operations:

```
 is-magma-hom : (M N : Magma 𝓤) → (⟨ M ⟩ → ⟨ N ⟩) → 𝓤 ̇
 is-magma-hom M N f = (x y : ⟨ M ⟩) → f (x ·⟨ M ⟩ y) ≡ f x ·⟨ N ⟩ f y

 id-is-magma-hom : (M : Magma 𝓤) → is-magma-hom M M (id ⟨ M ⟩)
 id-is-magma-hom M = λ (x y : ⟨ M ⟩) → refl (x ·⟨ M ⟩ y)

 is-magma-iso : (M N : Magma 𝓤) → (⟨ M ⟩ → ⟨ N ⟩) → 𝓤 ̇
 is-magma-iso M N f = is-magma-hom M N f
                    × (Σ g ꞉ (⟨ N ⟩ → ⟨ M ⟩), is-magma-hom N M g
                                            × (g ∘ f ∼ id ⟨ M ⟩)
                                            × (f ∘ g ∼ id ⟨ N ⟩))

 id-is-magma-iso : (M : Magma 𝓤) → is-magma-iso M M (id ⟨ M ⟩)
 id-is-magma-iso M = id-is-magma-hom M ,
                     id ⟨ M ⟩ ,
                     id-is-magma-hom M ,
                     refl ,
                     refl
```

Any identification of magmas gives rise to a magma isomorphism by transport:

```
 Id→iso : {M N : Magma 𝓤} → M ≡ N → ⟨ M ⟩ → ⟨ N ⟩
 Id→iso p = transport ⟨_⟩ p

 Id→iso-is-iso : {M N : Magma 𝓤} (p : M ≡ N) → is-magma-iso M N (Id→iso p)
 Id→iso-is-iso (refl M) = id-is-magma-iso M
```

The isomorphisms can be collected in a type:

```
 _≅ₘ_ : Magma 𝓤 → Magma 𝓤 → 𝓤 ̇
 M ≅ₘ N = Σ f ꞉ (⟨ M ⟩ → ⟨ N ⟩), is-magma-iso M N f
```

The following function will be a bijection in the presence of univalence, so that the identifications of magmas are in one-to-one correspondence with the magma isomorphisms:

```
magma-Id→iso : {M N : Magma 𝓤} → M ≡ N → M ≅ₘ N
magma-Id→iso p = (Id→iso p , Id→iso-is-iso p)
```

If we omit the sethood condition in the definition of the type of magmas, we get the type of what we could call ∞-magmas (then the type of magmas could be called `0-Magma`).

```
∞-Magma : (𝓤 : Universe) → 𝓤 ⁺ ̇
∞-Magma 𝓤 = Σ X ꞉ 𝓤 ̇ , (X → X → X)
```

A monoid is a set equipped with an associative binary operation and with a two-sided neutral element, and so we get the type of monoids as follows.

We first define the three laws:

```
module monoids where

 left-neutral : {X : 𝓤 ̇ } → X → (X → X → X) → 𝓤 ̇
 left-neutral e _·_ = ∀ x → e · x ≡ x

 right-neutral : {X : 𝓤 ̇ } → X → (X → X → X) → 𝓤 ̇
 right-neutral e _·_ = ∀ x → x · e ≡ x

 associative : {X : 𝓤 ̇ } → (X → X → X) → 𝓤 ̇
 associative _·_ = ∀ x y z → (x · y) · z ≡ x · (y · z)
```

Then a monoid is a set equipped with such `e` and `_·_` satisfying these three laws:

```
 Monoid : (𝓤 : Universe) → 𝓤 ⁺ ̇
 Monoid 𝓤 = Σ X ꞉ 𝓤 ̇ , is-set X
                      × (Σ · ꞉ (X → X → X) , (Σ e ꞉ X , (left-neutral e ·)
                                                        × (right-neutral e ·)
                                                        × (associative ·)))
```

*Remark.* People are more likely to use records in Agda rather than iterated `Σ`s as above (recall that we defined `Σ` using a record). This is fine, because records amount to iterated `Σ` types (recall that also `_×_` is a `Σ` type, by definition). Here, however, we are being deliberately spartan. Once we have defined our Agda notation for MLTT, we want to stick to it. This is for teaching purposes (of MLTT, encoded in Agda, not of Agda itself in its full glory).

We could drop the `is-set X` condition, but then we wouldn't get ∞-monoids in any reasonable sense. We would instead get "wild ∞-monoids" or "incoherent ∞-monoids". The reason is that in monoids (with sets as carriers) the neutrality and associativity equations can hold in at most one way, by definition of set. But if we drop the sethood requirement, then the equations can hold in multiple ways. And then one is forced to consider equations between the identifications (all the way up in the ∞-case), which is what "coherence data" means. The wildness or incoherence amounts to the absence of such data.

Similarly to the situation with magmas, identifications of monoids are in bijection with monoid isomorphisms, assuming univalence. For this to be the case, it is absolutely necessary that the carrier of a monoid is a set rather than an arbitrary type, for otherwise the monoid equations can hold in many possible ways, and we would need to consider a notion of monoid isomorphism that in addition to preserving the neutral element and the multiplication, preserves the associativity identifications.

*Exercise.* Define the type of groups (with sets as carriers).

*Exercise.* Write down the various types of categories defined in the HoTT book in Agda.

*Exercise.* Try to define a type of topological spaces.

## The identity type in univalent mathematics

We can view a type `X` as a sort of category with hom-types rather than hom-sets, with the identifications between points as the arrows.

We have that `refl` provides a neutral element for composition of identifications:

```
refl-left : {X : 𝓤 ̇ } {x y : X} {p : x ＝ y} → refl x ∙ p ＝ p
refl-left {𝓤} {X} {x} {x} {refl x} = refl (refl x)

refl-right : {X : 𝓤 ̇ } {x y : X} {p : x ＝ y} → p ∙ refl y ＝ p
refl-right {𝓤} {X} {x} {y} {p} = refl p
```

And composition is associative:

```
∙assoc : {X : 𝓤 ̇ } {x y z t : X} (p : x ＝ y) (q : y ＝ z) (r : z ＝ t)
       → (p ∙ q) ∙ r ＝ p ∙ (q ∙ r)

∙assoc p q (refl z) = refl (p ∙ q)
```

If we wanted to prove the above without pattern matching, this time we would need the dependent version `𝕁` of induction on `_＝_`.

*Exercise*. Try to do this with `𝕁` and with `ℍ`.

But all arrows, the identifications, are invertible:

```
⁻¹-left∙ : {X : 𝓤 ̇ } {x y : X} (p : x ＝ y)
         → p ⁻¹ ∙ p ＝ refl y

⁻¹-left∙ (refl x) = refl (refl x)

⁻¹-right∙ : {X : 𝓤 ̇ } {x y : X} (p : x ＝ y)
          → p ∙ p ⁻¹ ＝ refl x

⁻¹-right∙ (refl x) = refl (refl x)
```

A category in which all arrows are invertible is called a groupoid. The above is the basis for the Hofmann–Streicher groupoid model of type theory.

But we actually get higher groupoids, because given identifications

```
p q : x ＝ y
```

we can consider the identity type `p ＝ q`, and given

```
u v : p ＝ q
```

we can consider the type `u ＝ v`, and so on. See [van den Berg and Garner] and [Lumsdaine].

For some types, such as the natural numbers, we can prove that this process trivializes after the first step, because the type `x ＝ y` has at most one element. Such types are the sets as defined above.

Voevodsky defined the notion of *hlevel* to measure how long it takes for the process to trivialize.

Here are some more constructions with identifications:

```
⁻¹-involutive : {X : 𝓤 ̇ } {x y : X} (p : x ＝ y)
              → (p ⁻¹)⁻¹ ＝ p

⁻¹-involutive (refl x) = refl (refl x)
```

The application operation on identifications is functorial, in the sense that it preserves the neutral element and commutes with composition:

```
ap-refl : {X : 𝓤 ̇ } {Y : 𝓥 ̇ } (f : X → Y) (x : X)
        → ap f (refl x) ≡ refl (f x)

ap-refl f x = refl (refl (f x))

ap-∙ : {X : 𝓤 ̇ } {Y : 𝓥 ̇ } (f : X → Y) {x y z : X} (p : x ≡ y) (q : y ≡ z)
     → ap f (p ∙ q) ≡ ap f p ∙ ap f q

ap-∙ f p (refl y) = refl (ap f p)
```

Notice that we also have

```
ap⁻¹ : {X : 𝓤 ̇ } {Y : 𝓥 ̇ } (f : X → Y) {x y : X} (p : x ≡ y)
     → (ap f p)⁻¹ ≡ ap f (p ⁻¹)

ap⁻¹ f (refl x) = refl (refl (f x))
```

The above functions `ap-refl` and `ap-∙` constitute functoriality in the second argument. We also have functoriality in the first argument, in the following sense:

```
ap-id : {X : 𝓤 ̇ } {x y : X} (p : x ≡ y)
      → ap id p ≡ p

ap-id (refl x) = refl (refl x)

ap-∘ : {X : 𝓤 ̇ } {Y : 𝓥 ̇ } {Z : 𝓦 ̇ }
       (f : X → Y) (g : Y → Z) {x y : X} (p : x ≡ y)
     → ap (g ∘ f) p ≡ (ap g ∘ ap f) p

ap-∘ f g (refl x) = refl (refl (g (f x)))
```

Transport is also functorial with respect to identification composition and function composition. By construction, it maps the neutral element to the identity function. The apparent contravariance takes place because we have defined function composition in the usual order, but identification composition in the diagramatic order (as is customary in each case).

```
transport∙ : {X : 𝓤 ̇ } (A : X → 𝓥 ̇ ) {x y z : X} (p : x ≡ y) (q : y ≡ z)
           → transport A (p ∙ q) ≡ transport A q ∘ transport A p

transport∙ A p (refl y) = refl (transport A p)
```

Functions of a type into a universe can be considered as generalized presheaves, which we could perhaps call ∞-presheaves. Their morphisms are natural transformations:

```
Nat : {X : 𝓤 ̇ } → (X → 𝓥 ̇ ) → (X → 𝓦 ̇ ) → 𝓤 ⊔ 𝓥 ⊔ 𝓦 ̇
Nat A B = (x : domain A) → A x → B x
```

We don’t need to specify the naturality condition, because it is automatic:

```
Nats-are-natural : {X : 𝓤 ̇ } (A : X → 𝓥 ̇ ) (B : X → 𝓦 ̇ ) (τ : Nat A B)
                 → {x y : X} (p : x ≡ y)
                 → τ y ∘ transport A p ≡ transport B p ∘ τ x

Nats-are-natural A B τ (refl x) = refl (τ x)
```

We will use the following constructions a number of times:

```
NatΣ : {X : 𝓤 ̇ } {A : X → 𝓥 ̇ } {B : X → 𝓦 ̇ } → Nat A B → Σ A → Σ B
NatΣ τ (x , a) = (x , τ x a)

transport-ap : {X : 𝓤 ̇ } {Y : 𝓥 ̇ } (A : Y → 𝓦 ̇ )
               (f : X → Y) {x x' : X} (p : x ≡ x') (a : A (f x))
             → transport (A ∘ f) p a ≡ transport A (ap f p) a
```

```
transport-ap A f (refl x) a = refl a
```

### Identifications that depend on identifications

If we have an identification `p : A ＝ B` of two types `A` and `B`, and elements `a : A` and `b : B`, we cannot ask directly whether `a ＝ b`, because although the types are identified by `p`, they are not necessarily the same, in the sense of definitional equality. This is not merely a syntactical restriction of our formal system, but instead a fundamental fact that reflects the philosophy of univalent mathematics. For instance, consider the type

```
data Color : 𝓤₀ ̇  where
 Black White : Color
```

With univalence, we will have that `Color ＝ 𝟚` where `𝟚` is the two-point type `𝟙 + 𝟙` with elements `₀` and `₁`. But there will be two identifications `p₀ p₁ : Color ＝ 𝟚`, one that identifies `Black` with `₀` and `White` with `₁`, and another one that identifies `Black` with `₁` and `White` with `₀`. There is no preferred coding of binary colors as bits. And, precisely because of that, even if univalence does give inhabitants of the type `Color ＝ 𝟚`, it doesn't make sense to ask whether `Black ＝ ₀` holds without specifying one of the possible inhabitants `p₀` and `p₁`.

What we will have is that the functions `transport id p₀` and `transport id p₁` are the two possible bijections `Color → 𝟚` that identify colors with bits. So, it is not enough to have `Color ＝ 𝟚` to be able to compare a color `c : Color` with a bit `b : 𝟚`. We need to specify which identification `p : Color ＝ 𝟚` we want to consider for the comparison. The same considerations apply when we consider identifications `p : 𝟚 ＝ 𝟚`.

So the meaningful comparison in the more general situation is

```
transport id p a ＝ b
```

for a specific

```
p : A ＝ B,
```

where `id` is the identity function of the universe where the types `A` and `B` live, and hence

```
transport id : A ＝ B → (A → B)
```

is the function that transforms identifications into functions, which has already occurred above.

More generally, we want to consider the situation in which we replace the identity function `id` of the universe where `A` and `B` live by an arbitrary type family, which is what we do now.

If we have a type

```
X : 𝓤 ̇ ,
```

and a type family

```
A : X → 𝓥 ̇
```

and points

```
x y : X
```

and an identification

```
p : x ＝ y,
```

then we get the identification

```
ap A p : A x ＝ A y.
```

However, if we have

```
a : A x,

b : A y,
```

we again cannot write down the identity type

~~`a ＝ b`~~.

This is again a non-sensical mathematical statement, because the types `A x` and `A y` are not the same, but only identified, and in general there can be many identifications, not just `ap A p`, and so any identification between elements of `A x` and `A y` has to be with respect to a specific identification, as in the above particular case.

This time, the meaningful comparison, given `p : x ＝ y`, is

```
transport A p a ＝ b,
```

For example, this idea applies when comparing the values of a dependent function:

```
apd : {X : 𝓤 ̇ } {A : X → 𝓥 ̇ } (f : (x : X) → A x) {x y : X}
      (p : x ＝ y) → transport A p (f x) ＝ f y

apd f (refl x) = refl (f x)
```

## Equality in Σ types

With the above notion of dependent equality, we can characterize equality in `Σ` types as follows.

```
to-Σ-＝ : {X : 𝓤 ̇ } {A : X → 𝓥 ̇ } {σ τ : Σ A}
        → (Σ p ꞉ pr₁ σ ＝ pr₁ τ , transport A p (pr₂ σ) ＝ pr₂ τ)
        → σ ＝ τ

to-Σ-＝ (refl x , refl a) = refl (x , a)

from-Σ-＝ : {X : 𝓤 ̇ } {A : X → 𝓥 ̇ } {σ τ : Σ A}
          → σ ＝ τ
          → Σ p ꞉ pr₁ σ ＝ pr₁ τ , transport A p (pr₂ σ) ＝ pr₂ τ

from-Σ-＝ (refl (x , a)) = (refl x , refl a)
```

For the sake of readability, the above two definitions can be equivalently written as follows.

```
to-Σ-＝-again : {X : 𝓤 ̇ } {A : X → 𝓥 ̇ } {(x , a) (y , b) : Σ A}
              → Σ p ꞉ x ＝ y , transport A p a ＝ b
              → (x , a) ＝ (y , b)

to-Σ-＝-again (refl x , refl a) = refl (x , a)

from-Σ-＝-again : {X : 𝓤 ̇ } {A : X → 𝓥 ̇ } {(x , a) (y , b) : Σ A}
                → (x , a) ＝ (y , b)
                → Σ p ꞉ x ＝ y , transport A p a ＝ b

from-Σ-＝-again (refl (x , a)) = (refl x , refl a)
```

The above gives the logical equivalence

```
(σ ＝ τ) ⇔ (Σ p ꞉ pr₁ σ ＝ pr₁ τ , transport A p (pr₂ σ) ＝ pr₂ τ).
```

But this is a very weak statement when the left- and right-hand identity types have multiple elements, which is precisely the point of univalent mathematics.

What we want is the lhs and the rhs to be isomorphic, or more precisely, equivalent in the sense of Voevodsky.

Once we have defined this notion `_≃_` of type equivalence, this characterization will become an equivalence

```
(σ ≡ τ) ≃ (Σ p ꞉ pr₁ σ ≡ pr₁ τ , transport A p pr₂ σ ≡ pr₂ τ).
```

But even this is not sufficiently precise, because in general there are multiple equivalences between two types. For example, there are precisely two equivalences

```
𝟙 + 𝟙 ≃ 𝟙 + 𝟙,
```

namely the identity function and the function that flips left and right. What we want to say is that a *specific map* is an equivalence. In our case, it is the function `from-Σ-≡` defined above.

Voevodsky came up with a definition of a type "`f` is an equivalence" which is always a subsingleton: a given function `f` can be an equivalence in at most one way. In other words, being an equivalence is property of `f`, rather than data for `f`.

The following special case of `to-Σ-≡` is often useful:

```
to-Σ-≡' : {X : 𝓤 ̇ } {A : X → 𝓥 ̇ } {x : X} {a a' : A x}
        → a ≡ a' → Id (Σ A) (x , a) (x , a')

to-Σ-≡' {𝓤} {𝓥} {X} {A} {x} = ap (λ - → (x , -))
```

We take the opportunity to establish more equations for transport and to define a dependent version of transport:

```
transport-× : {X : 𝓤 ̇ } (A : X → 𝓥 ̇ ) (B : X → 𝓦 ̇ )
              {x y : X} (p : x ≡ y) {(a , b) : A x × B x}

             → transport (λ x → A x × B x) p (a , b)
             ≡ (transport A p a , transport B p b)

transport-× A B (refl _) = refl _

transportd : {X : 𝓤 ̇ } (A : X → 𝓥 ̇ ) (B : (x : X) → A x → 𝓦 ̇ )
             {x : X} ((a , b) : Σ a ꞉ A x , B x a) {y : X} (p : x ≡ y)
           → B x a → B y (transport A p a)

transportd A B (a , b) (refl x) = id

transport-Σ : {X : 𝓤 ̇ } (A : X → 𝓥 ̇ ) (B : (x : X) → A x → 𝓦 ̇ )
              {x : X} {y : X} (p : x ≡ y) {(a , b) : Σ a ꞉ A x , B x a}

             → transport (λ - → Σ (B -)) p (a , b)
             ≡ transport A p a , transportd A B (a , b) p b

transport-Σ A B (refl x) {a , b} = refl (a , b)
```

## Voevodsky's notion of hlevel

Voevodsky's hlevels `0,1,2,3,...` are shifted by `2` with respect to the `n`-groupoid numbering convention, and correspond to `-2`-groupoids (singletons), `-1`-groupoids (subsingletons), `0`-groupoids (sets),...

The `hlevel` relation is defined by induction on `ℕ`, with the induction step working with the identity types of the elements of the type in question:

```
_is-of-hlevel_ : 𝓤 ̇ → ℕ → 𝓤 ̇
X is-of-hlevel 0        = is-singleton X
X is-of-hlevel (succ n) = (x x' : X) → ((x ＝ x') is-of-hlevel n)
```

It is often convenient in practice to have equivalent formulations of the types of hlevel `1` (as subsingletons) and `2` (as sets), which we will develop soon.

### Hedberg's Theorem

To characterize sets as the types of hlevel 2, we first need to show that subsingletons are sets, and this is not easy. We use an argument due to Hedberg. This argument also shows that Voevodsky's hlevels are upper closed.

We choose to present an alternative formulation of Hedberg's Theorem, but we stress that the method of proof is essentially the same.

We first define a notion of constant map:

```
wconstant : {X : 𝓤 ̇ } {Y : 𝓥 ̇ } → (X → Y) → 𝓤 ⊔ 𝓥 ̇
wconstant f = (x x' : domain f) → f x ＝ f x'
```

The prefix "`w`" officially stands for "weakly". Perhaps *incoherently constant* or *wildly constant* would be better terminologies, with *coherence* understood in the ∞-categorical sense. We prefer to stick to *wildly* rather than *weakly*, and luckily both start with the letter "`w`".

We first define the type of constant endomaps of a given type:

```
wconstant-endomap : 𝓤 ̇ → 𝓤 ̇
wconstant-endomap X = Σ f ꞉ (X → X), wconstant f

wcmap : (X : 𝓤 ̇ ) → wconstant-endomap X → (X → X)
wcmap X (f , w) = f

wcmap-constancy : (X : 𝓤 ̇ ) (c : wconstant-endomap X)
                → wconstant (wcmap X c)
wcmap-constancy X (f , w) = w
```

The point is that a type is a set if and only if its identity types all have designated `wconstant` endomaps:

```
Hedberg : {X : 𝓤 ̇ } (x : X)
        → ((y : X) → wconstant-endomap (x ＝ y))
        → (y : X) → is-subsingleton (x ＝ y)

Hedberg {𝓤} {X} x c y p q =

 p                       ＝⟨ a y p                                   ⟩
 (f x (refl x))⁻¹ ∙ f y p ＝⟨ ap (λ - → (f x (refl x))⁻¹ ∙ -) (κ y p q) ⟩
 (f x (refl x))⁻¹ ∙ f y q ＝⟨ (a y q)⁻¹                              ⟩
 q                       ∎

 where
  f : (y : X) → x ＝ y → x ＝ y
  f y = wcmap (x ＝ y) (c y)

  κ : (y : X) (p q : x ＝ y) → f y p ＝ f y q
  κ y = wcmap-constancy (x ＝ y) (c y)

  a : (y : X) (p : x ＝ y) → p ＝ (f x (refl x))⁻¹ ∙ f y p
  a x (refl x) = (⁻¹-left∙ (f x (refl x)))⁻¹
```

### A characterization of sets

We consider types whose identity types all have designated constant endomaps:

```
wconstant-≡-endomaps : 𝓤 ̇ → 𝓤 ̇
wconstant-≡-endomaps X = (x y : X) → wconstant-endomap (x ≡ y)
```

The following is immediate from the definitions.

```
sets-have-wconstant-≡-endomaps : (X : 𝓤 ̇ ) → is-set X → wconstant-≡-endomaps X
sets-have-wconstant-≡-endomaps X s x y = (f , κ)
 where
  f : x ≡ y → x ≡ y
  f p = p

  κ : (p q : x ≡ y) → f p ≡ f q
  κ p q = s x y p q
```

And the converse is the content of Hedberg's Theorem.

```
types-with-wconstant-≡-endomaps-are-sets : (X : 𝓤 ̇ )
                                         → wconstant-≡-endomaps X → is-set X

types-with-wconstant-≡-endomaps-are-sets X c x = Hedberg x
                                                  (λ y → wcmap (x ≡ y) (c x y) ,
                                                   wcmap-constancy (x ≡ y) (c x y))
```

### Subsingletons are sets

In the following definition of the auxiliary function `f`, we ignore the argument `p`, using the fact that `X` is a subsingleton instead, to get a `wconstant` function:

```
subsingletons-have-wconstant-≡-endomaps : (X : 𝓤 ̇ )
                                        → is-subsingleton X
                                        → wconstant-≡-endomaps X

subsingletons-have-wconstant-≡-endomaps X s x y = (f , κ)
 where
  f : x ≡ y → x ≡ y
  f p = s x y

  κ : (p q : x ≡ y) → f p ≡ f q
  κ p q = refl (s x y)
```

And the corollary is that (sub)singleton types are sets.

```
subsingletons-are-sets : (X : 𝓤 ̇ ) → is-subsingleton X → is-set X
subsingletons-are-sets X s = types-with-wconstant-≡-endomaps-are-sets X
                               (subsingletons-have-wconstant-≡-endomaps X s)

singletons-are-sets : (X : 𝓤 ̇ ) → is-singleton X → is-set X
singletons-are-sets X = subsingletons-are-sets X ∘ singletons-are-subsingletons X
```

In particular, the types `𝟘` and `𝟙` are sets.

```
𝟘-is-set : is-set 𝟘
𝟘-is-set = subsingletons-are-sets 𝟘 𝟘-is-subsingleton

𝟙-is-set : is-set 𝟙
𝟙-is-set = subsingletons-are-sets 𝟙 𝟙-is-subsingleton
```

### The types of hlevel 1 are the subsingletons

Then with the above we get our desired characterization of the types of hlevel `1` as an immediate consequence:

```
subsingletons-are-of-hlevel-1 : (X : 𝓤 ̇ )
                              → is-subsingleton X
                              → X is-of-hlevel 1

subsingletons-are-of-hlevel-1 X = g
 where
  g : ((x y : X) → x ＝ y) → (x y : X) → is-singleton (x ＝ y)
  g t x y = t x y , subsingletons-are-sets X t x y (t x y)

types-of-hlevel-1-are-subsingletons : (X : 𝓤 ̇ )
                                    → X is-of-hlevel 1
                                    → is-subsingleton X

types-of-hlevel-1-are-subsingletons X = f
 where
  f : ((x y : X) → is-singleton (x ＝ y)) → (x y : X) → x ＝ y
  f s x y = center (x ＝ y) (s x y)
```

This is an "if and only if" characterization, but, under univalence, it becomes an equality because the types under consideration are subsingletons.

### The types of hlevel 2 are the sets

The same comments as for the previous section apply.

```
sets-are-of-hlevel-2 : (X : 𝓤 ̇ ) → is-set X → X is-of-hlevel 2
sets-are-of-hlevel-2 X = g
 where
  g : ((x y : X) → is-subsingleton (x ＝ y)) → (x y : X) → (x ＝ y) is-of-hlevel 1
  g t x y = subsingletons-are-of-hlevel-1 (x ＝ y) (t x y)

types-of-hlevel-2-are-sets : (X : 𝓤 ̇ ) → X is-of-hlevel 2 → is-set X
types-of-hlevel-2-are-sets X = f
 where
  f : ((x y : X) → (x ＝ y) is-of-hlevel 1) → (x y : X) → is-subsingleton (x ＝ y)
  f s x y = types-of-hlevel-1-are-subsingletons (x ＝ y) (s x y)
```

### The hlevels are upper closed

A singleton is a subsingleton, a subsingleton is a set, … , a type of hlevel `n` is of hlevel `n+1` too, …

Again we can conclude this almost immediately from the above development:

```
hlevel-upper : (X : 𝓤 ̇ ) (n : ℕ) → X is-of-hlevel n → X is-of-hlevel (succ n)
hlevel-upper X zero = γ
 where
  γ : is-singleton X → (x y : X) → is-singleton (x ＝ y)
  γ h x y = p , subsingletons-are-sets X k x y p
   where
    k : is-subsingleton X
    k = singletons-are-subsingletons X h
```

```
    p : x ≡ y
    p = k x y
```

```
hlevel-upper X (succ n) = λ h x y → hlevel-upper (x ≡ y) n (h x y)
```

To be consistent with the above terminology, we have to stipulate that all types have hlevel `∞`. We don't need a definition for this notion. But what may happen (and it does with univalence) is that there are types which don't have any finite hlevel. We can say that such types then have minimal hlevel `∞`.

*Exercise*. Formulate and prove the following. The type `𝟙` has minimal hlevel `0`.

```
_has-minimal-hlevel_ : 𝓤 ̇ → ℕ → 𝓤 ̇
X has-minimal-hlevel 0        = X is-of-hlevel 0
X has-minimal-hlevel (succ n) = (X is-of-hlevel (succ n)) × ¬(X is-of-hlevel n)

_has-minimal-hlevel-∞ : 𝓤 ̇ → 𝓤 ̇
X has-minimal-hlevel-∞ = (n : ℕ) → ¬(X is-of-hlevel n)
```

The type `𝟘` has minimal hlevel `1`, the type `ℕ` has minimal hlevel `2`. The solution to the fact that `ℕ` has hlevel 2 is given in the next section. More ambitiously, after univalence is available, show that the type of monoids has minimal hlevel `3`.

### ℕ and 𝟚 are sets

If a type has decidable equality we can define a `wconstant` function `x ≡ y → x ≡ y` and hence conclude that it is a set. This argument is due to Hedberg.

```
pointed-types-have-wconstant-endomap : {X : 𝓤 ̇ } → X → wconstant-endomap X
pointed-types-have-wconstant-endomap x = ((λ y → x) , (λ y y' → refl x))

empty-types-have-wconstant-endomap : {X : 𝓤 ̇ } → is-empty X → wconstant-endomap X
empty-types-have-wconstant-endomap e = (id , (λ x x' → !𝟘 (x ≡ x') (e x)))

decidable-has-wconstant-endomap : {X : 𝓤 ̇ } → decidable X → wconstant-endomap X
decidable-has-wconstant-endomap (inl x) = pointed-types-have-wconstant-endomap x
decidable-has-wconstant-endomap (inr e) = empty-types-have-wconstant-endomap e

hedberg-lemma : {X : 𝓤 ̇ } → has-decidable-equality X → wconstant-≡-endomaps X
hedberg-lemma {𝓤} {X} d x y = decidable-has-wconstant-endomap (d x y)

hedberg : {X : 𝓤 ̇ } → has-decidable-equality X → is-set X
hedberg {𝓤} {X} d = types-with-wconstant-≡-endomaps-are-sets X (hedberg-lemma d)

ℕ-is-set : is-set ℕ
ℕ-is-set = hedberg ℕ-has-decidable-equality

𝟚-is-set : is-set 𝟚
𝟚-is-set = hedberg 𝟚-has-decidable-equality
```

Notice that excluded middle implies directly that all sets have decidable equality, so that in its presence a type is a set if and only if it has decidable equality.

## Retracts

We use retracts as a mathematical technique to transfer properties between types. For instance, retracts of singletons are singletons. Showing that a particular type `X` is a singleton may be rather difficult to do directly by applying the definition of singleton and the definition of the particular type, but it may be easy to show that `X` is a retract of `Y` for a type `Y` that is already known to be a singleton. In these notes, a major application will be to get a simple proof of the known fact that invertible maps are equivalences in the sense of Voevodsky.

A *section* of a function is simply a right inverse, by definition:

```
has-section : {X : 𝓤 ̇ } {Y : 𝓥 ̇ } → (X → Y) → 𝓤 ⊔ 𝓥 ̇
has-section r = Σ s ꞉ (codomain r → domain r), r ∘ s ∼ id
```

Notice that `has-section r` is the type of all sections `(s , η)` of `r`, which may well be empty. So a point of this type is a designated section `s` of `r`, together with the datum `η`. Unless the domain of `r` is a set, this datum is not property, and we may well have an element `(s , η')` of the type `has-section r` with `η'` distinct from `η` for the same `s`.

We say that `X` *is a retract of* `Y`, written `X ◁ Y`, if we have a function `Y → X` which has a section:

```
_◁_ : 𝓤 ̇ → 𝓥 ̇ → 𝓤 ⊔ 𝓥 ̇
X ◁ Y = Σ r ꞉ (Y → X), has-section r
```

This type actually collects *all* the ways in which the type `X` can be a retract of the type `Y`, and so is data or structure on `X` and `Y`, rather than a property of them.

A function that has a section is called a retraction. We use this terminology, ambiguously, also for the function that projects out the retraction:

```
retraction : {X : 𝓤 ̇ } {Y : 𝓥 ̇ } → X ◁ Y → Y → X
retraction (r , s , η) = r

section : {X : 𝓤 ̇ } {Y : 𝓥 ̇ } → X ◁ Y → X → Y
section (r , s , η) = s

retract-equation : {X : 𝓤 ̇ } {Y : 𝓥 ̇ } (ρ : X ◁ Y)
                 → retraction ρ ∘ section ρ ∼ id X

retract-equation (r , s , η) = η

retraction-has-section : {X : 𝓤 ̇ } {Y : 𝓥 ̇ } (ρ : X ◁ Y)
                       → has-section (retraction ρ)

retraction-has-section (r , h) = h
```

We have an identity retraction:

```
id-◁ : (X : 𝓤 ̇ ) → X ◁ X
id-◁ X = id X , id X , refl
```

*Exercise*. The identity retraction is by no means the only retraction of a type onto itself in general, of course. Prove that we have (that is, produce an element of the type) `ℕ ◁ ℕ` with the function `pred : ℕ → ℕ` defined above as the retraction. Try to produce more inhabitants of this type.

We can define the composition of two retractions as follows:

```
_◁∘_ : {X : 𝓤 ̇ } {Y : 𝓥 ̇ } {Z : 𝓦 ̇ } → X ◁ Y → Y ◁ Z → X ◁ Z

(r , s , η) ◁∘ (r' , s' , η') = (r ∘ r' , s' ∘ s , η'')
 where
  η'' = λ x → r (r' (s' (s x))) ＝⟨ ap r (η' (s x)) ⟩
              r (s x)           ＝⟨ η x             ⟩
              x                 ∎
```

We also define composition with an implicit argument made explicit:

```
_◁⟨_⟩_ : (X : 𝓤 ̇ ) {Y : 𝓥 ̇ } {Z : 𝓦 ̇ } → X ◁ Y → Y ◁ Z → X ◁ Z
X ◁⟨ ρ ⟩ σ = ρ ◁∘ σ
```

And we introduce the following postfix notation for the identity retraction:

```
_◀ : (X : 𝓤 ̇ ) → X ◁ X
X ◀ = id-◁ X
```

These last two definitions are for notational convenience. See below for examples of their use.

We conclude this section with some facts about retracts of `Σ` types, which are of general use, in particular for dealing with equivalences in the sense of Voevosky in comparison with invertible maps.

A pointwise retraction gives a retraction of the total spaces:

```
Σ-retract : {X : 𝓤 ̇ } {A : X → 𝓥 ̇ } {B : X → 𝓦 ̇ }
          → ((x : X) → A x ◁  B x) → Σ A ◁ Σ B

Σ-retract {𝓤} {𝓥} {𝓦} {X} {A} {B} ρ = NatΣ r , NatΣ s , η'
 where
  r : (x : X) → B x → A x
  r x = retraction (ρ x)

  s : (x : X) → A x → B x
  s x = section (ρ x)

  η : (x : X) (a : A x) → r x (s x a) ≡ a
  η x = retract-equation (ρ x)

  η' : (σ : Σ A) → NatΣ r (NatΣ s σ) ≡ σ
  η' (x , a) = x , r x (s x a) ≡⟨ to-Σ-≡' (η x a) ⟩
               x , a           ∎
```

We have that `transport A (p ⁻¹)` is a two-sided inverse of `transport A p` using the functoriality of `transport A`, or directly by induction on `p`:

```
transport-is-retraction : {X : 𝓤 ̇ } (A : X → 𝓥 ̇ ) {x y : X} (p : x ≡ y)
                        → transport A p ∘ transport A (p ⁻¹) ∼ id (A y)

transport-is-retraction A (refl x) = refl

transport-is-section : {X : 𝓤 ̇ } (A : X → 𝓥 ̇ ) {x y : X} (p : x ≡ y)
                     → transport A (p ⁻¹) ∘ transport A p ∼ id (A x)

transport-is-section A (refl x) = refl
```

Using this, we have the following reindexing retraction of `Σ` types:

```
Σ-reindexing-retract : {X : 𝓤 ̇ } {Y : 𝓥 ̇ } {A : X → 𝓦 ̇ } (r : Y → X)
                     → has-section r
                     → (Σ x ꞉ X , A x) ◁ (Σ y ꞉ Y , A (r y))

Σ-reindexing-retract {𝓤} {𝓥} {𝓦} {X} {Y} {A} r (s , η) = γ , φ , γφ
 where
  γ : Σ (A ∘ r) → Σ A
  γ (y , a) = (r y , a)

  φ : Σ A → Σ (A ∘ r)
  φ (x , a) = (s x , transport A ((η x)⁻¹) a)

  γφ : (σ : Σ A) → γ (φ σ) ≡ σ
  γφ (x , a) = p
   where
    p : (r (s x) , transport A ((η x)⁻¹) a) ≡ (x , a)
    p = to-Σ-≡ (η x , transport-is-retraction A (η x) a)
```

We have defined the property of a type being a singleton. The singleton type `Σ y ꞉ X , x ≡ y` induced by a point `x : X` of a type `X` is denoted by `singleton-type x`. The terminology is justified by the fact that it is indeed a singleton in the sense defined above.

```
singleton-type : {X : 𝓤 ̇ } → X → 𝓤 ̇
singleton-type {𝓤} {X} x = Σ y ꞉ X , y ≡ x

singleton-type-center : {X : 𝓤 ̇ } (x : X) → singleton-type x
singleton-type-center x = (x , refl x)
```

```
singleton-type-centered : {X : 𝓤 ̇ } (x : X) (σ : singleton-type x)
                        → singleton-type-center x ≡ σ

singleton-type-centered x (x , refl x) = refl (x , refl x)

singleton-types-are-singletons : (X : 𝓤 ̇ ) (x : X)
                               → is-singleton (singleton-type x)

singleton-types-are-singletons X x = singleton-type-center x ,
                                     singleton-type-centered x
```

The following gives a technique for showing that some types are singletons:

```
retract-of-singleton : {X : 𝓤 ̇ } {Y : 𝓥 ̇ }
                     → Y ◁ X → is-singleton X → is-singleton Y

retract-of-singleton (r , s , η) (c , φ) = r c , γ
 where
  γ = λ y → r c     ≡⟨ ap r (φ (s y)) ⟩
            r (s y) ≡⟨ η y            ⟩
            y       ∎
```

Sometimes we need the following symmetric versions of the above:

```
singleton-type' : {X : 𝓤 ̇ } → X → 𝓤 ̇
singleton-type' {𝓤} {X} x = Σ y ꞉ X , x ≡ y

singleton-type'-center : {X : 𝓤 ̇ } (x : X) → singleton-type' x
singleton-type'-center x = (x , refl x)

singleton-type'-centered : {X : 𝓤 ̇ } (x : X) (σ : singleton-type' x)
                         → singleton-type'-center x ≡ σ

singleton-type'-centered x (x , refl x) = refl (x , refl x)

singleton-types'-are-singletons : (X : 𝓤 ̇ ) (x : X)
                                → is-singleton (singleton-type' x)

singleton-types'-are-singletons X x = singleton-type'-center x ,
                                      singleton-type'-centered x
```

## Voevodsky's notion of type equivalence

The main notions of univalent mathematics conceived by Voevodsky, with formulations in MLTT, are those of singleton type (or contractible type), hlevel (including the notions of subsingleton and set), and of type equivalence, which we define now.

We begin with a discussion of the notion of *invertible function*, whose only difference with the notion of equivalence is that it is data rather than property:

```
invertible : {X : 𝓤 ̇ } {Y : 𝓥 ̇ } → (X → Y) → 𝓤 ⊔ 𝓥 ̇
invertible f = Σ g ꞉ (codomain f → domain f) , (g ∘ f ∼ id) × (f ∘ g ∼ id)
```

The situation is that we will have a logical equivalence between

- *data* establishing invertibility of a given function, and
- the *property* of the function being an equivalence.

Mathematically, what happens is that the type

* `f` is an equivalence

is a retract of the type

* `f` is invertible.

This retraction property is not easy to show, and there are many approaches. We discuss an approach we came up with while developing these lecture notes, which is intended to be relatively simple and direct, but the reader should consult other approaches, such as that of the HoTT book, which has a well-established categorical pedigree.

The problem with the notion of invertibility of `f` is that, while we have that the inverse `g` is unique when it exists, we cannot in general prove that the identification data `g ∘ f ∼ id` and `f ∘ g ∼ id` are also unique, and, indeed, they are not in general.

The following is Voevodsky's proposed formulation of the notion of equivalence in MLTT, which relies on the concept of `fiber`:

```
fiber : {X : 𝓤 ̇ } {Y : 𝓥 ̇ } (f : X → Y) → Y → 𝓤 ⊔ 𝓥 ̇
fiber f y = Σ x ꞉ domain f , f x ＝ y

fiber-point : {X : 𝓤 ̇ } {Y : 𝓥 ̇ } {f : X → Y} {y : Y}
            → fiber f y → X

fiber-point (x , p) = x

fiber-identification : {X : 𝓤 ̇ } {Y : 𝓥 ̇ } {f : X → Y} {y : Y}
                     → (w : fiber f y) → f (fiber-point w) ＝ y

fiber-identification (x , p) = p
```

So the type `fiber f y` collects the points `x : X` which are mapped to a point identified with `y`, including the identification datum. Voevodsky's insight is that a general notion of equivalence can be formulated in MLTT by requiring the fibers to be singletons. It is important here that not only the `x : X` with `f x ＝ y` is unique, but also that the identification datum `p : f x ＝ y` is unique. This is similar to, or even a generalization, of the categorical notion of "uniqueness up to a unique isomorphism".

```
is-equiv : {X : 𝓤 ̇ } {Y : 𝓥 ̇ } → (X → Y) → 𝓤 ⊔ 𝓥 ̇
is-equiv f = (y : codomain f) → is-singleton (fiber f y)
```

We can read this as saying that for every `y : Y` there is a unique `x : X` with `f x ＝ y`, where the uniqueness refers not only to `x : X` but also to the identification datum `p : f x ＝ y`. More precisely, the *pair* `(x , p)` is required to be unique. It is easy to see that equivalences are invertible:

```
inverse : {X : 𝓤 ̇ } {Y : 𝓥 ̇ } (f : X → Y) → is-equiv f → (Y → X)
inverse f e y = fiber-point (center (fiber f y) (e y))

inverses-are-sections : {X : 𝓤 ̇ } {Y : 𝓥 ̇ } (f : X → Y) (e : is-equiv f)
                      → f ∘ inverse f e ∼ id

inverses-are-sections f e y = fiber-identification (center (fiber f y) (e y))

inverse-centrality : {X : 𝓤 ̇ } {Y : 𝓥 ̇ }
                     (f : X → Y) (e : is-equiv f) (y : Y) (t : fiber f y)
                   → (inverse f e y , inverses-are-sections f e y) ＝ t

inverse-centrality f e y = centrality (fiber f y) (e y)

inverses-are-retractions : {X : 𝓤 ̇ } {Y : 𝓥 ̇ } (f : X → Y) (e : is-equiv f)
                         → inverse f e ∘ f ∼ id

inverses-are-retractions f e x = ap fiber-point p
```

```
 where
  p : inverse f e (f x) , inverses-are-sections f e (f x) ＝ x , refl (f x)
  p = inverse-centrality f e (f x) (x , (refl (f x)))

equivs-are-invertible : {X : 𝓤 ̇ } {Y : 𝓥 ̇ } (f : X → Y)
                      → is-equiv f → invertible f

equivs-are-invertible f e = inverse f e ,
                            inverses-are-retractions f e ,
                            inverses-are-sections f e
```

The non-trivial direction derives the equivalence property from invertibility data, for which we use the retraction techniques explained above.

Suppose that invertibility data

`g : Y → X` ,

`η : (x : X) → g (f x) ＝ x`

`ε : (y : Y) → f (g y) ＝ y`

for a map `f : X → Y` is given, and that a point `y₀` in the codomain of `f` is given.

We need to show that the fiber `Σ x : X , f x ＝ y₀` of `y₀` is a singleton.

1. We first use the assumption `ε` to show that the type `f (g y) ＝ y₀` is a retract of the type `y ＝ y₀` for any given `y : Y`.

   To get the section `s : f (g y) ＝ y₀ → y ＝ y₀`, we transport along the identification `ε y : f (g y) ＝ y` over the family `A - = (- ＝ y₀)`, which can be abbreviated as `_＝ y₀`.

   To get the retraction `r` in the opposite direction, we transport along the inverse of the identification `ε y` over the same family.

   We already know that this gives a section-retraction pair by `transport-is-section`.

2. Next we have that the type `Σ x : X , f x ＝ y₀` is a retract of the type `Σ y : Y , f (g y) ＝ y₀` by `Σ-reindexing-retract` using the assumption that `η` exibits `g` as a section of `f`, which in turn is a retract of the type `Σ y : Y , y ＝ y₀` by applying `Σ` to both sides of the retraction `(f (g y) ＝ y₀) ◁ (y ＝ y₀)` of the previous step.

   This amounts to saying that the type `fiber f y₀` is a retract of `singleton-type y₀`.

3. But then we are done, because singleton types are singletons and retracts of singletons are singletons.

```
invertibles-are-equivs : {X : 𝓤 ̇ } {Y : 𝓥 ̇ } (f : X → Y)
                       → invertible f → is-equiv f

invertibles-are-equivs {𝓤} {𝓥} {X} {Y} f (g , η , ε) y₀ = iii
 where
  i : (y : Y) → (f (g y) ＝ y₀) ◁ (y ＝ y₀)
  i y =  r , s , transport-is-section (_＝ y₀) (ε y)
   where
    s : f (g y) ＝ y₀ → y ＝ y₀
    s = transport (_＝ y₀) (ε y)

    r : y ＝ y₀ → f (g y) ＝ y₀
    r = transport (_＝ y₀) ((ε y)⁻¹)

  ii : fiber f y₀ ◁ singleton-type y₀
  ii = (Σ x : X , f x ＝ y₀)      ◁⟨ Σ-reindexing-retract g (f , η) ⟩
       (Σ y : Y , f (g y) ＝ y₀) ◁⟨ Σ-retract i                    ⟩
       (Σ y : Y , y ＝ y₀)       ◀
```

```
  iii : is-singleton (fiber f y₀)
  iii = retract-of-singleton ii (singleton-types-are-singletons Y y₀)
```

An immediate consequence is that inverses of equivalences are themselves equivalences:

```
inverses-are-equivs : {X : 𝓤 ̇ } {Y : 𝓥 ̇ } (f : X → Y) (e : is-equiv f)
                    → is-equiv (inverse f e)

inverses-are-equivs f e = invertibles-are-equivs
                           (inverse f e)
                           (f , inverses-are-sections f e , inverses-are-retractions f e)
```

Notice that inversion is involutive on the nose:

```
inversion-involutive : {X : 𝓤 ̇ } {Y : 𝓥 ̇ } (f : X → Y) (e : is-equiv f)
                     → inverse (inverse f e) (inverses-are-equivs f e) ≡ f

inversion-involutive f e = refl f
```

To see that the above procedures do exhibit the type "`f` is an equivalence" as a retract of the type "`f` is invertible", it suffices to show that it is a subsingleton.

The identity function is invertible:

```
id-invertible : (X : 𝓤 ̇ ) → invertible (id X)
id-invertible X = id X , refl , refl
```

We can compose invertible maps:

```
∘-invertible : {X : 𝓤 ̇ } {Y : 𝓥 ̇ } {Z : 𝓦 ̇ } {f : X → Y} {f' : Y → Z}
             → invertible f' → invertible f → invertible (f' ∘ f)

∘-invertible {𝓤} {𝓥} {𝓦} {X} {Y} {Z} {f} {f'} (g' , gf' , fg') (g , gf , fg) =
  g ∘ g' , η , ε
 where
  η = λ x → g (g' (f' (f x))) ≡⟨ ap g (gf' (f x)) ⟩
            g (f x)           ≡⟨ gf x             ⟩
            x                 ∎

  ε = λ z → f' (f (g (g' z))) ≡⟨ ap f' (fg (g' z)) ⟩
            f' (g' z)         ≡⟨ fg' z             ⟩
            z                 ∎
```

There is an identity equivalence, and we get composition of equivalences by reduction to invertible maps:

```
id-is-equiv : (X : 𝓤 ̇ ) → is-equiv (id X)
id-is-equiv = singleton-types-are-singletons
```

An `abstract` definition is not expanded during type checking. One possible use of this is efficiency. In our case, it saves about half a minute in the checking of this file for correctness in the uses of `∘-is-equiv`:

```
∘-is-equiv : {X : 𝓤 ̇ } {Y : 𝓥 ̇ } {Z : 𝓦 ̇ } {f : X → Y} {g : Y → Z}
           → is-equiv g → is-equiv f → is-equiv (g ∘ f)

∘-is-equiv {𝓤} {𝓥} {𝓦} {X} {Y} {Z} {f} {g} i j = γ
 where
  abstract
   γ : is-equiv (g ∘ f)
   γ = invertibles-are-equivs (g ∘ f)
         (∘-invertible (equivs-are-invertible g i)
                       (equivs-are-invertible f j))
```

Because we have made the above definition abstract, we don't have access to the given construction when proving things involving `∘-is-equiv`, such as the contravariance of inversion:

```
inverse-of-∘ : {X : 𝓤 ̇ } {Y : 𝓥 ̇ } {Z : 𝓦 ̇ }
               (f : X → Y) (g : Y → Z)
```

```
             (i : is-equiv f) (j : is-equiv g)
           → inverse f i ∘ inverse g j ∼ inverse (g ∘ f) (∘-is-equiv j i)

inverse-of-∘ f g i j z =

  f' (g' z)             =⟨ (ap (f' ∘ g') (s z))⁻¹                       ⟩
  f' (g' (g (f (h z)))) =⟨ ap f' (inverses-are-retractions g j (f (h z))) ⟩
  f' (f (h z))          =⟨ inverses-are-retractions f i (h z)           ⟩
  h z                   ∎

 where
  f' = inverse f i
  g' = inverse g j
  h  = inverse (g ∘ f) (∘-is-equiv j i)

  s : g ∘ f ∘ h ∼ id
  s = inverses-are-sections (g ∘ f) (∘-is-equiv j i)
```

The type of equivalences is defined as follows:

```
_≃_ : 𝓤 ̇ → 𝓥 ̇ → 𝓤 ⊔ 𝓥 ̇
X ≃ Y = Σ f ꞉ (X → Y), is-equiv f
```

Notice that this doesn't just say that `X` and `Y` are equivalent: the type `X ≃ Y` collects all the ways in which the types `X` and `Y` are equivalent. For example, the two-point type `𝟚` is equivalent to itself in two ways, by the identity map, and by the map that interchanges its two points, and hence the type `𝟚 ≃ 𝟚` has two elements.

Again it is convenient to have special names for its first and second projections:

```
Eq→fun ⌜_⌝ : {X : 𝓤 ̇ } {Y : 𝓥 ̇ } → X ≃ Y → X → Y
Eq→fun (f , i) = f
⌜_⌝            = Eq→fun

Eq→fun-is-equiv ⌜⌝-is-equiv : {X : 𝓤 ̇ } {Y : 𝓥 ̇ } (e : X ≃ Y) → is-equiv (⌜ e ⌝)
Eq→fun-is-equiv (f , i) = i
⌜⌝-is-equiv             = Eq→fun-is-equiv

invertibility-gives-≃ : {X : 𝓤 ̇ } {Y : 𝓥 ̇ } (f : X → Y)
                      → invertible f → X ≃ Y

invertibility-gives-≃ f i = f , invertibles-are-equivs f i
```

Examples:

```
Σ-induction-≃ : {X : 𝓤 ̇ } {Y : X → 𝓥 ̇ } {A : Σ Y → 𝓦 ̇ }
              → ((x : X) (y : Y x) → A (x , y)) ≃ ((z : Σ Y) → A z)

Σ-induction-≃ = invertibility-gives-≃ Σ-induction (curry , refl , refl)

Σ-flip : {X : 𝓤 ̇ } {Y : 𝓥 ̇ } {A : X → Y → 𝓦 ̇ }
       → (Σ x ꞉ X , Σ y ꞉ Y , A x y) ≃ (Σ y ꞉ Y , Σ x ꞉ X , A x y)

Σ-flip = invertibility-gives-≃
          (λ (x , y , p) → (y , x , p))
          ((λ (y , x , p) → (x , y , p)) , refl , refl)

×-comm : {X : 𝓤 ̇ } {Y : 𝓥 ̇ }
       → (X × Y) ≃ (Y × X)

×-comm = invertibility-gives-≃
          (λ (x , y) → (y , x))
          ((λ (y , x) → (x , y)) , refl , refl)
```

The identity equivalence and the composition of two equivalences:

```
id-≃ : (X : 𝓤 ̇ ) → X ≃ X
id-≃ X = id X , id-is-equiv X
```

```
_●_ : {X : 𝓤 ̇ } {Y : 𝓥 ̇ } {Z : 𝓦 ̇ } → X ≃ Y → Y ≃ Z → X ≃ Z
(f , d) ● (f' , e) = f' ∘ f , ∘-is-equiv e d

≃-sym : {X : 𝓤 ̇ } {Y : 𝓥 ̇ } → X ≃ Y → Y ≃ X
≃-sym (f , e) = inverse f e , inverses-are-equivs f e
```

We can use the following notation for equational reasoning with equivalences:

```
_≃⟨_⟩_ : (X : 𝓤 ̇ ) {Y : 𝓥 ̇ } {Z : 𝓦 ̇ } → X ≃ Y → Y ≃ Z → X ≃ Z
_ ≃⟨ d ⟩ e = d ● e

_■ : (X : 𝓤 ̇ ) → X ≃ X
_■ = id-≃
```

We conclude this section with some important examples. The function `transport A p` is an equivalence.

```
transport-is-equiv : {X : 𝓤 ̇ } (A : X → 𝓥 ̇ ) {x y : X} (p : x ≡ y)
                   → is-equiv (transport A p)

transport-is-equiv A (refl x) = id-is-equiv (A x)
```

Alternatively, we could have used the fact that `transport A (p ⁻¹)` is an inverse of `transport A p`.

Here is the promised characterization of equality in `Σ` types:

```
Σ-≡-≃ : {X : 𝓤 ̇ } {A : X → 𝓥 ̇ } (σ τ : Σ A)
      → (σ ≡ τ) ≃ (Σ p ꞉ pr₁ σ ≡ pr₁ τ , transport A p (pr₂ σ) ≡ pr₂ τ)

Σ-≡-≃ {𝓤} {𝓥} {X} {A}  σ τ = invertibility-gives-≃ from-Σ-≡ (to-Σ-≡ , η , ε)
 where
  η : (q : σ ≡ τ) → to-Σ-≡ (from-Σ-≡ q) ≡ q
  η (refl σ) = refl (refl σ)

  ε : (w : Σ p ꞉ pr₁ σ ≡ pr₁ τ , transport A p (pr₂ σ) ≡ pr₂ τ)
    → from-Σ-≡ (to-Σ-≡ w) ≡ w

  ε (refl p , refl q) = refl (refl p , refl q)
```

Similarly we have:

```
to-×-≡ : {X : 𝓤 ̇ } {Y : 𝓥 ̇ } {z t : X × Y}
       → (pr₁ z ≡ pr₁ t) × (pr₂ z ≡ pr₂ t) → z ≡ t

to-×-≡ (refl x , refl y) = refl (x , y)

from-×-≡ : {X : 𝓤 ̇ } {Y : 𝓥 ̇ } {z t : X × Y}
         → z ≡ t → (pr₁ z ≡ pr₁ t) × (pr₂ z ≡ pr₂ t)

from-×-≡ {𝓤} {𝓥} {X} {Y} (refl (x , y)) = (refl x , refl y)

×-≡-≃ : {X : 𝓤 ̇ } {Y : 𝓥 ̇ } (z t : X × Y)
      → (z ≡ t) ≃ (pr₁ z ≡ pr₁ t) × (pr₂ z ≡ pr₂ t)

×-≡-≃ {𝓤} {𝓥} {X} {Y} z t = invertibility-gives-≃ from-×-≡ (to-×-≡ , η , ε)
 where
  η : (p : z ≡ t) → to-×-≡ (from-×-≡ p) ≡ p
  η (refl z) = refl (refl z)

  ε : (q : (pr₁ z ≡ pr₁ t) × (pr₂ z ≡ pr₂ t)) → from-×-≡ (to-×-≡ q) ≡ q
  ε (refl x , refl y) = refl (refl x , refl y)
```

The following are often useful:

```
ap-pr₁-to-×-= : {X : 𝓤 ̇ } {Y : 𝓥 ̇ } {z t : X × Y}
              → (p₁ : pr₁ z = pr₁ t)
              → (p₂ : pr₂ z = pr₂ t)
              → ap pr₁ (to-×-= (p₁ , p₂)) = p₁

ap-pr₁-to-×-= (refl x) (refl y) = refl (refl x)

ap-pr₂-to-×-= : {X : 𝓤 ̇ } {Y : 𝓥 ̇ } {z t : X × Y}
              → (p₁ : pr₁ z = pr₁ t)
              → (p₂ : pr₂ z = pr₂ t)
              → ap pr₂ (to-×-= (p₁ , p₂)) = p₂

ap-pr₂-to-×-= (refl x) (refl y) = refl (refl y)

Σ-cong : {X : 𝓤 ̇ } {A : X → 𝓥 ̇ } {B : X → 𝓦 ̇ }
       → ((x : X) → A x ≃ B x) → Σ A ≃ Σ B

Σ-cong {𝓤} {𝓥} {𝓦} {X} {A} {B} φ =
  invertibility-gives-≃ (NatΣ f) (NatΣ g , NatΣ-η , NatΣ-ε)
 where
  f : (x : X) → A x → B x
  f x = ⌜ φ x ⌝

  g : (x : X) → B x → A x
  g x = inverse (f x) (⌜⌝-is-equiv (φ x))

  η : (x : X) (a : A x) → g x (f x a) = a
  η x = inverses-are-retractions (f x) (⌜⌝-is-equiv (φ x))

  ε : (x : X) (b : B x) → f x (g x b) = b
  ε x = inverses-are-sections (f x) (⌜⌝-is-equiv (φ x))

  NatΣ-η : (w : Σ A) → NatΣ g (NatΣ f w) = w
  NatΣ-η (x , a) = x , g x (f x a) =⟨ to-Σ-=' (η x a) ⟩
                   x , a           ∎

  NatΣ-ε : (t : Σ B) → NatΣ f (NatΣ g t) = t
  NatΣ-ε (x , b) = x , f x (g x b) =⟨ to-Σ-=' (ε x b) ⟩
                   x , b           ∎

≃-gives-◁ : {X : 𝓤 ̇ } {Y : 𝓥 ̇ } → X ≃ Y → X ◁ Y
≃-gives-◁ (f , e) = (inverse f e , f , inverses-are-retractions f e)

≃-gives-▷ : {X : 𝓤 ̇ } {Y : 𝓥 ̇ } → X ≃ Y → Y ◁ X
≃-gives-▷ (f , e) = (f , inverse f e , inverses-are-sections f e)

equiv-to-singleton : {X : 𝓤 ̇ } {Y : 𝓥 ̇ }
                   → X ≃ Y → is-singleton Y → is-singleton X

equiv-to-singleton e = retract-of-singleton (≃-gives-◁ e)
```

## Voevodsky's univalence axiom

There is a canonical transformation `(X Y : 𝓤 ̇ ) → X = Y → X ≃ Y` that sends the identity identification `refl X : X = X` to the identity equivalence `id-≃ X : X ≃ X`. The univalence axiom, for the universe 𝓤, says that this canonical map is itself an equivalence.

```
Id→Eq : (X Y : 𝓤 ̇ ) → X = Y → X ≃ Y
Id→Eq X X (refl X) = id-≃ X
```

```
is-univalent : (𝓤 : Universe) → 𝓤 ⁺ ̇
is-univalent 𝓤 = (X Y : 𝓤 ̇ ) → is-equiv (Id→Eq X Y)
```

Thus, the univalence of the universe 𝓤 says that identifications `X ≡ Y` of types in 𝓤 are in canonical bijection with equivalences `X ≃ Y`, if by bijection we mean equivalence, where the canonical bijection is `Id→Eq`.

We emphasize that this doesn't posit that univalence holds. It says what univalence is (like the type that says what the twin-prime conjecture is).

```
univalence-≃ : is-univalent 𝓤 → (X Y : 𝓤 ̇ ) → (X ≡ Y) ≃ (X ≃ Y)
univalence-≃ ua X Y = Id→Eq X Y , ua X Y

Eq→Id : is-univalent 𝓤 → (X Y : 𝓤 ̇ ) → X ≃ Y → X ≡ Y
Eq→Id ua X Y = inverse (Id→Eq X Y) (ua X Y)
```

Here is a third way to convert a type identification into a function:

```
Id→fun : {X Y : 𝓤 ̇ } → X ≡ Y → X → Y
Id→fun {𝓤} {X} {Y} p = ⌜ Id→Eq X Y p ⌝

Id→funs-agree : {X Y : 𝓤 ̇ } (p : X ≡ Y)
              → Id→fun p ≡ Id→Fun p
Id→funs-agree (refl X) = refl (id X)
```

What characterizes univalent mathematics is not the univalence axiom. We have defined and studied the main concepts of univalent mathematics in a pure, spartan MLTT. It is the concepts of hlevel, including singleton, subsingleton and set, and the notion of equivalence that are at the heart of univalent mathematics. Univalence *is* a fundamental ingredient, but first we need the correct notion of equivalence to be able to formulate it.

*Remark*. If we formulate univalence with invertible maps instead of equivalences, we get a statement that is provably false in MLTT, and this is one of the reasons why Voevodsky's notion of equivalence is important. This is Exercise 4.6 of the HoTT book. There is a solution in Coq by Mike Shulman.

## Example of a type that is not a set under univalence

We have two automorphisms of `𝟚`, namely the identity function and the map that swaps `₀` and `₁`:

```
swap₂ : 𝟚 → 𝟚
swap₂ ₀ = ₁
swap₂ ₁ = ₀

swap₂-involutive : (n : 𝟚) → swap₂ (swap₂ n) ≡ n
swap₂-involutive ₀ = refl ₀
swap₂-involutive ₁ = refl ₁
```

That is, `swap₂` is its own inverse and hence is an equivalence:

```
swap₂-is-equiv : is-equiv swap₂
swap₂-is-equiv = invertibles-are-equivs
                  swap₂
                  (swap₂ , swap₂-involutive , swap₂-involutive)
```

We now use a local module to assume univalence of the first universe in the construction of our example:

```
module example-of-a-nonset (ua : is-univalent 𝓤₀) where
```

The above gives two distinct equivalences:

```
 e₀ e₁ : 𝟚 ≃ 𝟚
 e₀ = id-≃ 𝟚
 e₁ = swap₂ , swap₂-is-equiv
```

```
e₀-is-not-e₁ : e₀ ≢ e₁
e₀-is-not-e₁ p = ₁-is-not-₀ r
 where
  q : id ≡ swap₂
  q = ap ⌜_⌝ p

  r : ₁ ≡ ₀
  r = ap (λ - → - ₁) q
```

Using univalence, we get two different identifications of the type 𝟚 with itself:

```
p₀ p₁ : 𝟚 ≡ 𝟚
p₀ = Eq→Id ua 𝟚 𝟚 e₀
p₁ = Eq→Id ua 𝟚 𝟚 e₁

p₀-is-not-p₁ : p₀ ≢ p₁
p₀-is-not-p₁ q = e₀-is-not-e₁ r
 where
  r = e₀              ≡⟨ (inverses-are-sections (Id→Eq 𝟚 𝟚) (ua 𝟚 𝟚) e₀)⁻¹ ⟩
      Id→Eq 𝟚 𝟚 p₀    ≡⟨ ap (Id→Eq 𝟚 𝟚) q                                   ⟩
      Id→Eq 𝟚 𝟚 p₁    ≡⟨ inverses-are-sections (Id→Eq 𝟚 𝟚) (ua 𝟚 𝟚) e₁     ⟩
      e₁              ∎
```

If the universe $\mathcal{U}_0$ were a set, then the identifications $p_0$ and $p_1$ defined above would be equal, and therefore it is not a set.

```
𝓤₀-is-not-a-set : ¬(is-set (𝓤₀ ̇ ))
𝓤₀-is-not-a-set s = p₀-is-not-p₁ q
 where
  q : p₀ ≡ p₁
  q = s 𝟚 𝟚 p₀ p₁
```

For more examples, see [Kraus and Sattler].

## Exercises

Here are some facts whose proofs are left to the reader but that we will need from the next section onwards. Sample solutions are given below.

### Formulations

Define functions for the following type declarations. As a matter of procedure, we suggest to import this file in a solutions file and add another declaration with the same type and new name e.g. `sections-are-lc-solution`, because we already have solutions in this file. It is important not to forget to include the option `--without-K` in the solutions file that imports (the Agda version of) this file.

```
subsingleton-criterion : {X : 𝓤 ̇ }
                       → (X → is-singleton X)
                       → is-subsingleton X

subsingleton-criterion' : {X : 𝓤 ̇ }
                        → (X → is-subsingleton X)
                        → is-subsingleton X

retract-of-subsingleton : {X : 𝓤 ̇ } {Y : 𝓥 ̇ }
                        → Y ◁ X → is-subsingleton X → is-subsingleton Y

left-cancellable : {X : 𝓤 ̇ } {Y : 𝓥 ̇ } → (X → Y) → 𝓤 ⊔ 𝓥 ̇
left-cancellable f = {x x' : domain f} → f x ≡ f x' → x ≡ x'
```

```
lc-maps-reflect-subsingletons : {X : 𝓤 ̇ } {Y : 𝓥 ̇ } (f : X → Y)
                              → left-cancellable f
                              → is-subsingleton Y
                              → is-subsingleton X
```

```
has-retraction : {X : 𝓤 ̇ } {Y : 𝓥 ̇ } → (X → Y) → 𝓤 ⊔ 𝓥 ̇
has-retraction s = Σ r ꞉ (codomain s → domain s), r ∘ s ∼ id
```

```
sections-are-lc : {X : 𝓤 ̇ } {A : 𝓥 ̇ } (s : X → A)
                → has-retraction s → left-cancellable s
```

```
equivs-have-retractions : {X : 𝓤 ̇ } {Y : 𝓥 ̇ } (f : X → Y)
                        → is-equiv f → has-retraction f
```

```
equivs-have-sections : {X : 𝓤 ̇ } {Y : 𝓥 ̇ } (f : X → Y)
                     → is-equiv f → has-section f
```

```
equivs-are-lc : {X : 𝓤 ̇ } {Y : 𝓥 ̇ } (f : X → Y)
              → is-equiv f → left-cancellable f
```

```
equiv-to-subsingleton : {X : 𝓤 ̇ } {Y : 𝓥 ̇ }
                      → X ≃ Y
                      → is-subsingleton Y
                      → is-subsingleton X
```

```
comp-inverses : {X : 𝓤 ̇ } {Y : 𝓥 ̇ } {Z : 𝓦 ̇ }
                (f : X → Y) (g : Y → Z)
                (i : is-equiv f) (j : is-equiv g)
                (f' : Y → X) (g' : Z → Y)
              → f' ∼ inverse f i
              → g' ∼ inverse g j
              → f' ∘ g' ∼ inverse (g ∘ f) (∘-is-equiv j i)
```

```
equiv-to-set : {X : 𝓤 ̇ } {Y : 𝓥 ̇ }
             → X ≃ Y
             → is-set Y
             → is-set X
```

```
sections-closed-under-∼ : {X : 𝓤 ̇ } {Y : 𝓥 ̇ } (f g : X → Y)
                        → has-retraction f
                        → g ∼ f
                        → has-retraction g
```

```
retractions-closed-under-∼ : {X : 𝓤 ̇ } {Y : 𝓥 ̇ } (f g : X → Y)
                           → has-section f
                           → g ∼ f
                           → has-section g
```

An alternative notion of equivalence, equivalent to Voevodsky's, has been given by André Joyal:

```
is-joyal-equiv : {X : 𝓤 ̇ } {Y : 𝓥 ̇ } → (X → Y) → 𝓤 ⊔ 𝓥 ̇
is-joyal-equiv f = has-section f × has-retraction f
```

Provide definitions for the following type declarations:

```
one-inverse : (X : 𝓤 ̇ ) (Y : 𝓥 ̇ )
              (f : X → Y) (r s : Y → X)
            → (r ∘ f ∼ id)
```

```
         → (f ∘ s ∼ id)
         → r ∼ s

joyal-equivs-are-invertible : {X : 𝓤 ̇ } {Y : 𝓥 ̇ } (f : X → Y)
                            → is-joyal-equiv f → invertible f

joyal-equivs-are-equivs : {X : 𝓤 ̇ } {Y : 𝓥 ̇ } (f : X → Y)
                        → is-joyal-equiv f → is-equiv f

invertibles-are-joyal-equivs : {X : 𝓤 ̇ } {Y : 𝓥 ̇ } (f : X → Y)
                             → invertible f → is-joyal-equiv f

equivs-are-joyal-equivs : {X : 𝓤 ̇ } {Y : 𝓥 ̇ } (f : X → Y)
                        → is-equiv f → is-joyal-equiv f

equivs-closed-under-∼ : {X : 𝓤 ̇ } {Y : 𝓥 ̇ } {f g : X → Y}
                      → is-equiv f
                      → g ∼ f
                      → is-equiv g

equiv-to-singleton' : {X : 𝓤 ̇ } {Y : 𝓥 ̇ }
                    → X ≃ Y → is-singleton X → is-singleton Y

subtypes-of-sets-are-sets : {X : 𝓤 ̇ } {Y : 𝓥 ̇ } (m : X → Y)
                          → left-cancellable m → is-set Y → is-set X

pr₁-lc : {X : 𝓤 ̇ } {A : X → 𝓥 ̇ }
       → ((x : X) → is-subsingleton (A x))
       → left-cancellable (λ (t : Σ A) → pr₁ t)

subsets-of-sets-are-sets : (X : 𝓤 ̇ ) (A : X → 𝓥 ̇ )
                         → is-set X
                         → ((x : X) → is-subsingleton (A x))
                         → is-set (Σ x ꞉ X , A x)

to-subtype-≡ : {X : 𝓦 ̇ } {A : X → 𝓥 ̇ }
               {x y : X} {a : A x} {b : A y}
             → ((x : X) → is-subsingleton (A x))
             → x ≡ y
             → (x , a) ≡ (y , b)

pr₁-is-equiv : {X : 𝓤 ̇ } {A : X → 𝓥 ̇ }
             → ((x : X) → is-singleton (A x))
             → is-equiv (λ (t : Σ A) → pr₁ t)

pr₁-≃ : {X : 𝓤 ̇ } {A : X → 𝓥 ̇ }
      → ((x : X) → is-singleton (A x))
      → Σ A ≃ X

pr₁-≃ i = pr₁ , pr₁-is-equiv i

ΠΣ-distr-≃ : {X : 𝓤 ̇ } {A : X → 𝓥 ̇ } {P : (x : X) → A x → 𝓦 ̇ }
           → (Π x ꞉ X , Σ a ꞉ A x , P x a)
           ≃ (Σ f ꞉ Π A , Π x ꞉ X , P x (f x))

Σ-assoc : {X : 𝓤 ̇ } {Y : X → 𝓥 ̇ } {Z : Σ Y → 𝓦 ̇ }
        → Σ Z ≃ (Σ x ꞉ X , Σ y ꞉ Y x , Z (x , y))
```

```
⁻¹-≃ : {X : 𝓤 ̇ } (x y : X) → (x ≡ y) ≃ (y ≡ x)

singleton-types-≃ : {X : 𝓤 ̇ } (x : X) → singleton-type' x ≃ singleton-type x

singletons-≃ : {X : 𝓤 ̇ } {Y : 𝓥 ̇ }
             → is-singleton X → is-singleton Y → X ≃ Y

maps-of-singletons-are-equivs : {X : 𝓤 ̇ } {Y : 𝓥 ̇ } (f : X → Y)
                              → is-singleton X → is-singleton Y → is-equiv f

logically-equivalent-subsingletons-are-equivalent : (X : 𝓤 ̇ ) (Y : 𝓥 ̇ )
                                                  → is-subsingleton X
                                                  → is-subsingleton Y
                                                  → X ⇔ Y
                                                  → X ≃ Y

singletons-are-equivalent : (X : 𝓤 ̇ ) (Y : 𝓥 ̇ )
                          → is-singleton X
                          → is-singleton Y
                          → X ≃ Y

NatΣ-fiber-equiv : {X : 𝓤 ̇ } (A : X → 𝓥 ̇ ) (B : X → 𝓦 ̇ ) (φ : Nat A B)
                   (x : X) (b : B x)
                 → fiber (φ x) b ≃ fiber (NatΣ φ) (x , b)

NatΣ-equiv-gives-fiberwise-equiv : {X : 𝓤 ̇ } {A : X → 𝓥 ̇ } {B : X → 𝓦 ̇ }
                                   (φ : Nat A B)
                                 → is-equiv (NatΣ φ)
                                 → ((x : X) → is-equiv (φ x))

Σ-is-subsingleton : {X : 𝓤 ̇ } {A : X → 𝓥 ̇ }
                  → is-subsingleton X
                  → ((x : X) → is-subsingleton (A x))
                  → is-subsingleton (Σ A)

×-is-singleton : {X : 𝓤 ̇ } {Y : 𝓥 ̇ }
                  → is-singleton X
                  → is-singleton Y
                  → is-singleton (X × Y)

×-is-subsingleton : {X : 𝓤 ̇ } {Y : 𝓥 ̇ }
                  → is-subsingleton X
                  → is-subsingleton Y
                  → is-subsingleton (X × Y)

×-is-subsingleton' : {X : 𝓤 ̇ } {Y : 𝓥 ̇ }
                   → ((Y → is-subsingleton X) × (X → is-subsingleton Y))
                   → is-subsingleton (X × Y)

×-is-subsingleton'-back : {X : 𝓤 ̇ } {Y : 𝓥 ̇ }
                        → is-subsingleton (X × Y)
                        → (Y → is-subsingleton X) × (X → is-subsingleton Y)

ap₂ : {X : 𝓤 ̇ } {Y : 𝓥 ̇ } {Z : 𝓦 ̇ } (f : X → Y → Z) {x x' : X} {y y' : Y}
    → x ≡ x' → y ≡ y' → f x y ≡ f x' y'
```

## Solutions

For the sake of readability, we re-state the formulations of the exercises in the type of `sol` in a `where` clause for each exercise.

```
subsingleton-criterion = sol
 where
  sol : {X : 𝓤 ̇ } → (X → is-singleton X) → is-subsingleton X
  sol f x = singletons-are-subsingletons (domain f) (f x) x
```

```
subsingleton-criterion' = sol
 where
  sol : {X : 𝓤 ̇ } → (X → is-subsingleton X) → is-subsingleton X
  sol f x y = f x x y
```

```
retract-of-subsingleton = sol
 where
  sol : {X : 𝓤 ̇ } {Y : 𝓥 ̇ }
      → Y ◁ X → is-subsingleton X → is-subsingleton Y
  sol (r , s , η) i =  subsingleton-criterion
                        (λ x → retract-of-singleton (r , s , η)
                                (pointed-subsingletons-are-singletons
                                  (codomain s) (s x) i))
```

```
lc-maps-reflect-subsingletons = sol
 where
  sol : {X : 𝓤 ̇ } {Y : 𝓥 ̇ } (f : X → Y)
      → left-cancellable f → is-subsingleton Y → is-subsingleton X
  sol f l s x x' = l (s (f x) (f x'))
```

```
sections-are-lc = sol
 where
  sol : {X : 𝓤 ̇ } {A : 𝓥 ̇ } (s : X → A)
      → has-retraction s → left-cancellable s
  sol s (r , ε) {x} {y} p = x       =⟨ (ε x)⁻¹ ⟩
                            r (s x) =⟨ ap r p  ⟩
                            r (s y) =⟨ ε y     ⟩
                            y       ∎
```

```
equivs-have-retractions = sol
 where
  sol : {X : 𝓤 ̇ } {Y : 𝓥 ̇ } (f : X → Y) → is-equiv f → has-retraction f
  sol f e = (inverse f e , inverses-are-retractions f e)
```

```
equivs-have-sections = sol
 where
  sol : {X : 𝓤 ̇ } {Y : 𝓥 ̇ } (f : X → Y) → is-equiv f → has-section f
  sol f e = (inverse f e , inverses-are-sections f e)
```

```
equivs-are-lc = sol
 where
  sol : {X : 𝓤 ̇ } {Y : 𝓥 ̇ } (f : X → Y) → is-equiv f → left-cancellable f
  sol f e = sections-are-lc f (equivs-have-retractions f e)
```

```
equiv-to-subsingleton = sol
 where
  sol : {X : 𝓤 ̇ } {Y : 𝓥 ̇ } → X ≃ Y → is-subsingleton Y → is-subsingleton X
  sol (f , i) = lc-maps-reflect-subsingletons f (equivs-are-lc f i)
```

```
comp-inverses = sol
 where
```

```
  sol : {X : 𝓤 ̇ } {Y : 𝓥 ̇ } {Z : 𝓦 ̇ }
        (f : X → Y) (g : Y → Z)
        (i : is-equiv f) (j : is-equiv g)
        (f' : Y → X) (g' : Z → Y)
      → f' ∼ inverse f i
      → g' ∼ inverse g j
      → f' ∘ g' ∼ inverse (g ∘ f) (∘-is-equiv j i)
  sol f g i j f' g' h k z =
   f' (g' z)                          =⟨ h (g' z)               ⟩
   inverse f i (g' z)                 =⟨ ap (inverse f i) (k z) ⟩
   inverse f i (inverse g j z)        =⟨ inverse-of-∘ f g i j z ⟩
   inverse (g ∘ f) (∘-is-equiv j i) z ∎

equiv-to-set = sol
 where
  sol : {X : 𝓤 ̇ } {Y : 𝓥 ̇ } → X ≃ Y → is-set Y → is-set X
  sol e = subtypes-of-sets-are-sets ⌜ e ⌝ (equivs-are-lc ⌜ e ⌝ (⌜⌝-is-equiv e))

sections-closed-under-∼ = sol
 where
  sol : {X : 𝓤 ̇ } {Y : 𝓥 ̇ } (f g : X → Y)
      → has-retraction f → g ∼ f → has-retraction g
  sol f g (r , rf) h = (r ,
                        λ x → r (g x) =⟨ ap r (h x) ⟩
                              r (f x) =⟨ rf x       ⟩
                              x       ∎)

retractions-closed-under-∼ = sol
 where
  sol : {X : 𝓤 ̇ } {Y : 𝓥 ̇ } (f g : X → Y)
      → has-section f → g ∼ f → has-section g
  sol f g (s , fs) h = (s ,
                        λ y → g (s y) =⟨ h (s y) ⟩
                              f (s y) =⟨ fs y    ⟩
                              y ∎)

one-inverse = sol
 where
  sol : (X : 𝓤 ̇ ) (Y : 𝓥 ̇ )
        (f : X → Y) (r s : Y → X)
      → (r ∘ f ∼ id)
      → (f ∘ s ∼ id)
      → r ∼ s
  sol X Y f r s h k y = r y         =⟨ ap r ((k y)⁻¹) ⟩
                        r (f (s y)) =⟨ h (s y)        ⟩
                        s y         ∎

joyal-equivs-are-invertible = sol
 where
  sol : {X : 𝓤 ̇ } {Y : 𝓥 ̇ } (f : X → Y)
      → is-joyal-equiv f → invertible f
  sol f ((s , ε) , (r , η)) = (s , sf , ε)
   where
    sf = λ (x : domain f) → s(f x)       =⟨ (η (s (f x)))⁻¹ ⟩
                            r(f(s(f x))) =⟨ ap r (ε (f x))  ⟩
                            r(f x)       =⟨ η x             ⟩
                            x            ∎

joyal-equivs-are-equivs = sol
 where
  sol : {X : 𝓤 ̇ } {Y : 𝓥 ̇ } (f : X → Y)
      → is-joyal-equiv f → is-equiv f
  sol f j = invertibles-are-equivs f (joyal-equivs-are-invertible f j)
```

```
invertibles-are-joyal-equivs = sol
 where
  sol : {X : 𝓤 ̇ } {Y : 𝓥 ̇ } (f : X → Y)
      → invertible f → is-joyal-equiv f
  sol f (g , η , ε) = ((g , ε) , (g , η))

equivs-are-joyal-equivs = sol
 where
  sol : {X : 𝓤 ̇ } {Y : 𝓥 ̇ } (f : X → Y)
      → is-equiv f → is-joyal-equiv f
  sol f e = invertibles-are-joyal-equivs f (equivs-are-invertible f e)

equivs-closed-under-∼ = sol
 where
  sol : {X : 𝓤 ̇ } {Y : 𝓥 ̇ } {f g : X → Y}
      → is-equiv f → g ∼ f → is-equiv g
  sol {𝓤} {𝓥} {X} {Y} {f} {g} e h = joyal-equivs-are-equivs g
                                     (retractions-closed-under-∼ f g
                                      (equivs-have-sections f e) h ,
                                     sections-closed-under-∼ f g
                                      (equivs-have-retractions f e) h)

equiv-to-singleton' = sol
 where
  sol : {X : 𝓤 ̇ } {Y : 𝓥 ̇ }
      → X ≃ Y → is-singleton X → is-singleton Y
  sol e = retract-of-singleton (≃-gives-▷ e)

subtypes-of-sets-are-sets = sol
 where
  sol : {X : 𝓤 ̇ } {Y : 𝓥 ̇ } (m : X → Y)
      → left-cancellable m → is-set Y → is-set X
  sol {𝓤} {𝓥} {X} m i h = types-with-wconstant-＝-endomaps-are-sets X c
   where
    f : (x x' : X) → x ＝ x' → x ＝ x'
    f x x' r = i (ap m r)

    κ : (x x' : X) (r s : x ＝ x') → f x x' r ＝ f x x' s
    κ x x' r s = ap i (h (m x) (m x') (ap m r) (ap m s))

    c : wconstant-＝-endomaps X
    c x x' = f x x' , κ x x'

pr₁-lc = sol
 where
  sol : {X : 𝓤 ̇ } {A : X → 𝓥 ̇ }
      → ((x : X) → is-subsingleton (A x))
      → left-cancellable  (λ (t : Σ A) → pr₁ t)
  sol i p = to-Σ-＝ (p , i _ _ _)

subsets-of-sets-are-sets = sol
 where
  sol : (X : 𝓤 ̇ ) (A : X → 𝓥 ̇ )
     → is-set X
     → ((x : X) → is-subsingleton (A x))
     → is-set (Σ x ꞉ X , A x)
  sol X A h p = subtypes-of-sets-are-sets pr₁ (pr₁-lc p) h

to-subtype-＝ = sol
 where
  sol : {X : 𝓤 ̇ } {A : X → 𝓥 ̇ }
        {x y : X} {a : A x} {b : A y}
```

```
        → ((x : X) → is-subsingleton (A x))
        → x ≡ y
        → (x , a) ≡ (y , b)
    sol {𝓤} {𝓥} {X} {A} {x} {y} {a} {b} s p = to-Σ-≡ (p , s y (transport A p a) b)

  pr₁-is-equiv = sol
   where
    sol : {X : 𝓤 ̇ } {A : X → 𝓥 ̇ }
        → ((x : X) → is-singleton (A x))
        → is-equiv (λ (t : Σ A) → pr₁ t)
    sol {𝓤} {𝓥} {X} {A} s = invertibles-are-equivs pr₁ (g , η , ε)
     where
      g : X → Σ A
      g x = x , pr₁(s x)

      ε : (x : X) → pr₁ (g x) ≡ x
      ε x = refl (pr₁ (g x))

      η : (σ : Σ A) → g (pr₁ σ) ≡ σ
      η (x , a) = to-subtype-≡ (λ x → singletons-are-subsingletons (A x) (s x)) (ε x)

  ΠΣ-distr-≃ = sol
   where
    sol : {X : 𝓤 ̇ } {A : X → 𝓥 ̇ } {P : (x : X) → A x → 𝓦 ̇ }
        → (Π x ꞉ X , Σ a ꞉ A x , P x a)
        ≃ (Σ f ꞉ Π A , Π x ꞉ X , P x (f x))
    sol {𝓤} {𝓥} {𝓦} {X} {A} {P} = invertibility-gives-≃ φ (γ , η , ε)
     where
      φ : (Π x ꞉ X , Σ a ꞉ A x , P x a)
        → Σ f ꞉ Π A , Π x ꞉ X , P x (f x)
      φ g = ((λ x → pr₁ (g x)) , λ x → pr₂ (g x))

      γ : (Σ f ꞉ Π A , Π x ꞉ X , P x (f x))
        → Π x ꞉ X , Σ a ꞉ A x , P x a
      γ (f , φ) x = f x , φ x

      η : γ ∘ φ ∼ id
      η = refl

      ε : φ ∘ γ ∼ id
      ε = refl

  Σ-assoc = sol
   where
    sol : {X : 𝓤 ̇ } {Y : X → 𝓥 ̇ } {Z : Σ Y → 𝓦 ̇ }
        → Σ Z ≃ (Σ x ꞉ X , Σ y ꞉ Y x , Z (x , y))
    sol {𝓤} {𝓥} {𝓦} {X} {Y} {Z} = invertibility-gives-≃ f (g , refl , refl)
     where
      f : Σ Z → Σ x ꞉ X , Σ y ꞉ Y x , Z (x , y)
      f ((x , y) , z) = (x , (y , z))

      g : (Σ x ꞉ X , Σ y ꞉ Y x , Z (x , y)) → Σ Z
      g (x , (y , z)) = ((x , y) , z)

  ⁻¹-is-equiv : {X : 𝓤 ̇ } (x y : X)
              → is-equiv (λ (p : x ≡ y) → p ⁻¹)
  ⁻¹-is-equiv x y = invertibles-are-equivs _⁻¹
                     (_⁻¹ , ⁻¹-involutive , ⁻¹-involutive)

  ⁻¹-≃ = sol
   where
    sol : {X : 𝓤 ̇ } (x y : X) → (x ≡ y) ≃ (y ≡ x)
    sol x y = (_⁻¹ , ⁻¹-is-equiv x y)
```

```
singleton-types-≃ = sol
 where
  sol : {X : 𝓤 ̇ } (x : X) → singleton-type' x ≃ singleton-type x
  sol x = Σ-cong (λ y → ⁻¹-≃ x y)

singletons-≃ = sol
 where
  sol : {X : 𝓤 ̇ } {Y : 𝓥 ̇ }
      → is-singleton X → is-singleton Y → X ≃ Y
  sol {𝓤} {𝓥} {X} {Y} i j = invertibility-gives-≃ f (g , η , ε)
   where
    f : X → Y
    f x = center Y j

    g : Y → X
    g y = center X i

    η : (x : X) → g (f x) ≡ x
    η = centrality X i

    ε : (y : Y) → f (g y) ≡ y
    ε = centrality Y j

maps-of-singletons-are-equivs = sol
 where
  sol : {X : 𝓤 ̇ } {Y : 𝓥 ̇ } (f : X → Y)
      → is-singleton X → is-singleton Y → is-equiv f

  sol {𝓤} {𝓥} {X} {Y} f i j = invertibles-are-equivs f (g , η , ε)
   where
    g : Y → X
    g y = center X i

    η : (x : X) → g (f x) ≡ x
    η = centrality X i

    ε : (y : Y) → f (g y) ≡ y
    ε y = singletons-are-subsingletons Y j (f (g y)) y

logically-equivalent-subsingletons-are-equivalent = sol
 where
  sol : (X : 𝓤 ̇ ) (Y : 𝓥 ̇ )
      → is-subsingleton X → is-subsingleton Y → X ⇔ Y → X ≃ Y
  sol  X Y i j (f , g) = invertibility-gives-≃ f
                          (g ,
                           (λ x → i (g (f x)) x) ,
                           (λ y → j (f (g y)) y))

singletons-are-equivalent = sol
 where
  sol : (X : 𝓤 ̇ ) (Y : 𝓥 ̇ )
      → is-singleton X → is-singleton Y → X ≃ Y
  sol  X Y i j = invertibility-gives-≃ (λ _ → center Y j)
                  ((λ _ → center X i) ,
                   centrality X i ,
                   centrality Y j)

NatΣ-fiber-equiv = sol
 where
  sol : {X : 𝓤 ̇ } (A : X → 𝓥 ̇ ) (B : X → 𝓦 ̇ ) (φ : Nat A B)
        (x : X) (b : B x)
      → fiber (φ x) b ≃ fiber (NatΣ φ) (x , b)
  sol A B φ x b = invertibility-gives-≃ f (g , ε , η)
   where
```

```
    f : fiber (φ x) b → fiber (NatΣ φ) (x , b)
    f (a , refl _) = ((x , a) , refl (x , φ x a))

    g : fiber (NatΣ φ) (x , b) → fiber (φ x) b
    g ((x , a) , refl _) = (a , refl (φ x a))

    ε : (w : fiber (φ x) b) → g (f w) ≡ w
    ε (a , refl _) = refl (a , refl (φ x a))

    η : (t : fiber (NatΣ φ) (x , b)) → f (g t) ≡ t
    η ((x , a) , refl _) = refl ((x , a) , refl (NatΣ φ (x , a)))

NatΣ-equiv-gives-fiberwise-equiv = sol
 where
  sol : {X : 𝓤 ̇ } {A : X → 𝓥 ̇ } {B : X → 𝓦 ̇ } (φ : Nat A B)
      → is-equiv (NatΣ φ) → ((x : X) → is-equiv (φ x))
  sol {𝓤} {𝓥} {𝓦} {X} {A} {B} φ e x b = γ
   where
    d : fiber (φ x) b ≃ fiber (NatΣ φ) (x , b)
    d = NatΣ-fiber-equiv A B φ x b

    s : is-singleton (fiber (NatΣ φ) (x , b))
    s = e (x , b)

    γ : is-singleton (fiber (φ x) b)
    γ = equiv-to-singleton d s

Σ-is-subsingleton = sol
 where
  sol : {X : 𝓤 ̇ } {A : X → 𝓥 ̇ }
      → is-subsingleton X
      → ((x : X) → is-subsingleton (A x))
      → is-subsingleton (Σ A)
  sol i j (x , _) (y , _) = to-subtype-≡ j (i x y)

×-is-singleton = sol
 where
  sol : {X : 𝓤 ̇ } {Y : 𝓥 ̇ }
      → is-singleton X
      → is-singleton Y
      → is-singleton (X × Y)
  sol (x , φ) (y , γ) = (x , y) , δ
   where
    δ : ∀ z → (x , y) ≡ z
    δ (x' , y' ) = to-×-≡ (φ x' , γ y')

×-is-subsingleton = sol
 where
  sol : {X : 𝓤 ̇ } {Y : 𝓥 ̇ }
      → is-subsingleton X → is-subsingleton Y → is-subsingleton (X × Y)
  sol i j = Σ-is-subsingleton i (λ _ → j)

×-is-subsingleton' = sol
 where
  sol : {X : 𝓤 ̇ } {Y : 𝓥 ̇ }
      → ((Y → is-subsingleton X) × (X → is-subsingleton Y))
      → is-subsingleton (X × Y)
  sol {𝓤} {𝓥} {X} {Y} (i , j) = k
   where
    k : is-subsingleton (X × Y)
    k (x , y) (x' , y') = to-×-≡ (i y x x' , j x y y')

×-is-subsingleton'-back = sol
 where
  sol : {X : 𝓤 ̇ } {Y : 𝓥 ̇ }
```

```
       → is-subsingleton (X × Y)
       → (Y → is-subsingleton X) × (X → is-subsingleton Y)
  sol {𝓤} {𝓥} {X} {Y} k = i , j
   where
    i : Y → is-subsingleton X
    i y x x' = ap pr₁ (k (x , y) (x' , y))

    j : X → is-subsingleton Y
    j x y y' = ap pr₂ (k (x , y) (x , y'))

ap₂ = sol
 where
  sol : {X : 𝓤 ̇ } {Y : 𝓥 ̇ } {Z : 𝓦 ̇ } (f : X → Y → Z) {x x' : X} {y y' : Y}
      → x ≡ x' → y ≡ y' → f x y ≡ f x' y'
  sol f (refl x) (refl y) = refl (f x y)
```

## A characterization of univalence

We begin with two general results, which will be placed in a more general context later.

```
equiv-singleton-lemma : {X : 𝓤 ̇ } {A : X → 𝓥 ̇ } (x : X)
                        (f : (y : X) → x ≡ y → A y)
                      → ((y : X) → is-equiv (f y))
                      → is-singleton (Σ A)

equiv-singleton-lemma {𝓤} {𝓥} {X} {A} x f i = γ
 where
  abstract
   e : (y : X) → (x ≡ y) ≃ A y
   e y = (f y , i y)

   d : singleton-type' x ≃ Σ A
   d = Σ-cong e

   γ : is-singleton (Σ A)
   γ = equiv-to-singleton (≃-sym d) (singleton-types'-are-singletons X x)

singleton-equiv-lemma : {X : 𝓤 ̇ } {A : X → 𝓥 ̇ } (x : X)
                        (f : (y : X) → x ≡ y → A y)
                      → is-singleton (Σ A)
                      → (y : X) → is-equiv (f y)

singleton-equiv-lemma {𝓤} {𝓥} {X} {A} x f i = γ
 where
  abstract
   g : singleton-type' x → Σ A
   g = NatΣ f

   e : is-equiv g
   e = maps-of-singletons-are-equivs g (singleton-types'-are-singletons X x) i

   γ : (y : X) → is-equiv (f y)
   γ = NatΣ-equiv-gives-fiberwise-equiv f e
```

With this we can characterize univalence as follows:

```
univalence⇒ : is-univalent 𝓤
            → (X : 𝓤 ̇ ) → is-singleton (Σ Y ꞉ 𝓤 ̇ , X ≃ Y)

univalence⇒ ua X = equiv-singleton-lemma X (Id→Eq X) (ua X)

⇒univalence : ((X : 𝓤 ̇ ) → is-singleton (Σ Y ꞉ 𝓤 ̇ , X ≃ Y))
            → is-univalent 𝓤
```

```
⇒univalence i X = singleton-equiv-lemma X (Id→Eq X) (i X)
```

(Of course, this doesn't say that there is only one type `Y` equivalent to `X`, or only one equivalence from `X` to `Y`, because equality of `Σ` types is given by transport in the second component along an identification in the first component.)

We can replace *singleton* by *subsingleton* and still have a logical equivalence, and we sometimes need the characterization in this form:

```
univalence→ : is-univalent 𝓤
            → (X : 𝓤 ̇ ) → is-subsingleton (Σ Y ꞉ 𝓤 ̇ , X ≃ Y)

univalence→ ua X = singletons-are-subsingletons
                    (Σ (X ≃_)) (univalence⇒ ua X)

→univalence : ((X : 𝓤 ̇ ) → is-subsingleton (Σ Y ꞉ 𝓤 ̇ , X ≃ Y))
            → is-univalent 𝓤

→univalence i = ⇒univalence (λ X → pointed-subsingletons-are-singletons
                                    (Σ (X ≃_)) (X , id-≃ X) (i X))
```

## Equivalence induction

Under univalence, we get induction principles for type equivalences, corresponding to the induction principles `ℍ` and `𝕁` for identifications. To prove a property of equivalences, it is enough to prove it for the identity equivalence `id-≃ X` for all `X`.

Our objective is to get the induction principles `ℍ-≃` and `𝕁-≃` below and their corresponding equations. As above, it is easy to define `𝕁-≃` from `ℍ-≃`, and it is harder to define `ℍ-≃` directly, and it is much harder to prove the desired equation for `ℍ-≃` directly. In order to make this easy, we define an auxiliary induction principle `𝔾-≃`, from which we trivially derive `ℍ-≃`, and whose equation is easy to prove.

The reason the induction principle `𝔾-≃` and its equation are easy to construct and prove is that the type `Σ Y ꞉ 𝓤 ̇ , X ≃ Y` is a singleton by univalence, which considerably simplifies reasoning about transport. For `ℍ-≃` we consider `Y : 𝓤` and `e : X ≃ Y` separately, whereas for `𝔾-≃` we treat them as a pair `(Y , e)`. The point is that the type of such pairs is a singleton by univalence.

```
𝔾-≃ : is-univalent 𝓤
    → (X : 𝓤 ̇ ) (A : (Σ Y ꞉ 𝓤 ̇ , X ≃ Y) → 𝓥 ̇ )
    → A (X , id-≃ X) → (Y : 𝓤 ̇ ) (e : X ≃ Y) → A (Y , e)

𝔾-≃ {𝓤} ua X A a Y e = transport A p a
 where
  t : Σ Y ꞉ 𝓤 ̇ , X ≃ Y
  t = (X , id-≃ X)

  p : t ≡ (Y , e)
  p = univalence→ {𝓤} ua X t (Y , e)

𝔾-≃-equation : (ua : is-univalent 𝓤)
             → (X : 𝓤 ̇ ) (A : (Σ Y ꞉ 𝓤 ̇ , X ≃ Y) → 𝓥 ̇ ) (a : A (X , id-≃ X))
             → 𝔾-≃ ua X A a X (id-≃ X) ≡ a

𝔾-≃-equation {𝓤} {𝓥} ua X A a =

  𝔾-≃ ua X A a X (id-≃ X) ≡⟨ refl _                    ⟩
  transport A p a         ≡⟨ ap (λ - → transport A - a) q ⟩
  transport A (refl t) a  ≡⟨ refl _                   ⟩
  a                       ∎

 where
  t : Σ Y ꞉ 𝓤 ̇ , X ≃ Y
```

```
  t = (X , id-≃ X)

  p : t ≡ t
  p = univalence→ {𝓤} ua X t t

  q : p ≡ refl t
  q = subsingletons-are-sets (Σ Y ꞉ 𝓤 ̇ , X ≃ Y)
       (univalence→ {𝓤} ua X) t t p (refl t)
```

```
ℍ-≃ : is-univalent 𝓤
    → (X : 𝓤 ̇ ) (A : (Y : 𝓤 ̇ ) → X ≃ Y → 𝓥 ̇ )
    → A X (id-≃ X) → (Y : 𝓤 ̇ ) (e : X ≃ Y) → A Y e

ℍ-≃ ua X A = 𝔾-≃ ua X (Σ-induction A)
```

```
ℍ-≃-equation : (ua : is-univalent 𝓤)
             → (X : 𝓤 ̇ ) (A : (Y : 𝓤 ̇ ) → X ≃ Y → 𝓥 ̇ ) (a : A X  (id-≃ X))
             → ℍ-≃ ua X A a X (id-≃ X) ≡ a

ℍ-≃-equation ua X A = 𝔾-≃-equation ua X (Σ-induction A)
```

The induction principle `ℍ-≃` keeps `X` fixed and lets `Y` vary, while the induction principle `𝕁-≃` lets both vary:

```
𝕁-≃ : is-univalent 𝓤
    → (A : (X Y : 𝓤 ̇ ) → X ≃ Y → 𝓥 ̇ )
    → ((X : 𝓤 ̇ ) → A X X (id-≃ X))
    → (X Y : 𝓤 ̇ ) (e : X ≃ Y) → A X Y e

𝕁-≃ ua A φ X = ℍ-≃ ua X (A X) (φ X)
```

```
𝕁-≃-equation : (ua : is-univalent 𝓤)
             → (A : (X Y : 𝓤 ̇ ) → X ≃ Y → 𝓥 ̇ )
             → (φ : (X : 𝓤 ̇ ) → A X X (id-≃ X))
             → (X : 𝓤 ̇ ) → 𝕁-≃ ua A φ X X (id-≃ X) ≡ φ X

𝕁-≃-equation ua A φ X = ℍ-≃-equation ua X (A X) (φ X)
```

A second set of equivalence induction principles refer to `is-equiv` rather than ≃ and are proved by reduction to the first version `ℍ-≃`:

```
ℍ-equiv : is-univalent 𝓤
        → (X : 𝓤 ̇ ) (A : (Y : 𝓤 ̇ ) → (X → Y) → 𝓥 ̇ )
        → A X (𝑖𝑑 X) → (Y : 𝓤 ̇ ) (f : X → Y) → is-equiv f → A Y f

ℍ-equiv {𝓤} {𝓥} ua X A a Y f i = γ (f , i)
 where
  B : (Y : 𝓤 ̇ ) → X ≃ Y → 𝓥 ̇
  B Y (f , i) = A Y f

  b : B X (id-≃ X)
  b = a

  γ : (e : X ≃ Y) → B Y e
  γ = ℍ-≃ ua X B b Y
```

The above and the following say that to prove that a property of *functions* holds for all equivalences, it is enough to prove it for all identity functions:

```
𝕁-equiv : is-univalent 𝓤
        → (A : (X Y : 𝓤 ̇ ) → (X → Y) → 𝓥 ̇ )
        → ((X : 𝓤 ̇ ) → A X X (𝑖𝑑 X))
        → (X Y : 𝓤 ̇ ) (f : X → Y) → is-equiv f → A X Y f

𝕁-equiv ua A φ X = ℍ-equiv ua X (A X) (φ X)
```

And the following is an immediate consequence of the fact that invertible maps are equivalences:

```
𝕁-invertible : is-univalent 𝓤
             → (A : (X Y : 𝓤 ̇ ) → (X → Y) → 𝓥 ̇ )
             → ((X : 𝓤 ̇ ) → A X X (𝑖𝑑 X))
             → (X Y : 𝓤 ̇ ) (f : X → Y) → invertible f → A X Y f

𝕁-invertible ua A φ X Y f i = 𝕁-equiv ua A φ X Y f (invertibles-are-equivs f i)
```

For example, using `ℍ-equiv` we see that for any pair of functions

```
F : 𝓤 ̇ → 𝓤 ̇ ,
```

```
𝓕 : {X Y : 𝓤 ̇ } → (X → Y) → F X → F Y,
```

if 𝓕 preserves identities then it automatically preserves composition of equivalences. More generally, it is enough that at least one of the factors is an equivalence:

```
automatic-equiv-functoriality :

      (F : 𝓤 ̇ → 𝓤 ̇ )
      (𝓕 : {X Y : 𝓤 ̇ }  → (X → Y) → F X → F Y)
      (𝓕-id : {X : 𝓤 ̇ } → 𝓕 (𝑖𝑑 X) ≡ 𝑖𝑑 (F X))
      {X Y Z : 𝓤 ̇ }
      (f : X → Y)
      (g : Y → Z)

    → is-univalent 𝓤
    → is-equiv f + is-equiv g
    → 𝓕 (g ∘ f) ≡ 𝓕 g ∘ 𝓕 f

automatic-equiv-functoriality {𝓤} F 𝓕 𝓕-id {X} {Y} {Z} f g ua = γ
  where
   γ :  is-equiv f + is-equiv g → 𝓕 (g ∘ f) ≡ 𝓕 g ∘ 𝓕 f
   γ (inl i) = ℍ-equiv ua X A a Y f i g
    where
     A : (Y : 𝓤 ̇ ) → (X → Y) → 𝓤 ̇
     A Y f = (g : Y → Z) → 𝓕 (g ∘ f) ≡ 𝓕 g ∘ 𝓕 f

     a : (g : X → Z) → 𝓕 g ≡ 𝓕 g ∘ 𝓕 id
     a g = ap (𝓕 g ∘_) (𝓕-id ⁻¹)

   γ (inr j) = ℍ-equiv ua Y B b Z g j f
    where
     B : (Z : 𝓤 ̇ ) → (Y → Z) → 𝓤 ̇
     B Z g = (f : X → Y) → 𝓕 (g ∘ f) ≡ 𝓕 g ∘ 𝓕 f

     b : (f : X → Y) → 𝓕 f ≡ 𝓕 (𝑖𝑑 Y) ∘ 𝓕 f
     b f = ap (_∘ 𝓕 f) (𝓕-id ⁻¹)
```

Here is another example (see this for the terminology):

```
Σ-change-of-variable' : is-univalent 𝓤
                      → {X : 𝓤 ̇ } {Y : 𝓤 ̇ } (A : X → 𝓥 ̇ ) (f : X → Y)
                      → (i : is-equiv f)
                      → (Σ x ꞉ X , A x) ≡ (Σ y ꞉ Y , A (inverse f i y))

Σ-change-of-variable' {𝓤} {𝓥} ua {X} {Y} A f i = ℍ-≃ ua X B b Y (f , i)
 where
   B : (Y : 𝓤 ̇ ) → X ≃ Y →  (𝓤 ⊔ 𝓥)⁺ ̇
   B Y (f , i) = Σ A ≡ (Σ (A ∘ inverse f i))

   b : B X (id-≃ X)
   b = refl (Σ A)
```

An unprimed version of this is given below, after we study half adjoint equivalences.

The above version using the inverse of `f` can be proved directly by induction, but the following version is perhaps more natural.

```
Σ-change-of-variable'' : is-univalent 𝓤
                       → {X : 𝓤 ̇ } {Y : 𝓤 ̇ } (A : Y → 𝓥 ̇ ) (f : X → Y)
                       → is-equiv f
                       → (Σ y ꞉ Y , A y) ＝ (Σ x ꞉ X , A (f x))

Σ-change-of-variable'' ua A f i = Σ-change-of-variable' ua A
                                  (inverse f i)
                                  (inverses-are-equivs f i)
```

This particular proof works only because inversion is involutive on the nose.

As another example we have the following:

```
transport-map-along-＝ : {X Y Z : 𝓤 ̇ }
                          (p : X ＝ Y) (g : X → Z)
                        → transport (λ - → - → Z) p g
                        ＝ g ∘ Id→fun (p ⁻¹)

transport-map-along-＝ (refl X) = refl

transport-map-along-≃ : (ua : is-univalent 𝓤) {X Y Z : 𝓤 ̇ }
                          (e : X ≃ Y) (g : X → Z)
                        → transport (λ - → - → Z) (Eq→Id ua X Y e) g
                        ＝ g ∘ ⌜ ≃-sym e ⌝

transport-map-along-≃ {𝓤} ua {X} {Y} {Z} = 𝕁-≃ ua A a X Y
 where
  A : (X Y : 𝓤 ̇ ) → X ≃ Y → 𝓤 ̇
  A X Y e = (g : X → Z) → transport (λ - → - → Z) (Eq→Id ua X Y e) g
                        ＝ g ∘ ⌜ ≃-sym e ⌝
  a : (X : 𝓤 ̇ ) → A X X (id-≃ X)
  a X g = transport (λ - → - → Z) (Eq→Id ua X X (id-≃ X)) g ＝⟨ q       ⟩
          transport (λ - → - → Z) (refl X) g                 ＝⟨ refl _ ⟩
          g                                                  ∎
    where
     p : Eq→Id ua X X (id-≃ X) ＝ refl X
     p = inverses-are-retractions (Id→Eq X X) (ua X X) (refl X)

     q = ap (λ - → transport (λ - → - → Z) - g) p
```

An annoying feature of the use of `𝕁` (rather than pattern matching on `refl`) or `𝕁-≃` is that we have to repeat what we want to prove, as in the above examples.

## Half adjoint equivalences

An often useful alternative formulation of the notion of equivalence is that of half adjoint equivalence. If we have a function

```
f : X → Y
```

with invertibility data

```
g : Y → X ,

η : g ∘ f ∼ id,

ε : f ∘ g ∼ id,
```

then for any `x : X` we have that

```
ap f (η x)
```

and

```
     ε (f x)
```

are two identifications of type

```
     f (g (f x)) ＝ f x.
```

The half adjoint condition says that these two identifications are themselves identified. The addition of the constraint

```
     τ x : ap f (η x) ＝ ε (f x)
```

turns invertibility, which is data in general, into property of `f`.

```
is-hae : {X : 𝓤 ̇ } {Y : 𝓥 ̇ } → (X → Y) → 𝓤 ⊔ 𝓥 ̇
is-hae f = Σ g ꞉ (codomain f → domain f)
         , Σ η ꞉ g ∘ f ∼ id
         , Σ ε ꞉ f ∘ g ∼ id
         , ((x : domain f) → ap f (η x) ＝ ε (f x))
```

The following just forgets the constraint `τ`:

```
haes-are-invertible : {X : 𝓤 ̇ } {Y : 𝓥 ̇ } (f : X → Y)
                    → is-hae f → invertible f

haes-are-invertible f (g , η , ε , τ) = g , η , ε
```

Hence half adjoint equivalences are equivalences, because invertible maps are equivalences. But it is also easy to prove this directly, avoiding the detour via invertible maps. We begin with a construction which will be used a number of times in connection with half adjoint equivalences.

```
transport-ap-≃ : {X : 𝓤 ̇ } {Y : 𝓥 ̇ } (f : X → Y)
                 {x x' : X} (a : x' ＝ x) (b : f x' ＝ f x)
               → (transport (λ - → f - ＝ f x) a b ＝ refl (f x))
               ≃ (ap f a ＝ b)

transport-ap-≃ f (refl x) b = γ
 where
  γ : (b ＝ refl (f x)) ≃ (refl (f x) ＝ b)
  γ = ⁻¹-≃ b (refl (f x))

haes-are-equivs : {X : 𝓤 ̇ } {Y : 𝓥 ̇ } (f : X → Y)
                → is-hae f → is-equiv f

haes-are-equivs f (g , η , ε , τ) y = γ
 where
  c : (φ : fiber f y) → (g y , ε y) ＝ φ
  c (x , refl .(f x)) = q
   where
    p : transport (λ - → f - ＝ f x) (η x) (ε (f x)) ＝ refl (f x)
    p = ⌜ ≃-sym (transport-ap-≃ f (η x) (ε (f x))) ⌝ (τ x)

    q : (g (f x) , ε (f x)) ＝ (x , refl (f x))
    q = to-Σ-＝ (η x , p)

  γ : is-singleton (fiber f y)
  γ = (g y , ε y) , c
```

To recover the constraint for all equivalences, and hence for all invertible maps, under univalence, it is enough to give the constraint for identity maps:

```
id-is-hae : (X : 𝓤 ̇ ) → is-hae (id X)
id-is-hae X = id X , refl , refl , (λ x → refl (refl x))

ua-equivs-are-haes : is-univalent 𝓤
                   → {X Y : 𝓤 ̇ } (f : X → Y)
                   → is-equiv f → is-hae f
```

```
ua-equivs-are-haes ua {X} {Y} = 𝕁-equiv ua (λ X Y f → is-hae f) id-is-hae X Y
```

The above can be proved without univalence as follows. This argument also allows us to have `X` and `Y` in different universes. An example of an equivalence of types in different universes is `Id→Eq`, as stated by univalence.

```
equivs-are-haes : {X : 𝓤 ̇ } {Y : 𝓥 ̇ } (f : X → Y)
                → is-equiv f → is-hae f

equivs-are-haes {𝓤} {𝓥} {X} {Y} f i = (g , η , ε , τ)
 where
  g : Y → X
  g = inverse f i

  η : g ∘ f ∼ id
  η = inverses-are-retractions f i

  ε : f ∘ g ∼ id
  ε = inverses-are-sections f i

  τ : (x : X) → ap f (η x) = ε (f x)
  τ x = γ
   where
    φ : fiber f (f x)
    φ = center (fiber f (f x)) (i (f x))

    by-definition-of-g : g (f x) = fiber-point φ
    by-definition-of-g = refl _

    p : φ = (x , refl (f x))
    p = centrality (fiber f (f x)) (i (f x)) (x , refl (f x))

    a : g (f x) = x
    a = ap fiber-point p

    b : f (g (f x)) = f x
    b = fiber-identification φ

    by-definition-of-η : η x = a
    by-definition-of-η = refl _

    by-definition-of-ε : ε (f x) = b
    by-definition-of-ε = refl _

    q = transport (λ - → f - = f x)       a          b         =⟨ refl _    ⟩
        transport (λ - → f - = f x)       (ap pr₁ p) (pr₂ φ)   =⟨ t         ⟩
        transport (λ - → f (pr₁ -) = f x) p          (pr₂ φ)   =⟨ apd pr₂ p ⟩
        refl (f x)                                             ∎
     where
      t = (transport-ap (λ - → f - = f x) pr₁ p b)⁻¹

    γ : ap f (η x) = ε (f x)
    γ = ⌜ transport-ap-≃ f a b ⌝ q
```

We wrote the above proof of `equivs-are-haes` in a deliberately verbose form to aid understanding. Here is the same proof in a perversely reduced form:

```
equivs-are-haes' : {X : 𝓤 ̇ } {Y : 𝓥 ̇ } (f : X → Y)
                 → is-equiv f → is-hae f

equivs-are-haes' f e = (inverse f e ,
                        inverses-are-retractions f e ,
                        inverses-are-sections f e ,
                        τ)
 where
  τ : ∀ x → ap f (inverses-are-retractions f e x) = inverses-are-sections f e (f x)
  τ x = ⌜ transport-ap-≃ f (ap pr₁ p) (pr₂ φ) ⌝ q
   where
```

```
   φ : fiber f (f x)
   φ = pr₁ (e (f x))

   p : φ ≡ (x , refl (f x))
   p = pr₂ (e (f x)) (x , refl (f x))

   q : transport (λ - → f - ≡ f x) (ap pr₁ p) (pr₂ φ) ≡ refl (f x)
   q = (transport-ap (λ - → f - ≡ f x) pr₁ p ((pr₂ φ)))⁻¹ ∙ apd pr₂ p
```

Notice that we have the following factorization, on the nose, of the construction of invertibility data from the equivalence property:

```
equiv-invertible-hae-factorization : {X : 𝓤 ̇ } {Y : 𝓥 ̇ } (f : X → Y)
                                   → equivs-are-invertible f
                                   ∼ haes-are-invertible f ∘ equivs-are-haes f

equiv-invertible-hae-factorization f e = refl (equivs-are-invertible f e)
```

Instead of working with the notion of half adjoint equivalence, we can just work with Voevodsky's notion of equivalence, and use the fact that it satisfies the half adjoint condition:

```
half-adjoint-condition : {X : 𝓤 ̇ } {Y : 𝓥 ̇ } (f : X → Y) (e : is-equiv f) (x : X)
                       → ap f (inverses-are-retractions f e x) ≡ inverses-are-sections f e (f x)

half-adjoint-condition f e = pr₂ (pr₂ (pr₂ (equivs-are-haes f e)))
```

Here is an example, where, compared to `Σ-change-of-variable'`, we remove univalence from the hypothesis, generalize the universe of the type `Y`, and weaken equality to equivalence in the conclusion. Notice that the proof starts as that of `Σ-reindexing-retract`.

```
Σ-change-of-variable : {X : 𝓤 ̇ } {Y : 𝓥 ̇ } (A : Y → 𝓦 ̇ ) (f : X → Y)
                     → is-equiv f → (Σ y ꞉ Y , A y) ≃ (Σ x ꞉ X , A (f x))

Σ-change-of-variable {𝓤} {𝓥} {𝓦} {X} {Y} A f i = γ
 where
  g = inverse f i
  η = inverses-are-retractions f i
  ε = inverses-are-sections f i
  τ = half-adjoint-condition f i

  φ : Σ A → Σ (A ∘ f)
  φ (y , a) = (g y , transport A ((ε y)⁻¹) a)

  ψ : Σ (A ∘ f) → Σ A
  ψ (x , a) = (f x , a)

  ψφ : (z : Σ A) → ψ (φ z) ≡ z
  ψφ (y , a) = p
   where
    p : (f (g y) , transport A ((ε y)⁻¹) a) ≡ (y , a)
    p = to-Σ-≡ (ε y , transport-is-retraction A (ε y) a)

  φψ : (t : Σ (A ∘ f)) → φ (ψ t) ≡ t
  φψ (x , a) = p
   where
    a' : A (f (g (f x)))
    a' = transport A ((ε (f x))⁻¹) a

    q = transport (A ∘ f) (η x)  a' ≡⟨ transport-ap A f (η x) a'              ⟩
        transport A (ap f (η x)) a' ≡⟨ ap (λ - → transport A - a') (τ x)      ⟩
        transport A (ε (f x))    a' ≡⟨ transport-is-retraction A (ε (f x)) a ⟩
        a                           ∎

    p : (g (f x) , transport A ((ε (f x))⁻¹) a) ≡ (x , a)
    p = to-Σ-≡ (η x , q)
```

```
  γ : Σ A ≃ Σ (A ∘ f)
  γ = invertibility-gives-≃ φ (ψ , ψφ , φψ)
```

For the sake of completeness, we also include the proof from the HoTT book that invertible maps are half adjoint equivalences, which uses a standard argument coming from category theory.

We first need some naturality lemmas:

```
∼-naturality : {X : 𝓤 ̇ } {A : 𝓥 ̇ }
               (f g : X → A) (H : f ∼ g) {x y : X} {p : x ＝ y}
             → H x ∙ ap g p ＝ ap f p ∙ H y

∼-naturality f g H {x} {_} {refl a} = refl-left ⁻¹

∼-naturality' : {X : 𝓤 ̇ } {A : 𝓥 ̇ }
                (f g : X → A) (H : f ∼ g) {x y : X} {p : x ＝ y}
              → H x ∙ ap g p ∙ (H y)⁻¹ ＝ ap f p

∼-naturality' f g H {x} {x} {refl x} = ⁻¹-right∙ (H x)

∼-id-naturality : {X : 𝓤 ̇ }
                  (h : X → X) (η : h ∼ id) {x : X}
                → η (h x) ＝ ap h (η x)

∼-id-naturality h η {x} =

   η (h x)                         ＝⟨ refl _ ⟩
   η (h x) ∙ refl (h x)            ＝⟨ i      ⟩
   η (h x) ∙ (η x ∙ (η x)⁻¹)       ＝⟨ ii     ⟩
   η (h x) ∙ η x ∙ (η x)⁻¹         ＝⟨ iii    ⟩
   η (h x) ∙ ap id (η x) ∙ (η x)⁻¹ ＝⟨ iv     ⟩
   ap h (η x)                      ∎

 where
  i   = ap (η(h x) ∙_) ((⁻¹-right∙ (η x))⁻¹)
  ii  = (∙assoc (η (h x)) (η x) (η x ⁻¹))⁻¹
  iii = ap (λ - → η (h x) ∙ - ∙ η x ⁻¹) ((ap-id (η x))⁻¹)
  iv  = ∼-naturality' h id η {h x} {x} {η x}
```

The idea of the following proof is to improve `ε` to be able to give the required `τ`:

```
invertibles-are-haes : {X : 𝓤 ̇ } {Y : 𝓥 ̇ } (f : X → Y)
                     → invertible f → is-hae f

invertibles-are-haes f (g , η , ε) = g , η , ε' , τ
 where
  ε' = λ y → f (g y)         ＝⟨ (ε (f (g y)))⁻¹ ⟩
             f (g (f (g y))) ＝⟨ ap f (η (g y))  ⟩
             f (g y)         ＝⟨ ε y ⟩
             y               ∎

  module _ (x : domain f) where

   p = η (g (f x))       ＝⟨ ∼-id-naturality (g ∘ f) η  ⟩
       ap (g ∘ f) (η x)  ＝⟨ ap-∘ f g (η x)             ⟩
       ap g (ap f (η x)) ∎

   q = ap f (η (g (f x))) ∙ ε (f x)         ＝⟨ by-p             ⟩
       ap f (ap g (ap f (η x))) ∙ ε (f x)   ＝⟨ by-ap-∘          ⟩
       ap (f ∘ g) (ap f (η x))  ∙ ε (f x)   ＝⟨ by-∼-naturality ⟩
       ε (f (g (f x))) ∙ ap id (ap f (η x)) ＝⟨ by-ap-id         ⟩
       ε (f (g (f x))) ∙ ap f (η x)         ∎
    where
```

```
    by-p               = ap (λ - → ap f - · ε (f x)) p
    by-ap-∘            = ap (_· ε (f x)) ((ap-∘ g f (ap f (η x)))⁻¹)
    by-~-naturality = (~-naturality (f ∘ g) id ε {f (g (f x))} {f x} {ap f (η x)})⁻¹
    by-ap-id           = ap (ε (f (g (f x))) ·_) (ap-id (ap f (η x)))

  τ = ap f (η x)                                         ＝⟨ refl-left ⁻¹ ⟩
      refl (f (g (f x)))                    · ap f (η x)  ＝⟨ by-⁻¹-left·  ⟩
      (ε (f (g (f x))))⁻¹ ·  ε (f (g (f x))) · ap f (η x)  ＝⟨ by-·assoc    ⟩
      (ε (f (g (f x))))⁻¹ · (ε (f (g (f x))) · ap f (η x)) ＝⟨ by-q         ⟩
      (ε (f (g (f x))))⁻¹ · (ap f (η (g (f x))) · ε (f x)) ＝⟨ refl _       ⟩
      ε' (f x)                                             ∎
   where
    by-⁻¹-left· = ap (_· ap f (η x)) ((⁻¹-left· (ε (f (g (f x)))))⁻¹)
    by-·assoc   = ·assoc ((ε (f (g (f x))))⁻¹) (ε (f (g (f x)))) (ap f (η x))
    by-q        = ap ((ε (f (g (f x))))⁻¹ ·_) (q ⁻¹)
```

## Function extensionality from univalence

Function extensionality says that any two pointwise equal functions are equal. This is known to be not provable or disprovable in MLTT. It is an independent statement, which we abbreviate as `funext`.

```
funext : ∀ 𝓤 𝓥 → (𝓤 ⊔ 𝓥)⁺ ̇
funext 𝓤 𝓥 = {X : 𝓤 ̇ } {Y : 𝓥 ̇ } {f g : X → Y} → f ∼ g → f ＝ g
```

There will be two seemingly stronger statements, namely the generalization to dependent functions, and the requirement that the canonical map `f ＝ g → f ∼ g` is an equivalence.

*Exercise*. Assuming `funext`, prove that if a function `f : X → Y` is an equivalence then so is the precomposition map `_∘ f : (Y → Z) → (X → Z)`.

The crucial step in Voevodsky's proof that univalence implies `funext` is to establish the conclusion of the above exercise assuming univalence instead. We prove this by equivalence induction on `f`, which means that we only need to consider the case when `f` is an identity function, for which precomposition with `f` is itself an identity function (of a function type), and hence an equivalence:

```
precomp-is-equiv : is-univalent 𝓤
                 → (X Y : 𝓤 ̇ ) (f : X → Y)
                 → is-equiv f
                 → (Z : 𝓤 ̇ ) → is-equiv (λ (g : Y → Z) → g ∘ f)

precomp-is-equiv {𝓤} ua =
   𝕁-equiv ua
     (λ X Y (f : X → Y) → (Z : 𝓤 ̇ ) → is-equiv (λ g → g ∘ f))
     (λ X Z → id-is-equiv (X → Z))
```

With this we can prove the desired result as follows.

```
univalence-gives-funext : is-univalent 𝓤 → funext 𝓥 𝓤
univalence-gives-funext {𝓤} {𝓥} ua {X} {Y} {f₀} {f₁} = γ
 where
  Δ : 𝓤 ̇
  Δ = Σ y₀ ꞉ Y , Σ y₁ ꞉ Y , y₀ ＝ y₁

  δ : Y → Δ
  δ y = (y , y , refl y)

  π₀ π₁ : Δ → Y
  π₀ (y₀ , y₁ , p) = y₀
  π₁ (y₀ , y₁ , p) = y₁

  δ-is-equiv : is-equiv δ
  δ-is-equiv = invertibles-are-equivs δ (π₀ , η , ε)
   where
```

```
    η : (y : Y) → π₀ (δ y) ≡ y
    η y = refl y

    ε : (d : Δ) → δ (π₀ d) ≡ d
    ε (y , y , refl y) = refl (y , y , refl y)

  φ : (Δ → Y) → (Y → Y)
  φ π = π ∘ δ

  φ-is-equiv : is-equiv φ
  φ-is-equiv = precomp-is-equiv ua Y Δ δ δ-is-equiv Y

  p : φ π₀ ≡ φ π₁
  p = refl (id Y)

  q : π₀ ≡ π₁
  q = equivs-are-lc φ φ-is-equiv p

  γ : f₀ ∼ f₁ → f₀ ≡ f₁
  γ h = ap (λ π x → π (f₀ x , f₁ x , h x)) q
```

This definition of `γ` is probably too concise. Here are all the steps performed silently by Agda, by expanding judgmental equalities, indicated with `refl` here:

```
  γ' : f₀ ∼ f₁ → f₀ ≡ f₁
  γ' h = f₀                             ≡⟨ refl _ ⟩
         (λ x → f₀ x)                   ≡⟨ refl _ ⟩
         (λ x → π₀ (f₀ x , f₁ x , h x)) ≡⟨ ap (λ - x → - (f₀ x , f₁ x , h x)) q ⟩
         (λ x → π₁ (f₀ x , f₁ x , h x)) ≡⟨ refl _ ⟩
         (λ x → f₁ x)                   ≡⟨ refl _ ⟩
         f₁                             ∎
```

So notice that this relies on the so-called η-rule for judgmental equality of functions, namely

```
    f = λ x → f x.
```

Without it, we would only get that

```
    f₀ ∼ f₁ → (λ x → f₀ x) ≡ (λ x → f₁ x)
```

instead.

## Variations of function extensionality and their logical equivalence

Dependent function extensionality:

```
dfunext : ∀ 𝓤 𝓥 → (𝓤 ⊔ 𝓥)⁺ ̇
dfunext 𝓤 𝓥 = {X : 𝓤 ̇ } {A : X → 𝓥 ̇ } {f g : Π A} → f ∼ g → f ≡ g
```

The above says that there is some map `f ∼ g → f ≡ g`. The following instead says that the canonical map `happly` in the other direction is an equivalence:

```
happly : {X : 𝓤 ̇ } {A : X → 𝓥 ̇ } (f g : Π A) → f ≡ g → f ∼ g
happly f g p x = ap (λ - → - x) p

hfunext : ∀ 𝓤 𝓥 → (𝓤 ⊔ 𝓥)⁺ ̇
hfunext 𝓤 𝓥 = {X : 𝓤 ̇ } {A : X → 𝓥 ̇ } (f g : Π A) → is-equiv (happly f g)

hfunext-gives-dfunext : hfunext 𝓤 𝓥 → dfunext 𝓤 𝓥
hfunext-gives-dfunext hfe {X} {A} {f} {g} = inverse (happly f g) (hfe f g)
```

Voevodsky showed that all these notions of function extensionality are logically equivalent to saying that products of singletons are singletons:

```
vvfunext : ∀ 𝓤 𝓥 → (𝓤 ⊔ 𝓥)⁺ ̇
vvfunext 𝓤 𝓥 = {X : 𝓤 ̇ } {A : X → 𝓥 ̇ }
             → ((x : X) → is-singleton (A x))
             → is-singleton (Π A)

dfunext-gives-vvfunext : dfunext 𝓤 𝓥 → vvfunext 𝓤 𝓥
dfunext-gives-vvfunext fe {X} {A} i = γ
 where
  f : Π A
  f x = center (A x) (i x)

  c : (g : Π A) → f ≡ g
  c g = fe (λ (x : X) → centrality (A x) (i x) (g x))

  γ : is-singleton (Π A)
  γ = f , c

vvfunext-gives-hfunext : vvfunext 𝓤 𝓥 → hfunext 𝓤 𝓥
vvfunext-gives-hfunext vfe {X} {Y} f = γ
 where
  a : (x : X) → is-singleton (Σ y ꞉ Y x , f x ≡ y)
  a x = singleton-types'-are-singletons (Y x) (f x)

  c : is-singleton (Π x ꞉ X , Σ y ꞉ Y x , f x ≡ y)
  c = vfe a

  ρ : (Σ g ꞉ Π Y , f ∼ g) ◁ (Π x ꞉ X , Σ y ꞉ Y x , f x ≡ y)
  ρ = ≃-gives-▷ ΠΣ-distr-≃

  d : is-singleton (Σ g ꞉ Π Y , f ∼ g)
  d = retract-of-singleton ρ c

  e : (Σ g ꞉ Π Y , f ≡ g) → (Σ g ꞉ Π Y , f ∼ g)
  e = NatΣ (happly f)

  i : is-equiv e
  i = maps-of-singletons-are-equivs e (singleton-types'-are-singletons (Π Y) f) d

  γ : (g : Π Y) → is-equiv (happly f g)
  γ = NatΣ-equiv-gives-fiberwise-equiv (happly f) i
```

And finally the seemingly rather weak, non-dependent version `funext` implies the seemingly strongest version, which closes the circle of logical equivalences. We first need some lemmas.

```
postcomp-invertible : {X : 𝓤 ̇ } {Y : 𝓥 ̇ } {A : 𝓦 ̇ }
                    → funext 𝓦 𝓤
                    → funext 𝓦 𝓥
                    → (f : X → Y)
                    → invertible f
                    → invertible (λ (h : A → X) → f ∘ h)

postcomp-invertible {𝓤} {𝓥} {𝓦} {X} {Y} {A} fe fe' f (g , η , ε) = γ
 where
  f' : (A → X) → (A → Y)
  f' h = f ∘ h

  g' : (A → Y) → (A → X)
  g' k = g ∘ k

  η' : (h : A → X) → g' (f' h) ≡ h
  η' h = fe (η ∘ h)

  ε' : (k : A → Y) → f' (g' k) ≡ k
  ε' k = fe' (ε ∘ k)

  γ : invertible f'
```

```
  γ = (g' , η' , ε')

postcomp-is-equiv : {X : 𝓤 ̇ } {Y : 𝓥 ̇ } {A : 𝓦 ̇ }
                  → funext 𝓦 𝓤
                  → funext 𝓦 𝓥
                  → (f : X → Y)
                  → is-equiv f
                  → is-equiv (λ (h : A → X) → f ∘ h)

postcomp-is-equiv fe fe' f e =
 invertibles-are-equivs
  (λ h → f ∘ h)
  (postcomp-invertible fe fe' f (equivs-are-invertible f e))

funext-gives-vvfunext : funext 𝓤 (𝓤 ⊔ 𝓥) → funext 𝓤 𝓤 → vvfunext 𝓤 𝓥
funext-gives-vvfunext {𝓤} {𝓥} fe fe' {X} {A} φ = γ
 where
  f : Σ A → X
  f = pr₁

  f-is-equiv : is-equiv f
  f-is-equiv = pr₁-is-equiv φ

  g : (X → Σ A) → (X → X)
  g h = f ∘ h

  e : is-equiv g
  e = postcomp-is-equiv fe fe' f f-is-equiv

  i : is-singleton (Σ h ꞉ (X → Σ A), f ∘ h ≡ id X)
  i = e (id X)

  r : (Σ h ꞉ (X → Σ A), f ∘ h ≡ id X) → Π A
  r (h , p) x = transport A (happly (f ∘ h) (id X) p x) (pr₂ (h x))

  s : Π A → (Σ h ꞉ (X → Σ A), f ∘ h ≡ id X)
  s φ = (λ x → x , φ x) , refl (id X)

  η : ∀ φ → r (s φ) ≡ φ
  η φ = refl (r (s φ))

  γ : is-singleton (Π A)
  γ = retract-of-singleton (r , s , η) i
```

We have the following corollaries. We first formulate the types of some functions:

```
abstract
 funext-gives-hfunext       : funext 𝓤 (𝓤 ⊔ 𝓥) → funext 𝓤 𝓤 → hfunext 𝓤 𝓥
 dfunext-gives-hfunext      : dfunext 𝓤 𝓥 → hfunext 𝓤 𝓥
 funext-gives-dfunext       : funext 𝓤 (𝓤 ⊔ 𝓥) → funext 𝓤 𝓤 → dfunext 𝓤 𝓥
 univalence-gives-dfunext'  : is-univalent 𝓤 → is-univalent (𝓤 ⊔ 𝓥) → dfunext 𝓤 𝓥
 univalence-gives-hfunext'  : is-univalent 𝓤 → is-univalent (𝓤 ⊔ 𝓥) → hfunext 𝓤 𝓥
 univalence-gives-vvfunext' : is-univalent 𝓤 → is-univalent (𝓤 ⊔ 𝓥) → vvfunext 𝓤 𝓥
 univalence-gives-hfunext   : is-univalent 𝓤 → hfunext 𝓤 𝓤
 univalence-gives-dfunext   : is-univalent 𝓤 → dfunext 𝓤 𝓤
 univalence-gives-vvfunext  : is-univalent 𝓤 → vvfunext 𝓤 𝓤
```

And then we give their definitions (Agda makes sure there are no circularities):

```
 funext-gives-hfunext fe fe' = vvfunext-gives-hfunext (funext-gives-vvfunext fe fe')

 funext-gives-dfunext fe fe' = hfunext-gives-dfunext (funext-gives-hfunext fe fe')

 dfunext-gives-hfunext fe = vvfunext-gives-hfunext (dfunext-gives-vvfunext fe)

 univalence-gives-dfunext' ua ua' = funext-gives-dfunext
                                     (univalence-gives-funext ua')
                                     (univalence-gives-funext ua)
```

```
univalence-gives-hfunext' ua ua' = funext-gives-hfunext
                                     (univalence-gives-funext ua')
                                     (univalence-gives-funext ua)

univalence-gives-vvfunext' ua ua' = funext-gives-vvfunext
                                     (univalence-gives-funext ua')
                                     (univalence-gives-funext ua)

univalence-gives-hfunext ua = univalence-gives-hfunext' ua ua

univalence-gives-dfunext ua = univalence-gives-dfunext' ua ua

univalence-gives-vvfunext ua = univalence-gives-vvfunext' ua ua
```

## Universes are map classifiers

Under univalence, a universe `𝓤` becomes a map classifier, in the sense that maps from a type `X` in `𝓤` into a type `Y` in `𝓤` are in canonical bijection with functions `Y → 𝓤`. Using the following slice notation, this amounts to a bijection between `𝓤 / Y` and `Y → 𝓤`:

```
_/_ : (𝓤 : Universe) → 𝓥 ̇ → 𝓤 ⁺ ⊔ 𝓥 ̇
𝓤 / Y = Σ X ꞉ 𝓤 ̇ , (X → Y)
```

We need the following lemma, which has other uses:

```
total-fiber-is-domain : {X : 𝓤 ̇ } {Y : 𝓥 ̇ } (f : X → Y)
                      → Σ (fiber f) ≃ X

total-fiber-is-domain {𝓤} {𝓥} {X} {Y} f = invertibility-gives-≃ g (h , η , ε)
 where
  g : (Σ y ꞉ Y , Σ x ꞉ X , f x ≡ y) → X
  g (y , x , p) = x

  h : X → Σ y ꞉ Y , Σ x ꞉ X , f x ≡ y
  h x = (f x , x , refl (f x))

  η : ∀ t → h (g t) ≡ t
  η (_ , x , refl _) = refl (f x , x , refl _)

  ε : (x : X) → g (h x) ≡ x
  ε = refl
```

The function `χ` gives the *characteristic function* of a map into `Y`:

```
χ : (Y : 𝓤 ̇ ) → 𝓤 / Y  → (Y → 𝓤 ̇ )
χ Y (X , f) = fiber f
```

We say that a universe is a map classifier if the above function is an equivalence for every `Y` in the universe:

```
is-map-classifier : (𝓤 : Universe) → 𝓤 ⁺ ̇
is-map-classifier 𝓤 = (Y : 𝓤 ̇ ) → is-equiv (χ Y)
```

Any `Y → 𝓤` is the characteristic function of some map into `Y` by taking its total space and the first projection:

```
𝕋 : (Y : 𝓤 ̇ ) → (Y → 𝓤 ̇ ) → 𝓤 / Y
𝕋 Y A = Σ A , pr₁

χη : is-univalent 𝓤
   → (Y : 𝓤 ̇ ) (σ : 𝓤 / Y) → 𝕋 Y (χ Y σ) ≡ σ

χη ua Y (X , f) = r
 where
  e : Σ (fiber f) ≃ X
  e = total-fiber-is-domain f
```

```
  p : Σ (fiber f) ＝ X
  p = Eq→Id ua (Σ (fiber f)) X e

  observation : ⌜ ≃-sym e ⌝ ＝ (λ x → f x , x , refl (f x))
  observation = refl _

  q = transport (λ - → - → Y) p pr₁ ＝⟨ transport-map-along-≃ ua e pr₁ ⟩
      pr₁ ∘ ⌜ ≃-sym e ⌝             ＝⟨ refl _                          ⟩
      f                             ∎

  r : (Σ (fiber f) , pr₁) ＝ (X , f)
  r = to-Σ-＝ (p , q)

χε : is-univalent 𝓤
   → dfunext 𝓤 (𝓤 ⁺)
   → (Y : 𝓤 ̇ ) (A : Y → 𝓤 ̇ ) → χ Y (𝕋 Y A) ＝ A

χε ua fe Y A = fe γ
 where
  f : ∀ y → fiber pr₁ y → A y
  f y ((y , a) , refl p) = a

  g : ∀ y → A y → fiber pr₁ y
  g y a = (y , a) , refl y

  η : ∀ y σ → g y (f y σ) ＝ σ
  η y ((y , a) , refl p) = refl ((y , a) , refl p)

  ε : ∀ y a → f y (g y a) ＝ a
  ε y a = refl a

  γ : ∀ y → fiber pr₁ y ＝ A y
  γ y = Eq→Id ua _ _ (invertibility-gives-≃ (f y) (g y , η y , ε y))

universes-are-map-classifiers : is-univalent 𝓤
                              → dfunext 𝓤 (𝓤 ⁺)
                              → is-map-classifier 𝓤

universes-are-map-classifiers ua fe Y = invertibles-are-equivs (χ Y)
                                         (𝕋 Y , χη ua Y , χε ua fe Y)
```

Therefore we have the following canonical equivalence:

```
map-classification : is-univalent 𝓤
                   → dfunext 𝓤 (𝓤 ⁺)
                   → (Y : 𝓤 ̇ ) → 𝓤 / Y ≃ (Y → 𝓤 ̇ )

map-classification ua fe Y = χ Y , universes-are-map-classifiers ua fe Y
```

## The univalence axiom is a (sub)singleton

If we use a type as an axiom, it should better have at most one element. We prove some generally useful lemmas first.

```
Π-is-subsingleton : dfunext 𝓤 𝓥
                  → {X : 𝓤 ̇ } {A : X → 𝓥 ̇ }
                  → ((x : X) → is-subsingleton (A x))
                  → is-subsingleton (Π A)

Π-is-subsingleton fe i f g = fe (λ x → i x (f x) (g x))

being-singleton-is-subsingleton : dfunext 𝓤 𝓤
```

```
                                     → {X : 𝓤 ̇ } → is-subsingleton (is-singleton X)

being-singleton-is-subsingleton fe {X} (x , φ) (y , γ) = p
 where
  i : is-subsingleton X
  i = singletons-are-subsingletons X (y , γ)

  s : is-set X
  s = subsingletons-are-sets X i

  a : (z : X) → is-subsingleton ((t : X) → z ≡ t)
  a z = Π-is-subsingleton fe (s z)

  b : x ≡ y
  b = φ y

  p : (x , φ) ≡ (y , γ)
  p = to-subtype-≡ a b

being-equiv-is-subsingleton : dfunext 𝓥 (𝓤 ⊔ 𝓥)
                            → dfunext (𝓤 ⊔ 𝓥) (𝓤 ⊔ 𝓥)
                            → {X : 𝓤 ̇ } {Y : 𝓥 ̇ } (f : X → Y)
                            → is-subsingleton (is-equiv f)

being-equiv-is-subsingleton fe fe' f = Π-is-subsingleton fe
                                        (λ x → being-singleton-is-subsingleton fe')
```

In passing, we fulfill a promise made above:

```
subsingletons-are-retracts-of-logically-equivalent-types : {X : 𝓤 ̇ } {Y : 𝓥 ̇ }
                                                         → is-subsingleton X
                                                         → (X ⇔ Y)
                                                         → X ◁ Y

subsingletons-are-retracts-of-logically-equivalent-types i (f , g) = g , f , η
 where
  η : g ∘ f ∼ id
  η x = i (g (f x)) x

equivalence-property-is-retract-of-invertibility-data : dfunext 𝓥 (𝓤 ⊔ 𝓥)
                                                      → dfunext (𝓤 ⊔ 𝓥) (𝓤 ⊔ 𝓥)
                                                      → {X : 𝓤 ̇ } {Y : 𝓥 ̇ } (f : X → Y)
                                                      → is-equiv f ◁ invertible f

equivalence-property-is-retract-of-invertibility-data fe fe' f =
  subsingletons-are-retracts-of-logically-equivalent-types
   (being-equiv-is-subsingleton fe fe' f)
   (equivs-are-invertible f , invertibles-are-equivs f)
```

We now return to our main concern in this section.

```
univalence-is-subsingleton : is-univalent (𝓤 ⁺)
                           → is-subsingleton (is-univalent 𝓤)

univalence-is-subsingleton {𝓤} ua⁺ = subsingleton-criterion' γ
 where
  γ : is-univalent 𝓤 → is-subsingleton (is-univalent 𝓤)
  γ ua = i
   where
    dfe₁ : dfunext  𝓤    (𝓤 ⁺)
    dfe₂ : dfunext (𝓤 ⁺) (𝓤 ⁺)

    dfe₁ = univalence-gives-dfunext' ua ua⁺
    dfe₂ = univalence-gives-dfunext ua⁺

    i : is-subsingleton (is-univalent 𝓤)
    i = Π-is-subsingleton dfe₂
```

```
         (λ X → Π-is-subsingleton dfe₂
         (λ Y → being-equiv-is-subsingleton dfe₁ dfe₂ (Id→Eq X Y)))
```

So if all universes are univalent then "being univalent" is a subsingleton, and hence a singleton. This hypothesis of global univalence cannot be expressed in our MLTT that only has $\omega$ many universes, because global univalence would have to live in the first universe after them. Agda does have such a universe $\mathcal{U}\omega$, and so we can formulate it here. There would be no problem in extending our MLTT to have such a universe if we so wished, in which case we would be able to formulate and prove:

```
Univalence : 𝓤ω
Univalence = ∀ 𝓤 → is-univalent 𝓤

univalence-is-subsingletonω : Univalence → is-subsingleton (is-univalent 𝓤)
univalence-is-subsingletonω {𝓤} γ = univalence-is-subsingleton (γ (𝓤 ⁺))

univalence-is-a-singleton : Univalence → is-singleton (is-univalent 𝓤)
univalence-is-a-singleton {𝓤} γ = pointed-subsingletons-are-singletons
                                    (is-univalent 𝓤)
                                    (γ 𝓤)
                                    (univalence-is-subsingletonω γ)
```

That the type `Univalence` would be a subsingleton can't even be formulated in the absence of a successor $\mathcal{U}\omega^{+}$ of $\mathcal{U}\omega$, and Agda doesn't have such a successor universe (but there isn't any fundamental reason why it couldn't have it).

In the absence of a universe $\mathcal{U}\omega$ in our MLTT, we can simply have an axiom schema, consisting of $\omega$-many axioms, stating that each universe is univalent. Then we can prove in our MLTT that the univalence property for each universe is a (sub)singleton, with $\omega$-many proofs (or just one schematic proof with a free variable for a universe $\mathcal{U}_n$).

It follows immediately from the above that global univalence gives global function extensionality.

```
global-dfunext : 𝓤ω
global-dfunext = ∀ {𝓤 𝓥} → dfunext 𝓤 𝓥

univalence-gives-global-dfunext : Univalence → global-dfunext
univalence-gives-global-dfunext ua {𝓤} {𝓥} = univalence-gives-dfunext'
                                                (ua 𝓤) (ua (𝓤 ⊔ 𝓥))

global-hfunext : 𝓤ω
global-hfunext = ∀ {𝓤 𝓥} → hfunext 𝓤 𝓥

univalence-gives-global-hfunext : Univalence → global-hfunext
univalence-gives-global-hfunext ua {𝓤} {𝓥} = univalence-gives-hfunext'
                                                (ua 𝓤) (ua (𝓤 ⊔ 𝓥))
```

### `vvfunext 𝓤 𝓥` and `hfunext 𝓤 𝓥` are (sub)singletons

We need a version of `Π-is-subsingleton` for dependent functions with implicit arguments.

```
Π-is-subsingleton' : dfunext 𝓤 𝓥
                   → {X : 𝓤 ̇ } {A : X → 𝓥 ̇ }
                   → ((x : X) → is-subsingleton (A x))
                   → is-subsingleton ({x : X} → A x)

Π-is-subsingleton' fe {X} {A} i = γ
 where
  ρ : ({x : X} → A x) ◁ Π A
  ρ = (λ f {x} → f x) , (λ g x → g {x}) , refl

  γ : is-subsingleton ({x : X} → A x)
  γ = retract-of-subsingleton ρ (Π-is-subsingleton fe i)
```

To show that `vvfunext 𝓤 𝓥` and `hfunext 𝓤 𝓥` are subsingletons, we need assumptions of function extensionality for higher universes:

```
vv-and-hfunext-are-subsingletons-lemma  : dfunext (𝓤 ⁺)       (𝓤 ⊔ (𝓥 ⁺))
                                        → dfunext (𝓤 ⊔ (𝓥 ⁺)) (𝓤 ⊔ 𝓥)
                                        → dfunext (𝓤 ⊔ 𝓥)     (𝓤 ⊔ 𝓥)
```

```
                                                    → is-subsingleton (vvfunext 𝓤 𝓥)
                                                    × is-subsingleton (hfunext  𝓤 𝓥)

vv-and-hfunext-are-subsingletons-lemma {𝓤} {𝓥} dfe dfe' dfe'' = φ , γ
 where
  φ : is-subsingleton (vvfunext 𝓤 𝓥)
  φ = Π-is-subsingleton' dfe
       (λ X → Π-is-subsingleton' dfe'
       (λ A → Π-is-subsingleton dfe''
       (λ i → being-singleton-is-subsingleton dfe'')))

  γ : is-subsingleton (hfunext 𝓤 𝓥)
  γ = Π-is-subsingleton' dfe
       (λ X → Π-is-subsingleton' dfe'
       (λ A → Π-is-subsingleton dfe''
       (λ f → Π-is-subsingleton dfe''
       (λ g → being-equiv-is-subsingleton dfe'' dfe''
               (happly f g)))))
```

Hence they are singletons assuming global univalence (or just global function extensionality, of any kind):

```
vv-and-hfunext-are-singletons : Univalence
                              → is-singleton (vvfunext 𝓤 𝓥)
                              × is-singleton (hfunext  𝓤 𝓥)

vv-and-hfunext-are-singletons {𝓤} {𝓥} ua =

 f (vv-and-hfunext-are-subsingletons-lemma
     (univalence-gives-dfunext' (ua (𝓤 ⁺))       (ua ((𝓤 ⁺) ⊔ (𝓥 ⁺))))
     (univalence-gives-dfunext' (ua (𝓤 ⊔ (𝓥 ⁺))) (ua (𝓤 ⊔ (𝓥 ⁺))))
     (univalence-gives-dfunext' (ua (𝓤 ⊔ 𝓥))     (ua (𝓤 ⊔ 𝓥))))

 where
  f : is-subsingleton (vvfunext 𝓤 𝓥) × is-subsingleton (hfunext 𝓤 𝓥)
    → is-singleton (vvfunext 𝓤 𝓥) × is-singleton (hfunext 𝓤 𝓥)

  f (i , j) = pointed-subsingletons-are-singletons (vvfunext 𝓤 𝓥)
                (univalence-gives-vvfunext' (ua 𝓤) (ua (𝓤 ⊔ 𝓥))) i ,

              pointed-subsingletons-are-singletons (hfunext 𝓤 𝓥)
                (univalence-gives-hfunext' (ua 𝓤) (ua (𝓤 ⊔ 𝓥))) j
```

However, `funext 𝓤 𝓤` and `dfunext 𝓤 𝓤` are not subsingletons (see the HoTT book).

## Unique existence in univalent mathematics

Unique existence of `x : X` with `A x` in univalent mathematics, written

```
∃! x : X , A x
```

or simply

```
∃! A,
```

requires that not only the point `x : X` but also the datum `a : A x` to be unique. More precisely, we require that there is a unique *pair* `(x , a) : Σ A`.

This is particularly important in the formulation of universal properties involving types that are not necessarily sets, where it generalizes the categorical notion of uniqueness up to unique isomorphism.

```
∃! : {X : 𝓤 ̇ } → (X → 𝓥 ̇ ) → 𝓤 ⊔ 𝓥 ̇
∃! A = is-singleton (Σ A)

-∃! : {𝓤 𝓥 : Universe} (X : 𝓤 ̇ ) (Y : X → 𝓥 ̇ ) → 𝓤 ⊔ 𝓥 ̇
-∃! X Y = ∃! Y
```

```
syntax -∃! A (λ x → b) = ∃! x ꞉ A , b

∃!-is-subsingleton : {X : 𝓤 ̇ } (A : X → 𝓥 ̇ )
                   → dfunext (𝓤 ⊔ 𝓥) (𝓤 ⊔ 𝓥)
                   → is-subsingleton (∃! A)

∃!-is-subsingleton A fe = being-singleton-is-subsingleton fe

unique-existence-gives-weak-unique-existence : {X : 𝓤 ̇ } (A : X → 𝓥 ̇ )

  → (∃! x ꞉ X , A x)
  → (Σ x ꞉ X , A x) × ((x y : X) → A x → A y → x ≡ y)

unique-existence-gives-weak-unique-existence A s = center (Σ A) s , u
 where
  u : ∀ x y → A x → A y → x ≡ y
  u x y a b = ap pr₁ (singletons-are-subsingletons (Σ A) s (x , a) (y , b))
```

The converse holds if each `A x` is a subsingleton:

```
weak-unique-existence-gives-unique-existence-sometimes : {X : 𝓤 ̇ } (A : X → 𝓥 ̇ )

  →  ((x : X) → is-subsingleton (A x))
  → ((Σ x ꞉ X , A x) × ((x y : X) → A x → A y → x ≡ y))
  → (∃! x ꞉ X , A x)

weak-unique-existence-gives-unique-existence-sometimes A i ((x , a) , u) = (x , a) , φ
 where
  φ : (σ : Σ A) → x , a ≡ σ
  φ (y , b) = to-subtype-≡ i (u x y a b)
```

*Exercise*. Find a counter-example in the absence of the requirement that all types `A x` are subsingletons.

Similarly, the existence of at most one `x : X` with `A x` should be understood as

```
is-subsingleton (Σ A),
```

but we will not introduce special notation for this concept, although it will occur often.

## Universal property of the natural numbers

The natural numbers have the following universal property. What is noteworthy here is that the type `Y` need not be a set, so that the two equations can hold in multiple ways, but nevertheless we have unique existence in the sense of the previous section. Moreover, univalence is not needed. Function extensionality suffices.

```
ℕ-is-nno : hfunext 𝓤₀ 𝓤
         → (Y : 𝓤 ̇ ) (y₀ : Y) (g : Y → Y)
         → ∃! h ꞉ (ℕ → Y), (h 0 ≡ y₀) × (h ∘ succ ∼ g ∘ h)

ℕ-is-nno {𝓤} hfe Y y₀ g = γ
 where
```

We apply the same retraction techniques we used in order to prove that invertible maps are equivalences. We first show that, for any `h : ℕ → Y`, the type

```
(h 0 ≡ y₀) × (h ∘ succ ∼ g ∘ h)
```

is a retract of the type

```
h ∼ ℕ-iteration Y y₀ g
```

and hence, by function extensionality, we also have a retraction if we replace pointwise equality ∼ by equality ≡. Thus the type

```
Σ h : ℕ → Y , (h 0 ≡ y₀) × (h ∘ succ ∼ g ∘ h)
```

is a retract of the singleton type

```
Σ h : ℕ → Y , h ≡ ℕ-iteration Y y₀ g,
```

and therefore is itself a singleton, as required.

Now we do this in Agda. We first establish the following retraction (which is automatically an equivalence, but we don't need this fact).

```
lemma₀ : (h : ℕ → Y) → ((h 0 ≡ y₀) × (h ∘ succ ∼ g ∘ h)) ◁ (h ∼ ℕ-iteration Y y₀ g)
lemma₀ h = r , s , η
 where
  s : (h 0 ≡ y₀) × (h ∘ succ ∼ g ∘ h) → h ∼ ℕ-iteration Y y₀ g
  s (p , K) 0        = p
  s (p , K) (succ n) = h (succ n)                   ≡⟨ K n               ⟩
                       g (h n)                      ≡⟨ ap g (s (p , K) n) ⟩
                       ℕ-iteration Y y₀ g (succ n) ∎
```

The above section `s` is defined by induction on natural numbers, but the following retraction `r` is defined directly.

```
  r : codomain s → domain s
  r H = H 0 , (λ n → h (succ n)                ≡⟨ H (succ n)     ⟩
                     g (ℕ-iteration Y y₀ g n)   ≡⟨ ap g ((H n)⁻¹) ⟩
                     g (h n )                  ∎)
```

The retraction property doesn't need induction on natural numbers:

```
  η : (z : (h 0 ≡ y₀) × (h ∘ succ ∼ g ∘ h)) → r (s z) ≡ z
  η (p , K) = v
   where
    i = λ n →
     K n ∙  ap g (s (p , K) n) ∙  ap g ((s (p , K) n) ⁻¹)                  ≡⟨ ii  n ⟩
     K n ∙ (ap g (s (p , K) n) ∙  ap g ((s (p , K) n) ⁻¹))                 ≡⟨ iii n ⟩
     K n ∙ (ap g (s (p , K) n) ∙ (ap g  (s (p , K) n))⁻¹)                  ≡⟨ iv  n ⟩
     K n                                                                   ∎
      where
       ii  = λ n → ∙assoc (K n) (ap g (s (p , K) n)) (ap g ((s (p , K) n)⁻¹))
       iii = λ n → ap (λ - → K n ∙ (ap g (s (p , K) n) ∙ -)) (ap⁻¹ g (s (p , K) n) ⁻¹)
       iv  = λ n → ap (K n ∙_) (⁻¹-right∙ (ap g (s (p , K) n)))

    v : (p , (λ n → s (p , K) (succ n) ∙ ap g ((s (p , K) n)⁻¹))) ≡ (p , K)
    v = ap (p ,_) (hfunext-gives-dfunext hfe i)

lemma₁ = λ h → (h 0 ≡ y₀) × (h ∘ succ ∼ g ∘ h) ◁⟨ lemma₀ h ⟩
               (h ∼ ℕ-iteration Y y₀ g)        ◁⟨ i h      ⟩
               (h ≡ ℕ-iteration Y y₀ g)        ◀
 where
  i = λ h → ≃-gives-▷ (happly h (ℕ-iteration Y y₀ g) , hfe _ _)

lemma₂ : (Σ h : (ℕ → Y), (h 0 ≡ y₀) × (h ∘ succ ∼ g ∘ h))
       ◁ (Σ h : (ℕ → Y), h ≡ ℕ-iteration Y y₀ g)

lemma₂ = Σ-retract lemma₁

γ : is-singleton (Σ h : (ℕ → Y), (h 0 ≡ y₀) × (h ∘ succ ∼ g ∘ h))
γ = retract-of-singleton
     lemma₂
     (singleton-types-are-singletons (ℕ → Y) (ℕ-iteration Y y₀ g))
```

This concludes the proof of `ℕ-is-nno`. We say that `ℕ` is a natural numbers object, or, more precisely, the triple `(ℕ , 0 , succ)` is a natural numbers object.

Here is an important example, which given any `n : ℕ` constructs a type with `n` elements (and decidable equality):

```
module finite-types (hfe : hfunext 𝓤₀ 𝓤₁) where

 fin :  ∃! Fin ꞉ (ℕ → 𝓤₀ ̇ )
              , (Fin 0 ≡ 𝟘)
              × ((n : ℕ) → Fin (succ  n) ≡ Fin n + 𝟙)

 fin = ℕ-is-nno hfe (𝓤₀ ̇ ) 𝟘 (_+ 𝟙)

 Fin : ℕ → 𝓤₀ ̇
 Fin = pr₁ (center _ fin)
```

We could use the other projections to derive the following two equations, but they hold definitionally:

```
 Fin-equation₀ : Fin 0 ≡ 𝟘
 Fin-equation₀ = refl _

 Fin-equation-succ : Fin ∘ succ ≡ λ n → Fin n + 𝟙
 Fin-equation-succ = refl _
```

We also have

```
 Fin-equation-succ' : (n : ℕ) → Fin (succ n) ≡ Fin n + 𝟙
 Fin-equation-succ' n = refl _
```

and the examples

```
 Fin-equation₁ : Fin 1 ≡ 𝟘 + 𝟙
 Fin-equation₁ = refl _

 Fin-equation₂ : Fin 2 ≡ (𝟘 + 𝟙) + 𝟙
 Fin-equation₂ = refl _

 Fin-equation₃ : Fin 3 ≡ ((𝟘 + 𝟙) + 𝟙) + 𝟙
 Fin-equation₃ = refl _
```

*Exercise*. Assume univalence. The equation

```
Fin (succ n) ≡ Fin n + 𝟙
```

holds in multiple ways, because `Fin n` has `n!` automorphisms. Construct an involutive fiberwise equivalence

```
mirror : (n : ℕ) → Fin n → Fin n
```

different from the identity and hence an identification `Fin ≡ Fin` different from `refl Fin`. Consider

```
∃! Fin' ꞉ ℕ → 𝓤₀ ̇ , (Fin' 0 ≡ 𝟘) × (Fin' ∘ succ ≡ λ n → 𝟙 + Fin' n)
```

and show that `Fin' (succ n) ≡ Fin' n + 𝟙` so that `Fin'` satisfies the defining equations of `Fin`, although not judgmentally, and hence `Fin' ≡ Fin` by the univeral property of `Fin`. But there are many equalities `Fin' n ≡ Fin n`. Which one do we get by the universal property? Show that the type `Fin n` has decidable equality and hence is a set.

## More consequences of function extensionality

```
being-subsingleton-is-subsingleton : dfunext 𝓤 𝓤
                                   → {X : 𝓤 ̇ }
                                   → is-subsingleton (is-subsingleton X)
```

```
being-subsingleton-is-subsingleton fe {X} i j = c
 where
  l : is-set X
  l = subsingletons-are-sets X i

  a : (x y : X) → i x y ≡ j x y
  a x y = l x y (i x y) (j x y)

  b : (x : X) → i x ≡ j x
  b x = fe (a x)

  c : i ≡ j
  c = fe b

being-center-is-subsingleton : dfunext 𝓤 𝓤
                              → {X : 𝓤 ̇ } (c : X)
                              → is-subsingleton (is-center X c)

being-center-is-subsingleton fe {X} c φ γ = k
 where
  i : is-singleton X
  i = c , φ

  j : (x : X) → is-subsingleton (c ≡ x)
  j x = singletons-are-sets X i c x

  k : φ ≡ γ
  k = fe (λ x → j x (φ x) (γ x))
```

Here the version hfunext of function extensionality is what is needed:

```
Π-is-set : hfunext 𝓤 𝓥
         → {X : 𝓤 ̇ } {A : X → 𝓥 ̇ }
         → ((x : X) → is-set (A x))
         → is-set (Π A)

Π-is-set hfe s f g = b
 where
  a : is-subsingleton (f ∼ g)
  a p q = γ
   where
    h : ∀ x →  p x ≡ q x
    h x = s x (f x) (g x) (p x) (q x)
    γ : p ≡  q
    γ = hfunext-gives-dfunext hfe h

  e : (f ≡ g) ≃ (f ∼ g)
  e = (happly f g , hfe f g)

  b : is-subsingleton (f ≡ g)
  b = equiv-to-subsingleton e a

being-set-is-subsingleton : dfunext 𝓤 𝓤
                          → {X : 𝓤 ̇ }
                          → is-subsingleton (is-set X)

being-set-is-subsingleton fe = Π-is-subsingleton fe
                                (λ x → Π-is-subsingleton fe
                                (λ y → being-subsingleton-is-subsingleton fe))
```

More generally:

```
hlevel-relation-is-subsingleton : dfunext 𝓤 𝓤
                                → (n : ℕ) (X : 𝓤 ̇ )
                                → is-subsingleton (X is-of-hlevel n)

hlevel-relation-is-subsingleton {𝓤} fe zero X =
 being-singleton-is-subsingleton fe
```

```
hlevel-relation-is-subsingleton fe (succ n) X =
 Π-is-subsingleton fe
  (λ x → Π-is-subsingleton fe
  (λ x' → hlevel-relation-is-subsingleton fe n (x ＝ x')))
```

Composition of equivalences is associative:

```
●-assoc : dfunext 𝓣 (𝓤 ⊔ 𝓣)
        → dfunext (𝓤 ⊔ 𝓣) (𝓤 ⊔ 𝓣)
        → {X : 𝓤 ̇ } {Y : 𝓥 ̇ } {Z : 𝓦 ̇ } {T : 𝓣 ̇ }
          (α : X ≃ Y) (β : Y ≃ Z) (γ : Z ≃ T)
        → α ● (β ● γ) ＝ (α ● β) ● γ

●-assoc fe fe' (f , a) (g , b) (h , c) = ap (h ∘ g ∘ f ,_) q
 where
  d e : is-equiv (h ∘ g ∘ f)
  d = ∘-is-equiv (∘-is-equiv c b) a
  e = ∘-is-equiv c (∘-is-equiv b a)

  q : d ＝ e
  q = being-equiv-is-subsingleton fe fe' (h ∘ g ∘ f) _ _

≃-sym-involutive : dfunext 𝓥 (𝓤 ⊔ 𝓥)
                 → dfunext (𝓤 ⊔ 𝓥) (𝓤 ⊔ 𝓥) →
                   {X : 𝓤 ̇ } {Y : 𝓥 ̇ } (α : X ≃ Y)
                 → ≃-sym (≃-sym α) ＝ α

≃-sym-involutive fe fe' (f , a) = to-subtype-＝
                                   (being-equiv-is-subsingleton fe fe')
                                   (inversion-involutive f a)

≃-sym-is-equiv : dfunext 𝓥 (𝓤 ⊔ 𝓥)
               → dfunext 𝓤 (𝓤 ⊔ 𝓥)
               → dfunext (𝓤 ⊔ 𝓥) (𝓤 ⊔ 𝓥)
               → {X : 𝓤 ̇ } {Y : 𝓥 ̇ }
               → is-equiv (≃-sym {𝓤} {𝓥} {X} {Y})

≃-sym-is-equiv fe₀ fe₁ fe₂ = invertibles-are-equivs ≃-sym
                                (≃-sym ,
                                 ≃-sym-involutive fe₀ fe₂ ,
                                 ≃-sym-involutive fe₁ fe₂)

≃-sym-≃ : dfunext 𝓥 (𝓤 ⊔ 𝓥)
        → dfunext 𝓤 (𝓤 ⊔ 𝓥)
        → dfunext (𝓤 ⊔ 𝓥) (𝓤 ⊔ 𝓥)
        → (X : 𝓤 ̇ ) (Y : 𝓥 ̇ )
        → (X ≃ Y) ≃ (Y ≃ X)

≃-sym-≃ fe₀ fe₁ fe₂ X Y = ≃-sym , ≃-sym-is-equiv fe₀ fe₁ fe₂
```

*Exercise*. The hlevels are closed under `Σ` and, using `hfunext`, also under `Π`. Univalence is not needed, but makes the proof easier. (Without univalence, we need to show that the hlevels are closed under equivalence first.)

```
Π-cong : dfunext 𝓤 𝓥
       → dfunext 𝓤 𝓦
       → {X : 𝓤 ̇ } {Y : X → 𝓥 ̇ } {Y' : X → 𝓦 ̇ }
       → ((x : X) → Y x ≃ Y' x) → Π Y ≃ Π Y'

Π-cong fe fe' {X} {Y} {Y'} φ = invertibility-gives-≃ F (G , GF , FG)
 where
  f : (x : X) → Y x → Y' x
  f x = ⌜ φ x ⌝

  e : (x : X) → is-equiv (f x)
  e x = ⌜⌝-is-equiv (φ x)
```

```
  g : (x : X) → Y' x → Y x
  g x = inverse (f x) (e x)

  fg : (x : X) (y' : Y' x) → f x (g x y') ≡ y'
  fg x = inverses-are-sections (f x) (e x)

  gf : (x : X) (y : Y x) → g x (f x y) ≡ y
  gf x = inverses-are-retractions (f x) (e x)

  F : ((x : X) → Y x) → ((x : X) → Y' x)
  F φ x = f x (φ x)

  G : ((x : X) → Y' x) → (x : X) → Y x
  G γ x = g x (γ x)

  FG : (γ : ((x : X) → Y' x)) → F(G γ) ≡ γ
  FG γ = fe' (λ x → fg x (γ x))

  GF : (φ : ((x : X) → Y x)) → G(F φ) ≡ φ
  GF φ = fe (λ x → gf x (φ x))
```

An application of `Π-cong` is `hfunext₂-≃`:

```
hfunext-≃ : hfunext 𝓤 𝓥
          → {X : 𝓤 ̇ } {A : X → 𝓥 ̇ } (f g : Π A)
          → (f ≡ g) ≃ (f ∼ g)

hfunext-≃ hfe f g = (happly f g , hfe f g)

hfunext₂-≃ : hfunext 𝓤 (𝓥 ⊔ 𝓦) → hfunext 𝓥 𝓦
           → {X : 𝓤 ̇ } {Y : X → 𝓥 ̇ } {A : (x : X) → Y x → 𝓦 ̇ }
             (f g : (x : X) (y : Y x) → A x y)
           → (f ≡ g) ≃ (∀ x y → f x y ≡ g x y)

hfunext₂-≃ fe fe' {X} f g =

 (f ≡ g)                  ≃⟨ i  ⟩
 (∀ x → f x ≡ g x)        ≃⟨ ii ⟩
 (∀ x y → f x y ≡ g x y)  ■

 where
  i  = hfunext-≃ fe f g
  ii = Π-cong
        (hfunext-gives-dfunext fe)
        (hfunext-gives-dfunext fe)
        (λ x → hfunext-≃ fe' (f x) (g x))

precomp-invertible : dfunext 𝓥 𝓦
                   → dfunext 𝓤 𝓦
                   → {X : 𝓤 ̇ } {Y : 𝓥 ̇ } {Z : 𝓦 ̇ } (f : X → Y)
                   → invertible f
                   → invertible (λ (h : Y → Z) → h ∘ f)

precomp-invertible fe fe' {X} {Y} {Z} f (g , η , ε) = (g' , η' , ε')
 where
  f' : (Y → Z) → (X → Z)
  f' h = h ∘ f

  g' : (X → Z) → (Y → Z)
  g' k = k ∘ g

  η' : (h : Y → Z) → g' (f' h) ≡ h
  η' h = fe (λ y → ap h (ε y))

  ε' : (k : X → Z) → f' (g' k) ≡ k
```

```
  ε' k = fe' (λ x → ap k (η x))
```

We have already proved the following assuming univalence, in order to derive function extensionality from univalence. Now we prove it assuming function extensionality instead.

```
precomp-is-equiv' : dfunext 𝓥 𝓦
                  → dfunext 𝓤 𝓦
                  → {X : 𝓤 ̇ } {Y : 𝓥 ̇ } {Z : 𝓦 ̇ } (f : X → Y)
                  → is-equiv f
                  → is-equiv (λ (h : Y → Z) → h ∘ f)

precomp-is-equiv' fe fe' {X} {Y} {Z} f i = invertibles-are-equivs (_∘ f)
                                            (precomp-invertible fe fe' f
                                              (equivs-are-invertible f i))
```

More generally, using the half-adjoint condition for equivalences, we have the following dependent version of precomposition, which is also an equivalence assuming function extensionality.

```
dprecomp : {X : 𝓤 ̇ } {Y : 𝓥 ̇ } (A : Y → 𝓦 ̇ ) (f : X → Y)
         → Π A → Π (A ∘ f)

dprecomp A f = _∘ f

dprecomp-is-equiv : dfunext 𝓤 𝓦
                  → dfunext 𝓥 𝓦
                  → {X : 𝓤 ̇ } {Y : 𝓥 ̇ } (A : Y → 𝓦 ̇ ) (f : X → Y)
                  → is-equiv f
                  → is-equiv (dprecomp A f)

dprecomp-is-equiv fe fe' {X} {Y} A f i = invertibles-are-equivs φ (ψ , ψφ , φψ)
 where
  g = inverse f i
  η = inverses-are-retractions f i
  ε = inverses-are-sections f i

  τ : (x : X) → ap f (η x) ≡ ε (f x)
  τ = half-adjoint-condition f i

  φ : Π A → Π (A ∘ f)
  φ = dprecomp A f

  ψ : Π (A ∘ f) → Π A
  ψ k y = transport A (ε y) (k (g y))

  φψ₀ : (k : Π (A ∘ f)) (x : X) → transport A (ε (f x)) (k (g (f x))) ≡ k x
  φψ₀ k x = transport A (ε (f x))    (k (g (f x))) ≡⟨ a ⟩
            transport A (ap f (η x))(k (g (f x))) ≡⟨ b ⟩
            transport (A ∘ f) (η x) (k (g (f x))) ≡⟨ c ⟩
            k x                                    ∎
    where
     a = ap (λ - → transport A - (k (g (f x)))) ((τ x)⁻¹)
     b = (transport-ap A f (η x) (k (g (f x))))⁻¹
     c = apd k (η x)

  φψ : φ ∘ ψ ∼ id
  φψ k = fe (φψ₀ k)

  ψφ₀ : (h : Π A) (y : Y) → transport A (ε y) (h (f (g y))) ≡ h y
  ψφ₀ h y = apd h (ε y)

  ψφ : ψ ∘ φ ∼ id
  ψφ h = fe' (ψφ₀ h)
```

This amounts to saying that we can also change variables in Π:

```
Π-change-of-variable : dfunext 𝓤 𝓦
                     → dfunext 𝓥 𝓦
```

```
                → {X : 𝓤 ̇ } {Y : 𝓥 ̇ } (A : Y → 𝓦 ̇ ) (f : X → Y)
                → is-equiv f
                → (Π y ꞉ Y , A y) ≃ (Π x ꞉ X , A (f x))

Π-change-of-variable fe fe' A f i = dprecomp A f , dprecomp-is-equiv fe fe' A f i
```

Recall that a function is a Joyal equivalence if it has a section and it has a retraction. We now show that this notion is a subsingleton. For that purpose, we first show that if a function has a retraction then it has at most one section, and that if it has a section then it has at most one retraction.

```
at-most-one-section : dfunext 𝓥 𝓤
                    → hfunext 𝓥 𝓥
                    → {X : 𝓤 ̇ } {Y : 𝓥 ̇ } (f : X → Y)
                    → has-retraction f
                    → is-subsingleton (has-section f)

at-most-one-section {𝓥} {𝓤} fe hfe {X} {Y} f (g , gf) (h , fh) = d
 where
  fe' : dfunext 𝓥 𝓥
  fe' = hfunext-gives-dfunext hfe

  a : invertible f
  a = joyal-equivs-are-invertible f ((h , fh) , (g , gf))

  b : is-singleton (fiber (λ h →  f ∘ h) id)
  b = invertibles-are-equivs (λ h → f ∘ h) (postcomp-invertible fe fe' f a) id

  r : fiber (λ h →  f ∘ h) id → has-section f
  r (h , p) = (h , happly (f ∘ h) id p)

  s : has-section f → fiber (λ h → f ∘ h) id
  s (h , η) = (h , fe' η)

  rs : (σ : has-section f) → r (s σ) ＝ σ
  rs (h , η) = to-Σ-＝' q
   where
    q : happly (f ∘ h) id (inverse (happly (f ∘ h) id) (hfe (f ∘ h) id) η) ＝ η
    q = inverses-are-sections (happly (f ∘ h) id) (hfe (f ∘ h) id) η

  c : is-singleton (has-section f)
  c = retract-of-singleton (r , s , rs) b

  d : (σ : has-section f) → h , fh ＝ σ
  d = singletons-are-subsingletons (has-section f) c (h , fh)

at-most-one-retraction : hfunext 𝓤 𝓤
                       → dfunext 𝓥 𝓤
                       → {X : 𝓤 ̇ } {Y : 𝓥 ̇ } (f : X → Y)
                       → has-section f
                       → is-subsingleton (has-retraction f)

at-most-one-retraction {𝓤} {𝓥} hfe fe' {X} {Y} f (g , fg) (h , hf) = d
 where
  fe : dfunext 𝓤 𝓤
  fe = hfunext-gives-dfunext hfe

  a : invertible f
  a = joyal-equivs-are-invertible f ((g , fg) , (h , hf))

  b : is-singleton (fiber (λ h →  h ∘ f) id)
  b = invertibles-are-equivs (λ h → h ∘ f) (precomp-invertible fe' fe f a) id

  r : fiber (λ h →  h ∘ f) id → has-retraction f
  r (h , p) = (h , happly (h ∘ f) id p)

  s : has-retraction f → fiber (λ h →  h ∘ f) id
  s (h , η) = (h , fe η)

  rs : (σ : has-retraction f) → r (s σ) ＝ σ
```

```
  rs (h , η) = to-Σ-＝' q
   where
    q : happly (h ∘ f) id (inverse (happly (h ∘ f) id) (hfe (h ∘ f) id) η) ＝ η
    q = inverses-are-sections (happly (h ∘ f) id) (hfe (h ∘ f) id) η

  c : is-singleton (has-retraction f)
  c = retract-of-singleton (r , s , rs) b

  d : (ρ : has-retraction f) → h , hf ＝ ρ
  d = singletons-are-subsingletons (has-retraction f) c (h , hf)

being-joyal-equiv-is-subsingleton : hfunext 𝓤 𝓤
                                  → hfunext 𝓥 𝓥
                                  → dfunext 𝓥 𝓤
                                  → {X : 𝓤 ̇ } {Y : 𝓥 ̇ }
                                  → (f : X → Y)
                                  → is-subsingleton (is-joyal-equiv f)

being-joyal-equiv-is-subsingleton fe₀ fe₁ fe₂ f = ×-is-subsingleton'
                                                   (at-most-one-section    fe₂ fe₁ f ,
                                                    at-most-one-retraction fe₀ fe₂ f)
```

The fact that a function with a retraction has at most one section can also be used to prove that the notion of half adjoint equivalence is property. This is because the type `is-hae f` is equivalent to the type

```
     Σ (g , ε) : has-section f , ∀ x → (g (f x) , ε (f x)) ＝ (x , refl (f x)),
```

where the equality is in the type `fiber f (f x)`.

```
being-hae-is-subsingleton : dfunext 𝓥 𝓤
                          → hfunext 𝓥 𝓥
                          → dfunext 𝓤 (𝓥 ⊔ 𝓤)
                          → {X : 𝓤 ̇ } {Y : 𝓥 ̇ } (f : X → Y)
                          → is-subsingleton (is-hae f)

being-hae-is-subsingleton fe₀ fe₁ fe₂ {X} {Y} f = subsingleton-criterion' γ
 where
  a = λ g ε x
    → ((g (f x) , ε (f x)) ＝ (x , refl (f x)))                          ≃⟨ i  g ε x ⟩
      (Σ p ꞉ g (f x) ＝ x , transport (λ - → f - ＝ f x) p (ε (f x)) ＝ refl (f x)) ≃⟨ ii g ε x ⟩
      (Σ p ꞉ g (f x) ＝ x , ap f p ＝ ε (f x))                             ■
   where
    i  = λ g ε x → Σ-＝-≃ (g (f x) , ε (f x)) (x , refl (f x))
    ii = λ g ε x → Σ-cong (λ p → transport-ap-≃ f p (ε (f x)))

  b = (Σ (g , ε) ꞉ has-section f , ∀ x → (g (f x) , ε (f x)) ＝ (x , refl (f x)))         ≃⟨ i   ⟩
      (Σ (g , ε) ꞉ has-section f , ∀ x → Σ  p ꞉ g (f x) ＝ x , ap f p ＝ ε (f x))          ≃⟨ ii  ⟩
      (Σ g ꞉ (Y → X) , Σ ε ꞉ f ∘ g ∼ id , ∀ x → Σ  p ꞉ g (f x) ＝ x , ap f p ＝ ε (f x))   ≃⟨ iii ⟩
      (Σ g ꞉ (Y → X) , Σ ε ꞉ f ∘ g ∼ id , Σ η ꞉ g ∘ f ∼ id , ∀ x → ap f (η x) ＝ ε (f x)) ≃⟨ iv  ⟩
      is-hae f                                                                            ■
   where
    i   = Σ-cong (λ (g , ε) → Π-cong fe₂ fe₂ (a g ε))
    ii  = Σ-assoc
    iii = Σ-cong (λ g → Σ-cong (λ ε → ΠΣ-distr-≃))
    iv  = Σ-cong (λ g → Σ-flip)

  γ : is-hae f → is-subsingleton (is-hae f)
  γ (g₀ , ε₀ , η₀ , τ₀) = i
   where
    c : (x : X) → is-set (fiber f (f x))
    c x = singletons-are-sets (fiber f (f x)) (haes-are-equivs f (g₀ , ε₀ , η₀ , τ₀) (f x))

    d : ((g , ε) : has-section f) → is-subsingleton (∀ x → (g (f x) , ε (f x)) ＝ (x , refl (f x)))
    d (g , ε) = Π-is-subsingleton fe₂ (λ x → c x (g (f x) , ε (f x)) (x , refl (f x)))

    e : is-subsingleton (Σ (g , ε) ꞉ has-section f , ∀ x → (g (f x) , ε (f x)) ＝ (x , refl (f x)))
```

```
    e = Σ-is-subsingleton (at-most-one-section fe₀ fe₁ f (g₀ , ε₀)) d

    i : is-subsingleton (is-hae f)
    i = equiv-to-subsingleton (≃-sym b) e
```

Another consequence of function extensionality is that emptiness is a subsingleton:

```
emptiness-is-subsingleton : dfunext 𝓤 𝓤₀ → (X : 𝓤 ̇ )
                          → is-subsingleton (is-empty X)

emptiness-is-subsingleton fe X f g = fe (λ x → !𝟘 (f x ≡ g x) (f x))
```

If `P` is a subsingleton, then so is `P + ¬ P`. More generally:

```
+-is-subsingleton : {P : 𝓤 ̇ } {Q : 𝓥 ̇ }
                  → is-subsingleton P
                  → is-subsingleton Q
                  → (P → Q → 𝟘) → is-subsingleton (P + Q)

+-is-subsingleton {𝓤} {𝓥} {P} {Q} i j f = γ
 where
  γ : (x y : P + Q) → x ≡ y
  γ (inl p) (inl p') = ap inl (i p p')
  γ (inl p) (inr q)  = !𝟘 (inl p ≡ inr q) (f p q)
  γ (inr q) (inl p)  = !𝟘 (inr q ≡ inl p) (f p q)
  γ (inr q) (inr q') = ap inr (j q q')

+-is-subsingleton' : dfunext 𝓤 𝓤₀
                   → {P : 𝓤 ̇ } → is-subsingleton P → is-subsingleton (P + ¬ P)

+-is-subsingleton' fe {P} i = +-is-subsingleton i
                               (emptiness-is-subsingleton fe P)
                               (λ p n → n p)

EM-is-subsingleton : dfunext (𝓤 ⁺) 𝓤
                   → dfunext 𝓤 𝓤
                   → dfunext 𝓤 𝓤₀
                   → is-subsingleton (EM 𝓤)

EM-is-subsingleton fe⁺ fe fe₀ = Π-is-subsingleton fe⁺
                                 (λ P → Π-is-subsingleton fe
                                         (λ i → +-is-subsingleton' fe₀ i))
```

## Propositional extensionality

We have been using the mathematical terminology "subsingleton", but tradition in the formulation of the next notion demands the logical terminology "proposition". We now discuss the propositional extensionality axiom, its relation to univalence for propositions, and its use for dealing with powersets.

## The propositional extensionality axiom

Propositional extensionality says that any two logically equivalent propositions are equal:

```
propext : ∀ 𝓤  → 𝓤 ⁺ ̇
propext 𝓤 = {P Q : 𝓤 ̇ } → is-prop P → is-prop Q → (P → Q) → (Q → P) → P ≡ Q

global-propext : 𝓤ω
global-propext = ∀ {𝓤} → propext 𝓤
```

This is directly implied by univalence:

```
univalence-gives-propext : is-univalent 𝓤 → propext 𝓤
univalence-gives-propext ua {P} {Q} i j f g = Eq→Id ua P Q γ
 where
  γ : P ≃ Q
  γ = logically-equivalent-subsingletons-are-equivalent P Q i j (f , g)
```

## Propositional extensionality and the powerset

We also need a version of propositional extensionality for the type Ω 𝓤 of subsingletons in a given universe 𝓤, which lives in the next universe:

```
Ω : (𝓤 : Universe) → 𝓤 ⁺ ̇
Ω 𝓤 = Σ P ꞉ 𝓤 ̇ , is-subsingleton P

_holds : Ω 𝓤 → 𝓤 ̇
_holds (P , i) = P

holds-is-subsingleton : (p : Ω 𝓤) → is-subsingleton (p holds)
holds-is-subsingleton (P , i) = i

Ω-ext : dfunext 𝓤 𝓤 → propext 𝓤 → {p q : Ω 𝓤}
      → (p holds → q holds) → (q holds → p holds) → p ≡ q

Ω-ext {𝓤} fe pe {p} {q} f g = to-subtype-≡
                                   (λ _ → being-subsingleton-is-subsingleton fe)
                                   (pe (holds-is-subsingleton p) (holds-is-subsingleton q) f g)
```

With this and Hedberg, we get that Ω is a set:

```
Ω-is-a-set : dfunext 𝓤 𝓤 → propext 𝓤 → is-set (Ω 𝓤)
Ω-is-a-set {𝓤} fe pe = types-with-wconstant-≡-endomaps-are-sets (Ω 𝓤) c
 where
  A : (p q : Ω 𝓤) → 𝓤 ̇
  A p q = (p holds → q holds) × (q holds → p holds)

  i : (p q : Ω 𝓤) → is-subsingleton (A p q)
  i p q = Σ-is-subsingleton
           (Π-is-subsingleton fe
             (λ _ → holds-is-subsingleton q))
             (λ _ → Π-is-subsingleton fe (λ _ → holds-is-subsingleton p))

  g : (p q : Ω 𝓤) → p ≡ q → A p q
  g p q e = (u , v)
   where
    a : p holds ≡ q holds
    a = ap _holds e

    u : p holds → q holds
    u = Id→fun a

    v : q holds → p holds
    v = Id→fun (a ⁻¹)

  h : (p q : Ω 𝓤) → A p q → p ≡ q
  h p q (u , v) = Ω-ext fe pe u v

  f : (p q : Ω 𝓤) → p ≡ q → p ≡ q
  f p q e = h p q (g p q e)

  k : (p q : Ω 𝓤) (d e : p ≡ q) → f p q d ≡ f p q e
  k p q d e = ap (h p q) (i p q (g p q d) (g p q e))

  c : (p q : Ω 𝓤) → Σ f ꞉ (p ≡ q → p ≡ q), wconstant f
  c p q = (f p q , k p q)
```

Hence powersets, even of types that are not sets, are always sets.

```
powersets-are-sets : hfunext 𝓤 (𝓥 ⁺)
                   → dfunext 𝓥 𝓥
                   → propext 𝓥
                   → {X : 𝓤 ̇ } → is-set (X → Ω 𝓥)

powersets-are-sets fe fe' pe = Π-is-set fe (λ x → Ω-is-a-set fe' pe)
```

The above considers `X : 𝓤` and `Ω 𝓥`. When the two universes `𝓤` and `𝓥` are the same, we adopt the usual notation `𝓟 X` for the powerset `X → Ω 𝓤` of `X`.

```
𝓟 : 𝓤 ̇ → 𝓤 ⁺ ̇
𝓟 {𝓤} X = X → Ω 𝓤

powersets-are-sets' : Univalence
                    → {X : 𝓤 ̇ } → is-set (𝓟 X)

powersets-are-sets' {𝓤} ua = powersets-are-sets
                               (univalence-gives-hfunext' (ua 𝓤) (ua (𝓤 ⁺)))
                               (univalence-gives-dfunext (ua 𝓤))
                               (univalence-gives-propext (ua 𝓤))
```

Notice also that both `Ω` and the powerset live in the next universe. With propositional resizing, we get equivalent copies in the same universe.

Membership and containment for elements of the powerset are defined as follows:

```
_∈_ : {X : 𝓤 ̇ } → X → 𝓟 X → 𝓤 ̇
x ∈ A = A x holds

_∉_ : {X : 𝓤 ̇ } → X → 𝓟 X → 𝓤 ̇
x ∉ A = ¬(x ∈ A)

_⊆_ : {X : 𝓤 ̇ } → 𝓟 X → 𝓟 X → 𝓤 ̇
A ⊆ B = ∀ x → x ∈ A → x ∈ B

∈-is-subsingleton : {X : 𝓤 ̇ } (A : 𝓟 X) (x : X) → is-subsingleton (x ∈ A)
∈-is-subsingleton A x = holds-is-subsingleton (A x)

⊆-is-subsingleton : dfunext 𝓤 𝓤
                  → {X : 𝓤 ̇ } (A B : 𝓟 X) → is-subsingleton (A ⊆ B)

⊆-is-subsingleton fe A B = Π-is-subsingleton fe
                            (λ x → Π-is-subsingleton fe
                            (λ _ → ∈-is-subsingleton B x))

⊆-refl : {X : 𝓤 ̇ } (A : 𝓟 X) → A ⊆ A
⊆-refl A x = id (x ∈ A)

⊆-refl-consequence : {X : 𝓤 ̇ } (A B : 𝓟 X)
                   → A ≡ B → (A ⊆ B) × (B ⊆ A)

⊆-refl-consequence {X} A A (refl A) = ⊆-refl A , ⊆-refl A
```

Although `𝓟 X` is a set even if `X` is not, the total space `Σ x : X , A x holds` of a member `A : 𝓟 X` of the powerset need not be a set. For instance, if `A x holds = 𝟙` for all `x : X`, then the total space is equivalent to `X`, which may not be a set.

Propositional and functional extensionality give the usual extensionality condition for the powerset:

```
subset-extensionality : propext 𝓤
                      → dfunext 𝓤 𝓤
                      → dfunext 𝓤 (𝓤 ⁺)
                      → {X : 𝓤 ̇ } {A B : 𝓟 X}
                      → A ⊆ B → B ⊆ A → A ≡ B
```

```
subset-extensionality pe fe fe' {X} {A} {B} h k = fe' φ
 where
  φ : (x : X) → A x ≡ B x
  φ x = to-subtype-≡
           (λ _ → being-subsingleton-is-subsingleton fe)
           (pe (holds-is-subsingleton (A x)) (holds-is-subsingleton (B x))
               (h x) (k x))
```

And hence so does univalence:

```
subset-extensionality' : Univalence
                       → {X : 𝓤 ̇ } {A B : 𝓟 X}
                       → A ⊆ B → B ⊆ A → A ≡ B

subset-extensionality' {𝓤} ua = subset-extensionality
                                 (univalence-gives-propext (ua 𝓤))
                                 (univalence-gives-dfunext (ua 𝓤))
                                 (univalence-gives-dfunext' (ua 𝓤) (ua (𝓤 ⁺)))
```

For set-level mathematics, function extensionality and propositional extensionality are often the only consequences of univalence that are needed. A noteworthy exception is the theorem that the type of ordinals in a universe is an ordinal in the next universe, which requires univalence for sets (see the HoTT book or this).

## A characterization of propositional univalence

This section is not needed anywhere else and can be safely skipped. We characterize propositional univalence in terms of propositional extensionality and a restricted form of function extensionality.

By propositional univalence mean any of the following two equivalent notions.

```
prop-univalence prop-univalence' : (𝓤 : Universe) → 𝓤 ⁺ ̇
prop-univalence  𝓤 = (A : 𝓤 ̇ ) → is-prop A → (X : 𝓤 ̇ ) → is-equiv (Id→Eq A X)
prop-univalence' 𝓤 = (A : 𝓤 ̇ ) → is-prop A → (X : 𝓤 ̇ ) → is-prop X → is-equiv (Id→Eq A X)

prop-univalence-agreement : prop-univalence' 𝓤 ⇔ prop-univalence 𝓤
prop-univalence-agreement = (λ pu' A i X e → pu' A i X (equiv-to-subsingleton (≃-sym e) i) e) ,
                            (λ pu  A i X _ → pu  A i X)
```

The restricted form of function extensionality is given by any of the following four notions, which are equivalent under propositional extensionality and have the following type:

```
props-form-exponential-ideal
 props-are-closed-under-Π
 prop-vvfunext
 prop-hfunext : ∀ 𝓤 → 𝓤 ⁺ ̇
```

They are defined as follows.

```
props-form-exponential-ideal 𝓤 = (X A : 𝓤 ̇ ) → is-prop A → is-prop (X → A)

props-are-closed-under-Π 𝓤 = {X : 𝓤 ̇ } {A : X → 𝓤 ̇ }
                           → is-prop X
                           → ((x : X) → is-prop (A x))
                           → is-prop (Π A)

prop-vvfunext 𝓤 = {X : 𝓤 ̇ } {A : X → 𝓤 ̇ }
                → is-prop X
                → ((x : X) → is-singleton (A x))
                → is-singleton (Π A)

prop-hfunext 𝓤 = {X : 𝓤 ̇ } {A : X → 𝓤 ̇ }
               → is-prop X
               → (f g : Π A) → is-equiv (happly f g)
```

The last three agree unconditionally:

```
first-propositional-function-extensionality-agreement :

    (props-are-closed-under-Π 𝓤 → prop-vvfunext 𝓤)
  × (prop-vvfunext 𝓤             → prop-hfunext 𝓤)
  × (prop-hfunext 𝓤              → props-are-closed-under-Π 𝓤)
```

The proof is postponed. Moreover, as already mentioned, the four of them agree under propositional extensionality.

```
second-propositional-function-extensionality-agreement :

    propext 𝓤 → (props-form-exponential-ideal 𝓤 ⇔ props-are-closed-under-Π 𝓤)
```

The proof is again postponed, and this time there is a twist: we use a detour via propositional univalence to prove this.

*Question.* Does propositional function extensionality, in any of the above four incarnations, perhaps in the presence of propositional extensionality, imply full function extensionality?

*Discussion.* Notice that `props-form-exponential-ideal` requires `A` to be a proposition, but not `X`, whereas `prop-hfunext` requires `X` to be a proposition but not `A x`, while the other two versions require both `X` and `A(x)` to be propositions, and yet all versions are equivalent under propositional extensionality. Given this, a positive answer to the above question is not unlikely. We leave this as an open problem.

The following is the main theorem of this section, where the notion `props-form-an-exponential-ideal` plays a crucial role in its proof.

```
characterization-of-propositional-univalence : prop-univalence 𝓤
                                             ⇔ (propext 𝓤 × props-are-closed-under-Π 𝓤)
```

The remainder of this section is devoted to prove this, which has the second propositional function extensionality agreement as a consequence, but relies on the first agreement.

It is direct that propositional univalence implies propositional extensionality.

```
prop-univalence-gives-propext : prop-univalence 𝓤 → propext 𝓤
prop-univalence-gives-propext pu {P} {Q} i j f g = δ
 where
  γ : P ≃ Q
  γ = logically-equivalent-subsingletons-are-equivalent P Q i j (f , g)

  δ : P ＝ Q
  δ = inverse (Id→Eq P Q) (pu P i Q) γ
```

To show that propositional univalence implies that the propositions form an exponential ideal, we first need some lemmas.

```
prop-≃-induction : (𝓤 𝓥 : Universe) → (𝓤 ⊔ 𝓥)⁺ ̇
prop-≃-induction 𝓤 𝓥 = (P : 𝓤 ̇ )
                      → is-prop P
                      → (A : (X : 𝓤 ̇ ) → P ≃ X → 𝓥 ̇ )
                      → A P (id-≃ P) → (X : 𝓤 ̇ ) (e : P ≃ X) → A X e

prop-J-equiv : prop-univalence 𝓤
             → (𝓥 : Universe) → prop-≃-induction 𝓤 𝓥
prop-J-equiv {𝓤} pu 𝓥 P i A a X e = γ
 where
  A' : (X : 𝓤 ̇ ) → P ＝ X → 𝓥 ̇
  A' X q = A X (Id→Eq P X q)

  f : (X : 𝓤 ̇ ) (q : P ＝ X) → A' X q
  f = ℍ P A' a

  r : P ＝ X
  r = inverse (Id→Eq P X) (pu P i X) e

  g : A X (Id→Eq P X r)
```

```
  g = f X r

  γ : A X (id e)
  γ = transport (A X) (inverses-are-sections (Id→Eq P X) (pu P i X) e) g

prop-precomp-is-equiv : prop-univalence 𝓤
                      → (X Y Z : 𝓤 ̇ )
                      → is-prop X
                      → (f : X → Y)
                      → is-equiv f
                      → is-equiv (λ (g : Y → Z) → g ∘ f)
prop-precomp-is-equiv {𝓤} pu X Y Z i f f-is-equiv =
   prop-J-equiv pu 𝓤 X i (λ _ e → is-equiv (λ g → g ∘ ⌜ e ⌝))
     (id-is-equiv (X → Z)) Y (f , f-is-equiv)
```

We now apply the above lemmas to adapt the above proof that univalence implies function extensionality in order to obtain the following.

```
prop-univalence-gives-props-form-exponential-ideal : prop-univalence 𝓤
                                                   → props-form-exponential-ideal 𝓤

prop-univalence-gives-props-form-exponential-ideal {𝓤} pu X A A-is-prop = γ
 where
  Δ : 𝓤 ̇
  Δ = Σ a₀ ꞉ A , Σ a₁ ꞉ A , a₀ ＝ a₁

  δ : A → Δ
  δ a = (a , a , refl a)

  π₀ π₁ : Δ → A
  π₀ (a₀ , a₁ , a) = a₀
  π₁ (a₀ , a₁ , a) = a₁

  δ-is-equiv : is-equiv δ
  δ-is-equiv = invertibles-are-equivs δ (π₀ , η , ε)
   where
    η : (a : A) → π₀ (δ a) ＝ a
    η a = refl a

    ε : (d : Δ) → δ (π₀ d) ＝ d
    ε (a , a , refl a) = refl (a , a , refl a)

  φ : (Δ → A) → (A → A)
  φ π = π ∘ δ

  φ-is-equiv : is-equiv φ
  φ-is-equiv = prop-precomp-is-equiv pu A Δ A A-is-prop δ δ-is-equiv

  p : φ π₀ ＝ φ π₁
  p = refl (id A)

  q : π₀ ＝ π₁
  q = equivs-are-lc φ φ-is-equiv p

  h : (f₀ f₁ : X → A) → f₀ ∼ f₁
  h f₀ f₁ x = A-is-prop (f₀ x) (f₁ x)

  γ : (f₀ f₁ : X → A) → f₀ ＝ f₁
  γ f₀ f₁ = ap (λ π x → π (f₀ x , f₁ x , h f₀ f₁ x)) q
```

To prove the converse, we first establish the first propositional function extensionality agreement, in a number of steps. For this purpose we adapt some proofs regarding the various notions of (full) function extensionality.

```
props-are-closed-under-Π-gives-prop-vvfunext : props-are-closed-under-Π 𝓤 → prop-vvfunext 𝓤
props-are-closed-under-Π-gives-prop-vvfunext c {X} {A} X-is-prop A-is-prop-valued = γ
 where
  f : Π A
  f x = center (A x) (A-is-prop-valued x)
```

```
  d : (g : Π A) → f ≡ g
  d = c X-is-prop (λ (x : X) → singletons-are-subsingletons (A x) (A-is-prop-valued x)) f

  γ : is-singleton (Π A)
  γ = f , d

prop-vvfunext-gives-prop-hfunext : prop-vvfunext 𝓤 → prop-hfunext 𝓤
prop-vvfunext-gives-prop-hfunext vfe {X} {Y} X-is-prop f = γ
 where
  a : (x : X) → is-singleton (Σ y ꞉ Y x , f x ≡ y)
  a x = singleton-types'-are-singletons (Y x) (f x)

  c : is-singleton (Π x ꞉ X , Σ y ꞉ Y x , f x ≡ y)
  c = vfe X-is-prop a

  ρ : (Σ g ꞉ Π Y , f ∼ g) ◁ (Π x ꞉ X , Σ y ꞉ Y x , f x ≡ y)
  ρ = ≃-gives-▷ ΠΣ-distr-≃

  d : is-singleton (Σ g ꞉ Π Y , f ∼ g)
  d = retract-of-singleton ρ c

  e : (Σ g ꞉ Π Y , f ≡ g) → (Σ g ꞉ Π Y , f ∼ g)
  e = NatΣ (happly f)

  i : is-equiv e
  i = maps-of-singletons-are-equivs e (singleton-types'-are-singletons (Π Y) f) d

  γ : (g : Π Y) → is-equiv (happly f g)
  γ = NatΣ-equiv-gives-fiberwise-equiv (happly f) i

prop-hfunext-gives-props-are-closed-under-Π : prop-hfunext 𝓤 → props-are-closed-under-Π 𝓤
prop-hfunext-gives-props-are-closed-under-Π hfe {X} {A} X-is-prop A-is-prop-valued f g = γ
 where
  γ : f ≡ g
  γ = inverse (happly f g) (hfe X-is-prop f g) (λ x → A-is-prop-valued x (f x) (g x))
```

With this we can now fulfill the first promise:

```
first-propositional-function-extensionality-agreement =
  props-are-closed-under-Π-gives-prop-vvfunext ,
  prop-vvfunext-gives-prop-hfunext ,
  prop-hfunext-gives-props-are-closed-under-Π
```

We have already proved the following lemma `being-prop-is-prop`, as `being-subsingleton-is-subsingleton`, but under the stronger assumption `vvfunext`. The proof is almost, but not literally, the same, and needs some of the above lemmas.

```
prop-vvfunext-gives-props-are-closed-under-Π : prop-vvfunext 𝓤 → props-are-closed-under-Π 𝓤
prop-vvfunext-gives-props-are-closed-under-Π vfe =
    prop-hfunext-gives-props-are-closed-under-Π (prop-vvfunext-gives-prop-hfunext vfe)

being-prop-is-prop : prop-vvfunext 𝓤
                   → {X : 𝓤 ̇ } → is-prop (is-prop X)

being-prop-is-prop vfe {X} i j = γ
 where
  k : is-set X
  k = subsingletons-are-sets X i

  a : (x y : X) → i x y ≡ j x y
  a x y = k x y (i x y) (j x y)

  b : (x : X) → i x ≡ j x
  b x = prop-vvfunext-gives-props-are-closed-under-Π vfe i
          (subsingletons-are-sets X i x) (i x) (j x)
```

```
  c : (x : X) → is-prop ((y : X) → x ＝ y)
  c x = singletons-are-subsingletons ((y : X) → x ＝ y)
          (vfe i (λ y → pointed-subsingletons-are-singletons (x ＝ y) (i x y) (k x y)))

  γ : i ＝ j
  γ = prop-vvfunext-gives-props-are-closed-under-Π vfe i c i j
```

Similarly, we have already proved the following, as `being-singleton-is-subsingleton`, under the stronger assumption `dfunext`.

```
being-singleton-is-prop : props-are-closed-under-Π 𝓤
                        → {X : 𝓤 ̇ } → is-prop (is-singleton X)

being-singleton-is-prop c {X} (x , φ) (y , γ) = p
 where
  i : is-subsingleton X
  i = singletons-are-subsingletons X (y , γ)

  s : is-set X
  s = subsingletons-are-sets X i

  a : (z : X) → is-subsingleton ((t : X) → z ＝ t)
  a z = c i (λ x → s z x)

  b : x ＝ y
  b = φ y

  p : (x , φ) ＝ (y , γ)
  p = to-subtype-＝ a b
```

The following hasn't occurred above in any form, and is crucial for the main theorem of this section.

```
Id-of-props-is-prop : propext 𝓤
                    → prop-vvfunext 𝓤
                    → (P : 𝓤 ̇ )
                    → is-prop P
                    → (X : 𝓤 ̇ ) → is-prop (P ＝ X)

Id-of-props-is-prop {𝓤} pe vfe P i = Hedberg P (λ X → h X , k X)
 where
  module _ (X : 𝓤 ̇ ) where
   f : P ＝ X → is-prop X × (P ⇔ X)
   f p = transport is-prop p i , Id→fun p , (Id→fun (p ⁻¹))

   g : is-prop X × (P ⇔ X) → P ＝ X
   g (l , φ , ψ) = pe i l φ ψ

   h : P ＝ X → P ＝ X
   h = g ∘ f

   j : is-prop (is-prop X × (P ⇔ X))
   j = ×-is-subsingleton'
        ((λ (_ : P ⇔ X) → being-prop-is-prop vfe) ,
         (λ (l : is-prop X)
              → ×-is-subsingleton
                 (prop-vvfunext-gives-props-are-closed-under-Π vfe i (λ _ → l))
                 (prop-vvfunext-gives-props-are-closed-under-Π vfe l (λ _ → i))))

   k : wconstant h
   k p q = ap g (j (f p) (f q))

being-equiv-of-props-is-prop : props-are-closed-under-Π 𝓤
                             → {X Y : 𝓤 ̇ }
                             → is-prop X
                             → is-prop Y
                             → (f : X → Y) → is-prop (is-equiv f)
```

```
being-equiv-of-props-is-prop c i j f = c j (λ y → being-singleton-is-prop c)
```

Armed with the above lemmas, we are now in a position to show that propositional extensionality and propositional function extensionality together imply propositional univalence.

```
propext-and-props-are-closed-under-Π-give-prop-univalence : propext 𝓤
                                                          → props-are-closed-under-Π 𝓤
                                                          → prop-univalence 𝓤

propext-and-props-are-closed-under-Π-give-prop-univalence pe c A i X = γ
 where
  l : A ≃ X → is-subsingleton X
  l e = equiv-to-subsingleton (≃-sym e) i

  eqtoid : A ≃ X → A ＝ X
  eqtoid e = pe i
              (equiv-to-subsingleton (≃-sym e) i)
             ⌜ e ⌝
             ⌜ ≃-sym e ⌝

  m : is-subsingleton (A ≃ X)
  m (f₀ , k₀) (f₁ , k₁) = δ
   where
    j : (f : A → X) → is-prop (is-equiv f)
    j = being-equiv-of-props-is-prop c i
          (equiv-to-subsingleton (≃-sym (f₀ , k₀)) i)

    p : f₀ ＝ f₁
    p = c i (λ (a : A) → l (f₁ , k₁)) f₀ f₁

    δ : (f₀ , k₀) ＝ (f₁ , k₁)
    δ = to-subtype-＝ j p

  ε : (e : A ≃ X) → Id→Eq A X (eqtoid e) ＝ e
  ε e = m (Id→Eq A X (eqtoid e)) e

  η : (q : A ＝ X) → eqtoid (Id→Eq A X q) ＝ q
  η q = Id-of-props-is-prop pe
         (props-are-closed-under-Π-gives-prop-vvfunext c)
         A i X
         (eqtoid (Id→Eq A X q)) q

  γ : is-equiv (Id→Eq A X)
  γ = invertibles-are-equivs (Id→Eq A X) (eqtoid , η , ε)
```

We now need some lemmas to prove the converse. The next two have already been proved under different hypotheses.

```
prop-postcomp-invertible : {X Y A : 𝓤 ̇ }
                         → props-form-exponential-ideal 𝓤
                         → is-prop X
                         → is-prop Y
                         → (f : X → Y)
                         → invertible f
                         → invertible (λ (h : A → X) → f ∘ h)

prop-postcomp-invertible {𝓤} {X} {Y} {A} pei i j f (g , η , ε) = γ
 where
  f' : (A → X) → (A → Y)
  f' h = f ∘ h

  g' : (A → Y) → (A → X)
  g' k = g ∘ k

  η' : (h : A → X) → g' (f' h) ＝ h
  η' h = pei A X i (g' (f' h)) h

  ε' : (k : A → Y) → f' (g' k) ＝ k
  ε' k = pei A Y j (f' (g' k)) k
```

```
  γ : invertible f'
  γ = (g' , η' , ε')

prop-postcomp-is-equiv : {X Y A : 𝓤 ̇ }
                       → props-form-exponential-ideal 𝓤
                       → is-prop X
                       → is-prop Y
                       → (f : X → Y)
                       → is-equiv f
                       → is-equiv (λ (h : A → X) → f ∘ h)

prop-postcomp-is-equiv pei i j f e =
 invertibles-are-equivs
  (λ h → f ∘ h)
  (prop-postcomp-invertible pei i j f (equivs-are-invertible f e))
```

The following shows that `props-form-exponential-ideal` is stronger than `prop-vvfunext` (and perhaps properly stronger in the absence of propositional extensionality).

```
props-form-exponential-ideal-gives-vvfunext : props-form-exponential-ideal 𝓤 → prop-vvfunext 𝓤
props-form-exponential-ideal-gives-vvfunext {𝓤} pei {X} {A} X-is-prop φ = γ
 where
  f : Σ A → X
  f = pr₁

  A-is-prop-valued : (x : X) → is-prop (A x)
  A-is-prop-valued x = singletons-are-subsingletons (A x) (φ x)

  k : is-prop (Σ A)
  k = Σ-is-subsingleton X-is-prop A-is-prop-valued

  f-is-equiv : is-equiv f
  f-is-equiv = pr₁-is-equiv φ

  g : (X → Σ A) → (X → X)
  g h = f ∘ h

  e : is-equiv g
  e = prop-postcomp-is-equiv pei k X-is-prop f f-is-equiv

  i : is-singleton (Σ h ꞉ (X → Σ A), f ∘ h ＝ id X)
  i = e (id X)

  r : (Σ h ꞉ (X → Σ A), f ∘ h ＝ id X) → Π A
  r (h , p) x = transport A (happly (f ∘ h) (id X) p x) (pr₂ (h x))

  s : Π A → (Σ h ꞉ (X → Σ A), f ∘ h ＝ id X)
  s ψ = (λ x → x , ψ x) , refl (id X)

  η : ∀ ψ → r (s ψ) ＝ ψ
  η ψ = refl (r (s ψ))

  γ : is-singleton (Π A)
  γ = retract-of-singleton (r , s , η) i
```

Here are some corollaries of the above.

```
prop-univalence-gives-props-are-closed-under-Π : prop-univalence 𝓤 → props-are-closed-under-Π 𝓤
prop-univalence-gives-props-are-closed-under-Π pu =
    prop-vvfunext-gives-props-are-closed-under-Π
        (props-form-exponential-ideal-gives-vvfunext
              (prop-univalence-gives-props-form-exponential-ideal pu))

propext-and-props-closed-under-Π-give-props-form-exponential-ideal :

    propext 𝓤
  → props-are-closed-under-Π 𝓤
```

```
  → props-form-exponential-ideal 𝓤

propext-and-props-closed-under-Π-give-props-form-exponential-ideal pe c =
  prop-univalence-gives-props-form-exponential-ideal
      (propext-and-props-are-closed-under-Π-give-prop-univalence pe c)

props-form-exponential-ideal-gives-props-are-closed-under-Π :

    props-form-exponential-ideal 𝓤
  → props-are-closed-under-Π 𝓤

props-form-exponential-ideal-gives-props-are-closed-under-Π pei =
     prop-vvfunext-gives-props-are-closed-under-Π
         (props-form-exponential-ideal-gives-vvfunext pei)
```

And with this we can complete the proof of the main theorem of this section, formulated above in Agda.

```
characterization-of-propositional-univalence {𝓤} = α , β
 where
  α : prop-univalence 𝓤 → propext 𝓤 × props-are-closed-under-Π 𝓤
  α pu =  prop-univalence-gives-propext pu , prop-univalence-gives-props-are-closed-under-Π pu

  β : propext 𝓤 × props-are-closed-under-Π 𝓤 → prop-univalence 𝓤
  β (pe , fe) = propext-and-props-are-closed-under-Π-give-prop-univalence pe fe
```

And we can also fulfill the second promise again using `prop-univalence-gives-props-are-exponential-ideal`.

```
second-propositional-function-extensionality-agreement {𝓤} pe =
  props-form-exponential-ideal-gives-props-are-closed-under-Π ,
  propext-and-props-closed-under-Π-give-props-form-exponential-ideal pe
```

## Some constructions with types of equivalences

We first prove some properties of equivalence symmetrization and composition:

```
id-≃-left : dfunext 𝓥 (𝓤 ⊔ 𝓥)
          → dfunext (𝓤 ⊔ 𝓥) (𝓤 ⊔ 𝓥)
          → {X : 𝓤 ̇ } {Y : 𝓥 ̇ } (α : X ≃ Y)
          → id-≃ X ● α ≡ α

id-≃-left fe fe' α = to-subtype-≡ (being-equiv-is-subsingleton fe fe') (refl _)

≃-sym-left-inverse : dfunext 𝓥 𝓥
                   → {X : 𝓤 ̇ } {Y : 𝓥 ̇ } (α : X ≃ Y)
                   → ≃-sym α ● α ≡ id-≃ Y

≃-sym-left-inverse fe (f , e) = to-subtype-≡ (being-equiv-is-subsingleton fe fe) p
 where
  p : f ∘ inverse f e ≡ id
  p = fe (inverses-are-sections f e)

≃-sym-right-inverse : dfunext 𝓤 𝓤
                    → {X : 𝓤 ̇ } {Y : 𝓥 ̇ } (α : X ≃ Y)
                    → α ● ≃-sym α ≡ id-≃ X

≃-sym-right-inverse fe (f , e) = to-subtype-≡ (being-equiv-is-subsingleton fe fe) p
 where
  p : inverse f e ∘ f ≡ id
  p = fe (inverses-are-retractions f e)
```

We then transfer the above to equivalence types:

```
≃-Sym : dfunext 𝓥 (𝓤 ⊔ 𝓥)
      → dfunext 𝓤 (𝓤 ⊔ 𝓥)
      → dfunext (𝓤 ⊔ 𝓥) (𝓤 ⊔ 𝓥)
      → {X : 𝓤 ̇ } {Y : 𝓥 ̇ }
      → (X ≃ Y) ≃ (Y ≃ X)

≃-Sym fe₀ fe₁ fe₂ = ≃-sym , ≃-sym-is-equiv fe₀ fe₁ fe₂

≃-Comp : dfunext 𝓦 (𝓥 ⊔ 𝓦)
       → dfunext (𝓥 ⊔ 𝓦) (𝓥 ⊔ 𝓦 )
       → dfunext 𝓥 𝓥
       → dfunext 𝓦 (𝓤 ⊔ 𝓦)
       → dfunext (𝓤 ⊔ 𝓦) (𝓤 ⊔ 𝓦 )
       → dfunext 𝓤 𝓤
       → {X : 𝓤 ̇ } {Y : 𝓥 ̇ } (Z : 𝓦 ̇ )
       → X ≃ Y → (Y ≃ Z) ≃ (X ≃ Z)

≃-Comp fe₀ fe₁ fe₂ fe₃ fe₄ fe₅ Z α = invertibility-gives-≃ (α ●_)
                                      ((≃-sym α ●_) , p , q)
 where
  p = λ β → ≃-sym α ● (α ● β) ≡⟨ ●-assoc fe₀ fe₁ (≃-sym α) α β        ⟩
            (≃-sym α ● α) ● β ≡⟨ ap (_● β) (≃-sym-left-inverse fe₂ α) ⟩
            id-≃ _ ● β        ≡⟨ id-≃-left fe₀ fe₁ _                  ⟩
            β                 ∎

  q = λ γ → α ● (≃-sym α ● γ) ≡⟨ ●-assoc fe₃ fe₄ α (≃-sym α) γ         ⟩
            (α ● ≃-sym α) ● γ ≡⟨ ap (_● γ) (≃-sym-right-inverse fe₅ α) ⟩
            id-≃ _ ● γ        ≡⟨ id-≃-left fe₃ fe₄ _                   ⟩
            γ                 ∎
```

Using this we get the following self-congruence property of equivalences:

```
Eq-Eq-cong' : dfunext 𝓥 (𝓤 ⊔ 𝓥)
            → dfunext (𝓤 ⊔ 𝓥) (𝓤 ⊔ 𝓥)
            → dfunext 𝓤 𝓤
            → dfunext 𝓥 (𝓥 ⊔ 𝓦)
            → dfunext (𝓥 ⊔ 𝓦) (𝓥 ⊔ 𝓦)
            → dfunext 𝓦 𝓦
            → dfunext 𝓦 (𝓥 ⊔ 𝓦)
            → dfunext 𝓥 𝓥
            → dfunext 𝓦 (𝓦 ⊔ 𝓣)
            → dfunext (𝓦 ⊔ 𝓣) (𝓦 ⊔ 𝓣)
            → dfunext 𝓣 𝓣
            → dfunext 𝓣 (𝓦 ⊔ 𝓣)
            → {X : 𝓤 ̇ } {Y : 𝓥 ̇ } {A : 𝓦 ̇ } {B : 𝓣 ̇ }
            → X ≃ A → Y ≃ B → (X ≃ Y) ≃ (A ≃ B)

Eq-Eq-cong' fe₀ fe₁ fe₂ fe₃ fe₄ fe₅ fe₆ fe₇ fe₈ fe₉ fe₁₀ fe₁₁ {X} {Y} {A} {B} α β =

  X ≃ Y  ≃⟨ ≃-Comp fe₀ fe₁ fe₂ fe₃ fe₄ fe₅ Y (≃-sym α)  ⟩
  A ≃ Y  ≃⟨ ≃-Sym fe₃ fe₆ fe₄                          ⟩
  Y ≃ A  ≃⟨ ≃-Comp fe₆ fe₄ fe₇ fe₈ fe₉ fe₁₀ A (≃-sym β) ⟩
  B ≃ A  ≃⟨ ≃-Sym fe₈ fe₁₁ fe₉                         ⟩
  A ≃ B  ■
```

The above shows why global function extensionality would be a better assumption in practice, and so we define:

```
Eq-Eq-cong : global-dfunext
           → {X : 𝓤 ̇ } {Y : 𝓥 ̇ } {A : 𝓦 ̇ } {B : 𝓣 ̇ }
           → X ≃ A → Y ≃ B → (X ≃ Y) ≃ (A ≃ B)

Eq-Eq-cong fe = Eq-Eq-cong' fe fe fe fe fe fe fe fe fe fe fe fe
```

## Type embeddings

A function is called an embedding if its fibers are all subsingletons. In particular, equivalences are embeddings. However, sections of types more general than sets don't need to be embeddings.

```
is-embedding : {X : 𝓤 ̇ } {Y : 𝓥 ̇ } → (X → Y) → 𝓤 ⊔ 𝓥 ̇
is-embedding f = (y : codomain f) → is-subsingleton (fiber f y)
```

This says that for every `y : Y` there is at most one `x : X` with `f x ≡ y`, or, more precisely, there is at most one pair `(x , p)` with `x : X` and `p : f x ≡ y`.

```
being-embedding-is-subsingleton : global-dfunext
                                → {X : 𝓤 ̇ } {Y : 𝓥 ̇ } (f : X → Y)
                                → is-subsingleton (is-embedding f)

being-embedding-is-subsingleton fe f =
 Π-is-subsingleton fe
  (λ x → being-subsingleton-is-subsingleton fe)
```

For example, if `A` is a subsingleton, then the second projection `A × X → X` is an embedding:

```
pr₂-embedding : (A : 𝓤 ̇ ) (X : 𝓥 ̇ )
              → is-subsingleton A
              → is-embedding (λ (z : A × X) → pr₂ z)

pr₂-embedding A X i x ((a , x) , refl x) ((b , x) , refl x) = p
 where
  p : ((a , x) , refl x) ≡ ((b , x) , refl x)
  p = ap (λ - → ((- , x) , refl x)) (i a b)
```

*Exercise*. Show that the converse of `pr₂-embedding` holds.

More generally, with the arguments swapped, the projection `Σ A → X` is an embedding if `A x` is a subsingleton for every `x : X`:

```
pr₁-is-embedding : {X : 𝓤 ̇ } {A : X → 𝓥 ̇ }
                 → ((x : X) → is-subsingleton (A x))
                 → is-embedding (λ (σ : Σ A) → pr₁ σ)

pr₁-is-embedding i x ((x , a) , refl x) ((x , a') , refl x) = γ
 where
  p : a ≡ a'
  p = i x a a'

  γ : (x , a) , refl x ≡ (x , a') , refl x
  γ = ap (λ - → (x , -) , refl x) p
```

*Exercise*. Show that the converse of `pr₁-is-embedding` holds.

```
equivs-are-embeddings : {X : 𝓤 ̇ } {Y : 𝓥 ̇ }
                        (f : X → Y)
                      → is-equiv f
                      → is-embedding f

equivs-are-embeddings f i y = singletons-are-subsingletons (fiber f y) (i y)

id-is-embedding : {X : 𝓤 ̇ } → is-embedding (id X)
id-is-embedding {𝓤} {X} = equivs-are-embeddings id (id-is-equiv X)

∘-embedding : {X : 𝓤 ̇ } {Y : 𝓥 ̇ } {Z : 𝓦 ̇ }
              {f : X → Y} {g : Y → Z}
            → is-embedding g
            → is-embedding f
            → is-embedding (g ∘ f)

∘-embedding {𝓤} {𝓥} {𝓦} {X} {Y} {Z} {f} {g} d e = h
 where
  A : (z : Z) → 𝓤 ⊔ 𝓥 ⊔ 𝓦 ̇
```

```
  A z = Σ (y , p) ꞉ fiber g z , fiber f y

  i : (z : Z) → is-subsingleton (A z)
  i z = Σ-is-subsingleton (d z) (λ w → e (pr₁ w))

  φ : (z : Z) → fiber (g ∘ f) z → A z
  φ z (x , p) = (f x , p) , x , refl (f x)

  γ : (z : Z) → A z → fiber (g ∘ f) z
  γ z ((_ , p) , x , refl _) = x , p

  η : (z : Z) (t : fiber (g ∘ f) z) → γ z (φ z t) ＝ t
  η _ (x , refl _) = refl (x , refl ((g ∘ f) x))

  h : (z : Z) → is-subsingleton (fiber (g ∘ f) z)
  h z = lc-maps-reflect-subsingletons (φ z) (sections-are-lc (φ z) (γ z , η z)) (i z)
```

We can use the following criterion to prove that some maps are embeddings:

```
embedding-lemma : {X : 𝓤 ̇ } {Y : 𝓥 ̇ } (f : X → Y)
                → ((x : X) → is-singleton (fiber f (f x)))
                → is-embedding f

embedding-lemma f φ = γ
 where
  γ : (y : codomain f) (u v : fiber f y) → u ＝ v
  γ y (x , p) v = j (x , p) v
   where
    q : fiber f (f x) ＝ fiber f y
    q = ap (fiber f) p

    i : is-singleton (fiber f y)
    i = transport is-singleton q (φ x)

    j : is-subsingleton (fiber f y)
    j = singletons-are-subsingletons (fiber f y) i

embedding-criterion : {X : 𝓤 ̇ } {Y : 𝓥 ̇ } (f : X → Y)
                    → ((x x' : X) → (f x ＝ f x') ≃ (x ＝ x'))
                    → is-embedding f

embedding-criterion f e = embedding-lemma f b
 where
  X = domain f

  a : (x : X) → (Σ x' ꞉ X , f x' ＝ f x) ≃ (Σ x' ꞉ X , x' ＝ x)
  a x = Σ-cong (λ x' → e x' x)

  a' : (x : X) → fiber f (f x) ≃ singleton-type x
  a' = a

  b : (x : X) → is-singleton (fiber f (f x))
  b x = equiv-to-singleton (a' x) (singleton-types-are-singletons X x)
```

An equivalent formulation of `f` being an embedding is that the map

```
    ap f {x} {x'} : x ＝ x' → f x ＝ f x'
```

is an equivalence for all `x x' : X`.

```
ap-is-equiv-gives-embedding : {X : 𝓤 ̇ } {Y : 𝓥 ̇ } (f : X → Y)
                            → ((x x' : X) → is-equiv (ap f {x} {x'}))
                            → is-embedding f

ap-is-equiv-gives-embedding f i = embedding-criterion f
                                   (λ x' x → ≃-sym (ap f {x'} {x} , i x' x))
```

```
embedding-gives-ap-is-equiv : {X : 𝓤 ̇ } {Y : 𝓥 ̇ } (f : X → Y)
                            → is-embedding f
                            → (x x' : X) → is-equiv (ap f {x} {x'})

embedding-gives-ap-is-equiv {𝓤} {𝓥} {X} f e = γ
 where
  d : (x' : X) → (Σ x ꞉ X , f x' ＝ f x) ≃ (Σ x ꞉ X , f x ＝ f x')
  d x' = Σ-cong (λ x → ⁻¹-≃ (f x') (f x))

  s : (x' : X) → is-subsingleton (Σ x ꞉ X , f x' ＝ f x)
  s x' = equiv-to-subsingleton (d x') (e (f x'))

  γ : (x x' : X) → is-equiv (ap f {x} {x'})
  γ x = singleton-equiv-lemma x (λ x' → ap f {x} {x'})
         (pointed-subsingletons-are-singletons
           (Σ x' ꞉ X , f x ＝ f x') (x , (refl (f x))) (s x))

embedding-criterion-converse : {X : 𝓤 ̇ } {Y : 𝓥 ̇ } (f : X → Y)
                             → is-embedding f
                             → ((x' x : X) → (f x' ＝ f x) ≃ (x' ＝ x))

embedding-criterion-converse f e x' x = ≃-sym
                                         (ap f {x'} {x} ,
                                          embedding-gives-ap-is-equiv f e x' x)
```

Hence embeddings of arbitrary types are left cancellable, but the converse fails in general.

```
embeddings-are-lc : {X : 𝓤 ̇ } {Y : 𝓥 ̇ } (f : X → Y)
                  → is-embedding f
                  → left-cancellable f

embeddings-are-lc f e {x} {y} = ⌜ embedding-criterion-converse f e x y ⌝
```

Conversely, left cancellable maps into *sets* are always embeddings.

```
lc-maps-into-sets-are-embeddings : {X : 𝓤 ̇ } {Y : 𝓥 ̇ } (f : X → Y)
                                 → left-cancellable f
                                 → is-set Y
                                 → is-embedding f
lc-maps-into-sets-are-embeddings {𝓤} {𝓥} {X} {Y} f lc i y = γ
 where
  γ : is-subsingleton (Σ x ꞉ X , f x ＝ y)
  γ (x , p) (x' , p') = to-subtype-＝ j q
   where
    j : (x : X) → is-subsingleton (f x ＝ y)
    j x = i (f x) y

    q : x ＝ x'
    q = lc (f x  ＝⟨ p     ⟩
            y    ＝⟨ p' ⁻¹ ⟩
            f x' ∎)
```

If an embedding has a section, then it is an equivalence.

```
embedding-with-section-is-equiv : {X : 𝓤 ̇ } {Y : 𝓥 ̇ } (f : X → Y)
                                → is-embedding f
                                → has-section f
                                → is-equiv f
embedding-with-section-is-equiv f i (g , η) y = pointed-subsingletons-are-singletons
                                                 (fiber f y) (g y , η y) (i y)
```

Later we will see that a necessary and sufficient condition for an embedding to be an equivalence is that it is a surjection.

If a type `Y` is embedded into `Z`, then the function type `X → Y` is embedded into `X → Z`. More generally, if `A x` is embedded into `B x` for every `x : X`, then the dependent function type `Π A` is embedded into `Π B`.

```
NatΠ : {X : 𝓤 ̇ } {A : X → 𝓥 ̇ } {B : X → 𝓦 ̇ } → Nat A B → Π A → Π B
NatΠ τ f x = τ x (f x)
```

(Notice that `NatΠ` is a dependently typed version of the combinator `S` from combinatory logic. Its logical interpretation, here, is that if `A x` implies `B x` for all `x : X`, and `A x` holds for all `x : X`, then `B x` holds for all `x : X` too.)

```
NatΠ-is-embedding : hfunext 𝓤 𝓥
                  → hfunext 𝓤 𝓦
                  → {X : 𝓤 ̇ } {A : X → 𝓥 ̇ } {B : X → 𝓦 ̇ }
                  → (τ : Nat A B)
                  → ((x : X) → is-embedding (τ x))
                  → is-embedding (NatΠ τ)

NatΠ-is-embedding v w {X} {A} τ i = embedding-criterion (NatΠ τ) γ
 where
  γ : (f g : Π A) → (NatΠ τ f ≡ NatΠ τ g) ≃ (f ≡ g)
  γ f g = (NatΠ τ f ≡ NatΠ τ g) ≃⟨ hfunext-≃ w (NatΠ τ f) (NatΠ τ g) ⟩
          (NatΠ τ f ∼ NatΠ τ g) ≃⟨ b                                ⟩
          (f ∼ g)               ≃⟨ ≃-sym (hfunext-≃ v f g)            ⟩
          (f ≡ g)               ■

   where
    a : (x : X) → (NatΠ τ f x ≡ NatΠ τ g x) ≃ (f x ≡ g x)
    a x = embedding-criterion-converse (τ x) (i x) (f x) (g x)

    b : (NatΠ τ f ∼ NatΠ τ g) ≃ (f ∼ g)
    b = Π-cong (hfunext-gives-dfunext w) (hfunext-gives-dfunext v) a
```

Postcomposition with an embedding is itself an embedding (of a function type into another). This amounts to saying that any function `f : X → A` and any embedding `g : Y → A` can be completed to a commutative triangle in at most one way:

```
triangle-lemma : dfunext 𝓦 (𝓤 ⊔ 𝓥)
               → {Y : 𝓤 ̇ } {A : 𝓥 ̇ } (g : Y → A)
               → is-embedding g
               → {X : 𝓦 ̇ } (f : X → A) → is-subsingleton (Σ h ꞉ (X → Y) , g ∘ h ∼ f)

triangle-lemma fe {Y} {A} g i {X} f = iv
 where
  ii : (x : X) → is-subsingleton (Σ y ꞉ Y , g y ≡ f x)
  ii x = i (f x)

  iii : is-subsingleton (Π x ꞉ X , Σ y ꞉ Y , g y ≡ f x)
  iii = Π-is-subsingleton fe ii

  iv : is-subsingleton (Σ h ꞉ (X → Y) , g ∘ h ∼ f)
  iv = equiv-to-subsingleton (≃-sym ΠΣ-distr-≃) iii

postcomp-is-embedding : dfunext 𝓦 (𝓤 ⊔ 𝓥) → hfunext 𝓦 𝓥
                      → {Y : 𝓤 ̇ } {A : 𝓥 ̇ } (g : Y → A)
                      → is-embedding g
                      → (X : 𝓦 ̇ ) → is-embedding (λ (h : X → Y) → g ∘ h)

postcomp-is-embedding fe hfe {Y} {A} g i X = γ
 where
  γ : (f : X → A) → is-subsingleton (Σ h ꞉ (X → Y) , g ∘ h ≡ f)
  γ f = equiv-to-subsingleton u (triangle-lemma fe g i f)
   where
    u : (Σ h ꞉ (X → Y) , g ∘ h ≡ f) ≃ (Σ h ꞉ (X → Y) , g ∘ h ∼ f)
    u = Σ-cong (λ h → hfunext-≃ hfe (g ∘ h) f)
```

We conclude this section by introducing notation for the type of embeddings.

```
_↪_ : 𝓤 ̇ → 𝓥 ̇ → 𝓤 ⊔ 𝓥 ̇
X ↪ Y = Σ f ꞉ (X → Y), is-embedding f

Emb→fun : {X : 𝓤 ̇ } {Y : 𝓥 ̇ } → (X ↪ Y) → (X → Y)
Emb→fun (f , i) = f
```

The following justifies the terminology *subsingleton*:

*Exercise*. (1) Show that the type `is-subsingleton X` is logically equivalent to the type `X ↪ 𝟙`. (2) Hence assuming function extensionality and propositional extensionality, conclude that `is-subsingleton X ≡ (X ↪ 𝟙)`.

*Exercise*. Show that the map `Fin : ℕ → 𝓤₀` defined above is left-cancellable but not an embedding.

## The Yoneda Lemma for types

As we have seen, a type `X` can be seen as an ∞-groupoid and hence as an ∞-category, with identifications as the arrows. Likewise a universe `𝓤` can be seen as the ∞-generalization of the category of sets, with functions as the arrows. Hence a type family

```
A : X → 𝓤
```

can be seen as an ∞-presheaf, because groupoids are self-dual categories.

With this view, the identity type former `Id X : X → X → 𝓤` plays the role of the Yoneda embedding:

```
𝓨 : {X : 𝓤 ̇ } → X → (X → 𝓤 ̇ )
𝓨 {𝓤} {X} = Id X
```

Sometimes we want to make one of the parameters explicit:

```
𝑌 : (X : 𝓤 ̇ ) → X → (X → 𝓤 ̇ )
𝑌 {𝓤} X = 𝓨 {𝓤} {X}
```

By our definition of `Nat`, for any `A : X → 𝓥 ̇` and `x : X` we have

```
Nat (𝓨 x) A = (y : X) → x ≡ y → A y,
```

and, by `Nats-are-natural`, we have that `Nat (𝓨 x) A` is the type of natural transformations from the presheaf `𝓨 x` to the presheaf `A`. The starting point of the Yoneda Lemma, in our context, is that every such natural transformation is a transport.

```
transport-lemma : {X : 𝓤 ̇ } (A : X → 𝓥 ̇ ) (x : X)
                → (τ : Nat (𝓨 x) A)
                → (y : X) (p : x ≡ y) → τ y p ≡ transport A p (τ x (refl x))

transport-lemma A x τ x (refl x) = refl (τ x (refl x))
```

We denote the point `τ x (refl x)` in the above lemma by `𝓔 A x τ` as refer to it as the *Yoneda element* of the transformation `τ`.

```
𝓔 : {X : 𝓤 ̇ } (A : X → 𝓥 ̇ ) (x : X) → Nat (𝓨 x) A → A x
𝓔 A x τ = τ x (refl x)
```

The function

```
𝓔 A x : Nat (𝓨 x) A → A x
```

is an equivalence with inverse

```
𝓝 A x : A x → Nat (𝓨 x) A,
```

the transport natural transformation induced by `A` and `x`:

```
𝓝 : {X : 𝓤 ̇ } (A : X → 𝓥 ̇ ) (x : X) → A x → Nat (𝓨 x) A
𝓝 A x a y p = transport A p a
```

```
yoneda-η : dfunext 𝓤 (𝓤 ⊔ 𝓥)
         → dfunext 𝓤 𝓥
         → {X : 𝓤 ̇ } (A : X → 𝓥 ̇ ) (x : X)
```

```
         → 𝓝 A x ∘ 𝓔 A x ∼ id

yoneda-η fe fe' A x = γ
 where
  γ : (τ : Nat (𝓨 x) A) → (λ y p → transport A p (τ x (refl x))) ≡ τ
  γ τ = fe (λ y → fe' (λ p → (transport-lemma A x τ y p)⁻¹))

yoneda-ε : {X : 𝓤 ̇ } (A : X → 𝓥 ̇ ) (x : X)
         → 𝓔 A x ∘ 𝓝 A x ∼ id

yoneda-ε A x = γ
 where
  γ : (a : A x) → transport A (refl x) a ≡ a
  γ = refl
```

By a fiberwise equivalence we mean a natural transformation whose components are all equivalences:

```
is-fiberwise-equiv : {X : 𝓤 ̇ } {A : X → 𝓥 ̇ } {B : X → 𝓦 ̇ } → Nat A B → 𝓤 ⊔ 𝓥 ⊔ 𝓦 ̇
is-fiberwise-equiv τ = ∀ x → is-equiv (τ x)

𝓔-is-equiv : dfunext 𝓤 (𝓤 ⊔ 𝓥)
           → dfunext 𝓤 𝓥
           → {X : 𝓤 ̇ } (A : X → 𝓥 ̇ )
           → is-fiberwise-equiv (𝓔 A)

𝓔-is-equiv fe fe' A x = invertibles-are-equivs (𝓔 A x )
                         (𝓝 A x , yoneda-η fe fe' A x , yoneda-ε A x)

𝓝-is-equiv : dfunext 𝓤 (𝓤 ⊔ 𝓥)
           → dfunext 𝓤 𝓥
           → {X : 𝓤 ̇ } (A : X → 𝓥 ̇ )
           → is-fiberwise-equiv (𝓝 A)

𝓝-is-equiv fe fe' A x = invertibles-are-equivs (𝓝 A x)
                         (𝓔 A x , yoneda-ε A x , yoneda-η fe fe' A x)
```

This gives the Yoneda Lemma for types, which says that natural transformations from `𝓨 x` to `A` are in canonical bijection with elements of `A x`:

```
Yoneda-Lemma : dfunext 𝓤 (𝓤 ⊔ 𝓥)
             → dfunext 𝓤 𝓥
             → {X : 𝓤 ̇ } (A : X → 𝓥 ̇ ) (x : X)
             → Nat (𝓨 x) A ≃ A x

Yoneda-Lemma fe fe' A x = 𝓔 A x , 𝓔-is-equiv fe fe' A x
```

A universal element of a presheaf `A` corresponds in our context to an element of the type `is-singleton (Σ A)`, which can also be written `∃! A`.

If the transport transformation is a fiberwise equivalence, then `A` has a universal element. More generally, we have the following:

```
retract-universal-lemma : {X : 𝓤 ̇ } (A : X → 𝓥 ̇ ) (x : X)
                        → ((y : X) → A y ◁ (x ≡ y))
                        → ∃! A

retract-universal-lemma A x ρ = i
 where
  σ : Σ A ◁ singleton-type' x
  σ = Σ-retract ρ

  i : ∃! A
  i = retract-of-singleton σ (singleton-types'-are-singletons (domain A) x)

fiberwise-equiv-universal : {X : 𝓤 ̇ } (A : X → 𝓥 ̇ )
```

```
                                (x : X) (τ : Nat (𝓨 x) A)
                              → is-fiberwise-equiv τ
                              → ∃! A

fiberwise-equiv-universal A x τ e = retract-universal-lemma A x ρ
 where
  ρ : ∀ y → A y ◁ (x ＝ y)
  ρ y = ≃-gives-▷ ((τ y) , e y)
```

Conversely:

```
universal-fiberwise-equiv : {X : 𝓤 ̇ } (A : X → 𝓥 ̇ )
                          → ∃! A
                          → (x : X) (τ : Nat (𝓨 x) A) → is-fiberwise-equiv τ

universal-fiberwise-equiv {𝓤} {𝓥} {X} A u x τ = γ
 where
  g : singleton-type' x → Σ A
  g = NatΣ τ

  e : is-equiv g
  e = maps-of-singletons-are-equivs g (singleton-types'-are-singletons X x) u

  γ : is-fiberwise-equiv τ
  γ = NatΣ-equiv-gives-fiberwise-equiv τ e
```

In particular, the induced transport transformation `τ = 𝓝 A x a` is a fiberwise equivalence if and only if there is a unique `x : X` with `A x`, which we abbreviate as `∃! A`.

A corollary is the following characterization of function extensionality, similar to the above characterization of univalence, by taking the transformation `τ = happly f`, because `hfe f` says that `τ` is a fiberwise equivalence:

```
hfunext→ : hfunext 𝓤 𝓥
         → ((X : 𝓤 ̇ ) (A : X → 𝓥 ̇ ) (f : Π A) → ∃! g ꞉ Π A , f ∼ g)

hfunext→ hfe X A f = fiberwise-equiv-universal (f ∼_) f (happly f) (hfe f)

→hfunext : ((X : 𝓤 ̇ ) (A : X → 𝓥 ̇ ) (f : Π A) → ∃! g ꞉ Π A , f ∼ g)
         → hfunext 𝓤 𝓥

→hfunext φ {X} {A} f = universal-fiberwise-equiv (f ∼_) (φ X A f) f (happly f)
```

We also have the following general corollaries:

```
fiberwise-equiv-criterion : {X : 𝓤 ̇ } (A : X → 𝓥 ̇ )
                            (x : X)
                          → ((y : X) → A y ◁ (x ＝ y))
                          → (τ : Nat (𝓨 x) A) → is-fiberwise-equiv τ

fiberwise-equiv-criterion A x ρ τ = universal-fiberwise-equiv A
                                     (retract-universal-lemma A x ρ) x τ
```

This says that if we have a fiberwise retraction, then any natural transformation is an equivalence. And the following says that if we have a fiberwise equivalence, then any natural transformation is a fiberwise equivalence:

```
fiberwise-equiv-criterion' : {X : 𝓤 ̇ } (A : X → 𝓥 ̇ )
                             (x : X)
                           → ((y : X) → (x ＝ y) ≃ A y)
                           → (τ : Nat (𝓨 x) A) → is-fiberwise-equiv τ

fiberwise-equiv-criterion' A x e = fiberwise-equiv-criterion A x
                                    (λ y → ≃-gives-▷ (e y))
```

A presheaf is called *representable* if it is pointwise equivalent to a presheaf of the form `𝓨 x`:

```
_≃̇_ : {X : 𝓤 ̇ } → (X → 𝓥 ̇ ) → (X → 𝓦 ̇ ) → 𝓤 ⊔ 𝓥 ⊔ 𝓦 ̇
A ≃̇ B = ∀ x → A x ≃ B x
```

```
is-representable : {X : 𝓤 ̇ } → (X → 𝓥 ̇ ) → 𝓤 ⊔ 𝓥 ̇
is-representable A = Σ x ꞉ domain A , 𝓨 x ≃̇ A

representable-universal : {X : 𝓤 ̇ } (A : X → 𝓥 ̇ )
                        → is-representable A
                        → ∃! A

representable-universal A (x , e) = retract-universal-lemma A x
                                     (λ x → ≃-gives-▷ (e x))

universal-representable : {X : 𝓤 ̇ } {A : X → 𝓥 ̇ }
                        → ∃! A
                        → is-representable A

universal-representable {𝓤} {𝓥} {X} {A} ((x , a) , p) = x , φ
 where
  e : is-fiberwise-equiv (𝓝 A x a)
  e = universal-fiberwise-equiv A ((x , a) , p) x (𝓝 A x a)

  φ : (y : X) → (x ＝ y) ≃ A y
  φ y = (𝓝 A x a y , e y)
```

Combining `retract-universal-lemma` and `universal-fiberwise-equiv` we get the following:

```
fiberwise-retractions-are-equivs : {X : 𝓤 ̇ } (A : X → 𝓥 ̇ ) (x : X)
                                 → (τ : Nat (𝓨 x) A)
                                 → ((y : X) → has-section (τ y))
                                 → is-fiberwise-equiv τ

fiberwise-retractions-are-equivs {𝓤} {𝓥} {X} A x τ s = γ
 where
  ρ : (y : X) → A y ◁ (x ＝ y)
  ρ y = τ y , s y

  i : ∃! A
  i = retract-universal-lemma A x ρ

  γ : is-fiberwise-equiv τ
  γ = universal-fiberwise-equiv A i x τ
```

Perhaps the following formulation is more appealing:

```
fiberwise-◁-gives-≃ : (X : 𝓤 ̇ ) (A : X → 𝓥 ̇ ) (x : X)
                    → ((y : X) → A y ◁ (x ＝ y))
                    → ((y : X) → A y ≃ (x ＝ y))

fiberwise-◁-gives-≃ X A x ρ = γ
 where
  f : (y : X) → (x ＝ y) → A y
  f y = retraction (ρ y)

  e : is-fiberwise-equiv f
  e = fiberwise-retractions-are-equivs A x f (λ y → retraction-has-section (ρ y))

  γ : (y : X) → A y ≃ (x ＝ y)
  γ y = ≃-sym(f y , e y)
```

We have the following corollary:

```
embedding-criterion' : {X : 𝓤 ̇ } {Y : 𝓥 ̇ } (f : X → Y)
                     → ((x x' : X) → (f x ＝ f x') ◁ (x ＝ x'))
                     → is-embedding f

embedding-criterion' f ρ = embedding-criterion f
```

```
                    (λ x → fiberwise-◁-gives-≃ (domain f)
                             (λ - → f x ＝ f -) x (ρ x))
```

*Exercise.* It also follows that `f` is an embedding if and only if the map `ap f {x} {x'}` has a section.

To prove that `𝓨 {𝓤} {X}` is an embedding of `X` into `X → 𝓤` for any type `X : 𝓤`, we need the following two lemmas, which are interesting in their own right:

```
being-fiberwise-equiv-is-subsingleton : global-dfunext
                                       → {X : 𝓤 ̇ } {A : X → 𝓥 ̇ } {B : X → 𝓦 ̇ }
                                       → (τ : Nat A B)
                                       → is-subsingleton (is-fiberwise-equiv τ)

being-fiberwise-equiv-is-subsingleton fe τ =
 Π-is-subsingleton fe (λ y → being-equiv-is-subsingleton fe fe (τ y))

being-representable-is-subsingleton : global-dfunext
                                     → {X : 𝓤 ̇ } (A : X → 𝓥 ̇ )
                                     → is-subsingleton (is-representable A)

being-representable-is-subsingleton fe {X} A r₀ r₁ = γ
 where
  u : ∃! A
  u = representable-universal A r₀

  i : (x : X) (τ : Nat (𝓨 x) A) → is-singleton (is-fiberwise-equiv τ)
  i x τ = pointed-subsingletons-are-singletons
           (is-fiberwise-equiv τ)
           (universal-fiberwise-equiv A u x τ)
           (being-fiberwise-equiv-is-subsingleton fe τ)

  ε : (x : X) → (𝓨 x ≐ A) ≃ A x
  ε x = ((y : X) → 𝓨 x y ≃ A y)                         ≃⟨ ΠΣ-distr-≃          ⟩
        (Σ τ ꞉ Nat (𝓨 x) A , is-fiberwise-equiv τ)     ≃⟨ pr₁-≃ (i x)          ⟩
        Nat (𝓨 x) A                                     ≃⟨ Yoneda-Lemma fe fe A x ⟩
        A x                                             ■

  δ : is-representable A ≃ Σ A
  δ = Σ-cong ε

  v : is-singleton (is-representable A)
  v = equiv-to-singleton δ u

  γ : r₀ ＝ r₁
  γ = singletons-are-subsingletons (is-representable A) v r₀ r₁
```

With this it is almost immediate that the Yoneda map `𝑌 X` is an embedding of `X` into `X → 𝓤`:

```
𝓨-is-embedding : Univalence → (X : 𝓤 ̇ ) → is-embedding (𝑌 X)
𝓨-is-embedding {𝓤} ua X A = γ
 where
  hfe : global-hfunext
  hfe = univalence-gives-global-hfunext ua

  dfe : global-dfunext
  dfe = univalence-gives-global-dfunext ua

  p = λ x → (𝓨 x ＝ A)                  ≃⟨ i  x ⟩
            ((y : X) → 𝓨 x y ＝ A y)    ≃⟨ ii x ⟩
            ((y : X) → 𝓨 x y ≃ A y)     ■
    where
     i  = λ x → (happly (𝓨 x) A , hfe (𝓨 x) A)
     ii = λ x → Π-cong dfe dfe
                 (λ y → univalence-≃ (ua 𝓤)
                         (𝓨 x y) (A y))

  e : fiber 𝓨 A ≃ is-representable A
  e = Σ-cong p
```

```
  γ : is-subsingleton (fiber 𝓨 A)
  γ = equiv-to-subsingleton e (being-representable-is-subsingleton dfe A)
```

## What is a function?

In set theory, a function is a relation between two sets that associates to every element of the first set exactly one element of the second set. We say that the relation is *functional*.

In type theory, on the other hand, the notion of function is taken as primitive. However, we can show that the type of functions is equivalent to the type of functional relations. This relies on univalence.

When the types under consideration are sets, the corresponding relations are *truth valued*. But for the equivalence between functions and functional relations to hold for arbitrary types, we need to work with *type valued* relations.

More generally, we have a one-to-one corresponce between dependent functions `f : (x : X) → A x` and dependent type valued functional relations `R : (x : X) → A x → 𝓥`. We take the domain `X` and codomain `A` as parameters for a submodule:

```
module function-graphs
        (ua : Univalence)
        {𝓤 𝓥 : Universe}
        (X : 𝓤 ̇ )
        (A : X → 𝓥 ̇ )
       where

 hfe : global-hfunext
 hfe = univalence-gives-global-hfunext ua

 fe : global-dfunext
 fe = univalence-gives-global-dfunext ua
```

The type of dependent functions:

```
 Function : 𝓤 ⊔ 𝓥 ̇
 Function = (x : X) → A x
```

That of dependent relations:

```
 Relation : 𝓤 ⊔ (𝓥 ⁺) ̇
 Relation = (x : X) → A x → 𝓥 ̇
```

A relation `R` is said to be functional if for every `x : X` there is a unique `a : A x` with `R x a`:

```
 is-functional : Relation → 𝓤 ⊔ 𝓥 ̇
 is-functional R = (x : X) → ∃! a ꞉ A x , R x a
```

Although the relation is allowed to take values in arbitrary types, its functionality condition is a truth value:

```
 being-functional-is-subsingleton : (R : Relation)
                                  → is-subsingleton (is-functional R)

 being-functional-is-subsingleton R = Π-is-subsingleton fe
                                       (λ x → ∃!-is-subsingleton (R x) fe)
```

The type of functional relations:

```
 Functional-Relation : 𝓤 ⊔ (𝓥 ⁺) ̇
 Functional-Relation = Σ R ꞉ Relation , is-functional R
```

To a function `f` we associate the relation `R` defined by `R x a = (f x ≡ a)`. Notice that `R` is truth valued if the type `A x` is a set for every `x : X`, by definition of set.

```
 ρ : Function → Relation
 ρ f = λ x a → f x ≡ a
```

To show that the map `ρ` is an embedding we apply the Yoneda embedding and the fact that the map `NatΠ` transforms fiberwise embeddings into embeddings:

```
ρ-is-embedding : is-embedding ρ
ρ-is-embedding = NatΠ-is-embedding hfe hfe
                  (λ x → Y (A x))
                  (λ x → 𝓨-is-embedding ua (A x))
 where
```

This relies implicitly on the following remarks:

```
  τ : (x : X) → A x → (A x → 𝓥 ̇ )
  τ x a b = a ≡ b

  remark₀ : τ ≡ λ x → Y (A x)
  remark₀ = refl _

  remark₁ : ρ ≡ NatΠ τ
  remark₁ = refl _
```

The relation induced by a function is functional, of course:

```
ρ-is-functional : (f : Function) → is-functional (ρ f)
ρ-is-functional f = σ
 where
  σ : (x : X) → ∃! a ꞉ A x , f x ≡ a
  σ x = singleton-types'-are-singletons (A x) (f x)
```

The graph map associates functional relations to functions:

```
Γ : Function → Functional-Relation
Γ f = ρ f , ρ-is-functional f
```

The function `Γ` can be seen as the corestriction of `ρ` to its image.

We get a function from a functional relation by unique choice, which is just projection:

```
Φ : Functional-Relation → Function
Φ (R , σ) = λ x → pr₁ (center (Σ a ꞉ A x , R x a) (σ x))
```

To show that these two constructions are mutually inverse, we again apply the Yoneda machinery, but in a different way.

```
Γ-is-equiv : is-equiv Γ
Γ-is-equiv = invertibles-are-equivs Γ (Φ , η , ε)
 where
  η : Φ ∘ Γ ∼ id
  η = refl

  ε : Γ ∘ Φ ∼ id
  ε (R , σ) = a
   where
    f : Function
    f = Φ (R , σ)

    e : (x : X) → R x (f x)
    e x = pr₂ (center (Σ a ꞉ A x , R x a) (σ x))

    τ : (x : X) → Nat (𝓨 (f x)) (R x)
    τ x = 𝓝 (R x) (f x) (e x)

    τ-is-fiberwise-equiv : (x : X) → is-fiberwise-equiv (τ x)
    τ-is-fiberwise-equiv x = universal-fiberwise-equiv (R x) (σ x) (f x) (τ x)

    d : (x : X) (a : A x) → (f x ≡ a) ≃ R x a
    d x a = τ x a , τ-is-fiberwise-equiv x a

    c : (x : X) (a : A x) → (f x ≡ a) ≡ R x a
    c x a = Eq→Id (ua 𝓥) _ _ (d x a)
```

```
  b : ρ f ≡ R
  b = fe (λ x → fe (c x))

  a : (ρ f , ρ-is-functional f) ≡ (R , σ)
  a = to-subtype-≡ being-functional-is-subsingleton b
```

Therefore we have a bijection between functions and functional relations:

```
 functions-amount-to-functional-relations : Function ≃ Functional-Relation
 functions-amount-to-functional-relations = Γ , Γ-is-equiv
```

## Partial functions

Based on the previous section, we can define a *partial function* to be a relation `R` such that for every `x : X` there is *at most one* `a : A x` with `R x a`. We use `Πₚ` for the type of dependent partial functions and `⇀` for the type of partial functions.

```
Πₚ : {X : 𝓤 ̇ } → (X → 𝓥 ̇ ) → 𝓤 ⊔ (𝓥 ⁺) ̇
Πₚ {𝓤} {𝓥} {X} A = Σ R ꞉ ((x : X) → A x → 𝓥 ̇ )
                      , ((x : X) → is-subsingleton (Σ a ꞉ A x , R x a))

_⇀_ : 𝓤 ̇ → 𝓥 ̇ → 𝓤 ⊔ (𝓥 ⁺) ̇
X ⇀ Y = Πₚ (λ (_ : X) → Y)
```

*Exercise.* Define partial function composition, both in non-dependent and dependent versions.

```
is-defined : {X : 𝓤 ̇ } {A : X → 𝓥 ̇ } → Πₚ A → X → 𝓥 ̇
is-defined (R , σ) x = Σ a ꞉ _ , R x a

being-defined-is-subsingleton : {X : 𝓤 ̇ } {A : X → 𝓥 ̇ } (f : Πₚ A) (x : X)
                              → is-subsingleton (is-defined f x)

being-defined-is-subsingleton (R , σ) x = σ x
```

Notice that we have to write `is-defined f x`, and we say that `f` is defined at `x`, or that `x` is in the domain of definition of `f`, rather than `is-defined (f x)`. In fact, before being able to evaluate a partial function `f` at an argument `x`, we need to know that `f` is defined at `x`. However, in informal discussions we will say "`f x` is defined" by the usual abuse of notation and terminology.

We will write the application of a partial function `f` to `x`, under the information `i` that `f x` is defined, as `f [ x , i ]`.

```
_[_,_] :  {X : 𝓤 ̇ } {A : X → 𝓥 ̇ } (f : Πₚ A) (x : X) → is-defined f x → A x
(R , s) [ x , (a , r)] = a
```

*Exercise.* Define Kleene equality of two partial functions `f g : Πₚ A` by saying that for all `x : X`, if whenever one of `f x` and `g x` is defined then so is the other, and whenever they are both defined, then they are equal:

```
_≡ₖ_ : {X : 𝓤 ̇ } {A : X → 𝓥 ̇ } → Πₚ A → Πₚ A → 𝓤 ⊔ 𝓥 ̇
f ≡ₖ g = ∀ x → (is-defined f x ⇔ is-defined g x)
             × ((i : is-defined f x) (j : is-defined g x) → f [ x , i ] ≡ g [ x , j ])
```

Show that the equality of two partial functions, in the sense of the identity type, is equivalent to their Kleene equality. This needs univalence. If all types `A x` are sets, then functional and propositional extensionality suffice. In the general case, it is easier, or less hard, to approach this problem using the chapter on equality of mathematical structures.

*Example.* The famous μ-operator from recursion theory is a partial function.

```
module μ-operator (fe : dfunext 𝓤₀ 𝓤₀) where

 open basic-arithmetic-and-order
```

First we need to show that the property of being a minimal root is a truth value and that the type of minimal roots has at most one element. It is this that requires function extensionality.

```
being-minimal-root-is-subsingleton : (f : ℕ → ℕ) (m : ℕ)
                                   → is-subsingleton (is-minimal-root f m)

being-minimal-root-is-subsingleton f m = ×-is-subsingleton
                                          (ℕ-is-set (f m) 0)
                                          (Π-is-subsingleton fe
                                             (λ _ → Π-is-subsingleton fe
                                             (λ _ → Π-is-subsingleton fe
                                             (λ _ → 𝟘-is-subsingleton))))

minimal-root-is-subsingleton : (f : ℕ → ℕ)
                             → is-subsingleton (minimal-root f)

minimal-root-is-subsingleton f (m , p , φ) (m' , p' , φ') =
   to-subtype-≡
    (being-minimal-root-is-subsingleton f)
    (at-most-one-minimal-root f m m' (p , φ) (p' , φ'))
```

We now define `μ` so that if `f` has a root then `μ f` is defined, and, conversely, if `μ f` is defined then it is the minimal root of `f`. Most of the work has already been done in the module `basic-arithmetic-and-order`.

```
μ : (ℕ → ℕ) ⇀ ℕ
μ = is-minimal-root , minimal-root-is-subsingleton

μ-property₀ : (f : ℕ → ℕ) → (Σ n ꞉ ℕ , f n ≡ 0) → is-defined μ f
μ-property₀ = root-gives-minimal-root

μ-property₁ : (f : ℕ → ℕ) (i : is-defined μ f)
            → (f (μ [ f , i ]) ≡ 0)
            × ((n : ℕ) → n < μ [ f , i ] → f n ≢ 0)

μ-property₁ f = pr₂
```

*Exercise*. Define

```
is-total : {X : 𝓤 ̇ } {A : X → 𝓥 ̇ } → Πₚ A → 𝓤 ⊔ 𝓥 ̇
is-total f = ∀ x → is-defined f x
```

Show that the type `Σ f ꞉ Πₚ A , is-total f` of total partial functions is equivalent to the type `Π A` of functions. In particular, the type `Σ f ꞉ X ⇀ Y , is-total f` is equivalent to the type `X → Y`.

*Exercise*. Two other natural renderings of the notion of partial function, for `X Y : 𝓤`, are given by the equivalences

```
   (X ⇀ Y) ≃ (X → 𝓛 Y)
           ≃ Σ D ꞉ 𝓤 ̇ , (D ↪ X) × (D → Y)
```

where

```
       𝓛 Y = Σ P ꞉ 𝓤 ̇ , is-subsingleton P × (P → Y)
           ≃ (𝟙 ⇀ Y)
```

are two equivalent formulations of the type of partial elements of `Y`. Generalize the universes, and generalize these alternative descriptions of the type of partial functions to dependent partial functions, and prove them.

## Universe lifting

Universes are not cumulative on the nose in Agda, in the sense that from

```
      X : 𝓤
```

we would get that

```
X : 𝓤⁺
```

or that

```
X : 𝓤 ⊔ 𝓥.
```

Instead we work with embeddings of universes into larger universes.

The following together with its induction principle should be considered as part of the universe handling of our spartan Martin-Löf type theory:

```
record Lift {𝓤 : Universe} (𝓥 : Universe) (X : 𝓤 ̇ ) : 𝓤 ⊔ 𝓥 ̇  where
 constructor
  lift
 field
  lower : X

open Lift public
```

The functions `Lift`, `lift` and `lower` have the following types:

```
type-of-Lift  :             type-of (Lift  {𝓤} 𝓥)       ≡ (𝓤 ̇ → 𝓤 ⊔ 𝓥 ̇ )
type-of-lift  : {X : 𝓤 ̇ } → type-of (lift  {𝓤} {𝓥} {X}) ≡ (X → Lift 𝓥 X)
type-of-lower : {X : 𝓤 ̇ } → type-of (lower {𝓤} {𝓥} {X}) ≡ (Lift 𝓥 X → X)

type-of-Lift  = refl _
type-of-lift  = refl _
type-of-lower = refl _
```

The induction and recursion principles are as follows:

```
Lift-induction : ∀ {𝓤} 𝓥 (X : 𝓤 ̇ ) (A : Lift 𝓥 X → 𝓦 ̇ )
               → ((x : X) → A (lift x))
               → (l : Lift 𝓥 X) → A l

Lift-induction 𝓥 X A φ (lift x) = φ x

Lift-recursion : ∀ {𝓤} 𝓥 {X : 𝓤 ̇ } {B : 𝓦 ̇ }
               → (X → B) → Lift 𝓥 X → B

Lift-recursion 𝓥 {X} {B} = Lift-induction 𝓥 X (λ _ → B)
```

This gives an equivalence `lift : X → Lift 𝓥 X` and hence an embedding `Lift 𝓥 : 𝓤 ̇ → 𝓤 ⊔ 𝓥`. The following two constructions can be performed with induction, but actually hold on the nose by the so-called η rule for records:

```
lower-lift : {X : 𝓤 ̇ } (x : X) → lower {𝓤} {𝓥} (lift x) ≡ x
lower-lift = refl

lift-lower : {X : 𝓤 ̇ } (l : Lift 𝓥 X) → lift (lower l) ≡ l
lift-lower = refl

Lift-≃ : (X : 𝓤 ̇ ) → Lift 𝓥 X ≃ X
Lift-≃ {𝓤} {𝓥} X = invertibility-gives-≃ lower
                     (lift , lift-lower , lower-lift {𝓤} {𝓥})

≃-Lift : (X : 𝓤 ̇ ) → X ≃ Lift 𝓥 X
≃-Lift {𝓤} {𝓥} X = invertibility-gives-≃ lift
                     (lower , lower-lift {𝓤} {𝓥} , lift-lower)
```

With universe lifting, we can generalize equivalence induction as follows, in several steps.

Firstly, function extensionality for a pair of universes gives function extensionality for any pair of lower universes:

```
lower-dfunext : ∀ 𝓦 𝓣 𝓤 𝓥 → dfunext (𝓤 ⊔ 𝓦) (𝓥 ⊔ 𝓣) → dfunext 𝓤 𝓥
lower-dfunext 𝓦 𝓣 𝓤 𝓥 fe {X} {A} {f} {g} h = p
 where
  A' : Lift 𝓦 X → 𝓥 ⊔ 𝓣 ̇
  A' y = Lift 𝓣 (A (lower y))

  f' g' : Π A'
  f' y = lift (f (lower y))
  g' y = lift (g (lower y))

  h' : f' ∼ g'
  h' y = ap lift (h (lower y))

  p' : f' ≡ g'
  p' = fe h'

  p : f ≡ g
  p = ap (λ f' x → lower (f' (lift x))) p'
```

This could have been used above to remove a number of function extensionality hypotheses. But also it would have required to present the material of this section before we wanted to for learning purposes.

Secondly, a function from a universe to a higher universe is an embedding provided it maps any type to an equivalent type and the two universes are univalent:

```
universe-embedding-criterion : is-univalent 𝓤
                             → is-univalent (𝓤 ⊔ 𝓥)
                             → (f : 𝓤 ̇ → 𝓤 ⊔ 𝓥 ̇ )
                             → ((X : 𝓤 ̇ ) → f X ≃ X)
                             → is-embedding f

universe-embedding-criterion {𝓤} {𝓥} ua ua' f e = embedding-criterion f γ
 where
  fe : dfunext (𝓤 ⊔ 𝓥) (𝓤 ⊔ 𝓥)
  fe = univalence-gives-dfunext ua'

  fe₀ : dfunext 𝓤 𝓤
  fe₀ = lower-dfunext 𝓥 𝓥 𝓤 𝓤 fe

  fe₁ : dfunext 𝓤 (𝓤 ⊔ 𝓥)
  fe₁ = lower-dfunext 𝓥 𝓥 𝓤 (𝓤 ⊔ 𝓥) fe

  γ : (X X' : 𝓤 ̇ ) → (f X ≡ f X') ≃ (X ≡ X')
  γ X X' =  (f X ≡ f X')  ≃⟨ i   ⟩
            (f X ≃ f X')  ≃⟨ ii  ⟩
            (X ≃ X')      ≃⟨ iii ⟩
            (X ≡ X')      ■
   where
    i   = univalence-≃ ua' (f X) (f X')
    ii  = Eq-Eq-cong' fe fe fe fe fe fe₀ fe₁ fe fe₀ fe₀ fe₀ fe₀ (e X) (e X')
    iii = ≃-sym (univalence-≃ ua X X')
```

In particular, the function `Lift` is an embedding:

```
Lift-is-embedding : is-univalent 𝓤 → is-univalent (𝓤 ⊔ 𝓥)
                  → is-embedding (Lift {𝓤} 𝓥)

Lift-is-embedding {𝓤} {𝓥} ua ua' = universe-embedding-criterion {𝓤} {𝓥} ua ua'
                                      (Lift 𝓥) Lift-≃
```

Thirdly, we have a generalization of `univalence→` from a single universe to a pair of universes. We work with two symmetrical versions, where the second is derived from the first. We use an anonymous module to assume univalence in the following couple of constructions:

```
module _ {𝓤 𝓥 : Universe}
         (ua : is-univalent 𝓥)
```

```
          (ua' : is-univalent (𝓤 ⊔ 𝓥))
 where

 private
  fe : dfunext (𝓤 ⊔ 𝓥) (𝓤 ⊔ 𝓥)
  fe = univalence-gives-dfunext ua'

  fe₀ : dfunext 𝓥 (𝓤 ⊔ 𝓥)
  fe₀ = lower-dfunext 𝓤 𝓤 𝓥 (𝓤 ⊔ 𝓥) fe

  fe₁ : dfunext 𝓤 (𝓤 ⊔ 𝓥)
  fe₁ = lower-dfunext (𝓤 ⊔ 𝓥) 𝓤 𝓤 (𝓤 ⊔ 𝓥) fe

  fe₂ : dfunext 𝓥 𝓥
  fe₂ = lower-dfunext 𝓤 𝓤 𝓥 𝓥 fe

  fe₃ : dfunext 𝓤 𝓤
  fe₃ = lower-dfunext 𝓥 𝓥 𝓤 𝓤 fe

 univalence→' : (X : 𝓤 ̇ ) → is-subsingleton (Σ Y ꞉ 𝓥 ̇ , X ≃ Y)
 univalence→' X = s
  where
   e : (Y : 𝓥 ̇ ) → (X ≃ Y) ≃ (Lift 𝓤 Y ≡ Lift 𝓥 X)
   e Y = (X ≃ Y)                     ≃⟨ i   ⟩
         (Y ≃ X)                     ≃⟨ ii  ⟩
         (Lift 𝓤 Y ≃ Lift 𝓥 X)      ≃⟨ iii ⟩
         (Lift 𝓤 Y ≡ Lift 𝓥 X)      ■
    where
     i   = ≃-Sym fe₀ fe₁ fe
     ii  = Eq-Eq-cong' fe₁ fe fe₂ fe₁ fe fe fe fe₃
             fe fe fe fe (≃-Lift Y) (≃-Lift X)
     iii = ≃-sym (univalence-≃ ua' (Lift 𝓤 Y) (Lift 𝓥 X))

   d : (Σ Y ꞉ 𝓥 ̇ , X ≃ Y) ≃ (Σ Y ꞉ 𝓥 ̇ , Lift 𝓤 Y ≡ Lift 𝓥 X)
   d = Σ-cong e

   j : is-subsingleton (Σ Y ꞉ 𝓥 ̇ , Lift 𝓤 Y ≡ Lift 𝓥 X)
   j = Lift-is-embedding ua ua' (Lift 𝓥 X)

   abstract
    s : is-subsingleton (Σ Y ꞉ 𝓥 ̇ , X ≃ Y)
    s = equiv-to-subsingleton d j

 univalence→'-dual : (Y : 𝓤 ̇ ) → is-subsingleton (Σ X ꞉ 𝓥 ̇ , X ≃ Y)
 univalence→'-dual Y = equiv-to-subsingleton e i
  where
   e : (Σ X ꞉ 𝓥 ̇ , X ≃ Y) ≃ (Σ X ꞉ 𝓥 ̇ , Y ≃ X)
   e = Σ-cong (λ X → ≃-Sym fe₁ fe₀ fe)

   i : is-subsingleton (Σ X ꞉ 𝓥 ̇ , Y ≃ X)
   i = univalence→' Y
```

This is the end of the anonymous module. We are interested in these corollaries:

```
univalence→'' : is-univalent (𝓤 ⊔ 𝓥)
              → (X : 𝓤 ̇ ) → is-subsingleton (Σ Y ꞉ 𝓤 ⊔ 𝓥 ̇ , X ≃ Y)

univalence→'' ua = univalence→' ua ua

univalence→''-dual : is-univalent (𝓤 ⊔ 𝓥)
                   → (Y : 𝓤 ̇ ) → is-subsingleton (Σ X ꞉ 𝓤 ⊔ 𝓥 ̇ , X ≃ Y)

univalence→''-dual ua = univalence→'-dual ua ua
```

The first one is applied to get the following, where `Y` lives in a universe above that of `X`:

```
G↑-≃ : is-univalent (𝓤 ⊔ 𝓥)
     → (X : 𝓤 ̇ ) (A : (Σ Y ꞉ 𝓤 ⊔ 𝓥 ̇ , X ≃ Y) → 𝓦 ̇ )
     → A (Lift 𝓥 X , ≃-Lift X) → (Y : 𝓤 ⊔ 𝓥 ̇ ) (e : X ≃ Y) → A (Y , e)

G↑-≃ {𝓤} {𝓥} ua X A a Y e = transport A p a
 where
  t : Σ Y ꞉ 𝓤 ⊔ 𝓥 ̇ , X ≃ Y
  t = (Lift 𝓥 X , ≃-Lift X)

  p : t ≡ (Y , e)
  p = univalence→'' {𝓤} {𝓥} ua X t (Y , e)

H↑-≃ : is-univalent (𝓤 ⊔ 𝓥)
     → (X : 𝓤 ̇ ) (A : (Y : 𝓤 ⊔ 𝓥 ̇ ) → X ≃ Y → 𝓦 ̇ )
     → A (Lift 𝓥 X) (≃-Lift X) → (Y : 𝓤 ⊔ 𝓥 ̇ ) (e : X ≃ Y) → A Y e

H↑-≃ ua X A = G↑-≃ ua X (Σ-induction A)
```

*Exercise*. Formulate and prove the equations for `G↑-≃` and `H↑-≃` corresponding to those for `𝔾-≃` and `ℍ-≃`.

The difference with `ℍ-≃` is that here, to get the conclusion, we need to assume

```
A (Lift 𝓥 X) (≃-Lift X)
```

rather than

```
A X (id-≃).
```

And we have a similar development with a similar example:

```
J↑-≃ : is-univalent (𝓤 ⊔ 𝓥)
     → (A : (X : 𝓤 ̇ ) (Y : 𝓤 ⊔ 𝓥 ̇ ) → X ≃ Y → 𝓦 ̇ )
     → ((X : 𝓤 ̇ ) → A X (Lift 𝓥 X) (≃-Lift X))
     → (X : 𝓤 ̇ ) (Y : 𝓤 ⊔ 𝓥 ̇ ) (e : X ≃ Y) → A X Y e

J↑-≃ ua A φ X = H↑-≃ ua X (A X) (φ X)

H↑-equiv : is-univalent (𝓤 ⊔ 𝓥)
         → (X : 𝓤 ̇ ) (A : (Y : 𝓤 ⊔ 𝓥 ̇ ) → (X → Y) → 𝓦 ̇ )
         → A (Lift 𝓥 X) lift → (Y : 𝓤 ⊔ 𝓥 ̇ ) (f : X → Y) → is-equiv f → A Y f

H↑-equiv {𝓤} {𝓥} {𝓦} ua X A a Y f i = γ (f , i)
 where
  B : (Y : 𝓤 ⊔ 𝓥 ̇ ) → X ≃ Y → 𝓦 ̇
  B Y (f , i) = A Y f

  b : B (Lift 𝓥 X) (≃-Lift X)
  b = a

  γ : (e : X ≃ Y) → B Y e
  γ = H↑-≃ ua X B b Y

J↑-equiv : is-univalent (𝓤 ⊔ 𝓥)
         → (A : (X : 𝓤 ̇ ) (Y : 𝓤 ⊔ 𝓥 ̇ ) → (X → Y) → 𝓦 ̇ )
         → ((X : 𝓤 ̇ ) → A X (Lift 𝓥 X) lift)
         → (X : 𝓤 ̇ ) (Y : 𝓤 ⊔ 𝓥 ̇ ) (f : X → Y) → is-equiv f → A X Y f

J↑-equiv ua A φ X = H↑-equiv ua X (A X) (φ X)
```

All invertible functions from a type in a universe 𝓤 to a type in a higher universe `𝓤 ⊔ 𝓥` satisfy a given property if (and only if) the functions

```
lift {𝓤} {𝓥} {X} : X → Lift 𝓥 X
```

satisfy the property for all `X : 𝓤` (where we don't write the implicit arguments for `lift`):

```
J↑-invertible : is-univalent (𝓤 ⊔ 𝓥)
              → (A : (X : 𝓤 ̇ ) (Y : 𝓤 ⊔ 𝓥 ̇ ) → (X → Y) → 𝓦 ̇ )
              → ((X : 𝓤 ̇ ) → A X (Lift 𝓥 X) lift)
              → (X : 𝓤 ̇ ) (Y : 𝓤 ⊔ 𝓥 ̇ ) (f : X → Y) → invertible f → A X Y f

J↑-invertible ua A φ X Y f i = J↑-equiv ua A φ X Y f (invertibles-are-equivs f i)
```

Here is an example. First, `lift` is a half adjoint equivalence on the nose:

```
lift-is-hae : (X : 𝓤 ̇ ) → is-hae {𝓤} {𝓤 ⊔ 𝓥} {X} {Lift 𝓥 X} (lift {𝓤} {𝓥})
lift-is-hae {𝓤} {𝓥} X = lower ,
                        lower-lift {𝓤} {𝓥} ,
                        lift-lower ,
                        λ x → refl (refl (lift x))
```

Hence all invertible maps going up universe levels are half adjoint equivalences:

```
equivs-are-haes↑ : is-univalent (𝓤 ⊔ 𝓥)
                 → {X : 𝓤 ̇ } {Y : 𝓤 ⊔ 𝓥 ̇ } (f : X → Y)
                 → is-equiv f → is-hae f

equivs-are-haes↑ {𝓤} {𝓥} ua {X} {Y} = J↑-equiv {𝓤} {𝓥} ua (λ X Y f → is-hae f)
                                        lift-is-hae X Y
```

We have a dual development with the universes going down, where we consider `lower` in place of `lift`:

```
G↓-≃ : is-univalent (𝓤 ⊔ 𝓥)
     → (Y : 𝓤 ̇ ) (A : (Σ X ꞉ 𝓤 ⊔ 𝓥 ̇ , X ≃ Y) → 𝓦 ̇ )
     → A (Lift 𝓥 Y , Lift-≃ Y) → (X : 𝓤 ⊔ 𝓥 ̇ ) (e : X ≃ Y) → A (X , e)

G↓-≃ {𝓤} {𝓥} ua Y A a X e = transport A p a
 where
  t : Σ X ꞉ 𝓤 ⊔ 𝓥 ̇ , X ≃ Y
  t = (Lift 𝓥 Y , Lift-≃ Y)

  p : t ≡ (X , e)
  p = univalence→'-dual {𝓤} {𝓤 ⊔ 𝓥} ua ua Y t (X , e)

H↓-≃ : is-univalent (𝓤 ⊔ 𝓥)
     → (Y : 𝓤 ̇ ) (A : (X : 𝓤 ⊔ 𝓥 ̇ ) → X ≃ Y → 𝓦 ̇ )
     → A (Lift 𝓥 Y) (Lift-≃ Y) → (X : 𝓤 ⊔ 𝓥 ̇ ) (e : X ≃ Y) → A X e

H↓-≃ ua Y A = G↓-≃ ua Y (Σ-induction A)

J↓-≃ : is-univalent (𝓤 ⊔ 𝓥)
     → (A : (X : 𝓤 ⊔ 𝓥 ̇ ) (Y : 𝓤 ̇ ) → X ≃ Y → 𝓦 ̇ )
     → ((Y : 𝓤 ̇ ) → A (Lift 𝓥 Y) Y (Lift-≃ Y))
     → (X : 𝓤 ⊔ 𝓥 ̇ ) (Y : 𝓤 ̇ ) (e : X ≃ Y) → A X Y e

J↓-≃ ua A φ X Y = H↓-≃ ua Y (λ X → A X Y) (φ Y) X

H↓-equiv : is-univalent (𝓤 ⊔ 𝓥)
         → (Y : 𝓤 ̇ ) (A : (X : 𝓤 ⊔ 𝓥 ̇ ) → (X → Y) → 𝓦 ̇ )
         → A (Lift 𝓥 Y) lower → (X : 𝓤 ⊔ 𝓥 ̇ ) (f : X → Y) → is-equiv f → A X f

H↓-equiv {𝓤} {𝓥} {𝓦} ua Y A a X f i = γ (f , i)
 where
  B : (X : 𝓤 ⊔ 𝓥 ̇ ) → X ≃ Y → 𝓦 ̇
  B X (f , i) = A X f

  b : B (Lift 𝓥 Y) (Lift-≃ Y)
  b = a

  γ : (e : X ≃ Y) → B X e
  γ = H↓-≃ ua Y B b X
```

```
J↓-equiv : is-univalent (𝓤 ⊔ 𝓥)
         → (A : (X : 𝓤 ⊔ 𝓥 ̇ ) (Y : 𝓤 ̇ ) → (X → Y) → 𝓦 ̇ )
         → ((Y : 𝓤 ̇ ) → A (Lift 𝓥 Y) Y lower)
         → (X : 𝓤 ⊔ 𝓥 ̇ ) (Y : 𝓤 ̇ ) (f : X → Y) → is-equiv f → A X Y f

J↓-equiv ua A φ X Y = H↓-equiv ua Y (λ X → A X Y) (φ Y) X
```

All invertible functions from a type in a universe `𝓤 ⊔ 𝓥` to a type in the lower universe `𝓤` satisfy a given property if (and only if) the functions

```
    lower {𝓤} {𝓥} {Y} : Lift 𝓥 Y → Y
```

satisfy the property for all `Y : 𝓤`:

```
J↓-invertible : is-univalent (𝓤 ⊔ 𝓥)
              → (A : (X : 𝓤 ⊔ 𝓥 ̇ ) (Y : 𝓤 ̇ ) → (X → Y) → 𝓦 ̇ )
              → ((Y : 𝓤 ̇ ) → A (Lift 𝓥 Y) Y lower)
              → (X : 𝓤 ⊔ 𝓥 ̇ ) (Y : 𝓤 ̇ ) (f : X → Y) → invertible f → A X Y f

J↓-invertible ua A φ X Y f i = J↓-equiv ua A φ X Y f (invertibles-are-equivs f i)
```

And we have similar examples:

```
lower-is-hae : (X : 𝓤 ̇ ) → is-hae (lower {𝓤} {𝓥} {X})
lower-is-hae {𝓤} {𝓥} X = lift ,
                         lift-lower ,
                         lower-lift {𝓤} {𝓥} ,
                         (λ x → refl (refl (lower x)))

equivs-are-haes↓ : is-univalent (𝓤 ⊔ 𝓥)
                 → {X : 𝓤 ⊔ 𝓥 ̇ } {Y : 𝓤 ̇ } (f : X → Y)
                 → is-equiv f → is-hae f

equivs-are-haes↓ {𝓤} {𝓥} ua {X} {Y} = J↓-equiv {𝓤} {𝓥} ua (λ X Y f → is-hae f)
                                        lower-is-hae X Y
```

A crucial example of an equivalence "going down one universe" is `Id→Eq X Y`. This is because the identity type `X ≡ Y` lives in the successor universe `𝓤 ⁺` if `X` and `Y` live in `𝓤`, whereas the equivalence type `X ≃ Y` lives in the same universe as `X` and `Y`. Hence we can apply the above function `invertibles-are-haes↓` to `Id→Eq X Y` to conclude that it is a half adjoint equivalence:

```
Id→Eq-is-hae' : is-univalent 𝓤 → is-univalent (𝓤 ⁺)
              → {X Y : 𝓤 ̇ } → is-hae (Id→Eq X Y)

Id→Eq-is-hae' ua ua⁺ {X} {Y} = equivs-are-haes↓ ua⁺ (Id→Eq X Y) (ua X Y)
```

We can be parsimonious with the uses of univalence by instead using `invertibles-are-haes`, which doesn't require univalence. However, that `Id→Eq` is invertible of course requires univalence.

```
Id→Eq-is-hae : is-univalent 𝓤
             → {X Y : 𝓤 ̇ } → is-hae (Id→Eq X Y)

Id→Eq-is-hae ua {X} {Y} = invertibles-are-haes (Id→Eq X Y)
                           (equivs-are-invertible (Id→Eq X Y) (ua X Y))
```

The remainder of this section is not used anywhere else. Using the universe `𝓤ω` discussed above, we can consider global properties:

```
global-property-of-types : 𝓤ω
global-property-of-types = {𝓤 : Universe} → 𝓤 ̇ → 𝓤 ̇
```

We have already considered a few global properties, in fact, such as `is-singleton`, `is-subsingleton`, `is-set` and `_is-of-hlevel n`.

We may hope to have that if `A` is a global property of types, then, in the presence of univalence, for any `X : 𝓤` and `Y : 𝓥`, if `A X` holds then so does `A Y` if `X ≃ Y`. However, because we have a type of universes, or universe levels, we may define e.g.

`A {𝓤} X = (𝓤 ≡ 𝓤₀)`, which violates this hope. To get this conclusion, we need an assumption on `A`. We say that `A` is cumulative if it is preserved by the embedding `Lift` of universes into higher universes:

```
cumulative : global-property-of-types → 𝓤ω
cumulative A = {𝓤 𝓥 : Universe} (X : 𝓤 ̇ ) → A X ≃ A (Lift 𝓥 X)
```

We can prove the following:

```
global-≃-ap : Univalence
            → (A : global-property-of-types)
            → cumulative A
            → (X : 𝓤 ̇ ) (Y : 𝓥 ̇ ) → X ≃ Y → A X ≃ A Y
```

However, the above notion of global property is very restrictive. For example, `is-inhabited` defined below is a global property of type `{𝓤 : Universe} → 𝓤 ̇ → 𝓤 ⁺ ̇` . Hence we prove something more general, where in this example we take `F 𝓤 = 𝓤 ⁺`.

```
global-≃-ap' : Univalence
             → (F : Universe → Universe)
             → (A : {𝓤 : Universe} → 𝓤 ̇ → (F 𝓤) ̇ )
             → ({𝓤 𝓥 : Universe} (X : 𝓤 ̇ ) → A X ≃ A (Lift 𝓥 X))
             → (X : 𝓤 ̇ ) (Y : 𝓥 ̇ ) → X ≃ Y → A X ≃ A Y

global-≃-ap' {𝓤} {𝓥} ua F A φ X Y e =

  A X          ≃⟨ φ X                                  ⟩
  A (Lift 𝓥 X) ≃⟨ Id→Eq (A (Lift 𝓥 X)) (A (Lift 𝓤 Y)) q ⟩
  A (Lift 𝓤 Y) ≃⟨ ≃-sym (φ Y)                          ⟩
  A Y          ■
 where
  d : Lift 𝓥 X ≃ Lift 𝓤 Y
  d = Lift 𝓥 X ≃⟨ Lift-≃ X         ⟩
      X        ≃⟨ e                ⟩
      Y        ≃⟨ ≃-sym (Lift-≃ Y) ⟩
      Lift 𝓤 Y ■

  p : Lift 𝓥 X ≡ Lift 𝓤 Y
  p = Eq→Id (ua (𝓤 ⊔ 𝓥)) (Lift 𝓥 X) (Lift 𝓤 Y) d

  q : A (Lift 𝓥 X) ≡ A (Lift 𝓤 Y)
  q = ap A p
```

The first claim follows with `F = id`:

```
global-≃-ap ua = global-≃-ap' ua id
```

## The subtype classifier and other classifiers

A subtype of a type `Y` is a type `X` *together* with an embedding of `X` into `Y`:

```
Subtype : 𝓤 ̇ → 𝓤 ⁺ ̇
Subtype {𝓤} Y = Σ X ꞉ 𝓤 ̇ , X ↪ Y
```

The type `Ω 𝓤` of subsingletons in the universe `𝓤` is the subtype classifier of types in `𝓤`, in the sense that we have a canonical equivalence

```
Subtype Y ≃ (Y → Ω 𝓤)
```

for any type `Y : 𝓤`. We will derive this from something more general. We defined embeddings to be maps whose fibers are all subsingletons. We can replace `is-subsingleton` by an arbitrary property `P` of — or even structure on — types.

The following generalizes the slice constructor `_/_`:

```
_/[_]_ : (𝓤 : Universe) → (𝓤 ̇ → 𝓥 ̇ ) → 𝓤 ̇ → 𝓤 ⁺ ⊔ 𝓥 ̇
𝓤 /[ P ] Y = Σ X ꞉ 𝓤 ̇ , Σ f ꞉ (X → Y) , ((y : Y) → P (fiber f y))

χ-special : (P : 𝓤 ̇ → 𝓥 ̇ ) (Y : 𝓤 ̇ ) → 𝓤 /[ P ] Y  → (Y → Σ P)
χ-special P Y (X , f , φ) y = fiber f y , φ y

is-special-map-classifier : (𝓤 ̇ → 𝓥 ̇ ) → 𝓤 ⁺ ⊔ 𝓥 ̇
is-special-map-classifier {𝓤} P = (Y : 𝓤 ̇ ) → is-equiv (χ-special P Y)
```

If a universe is a map classifier then `Σ P` is the classifier of maps with `P`-fibers, for any `P : 𝓤 → 𝓥`:

```
mc-gives-sc : is-map-classifier 𝓤
            → (P : 𝓤 ̇ → 𝓥 ̇ ) → is-special-map-classifier P

mc-gives-sc {𝓤} s P Y = γ
 where
  e = (𝓤 /[ P ] Y)                                  ≃⟨ ≃-sym a ⟩
      (Σ σ ꞉ 𝓤 / Y , ((y : Y) → P ((χ Y) σ y)))     ≃⟨ ≃-sym b ⟩
      (Σ A ꞉ (Y → 𝓤 ̇ ), ((y : Y) → P (A y)))         ≃⟨ ≃-sym c ⟩
      (Y → Σ P)                                     ■
   where
    a = Σ-assoc
    b = Σ-change-of-variable (λ A → Π (P ∘ A)) (χ Y) (s Y)
    c = ΠΣ-distr-≃

  observation : χ-special P Y ≡ ⌜ e ⌝
  observation = refl _

  γ : is-equiv (χ-special P Y)
  γ = ⌜⌝-is-equiv e
```

Therefore we have the following canonical equivalence:

```
χ-special-is-equiv : is-univalent 𝓤
                   → dfunext 𝓤 (𝓤 ⁺)
                   → (P : 𝓤 ̇ → 𝓥 ̇ ) (Y : 𝓤 ̇ )
                   → is-equiv (χ-special P Y)

χ-special-is-equiv {𝓤} ua fe P Y = mc-gives-sc (universes-are-map-classifiers ua fe) P Y

special-map-classifier : is-univalent 𝓤
                       → dfunext 𝓤 (𝓤 ⁺)
                       → (P : 𝓤 ̇ → 𝓥 ̇ ) (Y : 𝓤 ̇ )
                       → 𝓤 /[ P ] Y ≃ (Y → Σ P)

special-map-classifier {𝓤} ua fe P Y = χ-special P Y , χ-special-is-equiv ua fe P Y
```

In particular, considering `P = is-subsingleton`, we get the promised fact that `Ω` is the subtype classifier:

```
Ω-is-subtype-classifier : Univalence
                        → (Y : 𝓤 ̇ ) → Subtype Y ≃ (Y → Ω 𝓤)

Ω-is-subtype-classifier {𝓤} ua = special-map-classifier (ua 𝓤)
                                  (univalence-gives-dfunext' (ua 𝓤) (ua (𝓤 ⁺)))
                                  is-subsingleton
```

It follows that the type of subtypes of `Y` is always a set, even if `Y` is not a set:

```
subtypes-form-set : Univalence → (Y : 𝓤 ̇ ) → is-set (Subtype Y)
subtypes-form-set {𝓤} ua Y = equiv-to-set
                               (Ω-is-subtype-classifier ua Y)
                               (powersets-are-sets' ua)
```

We now consider `P = is-singleton` and the type of singletons:

```
𝓢 : (𝓤 : Universe) → 𝓤 ⁺ ̇
𝓢 𝓤 = Σ S ꞉ 𝓤 ̇ , is-singleton S
```

```
equiv-classification : Univalence
                     → (Y : 𝓤 ̇ ) → (Σ X ꞉ 𝓤 ̇ , X ≃ Y) ≃ (Y → 𝓢 𝓤)

equiv-classification {𝓤} ua = special-map-classifier (ua 𝓤)
                                (univalence-gives-dfunext' (ua 𝓤) (ua (𝓤 ⁺)))
                                is-singleton
```

With this we can derive a fact we already know, as follows. First the type of singletons (in a universe) is itself a singleton (in the next universe):

```
the-singletons-form-a-singleton : propext 𝓤 → dfunext 𝓤 𝓤 → is-singleton (𝓢 𝓤)
the-singletons-form-a-singleton {𝓤} pe fe = c , φ
 where
  i : is-singleton (Lift 𝓤 𝟙)
  i = equiv-to-singleton (Lift-≃ 𝟙) 𝟙-is-singleton

  c : 𝓢 𝓤
  c = Lift 𝓤 𝟙 , i

  φ : (x : 𝓢 𝓤) → c ≡ x
  φ (S , s) = to-subtype-≡ (λ _ → being-singleton-is-subsingleton fe) p
   where
    p : Lift 𝓤 𝟙 ≡ S
    p = pe (singletons-are-subsingletons (Lift 𝓤 𝟙) i)
           (singletons-are-subsingletons S s)
           (λ _ → center S s) (λ _ → center (Lift 𝓤 𝟙) i)
```

What we already knew is this:

```
univalence→-again : Univalence
                  → (Y : 𝓤 ̇ ) → is-singleton (Σ X ꞉ 𝓤 ̇ , X ≃ Y)

univalence→-again {𝓤} ua Y = equiv-to-singleton (equiv-classification ua Y) i
 where
  i : is-singleton (Y → 𝓢 𝓤)
  i = univalence-gives-vvfunext' (ua 𝓤) (ua (𝓤 ⁺))
        (λ y → the-singletons-form-a-singleton
                (univalence-gives-propext (ua 𝓤))
                (univalence-gives-dfunext (ua 𝓤)))
```

*Exercise*. (1) Show that the retractions into `Y` are classified by the type `Σ A ꞉ 𝓤 ̇ , A` of pointed types. (2) After we have defined propositional truncation and surjections, show that the surjections into `Y` are classified by the type `Σ A ꞉ 𝓤 ̇ , ∥ A ∥` of inhabited types.

## Magma equivalences

We now define magma equivalences and show that the type of magma equivalences is identified with the type of magma isomorphisms. In the next section, which proves a *structure identity principles*, we apply this to characterize magma equality and equality of other mathematical structures in terms of equivalences of underlying types.

For simplicity we assume global univalence here.

```
module magma-equivalences (ua : Univalence) where

 open magmas

 dfe : global-dfunext
 dfe = univalence-gives-global-dfunext ua

 hfe : global-hfunext
 hfe = univalence-gives-global-hfunext ua
```

The magma homomorphism and isomorphism conditions are subsingleton types by virtue of the fact that the underlying type of a magma is a set by definition.

```
being-magma-hom-is-subsingleton : (M N : Magma 𝓤) (f : ⟨ M ⟩ → ⟨ N ⟩)
                                → is-subsingleton (is-magma-hom M N f)

being-magma-hom-is-subsingleton M N f =

 Π-is-subsingleton dfe
  (λ x → Π-is-subsingleton dfe
  (λ y → magma-is-set N (f (x ·⟨ M ⟩ y)) (f x ·⟨ N ⟩ f y)))

being-magma-iso-is-subsingleton : (M N : Magma 𝓤) (f : ⟨ M ⟩ → ⟨ N ⟩)
                                → is-subsingleton (is-magma-iso M N f)

being-magma-iso-is-subsingleton M N f (h , g , k , η , ε) (h' , g' , k' , η' , ε') = γ
 where
  p : h ≡ h'
  p = being-magma-hom-is-subsingleton M N f h h'

  q : g ≡ g'
  q = dfe (λ y → g y          ≡⟨ (ap g (ε' y))⁻¹ ⟩
                 g (f (g' y)) ≡⟨ η (g' y)        ⟩
                 g' y         ∎)

  i : is-subsingleton (is-magma-hom N M g' × (g' ∘ f ∼ id) × (f ∘ g' ∼ id))
  i = ×-is-subsingleton
       (being-magma-hom-is-subsingleton N M g')
       (×-is-subsingleton
          (Π-is-subsingleton dfe (λ x → magma-is-set M (g' (f x)) x))
          (Π-is-subsingleton dfe (λ y → magma-is-set N (f (g' y)) y)))

  γ : (h , g , k , η , ε) ≡ (h' , g' , k' , η' , ε')
  γ = to-×-≡ (p , to-Σ-≡ (q , i _ _))
```

By a magma equivalence we mean an equivalence which is a magma homomorphism. This notion is again a subsingleton type.

```
is-magma-equiv : (M N : Magma 𝓤) → (⟨ M ⟩ → ⟨ N ⟩) → 𝓤 ̇
is-magma-equiv M N f = is-equiv f × is-magma-hom M N f

being-magma-equiv-is-subsingleton : (M N : Magma 𝓤) (f : ⟨ M ⟩ → ⟨ N ⟩)
                                  → is-subsingleton (is-magma-equiv M N f)

being-magma-equiv-is-subsingleton M N f =
 ×-is-subsingleton
  (being-equiv-is-subsingleton dfe dfe f)
  (being-magma-hom-is-subsingleton M N f)
```

A function is a magma isomorphism if and only if it is a magma equivalence.

```
magma-isos-are-magma-equivs : (M N : Magma 𝓤) (f : ⟨ M ⟩ → ⟨ N ⟩)
                            → is-magma-iso M N f
                            → is-magma-equiv M N f

magma-isos-are-magma-equivs M N f (h , g , k , η , ε) = i , h
 where
  i : is-equiv f
  i = invertibles-are-equivs f (g , η , ε)

magma-equivs-are-magma-isos : (M N : Magma 𝓤) (f : ⟨ M ⟩ → ⟨ N ⟩)
                            → is-magma-equiv M N f
                            → is-magma-iso M N f

magma-equivs-are-magma-isos M N f (i , h) = h , g , k , η , ε
 where
```

```
  g : ⟨ N ⟩ → ⟨ M ⟩
  g = inverse f i

  η : g ∘ f ∼ id
  η = inverses-are-retractions f i

  ε : f ∘ g ∼ id
  ε = inverses-are-sections f i

  k : (a b : ⟨ N ⟩) → g (a ·⟨ N ⟩ b) ≡ g a ·⟨ M ⟩ g b
  k a b = g (a ·⟨ N ⟩ b)                 ≡⟨ ap₂ (λ a b → g (a ·⟨ N ⟩ b)) ((ε a)⁻¹)
                                                ((ε b)⁻¹)                        ⟩
          g (f (g a) ·⟨ N ⟩ f (g b))   ≡⟨ ap g ((h (g a) (g b))⁻¹)              ⟩
          g (f (g a ·⟨ M ⟩ g b))       ≡⟨ η (g a ·⟨ M ⟩ g b)                    ⟩
          g a ·⟨ M ⟩ g b               ∎
```

Because these two notions are subsingleton types, we conclude that they are equivalent.

```
magma-iso-charac : (M N : Magma 𝓤) (f : ⟨ M ⟩ → ⟨ N ⟩)
                 → is-magma-iso M N f ≃ is-magma-equiv M N f

magma-iso-charac M N f = logically-equivalent-subsingletons-are-equivalent
                          (is-magma-iso M N f)
                          (is-magma-equiv M N f)
                          (being-magma-iso-is-subsingleton M N f)
                          (being-magma-equiv-is-subsingleton M N f)
                          (magma-isos-are-magma-equivs M N f ,
                           magma-equivs-are-magma-isos M N f)
```

And hence they are equal by univalence.

```
magma-iso-charac' : (M N : Magma 𝓤) (f : ⟨ M ⟩ → ⟨ N ⟩)
                  → is-magma-iso M N f ≡ is-magma-equiv M N f

magma-iso-charac' M N f = Eq→Id (ua (universe-of ⟨ M ⟩))
                           (is-magma-iso M N f)
                           (is-magma-equiv M N f)
                           (magma-iso-charac M N f)
```

And by function extensionality the *properties* of being a magma isomorphism and a magma equivalence are the same:

```
magma-iso-charac'' : (M N : Magma 𝓤)
                   → is-magma-iso M N ≡ is-magma-equiv M N

magma-iso-charac'' M N = dfe (magma-iso-charac' M N)
```

Hence the type of magma equivalences is equivalent, and therefore equal, to the type of magma isomorphisms.

```
_≃ₘ_ : Magma 𝓤 → Magma 𝓤 → 𝓤 ̇
M ≃ₘ N = Σ f ꞉ (⟨ M ⟩ → ⟨ N ⟩), is-magma-equiv M N f

≅ₘ-charac : (M N : Magma 𝓤)
          → (M ≅ₘ N) ≃ (M ≃ₘ N)

≅ₘ-charac M N = Σ-cong (magma-iso-charac M N)

≅ₘ-charac' : (M N : Magma 𝓤)
           → (M ≅ₘ N) ≡ (M ≃ₘ N)

≅ₘ-charac' M N = ap Σ (magma-iso-charac'' M N)
```

It follows from the results of this and the next section that magma equality amounts to magma isomorphism.

## Equality of mathematical structures

Independently of any choice of foundation, we regard two groups to be the same, for all mathematical purposes, if they are isomorphic. Likewise, we consider two topological spaces to be the same if they are homeomorphic, two metric spaces to be the same if they are isometric, two categories to be the same if they are equivalent, and so on.

With Voevodsky's univalence axiom, we can *prove* that these notions of sameness are automatically captured by Martin-Löf's identity type. In particular, properties of groups are automatically invariant under isomorphism, properties of topological spaces are automatically invariant under homeomorphism, properties of metric spaces are automatically invariant under isometry, properties of categories are automatically invariant under equivalence, and so on, simply because, by design, properties are invariant under the notion of equality given by the identity type. In other foundations, the lack of such automatic invariance creates practical difficulties.

A *structure identity principle* describes the identity type of types of mathematical structures in terms of equivalences of underlying types, relying on univalence. The first published structure identity principle, for a large class of algebraic structures, is [Coquand and Danielsson]. The HoTT book (section 9.8) has a categorical version, whose formulation is attributed to Peter Aczel.

Here we formulate and prove a variation for types equipped with structure. We consider several versions:

1. One for raw structures subject to no axioms, such as ∞-magmas and pointed types.

2. One that adds axioms to a structure, so as to e.g. get an automatic characterization of magma identifications from a characterization of ∞-magma identifications.

3. One that joins two kinds of structure, so as to e.g. get an automatic characterization of identifications of pointed ∞-magmas from characterizations of identifications for pointed types and for ∞-magmas.

4. In particular, adding axioms to pointed ∞-magmas we get monoids with an automatic characterization of their identifications.

5. And then adding an axiom to monoids we get groups, again with an automatic characterization of their identitifications.

6. We also show that while two groups are equal precisely when they are isomorphic, two *subgroups* of a group are equal precisely when they have the same elements, if we define a subgroup to be a subset closed under the group operations.

We also apply these ideas to characterize identifications of metric spaces, topological spaces, graphs, partially ordered sets, categories and more.

### A structure identity principle for a standard notion of structure

```
module sip where
```

We consider mathematical structures specified by a function

```
S : 𝓤 → 𝓥
```

and we consider types `X : 𝓤` equipped with such structure `s : S X`, collected in the type

```
Σ X : 𝓤 , S X,
```

which, as we have seen, can be abbreviated as

```
Σ S.
```

For example, for the type of ∞-magmas we will take `𝓥 = 𝓤` and

```
S X = X → X → X.
```

Our objective is to describe the identity type `Id (Σ S) A B`, in favourable circumstances, in terms of equivalences of the underlying types `⟨ A ⟩` and `⟨ B ⟩` of `A B : Σ S`.

```
⟨_⟩ : {S : 𝓤 ̇ → 𝓥 ̇ } → Σ S → 𝓤 ̇
⟨ X , s ⟩ = X

structure : {S : 𝓤 ̇ → 𝓥 ̇ } (A : Σ S) → S ⟨ A ⟩
structure (X , s) = s
```

Our favourable circumstances will be given by data

```
ι : (A B : Σ S) → ⟨ A ⟩ ≃ ⟨ B ⟩ → 𝓦,

ρ : (A : Σ S) → ι A A (id-≃ ⟨ A ⟩).
```

The idea is that

- `ι` describes favourable equivalences, which will be called homomorphisms, and
- `ρ` then stipulates that all identity equivalences are homomorphisms.

We require that any two structures on the same type making the identity equivalence a homomorphism must be identified in a canonical way:

- The canonical map

  ```
  s ≡ t → ι (X , s) (X , t) (id-≃ X)
  ```

  defined by induction on identifications by

  ```
  refl s ↦ ρ (X , s)
  ```

  must be an equivalence for all `X : 𝓤` and `s t : S X` .

This may sound a bit abstract at this point, but in practical examples of interest it is easy to fulfill these requirements, as we will illustrate soon.

We first define the canonical map:

```
canonical-map : {S : 𝓤 ̇ → 𝓥 ̇ }
                (ι : (A B : Σ S) → ⟨ A ⟩ ≃ ⟨ B ⟩ → 𝓦 ̇ )
                (ρ : (A : Σ S) → ι A A (id-≃ ⟨ A ⟩))
                {X : 𝓤 ̇ }
                (s t : S X)

              → s ≡ t → ι (X , s) (X , t) (id-≃ X)

canonical-map ι ρ {X} s s (refl s) = ρ (X , s)
```

We refer to such favourable data as a *standard notion of structure* and collect them in the type

```
SNS S 𝓦
```

as follows:

```
SNS : (𝓤 ̇ → 𝓥 ̇ ) → (𝓦 : Universe) → 𝓤 ⁺ ⊔ 𝓥 ⊔ (𝓦 ⁺) ̇

SNS {𝓤} {𝓥} S 𝓦 = Σ ι ꞉ ((A B : Σ S) → (⟨ A ⟩ ≃ ⟨ B ⟩ → 𝓦 ̇ ))
                  , Σ ρ ꞉ ((A : Σ S) → ι A A (id-≃ ⟨ A ⟩))
                  , ({X : 𝓤 ̇ } (s t : S X) → is-equiv (canonical-map ι ρ s t))
```

We write `homomorphic` for the first projection (we don't need names for the other two projections):

```
homomorphic : {S : 𝓤 ̇ → 𝓥 ̇ } → SNS S 𝓦
            → (A B : Σ S) → ⟨ A ⟩ ≃ ⟨ B ⟩ → 𝓦 ̇

homomorphic (ι , ρ , θ) = ι
```

For example, when `S` specifies ∞-magma structure, we will have that `homomorphic σ A B (f , i)` amounts to `f` being a magma homomorphism.

We then collect the homomorphic equivalences of `A B : Σ S`, assuming that `S` is a standard notion of structure witnessed by `σ`, in a type

```
A ≃[ σ ] B.
```

Notice that only the first component of `σ`, namely `homomorphic σ`, is used in the definition. The other two components are used to prove properties of `A ≃[ σ ] B`.

```
_≃[_]_ : {S : 𝓤 ̇ → 𝓥 ̇ } → Σ S → SNS S 𝓦 → Σ S → 𝓤 ⊔ 𝓦 ̇

A ≃[ σ ] B = Σ f ꞉ (⟨ A ⟩ → ⟨ B ⟩)
           , Σ i ꞉ is-equiv f
           , homomorphic σ A B (f , i)

Id→homEq : {S : 𝓤 ̇ → 𝓥 ̇ } (σ : SNS S 𝓦)
         → (A B : Σ S) → (A ≡ B) → (A ≃[ σ ] B)

Id→homEq (_ , ρ , _) A A (refl A) = id , id-is-equiv ⟨ A ⟩ , ρ A
```

With this we are ready to prove the promised characterization of identity on `Σ S`:

```
homomorphism-lemma : {S : 𝓤 ̇ → 𝓥 ̇ } (σ : SNS S 𝓦)
                     (A B : Σ S) (p : ⟨ A ⟩ ≡ ⟨ B ⟩)
                   →
                     (transport S p (structure A) ≡ structure B)
                   ≃  homomorphic σ A B (Id→Eq ⟨ A ⟩ ⟨ B ⟩ p)

homomorphism-lemma (ι , ρ , θ) (X , s) (X , t) (refl X) = γ
 where
  γ : (s ≡ t) ≃ ι (X , s) (X , t) (id-≃ X)
  γ = (canonical-map ι ρ s t , θ s t)

characterization-of-≡ : is-univalent 𝓤
                      → {S : 𝓤 ̇ → 𝓥 ̇ } (σ : SNS S 𝓦)
                      → (A B : Σ S)

                      → (A ≡ B) ≃ (A ≃[ σ ] B)

characterization-of-≡ ua {S} σ A B =

   (A ≡ B)                                                              ≃⟨ i   ⟩
   (Σ p ꞉ ⟨ A ⟩ ≡ ⟨ B ⟩ , transport S p (structure A) ≡ structure B) ≃⟨ ii  ⟩
   (Σ p ꞉ ⟨ A ⟩ ≡ ⟨ B ⟩ , ι A B (Id→Eq ⟨ A ⟩ ⟨ B ⟩ p))               ≃⟨ iii ⟩
   (Σ e ꞉ ⟨ A ⟩ ≃ ⟨ B ⟩ , ι A B e)                                     ≃⟨ iv  ⟩
   (A ≃[ σ ] B)                                                         ■

 where
  ι   = homomorphic σ

  i   = Σ-≡-≃ A B
  ii  = Σ-cong (homomorphism-lemma σ A B)
  iii = ≃-sym (Σ-change-of-variable (ι A B) (Id→Eq ⟨ A ⟩ ⟨ B ⟩) (ua ⟨ A ⟩ ⟨ B ⟩))
  iv  = Σ-assoc
```

This equivalence is pointwise equal to `Id→homEq`, and hence `Id→homEq` is itself an equivalence:

```
Id→homEq-is-equiv : (ua : is-univalent 𝓤) {S : 𝓤 ̇ → 𝓥 ̇ } (σ : SNS S 𝓦)
                  → (A B : Σ S) → is-equiv (Id→homEq σ A B)

Id→homEq-is-equiv ua {S} σ A B = γ
 where
```

```
   h : (A B : Σ S) → Id→homEq σ A B ∼ ⌜ characterization-of-≡ ua σ A B ⌝
   h A A (refl A) = refl _

   γ : is-equiv (Id→homEq σ A B)
   γ = equivs-closed-under-∼
       (⌜⌝-is-equiv (characterization-of-≡ ua σ A B))
       (h A B)
```

We conclude this submodule with the following characterization of the canonical map and of when it is an equivalence, applying Yoneda.

```
 module _ {S : 𝓤 ̇ → 𝓥 ̇ }
          (ι : (A B : Σ S) → ⟨ A ⟩ ≃ ⟨ B ⟩ → 𝓦 ̇ )
          (ρ : (A : Σ S) → ι A A (id-≃ ⟨ A ⟩))
          {X : 𝓤 ̇ }

        where

  canonical-map-charac : (s t : S X) (p : s ≡ t)

                       → canonical-map ι ρ s t p
                       ≡ transport (λ - → ι (X , s) (X , -) (id-≃ X)) p (ρ (X , s))

  canonical-map-charac s = transport-lemma (λ t → ι (X , s) (X , t) (id-≃ X)) s
                            (canonical-map ι ρ s)

  when-canonical-map-is-equiv : ((s t : S X) → is-equiv (canonical-map ι ρ s t))
                              ⇔ ((s : S X) → ∃! t ꞉ S X , ι (X , s) (X , t) (id-≃ X))

  when-canonical-map-is-equiv = (λ e s → fiberwise-equiv-universal (A s) s (τ s) (e s)) ,
                                (λ φ s → universal-fiberwise-equiv (A s) (φ s) s (τ s))
   where
    A = λ s t → ι (X , s) (X , t) (id-≃ X)
    τ = canonical-map ι ρ
```

Another criterion is the following: It is enough to have any equivalence for the canonical map to be an equivalence:

```
  canonical-map-equiv-criterion : ((s t : S X) → (s ≡ t) ≃ ι (X , s) (X , t) (id-≃ X))
                                → (s t : S X) → is-equiv (canonical-map ι ρ s t)

  canonical-map-equiv-criterion φ s = fiberwise-equiv-criterion'
                                       (λ t → ι (X , s) (X , t) (id-≃ X))
                                       s (φ s) (canonical-map ι ρ s)
```

And in fact it is enough to have any retraction for the canonical map to be an equivalence:

```
  canonical-map-equiv-criterion' : ((s t : S X) → ι (X , s) (X , t) (id-≃ X) ◁ (s ≡ t))
                                 → (s t : S X) → is-equiv (canonical-map ι ρ s t)

  canonical-map-equiv-criterion' φ s = fiberwise-equiv-criterion
                                        (λ t → ι (X , s) (X , t) (id-≃ X))
                                        s (φ s) (canonical-map ι ρ s)
```

This concludes the module `sip`, and we now consider some examples of uses of this.

### ∞-Magmas

We now make precise the example outlined above:

```
module ∞-magma {𝓤 : Universe} where

 open sip

 ∞-magma-structure : 𝓤 ̇ → 𝓤 ̇
```

```
∞-magma-structure X = X → X → X

∞-Magma : 𝓤 ⁺ ̇
∞-Magma = Σ X ꞉ 𝓤 ̇ , ∞-magma-structure X

sns-data : SNS ∞-magma-structure 𝓤
sns-data = (ι , ρ , θ)
 where
  ι : (A B : ∞-Magma) → ⟨ A ⟩ ≃ ⟨ B ⟩ → 𝓤 ̇
  ι (X , _·_) (Y , _*_) (f , _) = (λ x x' → f (x · x')) ＝ (λ x x' → f x * f x')

  ρ : (A : ∞-Magma) → ι A A (id-≃ ⟨ A ⟩)
  ρ (X , _·_) = refl _·_

  h : {X : 𝓤 ̇ } {_·_ _*_ : ∞-magma-structure X}
    → canonical-map ι ρ _·_ _*_ ∼ id (_·_ ＝ _*_)

  h (refl _·_) = refl (refl _·_)

  θ : {X : 𝓤 ̇ } (_·_ _*_ : ∞-magma-structure X)
    → is-equiv (canonical-map ι ρ _·_ _*_)

  θ _·_ _*_ = equivs-closed-under-∼ (id-is-equiv (_·_ ＝ _*_)) h

 _≅_ : ∞-Magma → ∞-Magma → 𝓤 ̇

 (X , _·_) ≅ (Y , _*_) =

           Σ f ꞉ (X → Y), is-equiv f
                        × ((λ x x' → f (x · x')) ＝ (λ x x' → f x * f x'))

 characterization-of-∞-Magma-＝ : is-univalent 𝓤 → (A B : ∞-Magma) → (A ＝ B) ≃ (A ≅ B)
 characterization-of-∞-Magma-＝ ua = characterization-of-＝ ua sns-data
```

The above equivalence is characterized by induction on identifications as the function that maps the reflexive identification to the identity equivalence:

```
 characterization-of-characterization-of-∞-Magma-＝ :

    (ua : is-univalent 𝓤) (A : ∞-Magma)
  →
    ⌜ characterization-of-∞-Magma-＝ ua A A ⌝ (refl A)
  ＝
    (id ⟨ A ⟩ , id-is-equiv ⟨ A ⟩ , refl _)

 characterization-of-characterization-of-∞-Magma-＝ ua A = refl _
```

### Adding axioms

Next we want to account for situations in which axioms are considered, for example that the underlying type is a set, or that the monoid structure satisfies the unit and associativity laws. We do this in a submodule, by reduction to the characterization of identifications given in the module `sip`.

```
module sip-with-axioms where

 open sip
```

The first construction, given `S` as above, and given subsingleton-valued axioms for types equipped with structure specified by `S`, builds `SNS` data on `S'` defined by

```
    S' X = Σ s ꞉ S X , axioms X s
```

from given `SNS` data on `S`.

For that purpose we first define a forgetful map `Σ S' → Σ S` and an underlying-type function `Σ S → 𝓤`:

```
[_] : {S : 𝓤 ̇ → 𝓥 ̇ } {axioms : (X : 𝓤 ̇ ) → S X → 𝓦 ̇ }
    → (Σ X ꞉ 𝓤 ̇ , Σ s ꞉ S X , axioms X s) → Σ S

[ X , s , _ ] = (X , s)

⟪_⟫ : {S : 𝓤 ̇ → 𝓥 ̇ } {axioms : (X : 𝓤 ̇ ) → S X → 𝓦 ̇ }
    → (Σ X ꞉ 𝓤 ̇ , Σ s ꞉ S X , axioms X s) → 𝓤 ̇

⟪ X , _ , _ ⟫ = X
```

In the following construction:

- For `ι'` and `ρ'` we use `ι` and `ρ` ignoring the axioms.
- For `θ'` we need more work, but the essence of the construction is the fact that the projection

  ```
  S' X → S X
  ```

  that forgets the axioms is an embedding precisely because the axioms are subsingleton-valued.

```
add-axioms : {S : 𝓤 ̇ → 𝓥 ̇ }
             (axioms : (X : 𝓤 ̇ ) → S X → 𝓦 ̇ )
           → ((X : 𝓤 ̇ ) (s : S X) → is-subsingleton (axioms X s))

           → SNS S 𝓣
           → SNS (λ X → Σ s ꞉ S X , axioms X s) 𝓣

add-axioms {𝓤} {𝓥} {𝓦} {𝓣} {S} axioms i (ι , ρ , θ) = ι' , ρ' , θ'
 where
  S' : 𝓤 ̇ → 𝓥 ⊔ 𝓦 ̇
  S' X = Σ s ꞉ S X , axioms X s

  ι' : (A B : Σ S') → ⟨ A ⟩ ≃ ⟨ B ⟩ → 𝓣 ̇
  ι' A B = ι [ A ] [ B ]

  ρ' : (A : Σ S') → ι' A A (id-≃ ⟨ A ⟩)
  ρ' A = ρ [ A ]

  θ' : {X : 𝓤 ̇ } (s' t' : S' X) → is-equiv (canonical-map ι' ρ' s' t')
  θ' {X} (s , a) (t , b) = γ
   where
    π : S' X → S X
    π (s , _) = s

    j : is-embedding π
    j = pr₁-is-embedding (i X)

    k : {s' t' : S' X} → is-equiv (ap π {s'} {t'})
    k {s'} {t'} = embedding-gives-ap-is-equiv π j s' t'

    l : canonical-map ι' ρ' (s , a) (t , b)
      ∼ canonical-map ι ρ s t ∘ ap π {s , a} {t , b}

    l (refl (s , a)) = refl (ρ (X , s))

    e : is-equiv (canonical-map ι ρ s t ∘ ap π {s , a} {t , b})
    e = ∘-is-equiv (θ s t) k

    γ : is-equiv (canonical-map ι' ρ' (s , a) (t , b))
    γ = equivs-closed-under-∼ e l
```

And with this we can formulate and prove what `add-axioms` achieves, namely that the characterization of the identity type remains the same, ignoring the axioms:

```
 characterization-of-≡-with-axioms :

     is-univalent 𝓤
   → {S : 𝓤 ̇ → 𝓥 ̇ }
     (σ : SNS S 𝓣)
     (axioms : (X : 𝓤 ̇ ) → S X → 𝓦 ̇ )
   → ((X : 𝓤 ̇ ) (s : S X) → is-subsingleton (axioms X s))
   → (A B : Σ X ꞉ 𝓤 ̇ , Σ s ꞉ S X , axioms X s)

   → (A ≡ B) ≃ ([ A ] ≃[ σ ] [ B ])

 characterization-of-≡-with-axioms ua σ axioms i =
   characterization-of-≡ ua (add-axioms axioms i σ)
```

And this concludes the module `sip-with-axioms`. We now consider some examples.

### Magmas

As discussed above, we get magmas from ∞-magmas by adding an axiom:

```
module magma {𝓤 : Universe} where

 open sip-with-axioms

 Magma : 𝓤 ⁺ ̇
 Magma = Σ X ꞉ 𝓤 ̇ , (X → X → X) × is-set X

 _≅_ : Magma → Magma → 𝓤 ̇

 (X , _·_ , _) ≅ (Y , _*_ , _) =

               Σ f ꞉ (X → Y), is-equiv f
                            × ((λ x x' → f (x · x')) ≡ (λ x x' → f x * f x'))

 characterization-of-Magma-≡ : is-univalent 𝓤 → (A B : Magma ) → (A ≡ B) ≃ (A ≅ B)
 characterization-of-Magma-≡ ua =
   characterization-of-≡-with-axioms ua
     ∞-magma.sns-data
     (λ X s → is-set X)
     (λ X s → being-set-is-subsingleton (univalence-gives-dfunext ua))
```

*Exercise*. The above equivalence is characterized by induction on identifications as the function that maps the reflexive identification to the identity equivalence.

*Exercise*. Characterize identifications of monoids along the above lines. It is convenient to redefine the type of monoids to an equivalent type in the above format of structure with axioms. The following development solves this exercise.

### Pointed types

We now discuss equality of pointed types, where a pointed type is a type equipped with a designated point.

```
module pointed-type {𝓤 : Universe} where

 open sip

 Pointed : 𝓤 ̇ → 𝓤 ̇
 Pointed X = X

 sns-data : SNS Pointed 𝓤
 sns-data = (ι , ρ , θ)
  where
```

```
  ι : (A B : Σ Pointed) → ⟨ A ⟩ ≃ ⟨ B ⟩ → 𝓤 ̇
  ι (X , x₀) (Y , y₀) (f , _) = (f x₀ ≡ y₀)

  ρ : (A : Σ Pointed) → ι A A (id-≃ ⟨ A ⟩)
  ρ (X , x₀) = refl x₀

  θ : {X : 𝓤 ̇ } (x₀ x₁ : Pointed X) → is-equiv (canonical-map ι ρ x₀ x₁)
  θ x₀ x₁ = equivs-closed-under-∼ (id-is-equiv (x₀ ≡ x₁)) h
   where
    h : canonical-map ι ρ x₀ x₁ ∼ id (x₀ ≡ x₁)
    h (refl x₀) = refl (refl x₀)

 _≅_ : Σ Pointed → Σ Pointed → 𝓤 ̇
 (X , x₀) ≅ (Y , y₀) = Σ f ꞉ (X → Y), is-equiv f × (f x₀ ≡ y₀)

 characterization-of-pointed-type-≡ : is-univalent 𝓤
                                    → (A B : Σ Pointed) → (A ≡ B) ≃ (A ≅ B)

 characterization-of-pointed-type-≡ ua = characterization-of-≡ ua sns-data
```

*Exercise*. The above equivalence is characterized by induction on identifications as the function that maps the reflexive identification to the identity equivalence.

### Combining two mathematical structures

We now show how to join two mathematics structures so as to obtain a characterization of the identifications of the join from the characterization of the equalities of the structures. For example, we build the characterization of identifications of pointed ∞-magmas from the characterizations of the identifications of pointed types and the characterization of the identifications of magmas. Moreover, adding axioms, we get a characterization of identifications of monoids which amounts to the characterization of identifications of pointed ∞-magmas. Further adding an axiom, we get an automatic characterization of group identifications.

```
module sip-join where
```

We begin with the following technical lemma, whose proof uses the Yoneda machinery:

```
 technical-lemma :
     {X : 𝓤 ̇ } {A : X → X → 𝓥 ̇ }
     {Y : 𝓦 ̇ } {B : Y → Y → 𝓣 ̇ }
     (f : (x₀ x₁ : X) → x₀ ≡ x₁ → A x₀ x₁)
     (g : (y₀ y₁ : Y) → y₀ ≡ y₁ → B y₀ y₁)
   → ((x₀ x₁ : X) → is-equiv (f x₀ x₁))
   → ((y₀ y₁ : Y) → is-equiv (g y₀ y₁))

   → ((x₀ , y₀) (x₁ , y₁) : X × Y) → is-equiv (λ (p : (x₀ , y₀) ≡ (x₁ , y₁)) → f x₀ x₁ (ap pr₁ p) ,
                                                                                  g y₀ y₁ (ap pr₂ p))
 technical-lemma {𝓤} {𝓥} {𝓦} {𝓣} {X} {A} {Y} {B} f g i j (x₀ , y₀) = γ
  where
   u : ∃! x₁ ꞉ X , A x₀ x₁
   u = fiberwise-equiv-universal (A x₀) x₀ (f x₀) (i x₀)

   v : ∃! y₁ ꞉ Y , B y₀ y₁
   v = fiberwise-equiv-universal (B y₀) y₀ (g y₀) (j y₀)

   C : X × Y → 𝓥 ⊔ 𝓣 ̇
   C (x₁ , y₁) = A x₀ x₁ × B y₀ y₁

   w : (∃! x₁ ꞉ X , A x₀ x₁)
     → (∃! y₁ ꞉ Y , B y₀ y₁)
     →  ∃! (x₁ , y₁) ꞉ X × Y , C (x₁ , y₁)
   w ((x₁ , a₁) , φ) ((y₁ , b₁) , ψ) = ((x₁ , y₁) , (a₁ , b₁)) , δ
```

```
   where
    p : ∀ x y a b
      → (x₁ , a₁) ≡ (x , a)
      → (y₁ , b₁) ≡ (y , b)
      → (x₁ , y₁) , (a₁ , b₁) ≡ (x , y) , (a , b)
    p x₁ y₁ a₁ b₁ (refl (x₁ , a₁)) (refl (y₁ , b₁)) = refl ((x₁ , y₁) , (a₁ , b₁))

    δ : (((x , y) , (a , b)) : Σ C) → (x₁ , y₁) , (a₁ , b₁) ≡ ((x , y) , (a , b))
    δ ((x , y) , (a , b)) = p x y a b (φ (x , a)) (ψ (y , b))

  τ : Nat (𝓨 (x₀ , y₀)) C
  τ (x₁ , y₁) p = f x₀ x₁ (ap pr₁ p) , g y₀ y₁ (ap pr₂ p)

  γ : is-fiberwise-equiv τ
  γ = universal-fiberwise-equiv C (w u v) (x₀ , y₀) τ
```

We consider two given mathematical structures specified by $S_0$ and $S_1$, and work with structures specified by their combination `λ X → S₀ X × S₁ X`

```
variable
 𝓥₀ 𝓥₁ 𝓦₀ 𝓦₁ : Universe

open sip

⟪_⟫ : {S₀ : 𝓤 ̇ → 𝓥₀ ̇ } {S₁ : 𝓤 ̇ → 𝓥₁ ̇ }
    → (Σ X ꞉ 𝓤 ̇ , S₀ X × S₁ X) → 𝓤 ̇

⟪ X , s₀ , s₁ ⟫ = X

[_]₀ : {S₀ : 𝓤 ̇ → 𝓥₀ ̇ } {S₁ : 𝓤 ̇ → 𝓥₁ ̇ }
     → (Σ X ꞉ 𝓤 ̇ , S₀ X × S₁ X) → Σ S₀

[ X , s₀ , s₁ ]₀ = (X , s₀)

[_]₁ : {S₀ : 𝓤 ̇ → 𝓥₀ ̇ } {S₁ : 𝓤 ̇ → 𝓥₁ ̇ }
     → (Σ X ꞉ 𝓤 ̇ , S₀ X × S₁ X) → Σ S₁

[ X , s₀ , s₁ ]₁ = (X , s₁)
```

The main construction in this submodule is this:

```
join : {S₀ : 𝓤 ̇ → 𝓥₀ ̇ } {S₁ : 𝓤 ̇ → 𝓥₁ ̇ }
     → SNS S₀ 𝓦₀
     → SNS S₁ 𝓦₁
     → SNS (λ X → S₀ X × S₁ X) (𝓦₀ ⊔ 𝓦₁)

join {𝓤} {𝓥₀} {𝓥₁} {𝓦₀} {𝓦₁} {S₀} {S₁} (ι₀ , ρ₀ , θ₀) (ι₁ , ρ₁ , θ₁) = ι , ρ , θ
 where
  S : 𝓤 ̇ → 𝓥₀ ⊔ 𝓥₁ ̇
  S X = S₀ X × S₁ X

  ι : (A B : Σ S) → ⟨ A ⟩ ≃ ⟨ B ⟩ → 𝓦₀ ⊔ 𝓦₁ ̇
  ι A B e = ι₀ [ A ]₀ [ B ]₀ e  ×  ι₁ [ A ]₁ [ B ]₁ e

  ρ : (A : Σ S) → ι A A (id-≃ ⟨ A ⟩)
  ρ A = (ρ₀ [ A ]₀ , ρ₁ [ A ]₁)

  θ : {X : 𝓤 ̇ } (s t : S X) → is-equiv (canonical-map ι ρ s t)
  θ {X} (s₀ , s₁) (t₀ , t₁) = γ
   where
    c : (p : s₀ , s₁ ≡ t₀ , t₁) → ι₀ (X , s₀) (X , t₀) (id-≃ X)
                                × ι₁ (X , s₁) (X , t₁) (id-≃ X)

    c p = (canonical-map ι₀ ρ₀ s₀ t₀ (ap pr₁ p) ,
```

```
            canonical-map ι₁ ρ₁ s₁ t₁ (ap pr₂ p))

     i : is-equiv c
     i = technical-lemma
          (canonical-map ι₀ ρ₀)
          (canonical-map ι₁ ρ₁)
          θ₀ θ₁ (s₀ , s₁) (t₀ , t₁)

     e : canonical-map ι ρ (s₀ , s₁) (t₀ , t₁) ∼ c
     e (refl (s₀ , s₁)) = refl (ρ₀ (X , s₀) , ρ₁ (X , s₁))

     γ : is-equiv (canonical-map ι ρ (s₀ , s₁) (t₀ , t₁))
     γ = equivs-closed-under-∼ i e
```

We then can characterize the identity type of structures in the join by the following relation:

```
 _≃⟦_,_⟧_ : {S₀ : 𝓤 ̇ → 𝓥 ̇ } {S₁ : 𝓤 ̇ → 𝓥₁ ̇ }

          → (Σ X ꞉ 𝓤 ̇ , S₀ X × S₁ X)
          → SNS S₀ 𝓦₀
          → SNS S₁ 𝓦₁
          → (Σ X ꞉ 𝓤 ̇ , S₀ X × S₁ X)

          → 𝓤 ⊔ 𝓦₀ ⊔ 𝓦₁ ̇

 A ≃⟦ σ₀ , σ₁ ⟧ B = Σ f ꞉ (⟪ A ⟫ → ⟪ B ⟫)
                  , Σ i ꞉ is-equiv f , homomorphic σ₀ [ A ]₀ [ B ]₀ (f , i)
                                     × homomorphic σ₁ [ A ]₁ [ B ]₁ (f , i)
```

The following is then immediate from the join construction and the general structure identity principle:

```
 characterization-of-join-≡ : is-univalent 𝓤
                            → {S₀ : 𝓤 ̇ → 𝓥 ̇ } {S₁ : 𝓤 ̇ → 𝓥₁ ̇ }
                              (σ₀ : SNS S₀ 𝓦₀)   (σ₁ : SNS S₁ 𝓦₁)
                              (A B : Σ X ꞉ 𝓤 ̇ , S₀ X × S₁ X)

                            → (A ≡ B) ≃ (A ≃⟦ σ₀ , σ₁ ⟧ B)

 characterization-of-join-≡ ua σ₀ σ₁ = characterization-of-≡ ua (join σ₀ σ₁)
```

This concludes the submodule. Some examples of uses of this follow.

### Pointed ∞-magmas

Combining pointed types with ∞-magmas we get point ∞-magmas, and from the characterization of equality of pointed types and the characterization of equality of ∞-magmas we automatically get a characterization of the equality of pointed ∞-magmas.

```
module pointed-∞-magma {𝓤 : Universe} where

 open sip-join

 ∞-Magma· : 𝓤 ⁺ ̇
 ∞-Magma· = Σ X ꞉ 𝓤 ̇ , (X → X → X) × X

 _≅_ : ∞-Magma· → ∞-Magma· → 𝓤 ̇

 (X ,  _·_ , x₀) ≅ (Y ,  _*_ , y₀) =

                   Σ f ꞉ (X → Y), is-equiv f
                                × ((λ x x' → f (x · x')) ≡ (λ x x' → f x * f x'))
                                × (f x₀ ≡ y₀)
```

```
 characterization-of-pointed-magma-＝ : is-univalent 𝓤
                                      → (A B : ∞-Magma∙) → (A ＝ B) ≃ (A ≅ B)

 characterization-of-pointed-magma-＝ ua = characterization-of-join-＝ ua
                                             ∞-magma.sns-data
                                             pointed-type.sns-data
```

*Exercise*. The above equivalence is characterized by induction on identifications as the function that maps the reflexive identification to the identity equivalence.

### Monoids

In the following example, we combine joins and addition of axioms to get a characterization of the equality of monoids as monoid isomorphism.

```
module monoid {𝓤 : Universe} (ua : is-univalent 𝓤) where

 dfe : dfunext 𝓤 𝓤
 dfe = univalence-gives-dfunext ua

 open sip
 open sip-join
 open sip-with-axioms
 open monoids hiding (Monoid)

 monoid-structure : 𝓤 ̇ → 𝓤 ̇
 monoid-structure X = (X → X → X) × X

 monoid-axioms : (X : 𝓤 ̇ ) → monoid-structure X → 𝓤 ̇
 monoid-axioms X (_·_ , e) = is-set X
                           × left-neutral  e _·_
                           × right-neutral e _·_
                           × associative     _·_

 Monoid : 𝓤 ⁺ ̇
 Monoid = Σ X ꞉ 𝓤 ̇ , Σ s ꞉ monoid-structure X , monoid-axioms X s

 monoid-axioms-subsingleton : (X : 𝓤 ̇ ) (s : monoid-structure X)
                            → is-subsingleton (monoid-axioms X s)

 monoid-axioms-subsingleton X (_·_ , e) = subsingleton-criterion' γ
  where
   γ : monoid-axioms X (_·_ , e) → is-subsingleton (monoid-axioms X (_·_ , e))
   γ (i , _) = ×-is-subsingleton
                  (being-set-is-subsingleton dfe)
               (×-is-subsingleton
                  (Π-is-subsingleton dfe
                    (λ x → i (e · x) x))
               (×-is-subsingleton
                  (Π-is-subsingleton dfe
                    (λ x → i (x · e) x))
                  (Π-is-subsingleton dfe
                    (λ x → Π-is-subsingleton dfe
                    (λ y → Π-is-subsingleton dfe
                    (λ z → i ((x · y) · z) (x · (y · z))))))))

 sns-data : SNS (λ X → Σ s ꞉ monoid-structure X , monoid-axioms X s) 𝓤
 sns-data = add-axioms
              monoid-axioms monoid-axioms-subsingleton
              (join
                 ∞-magma.sns-data
                 pointed-type.sns-data)

 _≅_ : Monoid → Monoid → 𝓤 ̇
```

```
 (X , (_·_ , d) , _) ≅ (Y , (_*_ , e) , _) =

                    Σ f : (X → Y), is-equiv f
                                 × ((λ x x' → f (x · x')) ＝ (λ x x' → f x * f x'))
                                 × (f d ＝ e)

characterization-of-monoid-＝ : (A B : Monoid) → (A ＝ B) ≃ (A ≅ B)
characterization-of-monoid-＝ = characterization-of-＝ ua sns-data
```

*Exercise*. The above equivalence is characterized by induction on identifications as the function that maps the reflexive identification to the identity equivalence.

*Exercise.* A bijection that preserves the monoid multiplication automatically preserves the unit. We can say that the unit is property-like structure. This is because an associative magma, or semigroup, has at most one unit.

If we alternatively define monoids as associative magmas (that is, semigroups) with the property that a unit exists, then the structure identity principle automatically shows that identitications of monoids are equivalent to bijections that preserve the multiplication, without referring to the unit. However, functions that preserve the multiplication don't necessarily preserve the unit (exercise), that is, are not automatically monoid homomorphisms.

```
monoid-structure' : 𝓤 ̇ → 𝓤 ̇
monoid-structure' X = X → X → X

has-unit : {X : 𝓤 ̇ } → monoid-structure' X → 𝓤 ̇
has-unit {X} _·_ = Σ e ꞉ X , left-neutral  e _·_ × right-neutral e _·_
```

As discussed above, the difference is that now the unit is taken to be property rather than structure:

```
monoid-axioms' : (X : 𝓤 ̇ ) → monoid-structure' X → 𝓤 ̇
monoid-axioms' X _·_ = is-set X × has-unit _·_ × associative _·_

Monoid' : 𝓤 ⁺ ̇
Monoid' = Σ X ꞉ 𝓤 ̇  , Σ s ꞉ monoid-structure' X , monoid-axioms' X s
```

The equivalence of the alternative type `Monoid'` with the original type `Monoid` is just tuple reshuffling:

```
to-Monoid : Monoid' → Monoid
to-Monoid (X , _·_ , i , (e , l , r) , a) = (X , (_·_ , e) , (i , l , r , a))

from-Monoid : Monoid → Monoid'
from-Monoid (X , (_·_ , e) , (i , l , r , a)) = (X , _·_ , i , (e , l , r) , a)

to-Monoid-is-equiv : is-equiv to-Monoid
to-Monoid-is-equiv = invertibles-are-equivs to-Monoid (from-Monoid , refl , refl)

from-Monoid-is-equiv : is-equiv from-Monoid
from-Monoid-is-equiv = invertibles-are-equivs from-Monoid (to-Monoid , refl , refl)

the-two-types-of-monoids-coincide : Monoid' ≃ Monoid
the-two-types-of-monoids-coincide = to-Monoid , to-Monoid-is-equiv
```

And because the existence of a unit is property, the alternative monoid axioms are also property:

```
monoid-axioms'-subsingleton : (X : 𝓤 ̇ ) (s : monoid-structure' X)
                            → is-subsingleton (monoid-axioms' X s)

monoid-axioms'-subsingleton X _·_ = subsingleton-criterion' γ
 where
  γ : monoid-axioms' X _·_ → is-subsingleton (monoid-axioms' X _·_)
  γ (i , _ , _) =
    ×-is-subsingleton
     (being-set-is-subsingleton dfe)
   (×-is-subsingleton
     k
```

```
     (Π-is-subsingleton dfe (λ x →
      Π-is-subsingleton dfe (λ y →
      Π-is-subsingleton dfe (λ z → i ((x · y) · z) (x · (y · z)))))))
   where
    j : (e : X) → is-subsingleton (left-neutral e _·_ × right-neutral e _·_)
    j e = ×-is-subsingleton
           (Π-is-subsingleton dfe (λ x → i (e · x) x))
           (Π-is-subsingleton dfe (λ x → i (x · e) x))

    k : is-subsingleton (has-unit _·_)
    k (e , l , r) (e' , l' , r') = to-subtype-≡ j p
     where
      p = e        ≡⟨ (r' e)⁻¹ ⟩
         (e · e') ≡⟨ l e'     ⟩
         e'       ∎

sns-data' : SNS (λ X → Σ s ꞉ monoid-structure' X , monoid-axioms' X s) 𝓤
sns-data' = add-axioms
             monoid-axioms' monoid-axioms'-subsingleton
             ∞-magma.sns-data
```

As promised above, the characterization of equality doesn't refer to preservation of the unit:

```
_≅'_ : Monoid' → Monoid' → 𝓤 ̇
(X , _·_ , _) ≅' (Y , _*_ , _) =

              Σ f ꞉ (X → Y), is-equiv f
                          × ((λ x x' → f (x · x')) ≡ (λ x x' → f x * f x'))

characterization-of-monoid'-≡ : (A B : Monoid') → (A ≡ B) ≃ (A ≅' B)
characterization-of-monoid'-≡ = characterization-of-≡ ua sns-data'
```

We can define the type of semigroup isomorphisms of monoids as follows:

```
_≅ₛ_ : Monoid → Monoid → 𝓤 ̇

(X , (_·_ , _) , _) ≅ₛ (Y , (_*_ , _) , _) =

                    Σ f ꞉ (X → Y), is-equiv f
                                × ((λ x x' → f (x · x')) ≡ (λ x x' → f x * f x'))
```

We then get the following alternative characterization of monoid equality as semigroup-isomorphism, solving the above exercise, and a bit more.

```
2nd-characterization-of-monoid-≡ : (A B : Monoid) → (A ≡ B) ≃ A ≅ₛ B
2nd-characterization-of-monoid-≡ A B = (A ≡ B)                          ≃⟨ i   ⟩
                                       (from-Monoid A ≡ from-Monoid B)  ≃⟨ ii  ⟩
                                       (from-Monoid A ≅' from-Monoid B) ≃⟨ iii ⟩
                                       (A ≅ₛ B)                         ■

 where
  φ : A ≡ B → from-Monoid A ≡ from-Monoid B
  φ = ap from-Monoid

  from-Monoid-is-embedding : is-embedding from-Monoid
  from-Monoid-is-embedding = equivs-are-embeddings from-Monoid from-Monoid-is-equiv

  φ-is-equiv : is-equiv φ
  φ-is-equiv = embedding-gives-ap-is-equiv from-Monoid from-Monoid-is-embedding A B

  clearly : (from-Monoid A ≅' from-Monoid B) ≡ (A ≅ₛ B)
  clearly = refl _

  i   = (φ , φ-is-equiv)
  ii  = characterization-of-monoid'-≡ _ _
  iii = Id→Eq _ _ clearly
```

### Associative ∞-magmas

In the absence of the requirement that the underlying type is a set, the equivalences in the characterization of equality of associative ∞-magmas not only have to be homomorphic with respect to the magma operations but also need to respect the associativity data.

```
module associative-∞-magma
        {𝓤 : Universe}
        (ua : is-univalent 𝓤)
       where

 fe : dfunext 𝓤 𝓤
 fe = univalence-gives-dfunext ua

 associative : {X : 𝓤 ̇ } → (X → X → X) → 𝓤 ̇
 associative _·_ = ∀ x y z → (x · y) · z ≡ x · (y · z)

 ∞-amagma-structure : 𝓤 ̇ → 𝓤 ̇
 ∞-amagma-structure X = Σ _·_ ꞉ (X → X → X), (associative _·_)

 ∞-aMagma : 𝓤 ⁺ ̇
 ∞-aMagma = Σ X ꞉ 𝓤 ̇ , ∞-amagma-structure X

 homomorphic : {X Y : 𝓤 ̇ } → (X → X → X) → (Y → Y → Y) → (X → Y) → 𝓤 ̇
 homomorphic _·_ _*_ f = (λ x y → f (x · y)) ≡ (λ x y → f x * f y)
```

The notion of preservation of the associativity data depends not only on the homomorphism `f` but also on the homomorphism data `h` for `f`:

```
 respect-assoc : {X A : 𝓤 ̇ } (_·_ : X → X → X) (_*_ : A → A → A)
               → associative _·_
               → associative _*_
               → (f : X → A) → homomorphic _·_ _*_ f → 𝓤 ̇

 respect-assoc _·_ _*_ α β f h  =  fα ≡ βf
  where
   l = λ x y z → f ((x · y) · z)   ≡⟨ ap (λ - → - (x · y) z) h ⟩
                 f (x · y) * f z   ≡⟨ ap (λ - → - x y * f z) h ⟩
                 (f x * f y) * f z ∎

   r = λ x y z → f (x · (y · z))   ≡⟨ ap (λ - → - x (y · z)) h ⟩
                 f x * f (y · z)   ≡⟨ ap (λ - → f x * - y z) h ⟩
                 f x * (f y * f z) ∎

   fα βf : ∀ x y z → (f x * f y) * f z ≡ f x * (f y * f z)
   fα x y z = (l x y z)⁻¹ ∙ ap f (α x y z) ∙ r x y z
   βf x y z = β (f x) (f y) (f z)
```

The functions `l` and `r`, defined from the binary homomorphism condition `h`, give the homomorphism condition for the two induced ternary magma operations of each magma.

The following, which holds by construction, will be used implicitly:

```
 remark : {X : 𝓤 ̇ } (_·_ : X → X → X) (α β : associative _·_ )
        → respect-assoc _·_ _·_ α β id (refl _·_)
        ≡ ((λ x y z → refl ((x · y) · z) ∙ ap id (α x y z)) ≡ β)

 remark _·_ α β = refl _
```

The homomorphism condition `ι` is then defined as expected and the reflexivity condition `ρ` relies on the above remark.

```
 open sip hiding (homomorphic)

 sns-data : SNS ∞-amagma-structure 𝓤
```

```
sns-data = (ι , ρ , θ)
 where
  ι : (𝓧 𝓐 : ∞-aMagma) → ⟨ 𝓧 ⟩ ≃ ⟨ 𝓐 ⟩ → 𝓤 ̇
  ι (X , _·_ , α) (A , _*_ , β) (f , i) = Σ h ꞉ homomorphic _·_ _*_ f
                                              , respect-assoc _·_ _*_ α β f h

  ρ : (𝓧 : ∞-aMagma) → ι 𝓧 𝓧 (id-≃ ⟨ 𝓧 ⟩)
  ρ (X , _·_ , α) = h , p
   where
    h : homomorphic _·_ _·_ id
    h = refl _·_

    p : (λ x y z → refl ((x · y) · z) ∙ ap id (α x y z)) ≡ α
    p = fe (λ x → fe (λ y → fe (λ z → refl-left ∙ ap-id (α x y z))))
```

We prove the canonicity condition θ with the Yoneda machinery.

```
  u : (X : 𝓤 ̇ ) → ∀ s → ∃! t ꞉ ∞-amagma-structure X , ι (X , s) (X , t) (id-≃ X)
  u X (_·_ , α) = c , φ
   where
    c : Σ t ꞉ ∞-amagma-structure X , ι (X , _·_ , α) (X , t) (id-≃ X)
    c = (_·_ , α) , ρ (X , _·_ , α)

    φ : (σ : Σ t ꞉ ∞-amagma-structure X , ι (X , _·_ , α) (X , t) (id-≃ X)) → c ≡ σ
    φ ((_·_ , β) , refl _·_ , k) = γ
     where
      a : associative _·_
      a x y z = refl ((x · y) · z) ∙ ap id (α x y z)

      g : singleton-type' a → Σ t ꞉ ∞-amagma-structure X , ι (X , _·_ , α) (X , t) (id-≃ X)
      g (β , k) = (_·_ , β) , refl _·_ , k

      i : is-subsingleton (singleton-type' a)
      i = singletons-are-subsingletons _ (singleton-types'-are-singletons _ a)

      q : α , pr₂ (ρ (X , _·_ , α)) ≡ β , k
      q = i _ _

      γ : c ≡ (_·_ , β) , refl _·_ , k
      γ = ap g q

  θ : {X : 𝓤 ̇ } (s t : ∞-amagma-structure X) → is-equiv (canonical-map ι ρ s t)
  θ {X} s = universal-fiberwise-equiv (λ t → ι (X , s) (X , t) (id-≃ X))
             (u X s) s (canonical-map ι ρ s)
```

The promised characterization of associative ∞-magma equality then follows directly from the general structure of identity principle:

```
 _≅_ : ∞-aMagma → ∞-aMagma → 𝓤 ̇
 (X , _·_ , α) ≅ (Y , _*_ , β) = Σ f ꞉ (X → Y)
                                     , is-equiv f
                                     × (Σ h ꞉ homomorphic _·_ _*_ f
                                            , respect-assoc _·_ _*_ α β f h)

 characterization-of-∞-aMagma-≡ : (A B : ∞-aMagma) → (A ≡ B) ≃ (A ≅ B)
 characterization-of-∞-aMagma-≡ = characterization-of-≡ ua sns-data
```

### Groups

We add an axiom to monoids to get groups, so that we get a characterization of group equality from that of monoid equality.

```
module group {𝓤 : Universe} (ua : is-univalent 𝓤) where

 hfe : hfunext 𝓤 𝓤
```

```
hfe = univalence-gives-hfunext ua

open sip
open sip-with-axioms
open monoid {𝓤} ua hiding (sns-data ; _≅_ ; _≅'_)

group-structure : 𝓤 ̇ → 𝓤 ̇
group-structure X = Σ s ꞉ monoid-structure X , monoid-axioms X s

group-axiom : (X : 𝓤 ̇ ) → monoid-structure X → 𝓤 ̇
group-axiom X (_·_ , e) = (x : X) → Σ x' ꞉ X , (x · x' ≡ e) × (x' · x ≡ e)

Group : 𝓤 ⁺ ̇
Group = Σ X ꞉ 𝓤 ̇ , Σ ((_·_ , e) , a) ꞉ group-structure X , group-axiom X (_·_ , e)

inv-lemma : (X : 𝓤 ̇ ) (_·_ : X → X → X) (e : X)
          → monoid-axioms X (_·_ , e)
          → (x y z : X)
          → (y · x) ≡ e
          → (x · z) ≡ e
          → y ≡ z

inv-lemma X _·_  e (s , l , r , a) x y z q p =

   y             ≡⟨ (r y)⁻¹          ⟩
   (y · e)       ≡⟨ ap (y ·_) (p ⁻¹) ⟩
   (y · (x · z)) ≡⟨ (a y x z)⁻¹      ⟩
   ((y · x) · z) ≡⟨ ap (_· z) q      ⟩
   (e · z)       ≡⟨ l z              ⟩
   z             ∎

group-axiom-is-subsingleton : (X : 𝓤 ̇ )
                            → (s : group-structure X)
                            → is-subsingleton (group-axiom X (pr₁ s))

group-axiom-is-subsingleton X ((_·_ , e) , (s , l , r , a)) = γ
 where
  i : (x : X) → is-subsingleton (Σ x' ꞉ X , (x · x' ≡ e) × (x' · x ≡ e))
  i x (y , _ , q) (z , p , _) = u
   where
    t : y ≡ z
    t = inv-lemma X _·_ e (s , l , r , a) x y z q p

    u : (y , _ , q) ≡ (z , p , _)
    u = to-subtype-≡ (λ x' → ×-is-subsingleton (s (x · x') e) (s (x' · x) e)) t

  γ : is-subsingleton (group-axiom X (_·_ , e))
  γ = Π-is-subsingleton dfe i

sns-data : SNS (λ X → Σ s ꞉ group-structure X , group-axiom X (pr₁ s)) 𝓤
sns-data = add-axioms
            (λ X s → group-axiom X (pr₁ s)) group-axiom-is-subsingleton
            (monoid.sns-data ua)

_≅_ : Group → Group → 𝓤 ̇

(X , ((_·_ , d) , _) , _) ≅ (Y , ((_*_ , e) , _) , _) =

            Σ f ꞉ (X → Y), is-equiv f
                         × ((λ x x' → f (x · x')) ≡ (λ x x' → f x * f x'))
                         × (f d ≡ e)
```

```
characterization-of-group-≡ : (A B : Group) → (A ≡ B) ≃ (A ≅ B)
characterization-of-group-≡ = characterization-of-≡ ua sns-data
```

*Exercise*. The above equivalence is characterized by induction on identifications as the function that maps the reflexive identification to the identity equivalence.

*Exercise*. As in monoids, the preservation of the unit follows from the preservation of the multiplication, and hence one can remove `f d ≡ e` from the above definition. Prove that

```
(A ≅ B) ≃ (A ≅' B)
```

and hence, by transitivity,

```
(A ≡ B) ≃ (A ≅' B)
```

where

```
_≅'_ : Group → Group → 𝓤 ̇

(X , ((_·_ , d) , _) , _) ≅' (Y , ((_*_ , e) , _) , _) =

            Σ f ꞉ (X → Y), is-equiv f
                       × ((λ x x' → f (x · x')) ≡ (λ x x' → f x * f x'))
```

We now solve this exercise and do a bit more on the way, but in a different way as we did for monoids, for the sake of variation. We first name various projections and introduce notation.

```
group-structure-of : (G : Group) → group-structure ⟨ G ⟩
group-structure-of (X , ((_·_ , e) , i , l , r , a) , γ) = (_·_ , e) , i , l , r , a

monoid-structure-of : (G : Group) → monoid-structure ⟨ G ⟩
monoid-structure-of (X , ((_·_ , e) , i , l , r , a) , γ) = (_·_ , e)

monoid-axioms-of : (G : Group) → monoid-axioms ⟨ G ⟩ (monoid-structure-of G)
monoid-axioms-of (X , ((_·_ , e) , i , l , r , a) , γ) = i , l , r , a

multiplication : (G : Group) → ⟨ G ⟩ → ⟨ G ⟩ → ⟨ G ⟩
multiplication (X , ((_·_ , e) , i , l , r , a) , γ) = _·_

syntax multiplication G x y = x ·⟨ G ⟩ y

unit : (G : Group) → ⟨ G ⟩
unit (X , ((_·_ , e) , i , l , r , a) , γ) = e

group-is-set : (G : Group)
             → is-set ⟨ G ⟩

group-is-set (X , ((_·_ , e) , i , l , r , a) , γ) = i

unit-left : (G : Group) (x : ⟨ G ⟩)
          → unit G ·⟨ G ⟩ x ≡ x

unit-left (X , ((_·_ , e) , i , l , r , a) , γ) x = l x

unit-right : (G : Group) (x : ⟨ G ⟩)
           → x ·⟨ G ⟩ unit G ≡ x

unit-right (X , ((_·_ , e) , i , l , r , a) , γ) x = r x

assoc : (G : Group) (x y z : ⟨ G ⟩)
      → (x ·⟨ G ⟩ y) ·⟨ G ⟩ z ≡ x ·⟨ G ⟩ (y ·⟨ G ⟩ z)

assoc (X , ((_·_ , e) , i , l , r , a) , γ) = a
```

```
inv : (G : Group) → ⟨ G ⟩ → ⟨ G ⟩
inv (X , ((_·_ , e) , i , l , r , a) , γ) x = pr₁ (γ x)

inv-left : (G : Group) (x : ⟨ G ⟩)
         → inv G x ·⟨ G ⟩ x ≡ unit G

inv-left (X , ((_·_ , e) , i , l , r , a) , γ) x = pr₂ (pr₂ (γ x))

inv-right : (G : Group) (x : ⟨ G ⟩)
          → x ·⟨ G ⟩ inv G x ≡ unit G

inv-right (X , ((_·_ , e) , i , l , r , a) , γ) x = pr₁ (pr₂ (γ x))
```

We now solve the exercise.

```
preserves-multiplication : (G H : Group) → (⟨ G ⟩ → ⟨ H ⟩) → 𝓤 ̇
preserves-multiplication G H f = (λ (x y : ⟨ G ⟩) → f (x ·⟨ G ⟩ y))
                               ≡ (λ (x y : ⟨ G ⟩) → f x ·⟨ H ⟩ f y)

preserves-unit : (G H : Group) → (⟨ G ⟩ → ⟨ H ⟩) → 𝓤 ̇
preserves-unit G H f = f (unit G) ≡ unit H

idempotent-is-unit : (G : Group) (x : ⟨ G ⟩)
                   → x ·⟨ G ⟩ x ≡ x
                   → x ≡ unit G

idempotent-is-unit G x p = γ
 where
  x' = inv G x
  γ = x                        =⟨ (unit-left G x)⁻¹                        ⟩
      unit G ·⟨ G ⟩ x          =⟨ (ap (λ - → - ·⟨ G ⟩ x) (inv-left G x))⁻¹ ⟩
      (x' ·⟨ G ⟩ x) ·⟨ G ⟩ x   =⟨ assoc G x' x x                           ⟩
      x' ·⟨ G ⟩ (x ·⟨ G ⟩ x)   =⟨ ap (λ - → x' ·⟨ G ⟩ -) p                 ⟩
      x' ·⟨ G ⟩ x              =⟨ inv-left G x                             ⟩
      unit G                   ∎

unit-preservation-lemma : (G H : Group) (f : ⟨ G ⟩ → ⟨ H ⟩)
                        → preserves-multiplication G H f
                        → preserves-unit G H f

unit-preservation-lemma G H f m = idempotent-is-unit H e p
 where
  e  = f (unit G)

  p = e ·⟨ H ⟩ e               =⟨ ap (λ - → - (unit G) (unit G)) (m ⁻¹)   ⟩
      f (unit G ·⟨ G ⟩ unit G) =⟨ ap f (unit-left G (unit G))             ⟩
      e                        ∎
```

If a map preverves multiplication then it also preserves inverses:

```
inv-Lemma : (G : Group) (x y z : ⟨ G ⟩)
          → (y ·⟨ G ⟩ x) ≡ unit G
          → (x ·⟨ G ⟩ z) ≡ unit G
          → y ≡ z

inv-Lemma G = inv-lemma ⟨ G ⟩ (multiplication G) (unit G) (monoid-axioms-of G)

one-left-inv : (G : Group) (x x' : ⟨ G ⟩)
             → (x' ·⟨ G ⟩ x) ≡ unit G
             → x' ≡ inv G x
```

```
one-left-inv G x x' p = inv-Lemma G x x' (inv G x) p (inv-right G x)

one-right-inv : (G : Group) (x x' : ⟨ G ⟩)
              → (x ·⟨ G ⟩ x') ≡ unit G
              → x' ≡ inv G x

one-right-inv G x x' p = (inv-Lemma G x (inv G x) x' (inv-left G x) p)⁻¹

preserves-inv : (G H : Group) → (⟨ G ⟩ → ⟨ H ⟩) → 𝓤 ̇
preserves-inv G H f = (x : ⟨ G ⟩) → f (inv G x) ≡ inv H (f x)

inv-preservation-lemma : (G H : Group) (f : ⟨ G ⟩ → ⟨ H ⟩)
                       → preserves-multiplication G H f
                       → preserves-inv G H f

inv-preservation-lemma G H f m x = γ
 where
  p = f (inv G x) ·⟨ H ⟩ f x ≡⟨ (ap (λ - → - (inv G x) x) m)⁻¹  ⟩
      f (inv G x ·⟨ G ⟩ x)    ≡⟨ ap f (inv-left G x)             ⟩
      f (unit G)              ≡⟨ unit-preservation-lemma G H f m ⟩
      unit H                  ∎

  γ : f (inv G x) ≡ inv H (f x)
  γ = one-left-inv H (f x) (f (inv G x)) p
```

The usual notion of group homomorphism is that of multiplication-preserving function. But it is known that a group homomorphism in this sense is the same thing as a function that preserves both the multiplication and the unit.

```
is-homomorphism : (G H : Group) → (⟨ G ⟩ → ⟨ H ⟩) → 𝓤 ̇
is-homomorphism G H f = preserves-multiplication G H f
                      × preserves-unit G H f

preservation-of-mult-is-subsingleton : (G H : Group) (f : ⟨ G ⟩ → ⟨ H ⟩)
                                     → is-subsingleton (preserves-multiplication G H f)
preservation-of-mult-is-subsingleton G H f = j
 where
  j : is-subsingleton (preserves-multiplication G H f)
  j = Π-is-set hfe
       (λ _ → Π-is-set hfe
       (λ _ → group-is-set H))
       (λ (x y : ⟨ G ⟩) → f (x ·⟨ G ⟩ y))
       (λ (x y : ⟨ G ⟩) → f x ·⟨ H ⟩ f y)

being-homomorphism-is-subsingleton : (G H : Group) (f : ⟨ G ⟩ → ⟨ H ⟩)
                                   → is-subsingleton (is-homomorphism G H f)
being-homomorphism-is-subsingleton G H f = i
 where

  i : is-subsingleton (is-homomorphism G H f)
  i = ×-is-subsingleton
       (preservation-of-mult-is-subsingleton G H f)
       (group-is-set H (f (unit G)) (unit H))

notions-of-homomorphism-agree : (G H : Group) (f : ⟨ G ⟩ → ⟨ H ⟩)
                              → is-homomorphism G H f
                              ≃ preserves-multiplication G H f

notions-of-homomorphism-agree G H f = γ
 where
  α : is-homomorphism G H f → preserves-multiplication G H f
  α = pr₁

  β : preserves-multiplication G H f → is-homomorphism G H f
  β m = m , unit-preservation-lemma G H f m
```

```
  γ : is-homomorphism G H f ≃ preserves-multiplication G H f
  γ = logically-equivalent-subsingletons-are-equivalent _ _
       (being-homomorphism-is-subsingleton G H f)
       (preservation-of-mult-is-subsingleton G H f)
       (α , β)

 ≅-agreement : (G H : Group) → (G ≅ H) ≃ (G ≅' H)
 ≅-agreement G H = Σ-cong (λ f → Σ-cong (λ _ → notions-of-homomorphism-agree G H f))
```

This equivalence is that which forgets the preservation of the unit:

```
 forget-unit-preservation : (G H : Group) → (G ≅ H) → (G ≅' H)
 forget-unit-preservation G H (f , e , m , _) = f , e , m

 NB : (G H : Group) → ⌜ ≅-agreement G H ⌝ ≡ forget-unit-preservation G H
 NB G H = refl _

 forget-unit-preservation-is-equiv : (G H : Group)
                                   → is-equiv (forget-unit-preservation G H)

 forget-unit-preservation-is-equiv G H = ⌜⌝-is-equiv (≅-agreement G H)
```

This completes the solution of the exercise.

### The slice type

We now apply the above machinery to get a characterization of equality in the slice type.

```
module slice
        {𝓤 𝓥 : Universe}
        (R : 𝓥 ⁺ )
       where

 open sip

 S : 𝓤 ⁺ → 𝓤 ⊔ 𝓥 ⁺
 S X = X → R

 sns-data : SNS S (𝓤 ⊔ 𝓥)
 sns-data = (ι , ρ , θ)
  where
   ι : (A B : Σ S) → ⟨ A ⟩ ≃ ⟨ B ⟩ → 𝓤 ⊔ 𝓥 ⁺
   ι (X , g) (Y , h) (f , _) = (g ≡ h ∘ f)

   ρ : (A : Σ S) → ι A A (id-≃ ⟨ A ⟩)
   ρ (X , g) = refl g

   k : {X : 𝓤 ⁺ } {g h : S X} → canonical-map ι ρ g h ∼ id (g ≡ h)
   k (refl g) = refl (refl g)

   θ : {X : 𝓤 ⁺ } (g h : S X) → is-equiv (canonical-map ι ρ g h)
   θ g h = equivs-closed-under-∼ (id-is-equiv (g ≡ h)) k

 _≅_  : 𝓤 / R → 𝓤 / R → 𝓤 ⊔ 𝓥 ⁺
 (X , g) ≅ (Y , h) = Σ f ꞉ (X → Y), is-equiv f × (g ≡ h ∘ f)

 characterization-of-/-≡ : is-univalent 𝓤 → (A B : 𝓤 / R) → (A ≡ B) ≃ (A ≅ B)
 characterization-of-/-≡ ua = characterization-of-≡ ua sns-data
```

*Exercise*. The above equivalence is characterized by induction on identifications as the function that maps the reflexive identification to the identity equivalence.

### Subgroups

It is common mathematical practice to regard isomorphic groups as the same, which is a theorem in univalent mathematics, with the notion of sameness articulated by the identity type, as shown above. However, for some purposes, we may wish to consider two groups to be the same if they have the same elements. For example, in order to show that the subgroups of a group form an algebraic lattice with the finitely generated subgroups as the compact elements, it is this notion of equality that is used, with subgroup containment as the lattice order.

Asking whether two groups have the same elements in univalent mathematics doesn't make sense unless they are subgroups of the same ambient group. In the same way that in univalent mathematics two members of the powerset are equal iff they have the same elements, two subgroups are equal if and only if they have the same elements. The existence of an isomorphism of two subgroups does *not* imply their equality *in the type of subgroups*.

The notion of subgroup can be formulated in two equivalent ways.

1. A subgroup is an element of the powerset of the underlying set of the group that is closed under the group operations. So the type of subgroups of a given group is embedded as a subtype of the powerset of the underlying set and hence inherits the characterization of equality from the powerset.

2. A subgroup of a group `G` is a group `H` *together* with a homomorphic embedding `H → G`. With this second definition, two subgroups `H` and `H'` are equal iff the embeddings `H → G` and `H' → G` can be completed to a commutative triangle by a (necessarily unique) equivalence `H → H'`. So the type of subgroups of a group `G` with underlying set `⟨ G ⟩ : 𝓤` is embedded in the slice type `𝓤 / ⟨ G ⟩` and hence inherits the characterization of equality from the slice type.

```
module subgroup
        (𝓤  : Universe)
        (ua : Univalence)
       where

 gfe : global-dfunext
 gfe = univalence-gives-global-dfunext ua

 open sip
 open monoid {𝓤} (ua 𝓤) hiding (sns-data ; _≅_)
 open group {𝓤} (ua 𝓤)
```

We assume an arbitrary ambient group `G` in the following discussion.

```
 module ambient (G : Group) where

  _·_ : ⟨ G ⟩ → ⟨ G ⟩ → ⟨ G ⟩
  x · y = x ·⟨ G ⟩ y

  infixl 42 _·_
```

We abbreviate "closed under the group operations" by "group-closed":

```
  group-closed : (⟨ G ⟩ → 𝓥 ̇ ) → 𝓤 ⊔ 𝓥 ̇
  group-closed 𝓐 = 𝓐 (unit G)
                 × ((x y : ⟨ G ⟩) → 𝓐 x → 𝓐 y → 𝓐 (x · y))
                 × ((x : ⟨ G ⟩) → 𝓐 x → 𝓐 (inv G x))

  Subgroup : 𝓤 ⁺ ̇
  Subgroup = Σ A ꞉ 𝓟 ⟨ G ⟩ , group-closed (_∈ A)

  ⟪_⟫ : Subgroup → 𝓟 ⟨ G ⟩
  ⟪ A , u , c , ι ⟫ = A

  being-group-closed-subset-is-subsingleton : (A : 𝓟 ⟨ G ⟩) → is-subsingleton (group-closed (_∈ A))
  being-group-closed-subset-is-subsingleton A = ×-is-subsingleton
                                                  (∈-is-subsingleton A (unit G))
                                               (×-is-subsingleton
                                                  (Π-is-subsingleton dfe
                                                     (λ x → Π-is-subsingleton dfe
```

```
                                                   (λ y → Π-is-subsingleton dfe
                                                   (λ _ → Π-is-subsingleton dfe
                                                   (λ _ → ∈-is-subsingleton A (x · y))))))
                                                (Π-is-subsingleton dfe
                                                   (λ x → Π-is-subsingleton dfe
                                                   (λ _ → ∈-is-subsingleton A (inv G x)))))

⟪⟫-is-embedding : is-embedding ⟪_⟫
⟪⟫-is-embedding = pr₁-is-embedding being-group-closed-subset-is-subsingleton
```

Therefore equality of subgroups is equality of their underlying subsets in the powerset:

```
ap-⟪⟫ : (S T : Subgroup) → S ≡ T → ⟪ S ⟫ ≡ ⟪ T ⟫
ap-⟪⟫ S T = ap ⟪_⟫

ap-⟪⟫-is-equiv : (S T : Subgroup) → is-equiv (ap-⟪⟫ S T)
ap-⟪⟫-is-equiv = embedding-gives-ap-is-equiv ⟪_⟫ ⟪⟫-is-embedding

subgroups-form-a-set : is-set Subgroup
subgroups-form-a-set S T = equiv-to-subsingleton
                            (ap-⟪⟫ S T , ap-⟪⟫-is-equiv S T)
                            (powersets-are-sets' ua ⟪ S ⟫ ⟪ T ⟫)
```

It follows that two subgroups are equal if and only if they have the same elements:

```
subgroup-equality : (S T : Subgroup)
                  → (S ≡ T)
                  ≃ ((x : ⟨ G ⟩) → (x ∈ ⟪ S ⟫) ⇔ (x ∈ ⟪ T ⟫))

subgroup-equality S T = γ
 where
  f : S ≡ T → (x : ⟨ G ⟩) → x ∈ ⟪ S ⟫ ⇔ x ∈ ⟪ T ⟫
  f p x = transport (λ - → x ∈ ⟪ - ⟫) p , transport (λ - → x ∈ ⟪ - ⟫) (p ⁻¹)

  h : ((x : ⟨ G ⟩) → x ∈ ⟪ S ⟫ ⇔ x ∈ ⟪ T ⟫) → ⟪ S ⟫ ≡ ⟪ T ⟫
  h φ = subset-extensionality' ua α β
   where
    α : ⟪ S ⟫ ⊆ ⟪ T ⟫
    α x = lr-implication (φ x)

    β : ⟪ T ⟫ ⊆ ⟪ S ⟫
    β x = rl-implication (φ x)

  g : ((x : ⟨ G ⟩) → x ∈ ⟪ S ⟫ ⇔ x ∈ ⟪ T ⟫) → S ≡ T
  g = inverse (ap-⟪⟫ S T) (ap-⟪⟫-is-equiv S T) ∘ h

  γ : (S ≡ T) ≃ ((x : ⟨ G ⟩) → x ∈ ⟪ S ⟫ ⇔ x ∈ ⟪ T ⟫)
  γ = logically-equivalent-subsingletons-are-equivalent _ _
        (subgroups-form-a-set S T)
        (Π-is-subsingleton dfe
           (λ x → ×-is-subsingleton
                    (Π-is-subsingleton dfe (λ _ → ∈-is-subsingleton ⟪ T ⟫ x))
                    (Π-is-subsingleton dfe (λ _ → ∈-is-subsingleton ⟪ S ⟫ x))))
        (f , g)
```

We now show that the type of subgroups is equivalent to the following type, as an application of the subtype classifier.

```
Subgroup' : 𝓤 ⁺ ̇
Subgroup' = Σ H ꞉ Group
          , Σ h ꞉ (⟨ H ⟩ → ⟨ G ⟩)
          , is-embedding h
          × is-homomorphism H G h
```

It will be convenient to introduce notation for the type of group structures satisfying the group axioms:

```
T : 𝓤 ̇ → 𝓤 ̇
T X = Σ ((_·_ , e) , a) ꞉ group-structure X , group-axiom X (_·_ , e)
```

We use an anonymous module to give common assumptions for the following few lemmas:

```
 module _ {X : 𝓤 ̇ } (h : X → ⟨ G ⟩) (h-is-embedding : is-embedding h) where

  private
   h-lc : left-cancellable h
   h-lc = embeddings-are-lc h h-is-embedding

  having-group-closed-fiber-is-subsingleton : is-subsingleton (group-closed (fiber h))
  having-group-closed-fiber-is-subsingleton = being-group-closed-subset-is-subsingleton A
   where
    A : 𝓟 ⟨ G ⟩
    A y = (fiber h y , h-is-embedding y)

  at-most-one-homomorphic-structure : is-subsingleton (Σ τ ꞉ T X , is-homomorphism (X , τ) G h)
  at-most-one-homomorphic-structure
     ((((_*_ ,  unitH) ,  maxioms) ,  gaxiom) ,  (pmult ,  punit))
     ((((_*'_ , unitH') , maxioms') , gaxiom') , (pmult' , punit'))
   = γ
   where
    τ τ' : T X
    τ  = ((_*_ ,  unitH) ,  maxioms) ,  gaxiom
    τ' = ((_*'_ , unitH') , maxioms') , gaxiom'

    i :  is-homomorphism (X , τ)  G h
    i  = (pmult ,  punit)

    i' : is-homomorphism (X , τ') G h
    i' = (pmult' , punit')

    p : _*_ ＝ _*'_
    p = gfe (λ x → gfe (λ y → h-lc (h (x * y)  ＝⟨  ap (λ - → - x y) pmult      ⟩
                                    h x · h y  ＝⟨ (ap (λ - → - x y) pmult')⁻¹ ⟩
                                    h (x *' y) ∎)))

    q : unitH ＝ unitH'
    q = h-lc (h unitH  ＝⟨  punit     ⟩
              unit G   ＝⟨  punit' ⁻¹ ⟩
              h unitH' ∎)

    r : (_*_ , unitH) ＝ (_*'_ , unitH')
    r = to-×-＝ (p , q)

    δ : τ ＝ τ'
    δ = to-subtype-＝
          (group-axiom-is-subsingleton X)
          (to-subtype-＝
             (monoid-axioms-subsingleton X)
             r)

    γ : (τ  , i) ＝ (τ' , i')
    γ = to-subtype-＝ (λ τ → being-homomorphism-is-subsingleton (X , τ) G h) δ

  homomorphic-structure-gives-group-closed-fiber : (Σ τ ꞉ T X , is-homomorphism (X , τ) G h)
                                                 → group-closed (fiber h)

  homomorphic-structure-gives-group-closed-fiber
      ((((_*_ , unitH) , maxioms) , gaxiom) , (pmult , punit))
    = (unitc , mulc , invc)
   where
    H : Group
    H = X , ((_*_ , unitH) , maxioms) , gaxiom

    unitc : fiber h (unit G)
    unitc = unitH , punit

    mulc : ((x y : ⟨ G ⟩) → fiber h x → fiber h y → fiber h (x · y))
```

```
  mulc x y (a , p) (b , q) = (a * b) ,
                             (h (a * b) =⟨ ap (λ - → - a b) pmult    ⟩
                              h a · h b =⟨ ap₂ (λ - -' → - · -') p q ⟩
                              x · y     ∎)

  invc : ((x : ⟨ G ⟩) → fiber h x → fiber h (inv G x))
  invc x (a , p) = inv H a ,
                   (h (inv H a) =⟨ inv-preservation-lemma H G h pmult a ⟩
                    inv G (h a) =⟨ ap (inv G) p                         ⟩
                    inv G x     ∎)
```

Conversely:

```
group-closed-fiber-gives-homomorphic-structure : group-closed (fiber h)
                                               → (Σ τ ꞉ T X , is-homomorphism (X , τ) G h)

group-closed-fiber-gives-homomorphic-structure (unitc , mulc , invc) = τ , i
 where
  φ : (x : X) → fiber h (h x)
  φ x = (x , refl (h x))

  unitH : X
  unitH = fiber-point unitc

  _*_ : X → X → X
  x * y = fiber-point (mulc (h x) (h y) (φ x) (φ y))

  invH : X → X
  invH x = fiber-point (invc (h x) (φ x))

  pmul : (x y : X) → h (x * y) ≡ h x · h y
  pmul x y = fiber-identification (mulc (h x) (h y) (φ x) (φ y))

  punit : h unitH ≡ unit G
  punit = fiber-identification unitc

  pinv : (x : X) → h (invH x) ≡ inv G (h x)
  pinv x = fiber-identification (invc (h x) (φ x))

  unitH-left : (x : X) → unitH * x ≡ x
  unitH-left x = h-lc (h (unitH * x) =⟨ pmul unitH x       ⟩
                       h unitH · h x =⟨ ap (_· h x) punit ⟩
                       unit G · h x  =⟨ unit-left G (h x) ⟩
                       h x           ∎)

  unitH-right : (x : X) → x * unitH ≡ x
  unitH-right x = h-lc (h (x * unitH) =⟨ pmul x unitH        ⟩
                        h x · h unitH =⟨ ap (h x ·_) punit  ⟩
                        h x · unit G  =⟨ unit-right G (h x) ⟩
                        h x           ∎)

  assocH : (x y z : X) → ((x * y) * z) ≡ (x * (y * z))
  assocH x y z = h-lc (h ((x * y) * z)   =⟨ pmul (x * y) z             ⟩
                       h (x * y) · h z   =⟨ ap (_· h z) (pmul x y)     ⟩
                       (h x · h y) · h z =⟨ assoc G (h x) (h y) (h z)  ⟩
                       h x · (h y · h z) =⟨ (ap (h x ·_) (pmul y z))⁻¹ ⟩
                       h x · h (y * z)   =⟨ (pmul x (y * z))⁻¹         ⟩
                       h (x * (y * z))   ∎)

  group-axiomH : (x : X) → Σ x' ꞉ X , (x * x' ≡ unitH) × (x' * x ≡ unitH)
  group-axiomH x = invH x ,

                   h-lc (h (x * invH x)     =⟨ pmul x (invH x)      ⟩
                         h x · h (invH x)   =⟨ ap (h x ·_) (pinv x) ⟩
                         h x · inv G (h x)  =⟨ inv-right G (h x)    ⟩
```

```
                     unit G             =⟨ punit ⁻¹          ⟩
                     h unitH            ∎),

              h-lc ((h (invH x * x)     =⟨ pmul (invH x) x    ⟩
                      h (invH x) · h x  =⟨ ap (_· h x) (pinv x) ⟩
                      inv G (h x) · h x =⟨ inv-left G (h x)   ⟩
                      unit G            =⟨ punit ⁻¹           ⟩
                      h unitH           ∎))

  j : is-set X
  j = subtypes-of-sets-are-sets h h-lc (group-is-set G)

  τ : T X
  τ = ((_*_ , unitH) , (j , unitH-left , unitH-right , assocH)) , group-axiomH

  i : is-homomorphism (X , τ) G h
  i = gfe (λ x → gfe (pmul x)) , punit
```

What is important for our purposes is that this gives an equivalence:

```
 fiber-structure-lemma : group-closed (fiber h)
                       ≃ (Σ τ : T X , is-homomorphism (X , τ) G h)

 fiber-structure-lemma = logically-equivalent-subsingletons-are-equivalent _ _
                          having-group-closed-fiber-is-subsingleton
                          at-most-one-homomorphic-structure
                          (group-closed-fiber-gives-homomorphic-structure ,
                           homomorphic-structure-gives-group-closed-fiber)
```

This is the end of the anonymous submodule and we can now prove the desired result. We apply the material on the subtype classifier.

```
characterization-of-the-type-of-subgroups :  Subgroup ≃ Subgroup'
characterization-of-the-type-of-subgroups =

 Subgroup                                                                          ≃⟨ i
 (Σ A : 𝓟 ⟨ G ⟩ , group-closed (_∈ A))                                             ≃⟨ ii
 (Σ (X , h , e) : Subtype ⟨ G ⟩ , group-closed (fiber h))                          ≃⟨ iii
 (Σ X : 𝓤 ̇ , Σ (h , e) : X ↪ ⟨ G ⟩ , group-closed (fiber h))                       ≃⟨ iv
 (Σ X : 𝓤 ̇ , Σ (h , e) : X ↪ ⟨ G ⟩ , Σ τ : T X , is-homomorphism (X , τ) G h)      ≃⟨ v    ⟩
 (Σ X : 𝓤 ̇ , Σ h : (X → ⟨ G ⟩) , Σ e : is-embedding h , Σ τ : T X , is-homomorphism (X , τ) G h) ≃⟨ vi   ⟩
 (Σ X : 𝓤 ̇ , Σ h : (X → ⟨ G ⟩) , Σ τ : T X , Σ e : is-embedding h , is-homomorphism (X , τ) G h) ≃⟨ vii  ⟩
 (Σ X : 𝓤 ̇ , Σ τ : T X , Σ h : (X → ⟨ G ⟩) , is-embedding h × is-homomorphism (X , τ) G h)    ≃⟨ viii ⟩
 (Σ H : Group , Σ h : (⟨ H ⟩ → ⟨ G ⟩) , is-embedding h × is-homomorphism H G h)               ≃⟨ ix   ⟩
 Subgroup'                                                                         ■

   where
    φ : Subtype ⟨ G ⟩ → 𝓟 ⟨ G ⟩
    φ = χ-special is-subsingleton ⟨ G ⟩

    j : is-equiv φ
    j = χ-special-is-equiv (ua 𝓤) gfe is-subsingleton ⟨ G ⟩

    i    = Id→Eq _ _ (refl Subgroup)
    ii   = Σ-change-of-variable (λ (A : 𝓟 ⟨ G ⟩) → group-closed (_∈ A)) φ j
    iii  = Σ-assoc
    iv   = Σ-cong (λ X → Σ-cong (λ (h , e) → fiber-structure-lemma h e))
    v    = Σ-cong (λ X → Σ-assoc)
    vi   = Σ-cong (λ X → Σ-cong (λ h → Σ-flip))
    vii  = Σ-cong (λ X → Σ-flip)
    viii = ≃-sym Σ-assoc
    ix   = Id→Eq _ _ (refl Subgroup')
```

In particular, a subgroup induces a genuine group:

```
 induced-group : Subgroup → Group
 induced-group S = pr₁ (⌜ characterization-of-the-type-of-subgroups ⌝ S)
```

By applying the other projections, the induced group is homomorphically embedded into the ambient group.

The crucial tool to characterize equality in the alternative type of subgroups is the following embedding into the slice type.

```
forgetful-map : Subgroup' → 𝓤 / ⟨ G ⟩
forgetful-map ((X , _)  , h  , _) = (X , h)
```

To show that this map is an embedding, we express it as a composition of maps that are more easily seen to be embeddings.

```
forgetful-map-is-embedding : is-embedding forgetful-map
forgetful-map-is-embedding = γ
 where
  Subtype' : 𝓤 ̇ → 𝓤 ⁺ ̇
  Subtype' X = Σ (X , h) : 𝓤 / ⟨ G ⟩ , is-embedding h

  f₀ : Subgroup' → Subtype ⟨ G ⟩
  f₀ ((X , _)  , h  , e , _) = (X , h , e)

  f₁ : Subtype ⟨ G ⟩ → Subtype' ⟨ G ⟩
  f₁ (X , h , e) = ((X , h) , e)

  f₂ : Subtype' ⟨ G ⟩ → 𝓤 / ⟨ G ⟩
  f₂ ((X , h) , e) = (X , h)

  by-construction : forgetful-map ≡ f₂ ∘ f₁ ∘ f₀
  by-construction = refl _

  f₀-lc : left-cancellable f₀
  f₀-lc {(X , τ) , h , e , i} {(X , τ') , h , e , i'} (refl (X , h , e)) = δ
   where
    p : (τ , i) ≡ (τ' , i')
    p = at-most-one-homomorphic-structure h e (τ , i) (τ' , i')

    φ : (Σ τ : T X , is-homomorphism (X , τ) G h) → Subgroup'
    φ (τ , i) = ((X , τ) , h , e , i)

    δ : ((X , τ) , h , e , i) ≡ ((X , τ') , h , e , i')
    δ = ap φ p

  f₀-is-embedding : is-embedding f₀
  f₀-is-embedding = lc-maps-into-sets-are-embeddings f₀ f₀-lc (subtypes-form-set ua ⟨ G ⟩)

  f₁-is-equiv : is-equiv f₁
  f₁-is-equiv = invertibles-are-equivs f₁ ((λ ((X , h) , e) → (X , h , e)) , refl , refl)

  f₁-is-embedding : is-embedding f₁
  f₁-is-embedding = equivs-are-embeddings f₁ f₁-is-equiv

  f₂-is-embedding : is-embedding f₂
  f₂-is-embedding = pr₁-is-embedding (λ (X , h) → being-embedding-is-subsingleton gfe h)

  γ : is-embedding forgetful-map
  γ = ∘-embedding f₂-is-embedding (∘-embedding f₁-is-embedding f₀-is-embedding)
```

With this and the characterization of equality in the slice type, we get the promised characterization of equality of the alternative type of subgroups.

```
_≡ₛ_ : Subgroup' →  Subgroup' → 𝓤 ̇
(H , h , _ ) ≡ₛ (H' , h' , _ ) = Σ f : (⟨ H ⟩ → ⟨ H' ⟩) , is-equiv f × (h ≡ h' ∘ f)

subgroup'-equality : (S T : Subgroup') → (S ≡ T) ≃ (S ≡ₛ T)
subgroup'-equality S T = (S ≡ T)                               ≃⟨ i  ⟩
                         (forgetful-map S ≡ forgetful-map T)  ≃⟨ ii ⟩
                         (S ≡ₛ T)                              ■
 where
  open slice ⟨ G ⟩ hiding (S)
  i  = ≃-sym (embedding-criterion-converse forgetful-map forgetful-map-is-embedding S T)
  ii = characterization-of-/-≡ (ua 𝓤) (forgetful-map S) (forgetful-map T)
```

The equivalence `f` in the definition of the relation `＝ₛ` is unique when it exists, because `h'` is an embedding and hence is left-cancellable. Moreover, the type `S ＝ₛ T` has at most one element:

```
subgroups'-form-a-set : is-set Subgroup'
subgroups'-form-a-set = equiv-to-set
                         (≃-sym characterization-of-the-type-of-subgroups)
                         subgroups-form-a-set

＝ₛ-is-subsingleton-valued : (S T : Subgroup') → is-subsingleton (S ＝ₛ T)
＝ₛ-is-subsingleton-valued S T = γ
 where
  i : is-subsingleton (S ＝ T)
  i = subgroups'-form-a-set S T

  γ : is-subsingleton (S ＝ₛ T)
  γ = equiv-to-subsingleton (≃-sym (subgroup'-equality S T)) i
```

Here is an alternative proof that avoids the equivalence `Subgroup ≃ Subgroup'` used above to show that the alternative type of subgroups is a set:

```
＝ₛ-is-subsingleton-valued' : (S S' : Subgroup') → is-subsingleton (S ＝ₛ S')
＝ₛ-is-subsingleton-valued' (H , h , e , i) (H' , h' , e' , i') = γ
 where
  S  = (H  , h  , e  , i )
  S' = (H' , h' , e' , i')

  A = Σ f ꞉ (⟨ H ⟩ → ⟨ H' ⟩) , h' ∘ f ＝ h
  B = Σ (f , p) ꞉ A , is-equiv f

  A-is-subsingleton : is-subsingleton A
  A-is-subsingleton = postcomp-is-embedding gfe hfe h' e' ⟨ H ⟩ h

  B-is-subsingleton : is-subsingleton B
  B-is-subsingleton = Σ-is-subsingleton
                       A-is-subsingleton
                       (λ (f , p) → being-equiv-is-subsingleton gfe gfe f)

  δ : (S ＝ₛ S') ≃ B
  δ = invertibility-gives-≃ α (β , η , ε)
   where
    α = λ (f , i , p) → ((f , (p ⁻¹)) , i)
    β = λ ((f , p) , i) → (f , i , (p ⁻¹))
    η = λ (f , i , p) → ap (λ - → (f , i , -)) (⁻¹-involutive p)
    ε = λ ((f , p) , i) → ap (λ - → ((f , -) , i)) (⁻¹-involutive p)

  γ : is-subsingleton (S ＝ₛ S')
  γ = equiv-to-subsingleton δ B-is-subsingleton
```

### Rings

A mathematician asked us what a formalization of Noetherian local rings would look like in univalent type theory, in particular with respect to automatic preservation of theorems about rings by ring isomorphisms. In this section we consider rings, and we consider Noetherian local rings after we discuss unspecified existence.

We consider rings without unit, called *rngs*, and with unit, called *rings*.

There are several options to apply the above techniques to accomplish this. One way would be to add structure to Abelian groups. However, in order to avoid having to discuss the preservation of the neutral element of addition by homomorphisms separately as above in the case of groups, we proceed from scratch, where the neutral element is not part of the structure but instead its existence is part of the axioms. Cf. the discussions of equality of monoids and groups.

*Exercise.* Proceed using the alternative approach, which should be equally easy and short (and perhaps even shorter).

We consider r(i)ngs in a universe $\mathcal{U}$, and we assume univalence in their development. We hide the notation `⟨_⟩` from the module `sip` because we are going to use it for the underlying `Rng` of a `Ring`:

```
module ring {𝓤 : Universe} (ua : Univalence) where

 open sip hiding (⟨_⟩)
 open sip-with-axioms
 open sip-join
```

We derive function extensionality from univalence:

```
 fe : global-dfunext
 fe = univalence-gives-global-dfunext ua

 hfe : global-hfunext
 hfe = univalence-gives-global-hfunext ua
```

We take rng structure to be the product of two magma structures:

```
 rng-structure : 𝓤 ̇ → 𝓤 ̇
 rng-structure X = (X → X → X) × (X → X → X)
```

The axioms are the usual ones, with the additional requirement that the underlying type is a set, as opposed to an arbitrary ∞-groupoid:

```
 rng-axioms : (R : 𝓤 ̇ ) → rng-structure R → 𝓤 ̇
 rng-axioms R (_+_ , _·_) = I × II × III × IV × V × VI × VII
  where
    I   = is-set R
    II  = (x y z : R) → (x + y) + z ＝ x + (y + z)
    III = (x y : R) → x + y ＝ y + x
    IV  = Σ O ꞉ R , ((x : R) → x + O ＝ x) × ((x : R) → Σ x' ꞉ R , x + x' ＝ O)
    V   = (x y z : R) → (x · y) · z ＝ x · (y · z)
    VI  = (x y z : R) → x · (y + z) ＝ (x · y) + (x · z)
    VII = (x y z : R) → (y + z) · x ＝ (y · x) + (z · x)
```

The type of rngs in the universe $\mathcal{U}$, which lives in the universe after $\mathcal{U}$:

```
 Rng : 𝓤 ⁺ ̇
 Rng = Σ R ꞉ 𝓤 ̇ , Σ s ꞉ rng-structure R , rng-axioms R s
```

In order to be able to apply univalence to show that the identity type `𝓡 ＝ 𝓡'` of two rngs is in canonical bijection with the type `𝓡 ≅ 𝓡'` of ring isomorphisms, we need to show that the axioms constitute property rather than data, that is, they form a subsingleton, or a type with at most one element. The proof is a mix of algebra (to show that an additive semigroup has at most one zero element, and at most one additive inverse for each element) and general facts about subsingletons (e.g. they are closed under products) and is entirely routine.

```
 rng-axioms-is-subsingleton : (R : 𝓤 ̇ ) (s : rng-structure R)
                            → is-subsingleton (rng-axioms R s)

 rng-axioms-is-subsingleton R (_+_ , _·_) (i , ii , iii , iv-vii) = δ
  where
    A   = λ (O : R) → ((x : R) → x + O ＝ x)
                    × ((x : R) → Σ x' ꞉ R , x + x' ＝ O)

    IV  = Σ A

    a : (O O' : R) → ((x : R) → x + O ＝ x) → ((x : R) → x + O' ＝ x) → O ＝ O'
    a O O' f f' = O        ＝⟨ (f' O)⁻¹ ⟩
                 (O + O') ＝⟨ iii O O' ⟩
                 (O' + O) ＝⟨ f O'     ⟩
                  O'       ∎

    b : (O : R) → is-subsingleton ((x : R) → x + O ＝ x)
    b O = Π-is-subsingleton fe (λ x → i (x + O) x)
```

```
    c : (O : R)
      → ((x : R) → x + O ≡ x)
      → (x : R) → is-subsingleton (Σ x' ꞉ R , x + x' ≡ O)
    c O f x (x' , p') (x'' , p'') = to-subtype-≡ (λ x' → i (x + x') O) r
     where
      r : x' ≡ x''
      r = x'               ≡⟨ (f x')⁻¹               ⟩
          (x' + O)         ≡⟨ ap (x' +_) (p'' ⁻¹)    ⟩
          (x' + (x + x'')) ≡⟨ (ii x' x x'')⁻¹        ⟩
          ((x' + x) + x'') ≡⟨ ap (_+ x'') (iii x' x) ⟩
          ((x + x') + x'') ≡⟨ ap (_+ x'') p'         ⟩
          (O + x'')        ≡⟨ iii O x''              ⟩
          (x'' + O)        ≡⟨ f x''                  ⟩
          x''              ∎

    d : (O : R) → is-subsingleton (A O)
    d O (f , g) = φ (f , g)
     where
      φ : is-subsingleton (A O)
      φ = ×-is-subsingleton (b O) (Π-is-subsingleton fe (λ x → c O f x))

    IV-is-subsingleton : is-subsingleton IV
    IV-is-subsingleton (O , f , g) (O' , f' , g') = e
     where
      e : (O , f , g) ≡ (O' , f' , g')
      e = to-subtype-≡ d (a O O' f f')

    γ : is-subsingleton (rng-axioms R (_+_ , _·_))
    γ = ×-is-subsingleton
          (being-set-is-subsingleton fe)
       (×-is-subsingleton
          (Π-is-subsingleton fe
          (λ x → Π-is-subsingleton fe
          (λ y → Π-is-subsingleton fe
          (λ z → i ((x + y) + z) (x + (y + z))))))
       (×-is-subsingleton
          (Π-is-subsingleton fe
          (λ x → Π-is-subsingleton fe
          (λ y → i (x + y) (y + x))))
       (×-is-subsingleton
          IV-is-subsingleton
       (×-is-subsingleton
          (Π-is-subsingleton fe
          (λ x → Π-is-subsingleton fe
          (λ y → Π-is-subsingleton fe
          (λ z → i ((x · y) · z) (x · (y · z))))))
       (×-is-subsingleton
          (Π-is-subsingleton fe
          (λ x → Π-is-subsingleton fe
          (λ y → Π-is-subsingleton fe
          (λ z → i (x · (y + z)) ((x · y) + (x · z))))))

          (Π-is-subsingleton fe
          (λ x → Π-is-subsingleton fe
          (λ y → Π-is-subsingleton fe
          (λ z → i ((y + z) · x) ((y · x) + (z · x)))))))))))

    δ : (α : rng-axioms R (_+_ , _·_)) → (i , ii , iii , iv-vii) ≡ α
    δ = γ (i , ii , iii , iv-vii)
```

We define a rng isomorphism to be a bijection that preserves addition and multiplication, and collect all isomorphisms of two rngs 𝓡 and 𝓡' in a type 𝓡 ≅[Rng] 𝓡':

```
 _≅[Rng]_ : Rng → Rng → 𝓤 ̇

 (R , (_+_ , _·_) , _) ≅[Rng] (R' , (_+'_ , _·'_) , _) =

                       Σ f ꞉ (R → R')
```

```
                                            , is-equiv f
                                            × ((λ x y → f (x + y)) ≡ (λ x y → f x +' f y))
                                            × ((λ x y → f (x · y)) ≡ (λ x y → f x ·' f y))
```

Then the type of rng identities is in bijection with the type of ring isomorphisms by the above general machinery:

```
characterization-of-rng-≡ : (𝓡 𝓡' : Rng) → (𝓡 ≡ 𝓡') ≃ (𝓡 ≅[Rng] 𝓡')
characterization-of-rng-≡ = characterization-of-≡ (ua 𝓤)
                             (add-axioms
                               rng-axioms
                               rng-axioms-is-subsingleton
                               (join
                                 ∞-magma.sns-data
                                 ∞-magma.sns-data))
```

The underlying type of a rng:

```
⟨_⟩ : (𝓡 : Rng) → 𝓤 ̇
⟨ R , _ ⟩ = R
```

We now add units to rngs to get rings.

```
ring-structure : 𝓤 ̇ → 𝓤 ̇
ring-structure X = X × rng-structure X

ring-axioms : (R : 𝓤 ̇ ) → ring-structure R → 𝓤 ̇
ring-axioms R (𝟏 , _+_ , _·_) = rng-axioms R (_+_ , _·_) × VIII
 where
  VIII = (x : R) → (x · 𝟏 ≡ x) × (𝟏 · x ≡ x)

ring-axioms-is-subsingleton : (R : 𝓤 ̇ ) (s : ring-structure R)
                            → is-subsingleton (ring-axioms R s)

ring-axioms-is-subsingleton R (𝟏 , _+_ , _·_) ((i , ii-vii) , viii) = γ ((i , ii-vii) , viii)
 where
  γ : is-subsingleton (ring-axioms R (𝟏 , _+_ , _·_))
  γ = ×-is-subsingleton
        (rng-axioms-is-subsingleton R (_+_ , _·_))
        (Π-is-subsingleton fe (λ x → ×-is-subsingleton (i (x · 𝟏) x) (i (𝟏 · x) x)))
```

The type of rings with unit:

```
Ring : 𝓤 ⁺ ̇
Ring = Σ R ꞉ 𝓤 ̇ , Σ s ꞉ ring-structure R , ring-axioms R s

_≅[Ring]_ : Ring → Ring → 𝓤 ̇

(R , (𝟏 , _+_ , _·_) , _) ≅[Ring] (R' , (𝟏' , _+'_ , _·'_) , _) =

                                Σ f ꞉ (R → R')
                                    , is-equiv f
                                    × (f 𝟏 ≡ 𝟏')
                                    × ((λ x y → f (x + y)) ≡ (λ x y → f x +' f y))
                                    × ((λ x y → f (x · y)) ≡ (λ x y → f x ·' f y))

characterization-of-ring-≡ : (𝓡 𝓡' : Ring) → (𝓡 ≡ 𝓡') ≃ (𝓡 ≅[Ring] 𝓡')
characterization-of-ring-≡ = characterization-of-≡ (ua 𝓤)
                              (add-axioms
                                ring-axioms
                                ring-axioms-is-subsingleton
                                (join
                                  pointed-type.sns-data
                                    (join
```

```
                                      ∞-magma.sns-data
                                      ∞-magma.sns-data)))
```

### Metric spaces, graphs and ordered structures

We now apply the above machinery to get a characterization of equality of metric spaces, graphs and ordered structures/

```
module generalized-metric-space
        {𝓤 𝓥 : Universe}
        (R : 𝓥 ̇ )
        (axioms  : (X : 𝓤 ̇ ) → (X → X → R) → 𝓤 ⊔ 𝓥 ̇ )
        (axiomss : (X : 𝓤 ̇ ) (d : X → X → R) → is-subsingleton (axioms X d))
       where

 open sip
 open sip-with-axioms

 S : 𝓤 ̇ → 𝓤 ⊔ 𝓥 ̇
 S X = X → X → R

 sns-data : SNS S (𝓤 ⊔ 𝓥)
 sns-data = (ι , ρ , θ)
  where
   ι : (A B : Σ S) → ⟨ A ⟩ ≃ ⟨ B ⟩ → 𝓤 ⊔ 𝓥 ̇
   ι (X , d) (Y , e) (f , _) = (d ≡ λ x x' → e (f x) (f x'))

   ρ : (A : Σ S) → ι A A (id-≃ ⟨ A ⟩)
   ρ (X , d) = refl d

   h : {X : 𝓤 ̇ } {d e : S X} → canonical-map ι ρ d e ∼ id (d ≡ e)
   h (refl d) = refl (refl d)

   θ : {X : 𝓤 ̇ } (d e : S X) → is-equiv (canonical-map ι ρ d e)
   θ d e = equivs-closed-under-∼ (id-is-equiv (d ≡ e)) h

 M : 𝓤 ⁺ ⊔ 𝓥 ̇
 M = Σ X ꞉ 𝓤 ̇ , Σ d ꞉ (X → X → R) , axioms X d

 _≅_  : M → M → 𝓤 ⊔ 𝓥 ̇
 (X , d , _) ≅ (Y , e , _) = Σ f ꞉ (X → Y), is-equiv f
                                          × (d ≡ λ x x' → e (f x) (f x'))

 characterization-of-M-≡ : is-univalent 𝓤
                         → (A B : M) → (A ≡ B) ≃ (A ≅ B)

 characterization-of-M-≡ ua = characterization-of-≡-with-axioms ua
                                sns-data
                                axioms axiomss
```

*Exercise*. The above equivalence is characterized by induction on identifications as the function that maps the reflexive identification to the identity equivalence.

We have the following particular cases of interest:

1. *Metric spaces*. If `R` is a type of real numbers, then the axioms can be taken to be those for metric spaces, in which case `M` amounts to the type of metric spaces. Then the above characterizes metric space identification as isometry.

2. *Graphs*. If `R` is the type of truth values, and the `axioms` function is constant with value *true*, then `M` amounts to the type of directed graphs, and the above characterizes graph identification as graph isomorphism. We get undirected graphs by requiring the relation to be symmetric in the axioms.

3. *Partially ordered sets*. Again with `R` taken to be the type of truth values and suitable axioms, we get posets and other ordered structures, and the above says that their identifications amount to order isomorphisms.

### Topological spaces

We get a type of topological spaces when `R` is the type of truth values and the axioms are appropriately chosen.

```
module generalized-topological-space
        (𝓤 𝓥 : Universe)
        (R : 𝓥 ̇ )
        (axioms  : (X : 𝓤 ̇ ) → ((X → R) → R) → 𝓤 ⊔ 𝓥 ̇ )
        (axiomss : (X : 𝓤 ̇ ) (𝓞 : (X → R) → R) → is-subsingleton (axioms X 𝓞))
       where

 open sip
 open sip-with-axioms
```

When `R` is the type of truth values, the type `(X → R)` is the powerset of `X`, and membership amounts to function application:

```
 ℙ : 𝓦 ̇ → 𝓥 ⊔ 𝓦 ̇
 ℙ X = X → R

 _∈_ : {X : 𝓦 ̇ } → X → ℙ X → R
 x ∈ A = A x

 inverse-image : {X Y : 𝓤 ̇ } → (X → Y) → ℙ Y → ℙ X
 inverse-image f B = λ x → f x ∈ B

 ℙℙ : 𝓤 ̇ → 𝓤 ⊔ 𝓥 ̇
 ℙℙ X = ℙ (ℙ X)

 Space : 𝓤 ⁺ ⊔ 𝓥 ̇
 Space = Σ X ꞉ 𝓤 ̇ , Σ 𝓞 ꞉ ℙℙ X , axioms X 𝓞
```

If `(X , 𝓞X , a)` and `(Y , 𝓞Y , b)` are spaces, a homeomorphism can be described as a bijection `f : X → Y` such that the open sets of `Y` are precisely those whose inverse images are open in `X`, which can be written as

```
    (λ (V : ℙ Y) → inverse-image f V ∈ 𝓞X) ≡ 𝓞Y
```

Then `ι` defined below expresses the fact that a given bijection is a homeomorphism:

```
 sns-data : SNS ℙℙ (𝓤 ⊔ 𝓥)
 sns-data = (ι , ρ , θ)
  where
   ι : (A B : Σ ℙℙ) → ⟨ A ⟩ ≃ ⟨ B ⟩ → 𝓤 ⊔ 𝓥 ̇
   ι (X , 𝓞X) (Y , 𝓞Y) (f , _) = (λ (V : ℙ Y) → inverse-image f V ∈ 𝓞X) ≡ 𝓞Y
```

What `ρ` says is that identity function is a homeomorphism, trivially:

```
   ρ : (A : Σ ℙℙ) → ι A A (id-≃ ⟨ A ⟩)
   ρ (X , 𝓞) = refl 𝓞
```

Then `θ` amounts to the fact that two topologies on the same set must be the same if they make the identity function into a homeomorphism.

```
   h : {X : 𝓤 ̇ } {𝓞 𝓞' : ℙℙ X} → canonical-map ι ρ 𝓞 𝓞' ∼ id {𝓞 ≡ 𝓞'}
   h (refl 𝓞) = refl (refl 𝓞)

   θ : {X : 𝓤 ̇ } (𝓞 𝓞' : ℙℙ X) → is-equiv (canonical-map ι ρ 𝓞 𝓞')
   θ {X} 𝓞 𝓞' = equivs-closed-under-∼ (id-is-equiv (𝓞 ≡ 𝓞')) h
```

We introduce notation for the type of homeomorphisms:

```
 _≅_  : Space → Space → 𝓤 ⊔ 𝓥 ̇

 (X , 𝓞X , _) ≅ (Y , 𝓞Y , _) =
```

```
              Σ f ꞉ (X → Y), is-equiv f
                           × ((λ V → inverse-image f V ∈ 𝓞X) ＝ 𝓞Y)

 characterization-of-Space-＝ : is-univalent 𝓤
                              → (A B : Space) → (A ＝ B) ≃ (A ≅ B)

 characterization-of-Space-＝ ua = characterization-of-＝-with-axioms ua
                                    sns-data axioms axiomss
```

*Exercise*. The above equivalence is characterized by induction on identifications as the function that maps the reflexive identification to the identity equivalence.

But of course there are other choices for `R` that also make sense. For example, we can take `R` to be a type of real numbers, with the axioms for `X` and `F : (X → R) → R` saying that `F` is a linear functional. Then the above gives a characterization of the identity type of types equipped with linear functionals, in which case we may prefer to rephrase the above as

```
 _≅'_  : Space → Space → 𝓤 ⊔ 𝓥 ̇

 (X , F , _) ≅' (Y , G , _) =

             Σ f ꞉ (X → Y), is-equiv f
                          × ((λ (v : Y → R) → F (v ∘ f)) ＝ G)

 characterization-of-Space-＝' : is-univalent 𝓤
                               → (A B : Space) → (A ＝ B) ≃ (A ≅' B)

 characterization-of-Space-＝' = characterization-of-Space-＝
```

Linear functions on certain spaces correspond to special kinds of measures by the Riesz representation theorem, and hence in this case `Space` becomes a type of such kind of measure spaces by an appropriate choice of axioms.

### Selection spaces

By a selection space we mean a type `X` equipped with a selection function, where a selection function is a map `(X → R) → X`, where the type `R` is a parameter.

```
module selection-space
        (𝓤 𝓥 : Universe)
        (R : 𝓥 ̇ )
        (axioms  : (X : 𝓤 ̇ ) → ((X → R) → X) → 𝓤 ⊔ 𝓥 ̇ )
        (axiomss : (X : 𝓤 ̇ ) (ε : (X → R) → X) → is-subsingleton (axioms X ε))
       where

 open sip
 open sip-with-axioms

 S : 𝓤 ̇ → 𝓤 ⊔ 𝓥 ̇
 S X = (X → R) → X

 SelectionSpace : 𝓤 ⁺ ⊔ 𝓥 ̇
 SelectionSpace = Σ X ꞉ 𝓤 ̇ , Σ ε ꞉ S X , axioms X ε

 sns-data : SNS S (𝓤 ⊔ 𝓥)
 sns-data = (ι , ρ , θ)
  where
   ι : (A B : Σ S) → ⟨ A ⟩ ≃ ⟨ B ⟩ → 𝓤 ⊔ 𝓥 ̇
   ι (X , ε) (Y , δ) (f , _) = (λ (q : Y → R) → f (ε (q ∘ f))) ＝ δ

   ρ : (A : Σ S) → ι A A (id-≃ ⟨ A ⟩)
   ρ (X , ε) = refl ε

   θ : {X : 𝓤 ̇ } (ε δ : S X) → is-equiv (canonical-map ι ρ ε δ)
```

```
  θ {X} ε δ = γ
   where
    h : canonical-map ι ρ ε δ ∼ id (ε ≡ δ)
    h (refl ε) = refl (refl ε)

    γ : is-equiv (canonical-map ι ρ ε δ)
    γ = equivs-closed-under-∼ (id-is-equiv (ε ≡ δ)) h

 _≅_  :  SelectionSpace → SelectionSpace → 𝓤 ⊔ 𝓥 ̇

 (X , ε , _) ≅ (Y , δ , _) =

            Σ f ꞉ (X → Y), is-equiv f
                       × ((λ (q : Y → R) → f (ε (q ∘ f))) ≡ δ)

 characterization-of-selection-space-≡ : is-univalent 𝓤
                                       → (A B : SelectionSpace) → (A ≡ B) ≃ (A ≅ B)

 characterization-of-selection-space-≡ ua = characterization-of-≡-with-axioms ua
                                             sns-data
                                             axioms axiomss
```

*Exercise*. The above equivalence is characterized by induction on identifications as the function that maps the reflexive identification to the identity equivalence.

### A contrived example

Here is an example where we need to refer to the inverse of the equivalence under consideration.

We take the opportunity to illustrate how the above boiler-plate code can be avoided by defining `sns-data` on the fly, at the expense of readability:

```
module contrived-example (𝓤 : Universe) where

 open sip

 contrived-≡ : is-univalent 𝓤 →

    (X Y : 𝓤 ̇ ) (φ : (X → X) → X) (γ : (Y → Y) → Y)
  →
    ((X , φ) ≡ (Y , γ)) ≃ (Σ f ꞉ (X → Y)
                         , Σ i ꞉ is-equiv f
                         , (λ (g : Y → Y) → f (φ (inverse f i ∘ g ∘ f))) ≡ γ)

 contrived-≡ ua X Y φ γ =
   characterization-of-≡ ua
    ((λ (X , φ) (Y , γ) (f , i) → (λ (g : Y → Y) → f (φ (inverse f i ∘ g ∘ f))) ≡ γ) ,
     (λ (X , φ) → refl φ) ,
     (λ φ γ → equivs-closed-under-∼ (id-is-equiv (φ ≡ γ)) (λ {(refl φ) → refl (refl φ)})))
    (X , φ) (Y , γ)
```

Many of the above examples can be written in such a concise form.

### Functor algebras

We now characterize equality of functor algebras. In the following, we don't need to know that the functor preserves composition or to give coherence data for the identification `𝓕-id`.

```
module generalized-functor-algebra
         {𝓤 𝓥 : Universe}
         (F : 𝓤 ̇ → 𝓥 ̇ )
         (𝓕 : {X Y : 𝓤 ̇ } → (X → Y) → F X → F Y)
         (𝓕-id : {X : 𝓤 ̇ } → 𝓕 (id X) ≡ id (F X))
       where

 open sip

 S : 𝓤 ̇ → 𝓤 ⊔ 𝓥 ̇
 S X = F X → X

 sns-data : SNS S (𝓤 ⊔ 𝓥)
 sns-data = (ι , ρ , θ)
  where
   ι : (A B : Σ S) → ⟨ A ⟩ ≃ ⟨ B ⟩ → 𝓤 ⊔ 𝓥 ̇
   ι (X , α) (Y , β) (f , _) = f ∘ α ≡ β ∘ 𝓕 f

   ρ : (A : Σ S) → ι A A (id-≃ ⟨ A ⟩)
   ρ (X , α) = α          ≡⟨ ap (α ∘_) (𝓕-id ⁻¹) ⟩
               α ∘ 𝓕 id ∎

   θ : {X : 𝓤 ̇ } (α β : S X) → is-equiv (canonical-map ι ρ α β)
   θ {X} α β = γ
    where
     c : α ≡ β → α ≡ β ∘ 𝓕 id
     c = transport (α ≡_) (ρ (X , β))

     i : is-equiv c
     i = transport-is-equiv (α ≡_) (ρ (X , β))

     h : canonical-map ι ρ α β ∼ c
     h (refl _) = ρ (X , α)          ≡⟨ refl-left ⁻¹ ⟩
                  refl α ∙ ρ (X , α) ∎

     γ : is-equiv (canonical-map ι ρ α β)
     γ = equivs-closed-under-∼ i h

 characterization-of-functor-algebra-≡ : is-univalent 𝓤
   → (X Y : 𝓤 ̇ ) (α : F X → X) (β : F Y → Y)

   → ((X , α) ≡ (Y , β))  ≃  (Σ f ꞉ (X → Y), is-equiv f × (f ∘ α ≡ β ∘ 𝓕 f))

 characterization-of-functor-algebra-≡ ua X Y α β =
   characterization-of-≡ ua sns-data (X , α) (Y , β)
```

*Exercise*. The above equivalence is characterized by induction on identifications as the function that maps the reflexive identification to the identity equivalence.

### Type-valued preorders

We now characterize equality of generalized preordered sets. This example is harder than the previous ones.

A type-valued preorder on a type `X` is a type-valued relation which is reflexive and transitive. A type-valued, as opposed to a subsingleton-valued preorder, could also be called an ∞-preorder.

```
type-valued-preorder-S : 𝓤 ̇ → 𝓤 ⊔ (𝓥 ⁺) ̇
type-valued-preorder-S {𝓤} {𝓥} X = Σ _≤_ ꞉ (X → X → 𝓥 ̇ )
                                         , ((x : X) → x ≤ x)
                                         × ((x y z : X) → x ≤ y → y ≤ z → x ≤ z)
```

A category, also known as a `1`-category, is a type-valued preorder subject to suitable axioms, where the relation `_≤_` is traditionally written `hom`, and where identities are given by the reflexivity law, and composition is given by the transitivity law.

We choose to use categorical notation and terminology for type-valued preorders.

```
module type-valued-preorder
        (𝓤 𝓥 : Universe)
        (ua : Univalence)
       where

 open sip

 fe : global-dfunext
 fe = univalence-gives-global-dfunext ua

 hfe : global-hfunext
 hfe = univalence-gives-global-hfunext ua

 S : 𝓤 ̇ → 𝓤 ⊔ (𝓥 ⁺) ̇
 S = type-valued-preorder-S {𝓤} {𝓥}

 Type-valued-preorder : (𝓤 ⊔ 𝓥) ⁺ ̇
 Type-valued-preorder = Σ S
```

But we will use the shorter notation `Σ S` in this submodule.

The type of objects of a type-valued preorder:

```
 Ob : Σ S → 𝓤 ̇
 Ob (X , homX , idX , compX ) = X
```

Its hom-types (or preorder):

```
 hom : (𝓧 : Σ S) → Ob 𝓧 → Ob 𝓧 → 𝓥 ̇
 hom (X , homX , idX , compX) = homX
```

Its identities (or reflexivities):

```
 𝒾𝒹 : (𝓧 : Σ S) (x : Ob 𝓧) → hom 𝓧 x x
 𝒾𝒹 (X , homX , idX , compX) = idX
```

Its composition law (or transitivity):

```
 comp : (𝓧 : Σ S) (x y z : Ob 𝓧) → hom 𝓧 x y → hom 𝓧 y z → hom 𝓧 x z
 comp (X , homX , idX , compX) = compX
```

Notice that we have the so-called *diagramatic order* for composition.

The functoriality of a pair `F , 𝓕` (where in category theory the latter is also written `F`, by an abuse of notation) says that `𝓕` preserves identities and composition:

```
 functorial : (𝓧 𝓐 : Σ S)
            → (F : Ob 𝓧 → Ob 𝓐)
            → ((x y : Ob 𝓧) → hom 𝓧 x y → hom 𝓐 (F x) (F y))
            → 𝓤 ⊔ 𝓥 ̇

 functorial 𝓧 𝓐 F 𝓕' = pidentity × pcomposition
  where
```

In order to express the preservation of identities and composition in traditional form, we first define, locally, symbols for composition in applicative order, making the objects implicit:

```
   _o_ : {x y z : Ob 𝓧} → hom 𝓧 y z → hom 𝓧 x y → hom 𝓧 x z
   g o f = comp 𝓧 _ _ _ f g

   _□_ : {a b c : Ob 𝓐} → hom 𝓐 b c → hom 𝓐 a b → hom 𝓐 a c
   g □ f = comp 𝓐 _ _ _ f g
```

And we also make implicit the object parameters of the action of the functor on arrows:

```
  𝓕 : {x y : Ob 𝓧} → hom 𝓧 x y → hom 𝓐 (F x) (F y)
  𝓕 f = 𝓕' _ _ f
```

Preservation of identities:

```
  pidentity = (λ x → 𝓕 (𝑖𝑑 𝓧 x)) ≡ (λ x → 𝑖𝑑 𝓐 (F x))
```

Preservation of composition:

```
  pcomposition = (λ x y z (f : hom 𝓧 x y) (g : hom 𝓧 y z) → 𝓕 (g ∘ f))
               ≡ (λ x y z (f : hom 𝓧 x y) (g : hom 𝓧 y z) → 𝓕 g □ 𝓕 f)
```

This time we will need two steps to characterize equality of type-valued preorders. The first one is as above, by considering a standard notion of structure:

```
sns-data : SNS S (𝓤 ⊔ (𝓥 ⁺))
sns-data = (ι , ρ , θ)
 where
  ι : (𝓧 𝓐 : Σ S) → ⟨ 𝓧 ⟩ ≃ ⟨ 𝓐 ⟩ → 𝓤 ⊔ (𝓥 ⁺) ̇
  ι 𝓧 𝓐 (F , _) = Σ p ꞉ hom 𝓧 ≡ (λ x y → hom 𝓐 (F x) (F y))
                        , functorial 𝓧 𝓐 F (λ x y → transport (λ - → - x y) p)

  ρ : (𝓧 : Σ S) → ι 𝓧 𝓧 (id-≃ ⟨ 𝓧 ⟩)
  ρ 𝓧 = refl (hom 𝓧) , refl (𝑖𝑑 𝓧) , refl (comp 𝓧)

  θ : {X : 𝓤 ̇ } (s t : S X) → is-equiv (canonical-map ι ρ s t)
  θ {X} (homX , idX , compX) (homA , idA , compA) = g
   where
    φ = canonical-map ι ρ (homX , idX , compX) (homA , idA , compA)

    γ : codomain φ → domain φ
    γ (refl _ , refl _ , refl _) = refl _

    η : γ ∘ φ ∼ id
    η (refl _) = refl _

    ε : φ ∘ γ ∼ id
    ε (refl _ , refl _ , refl _) = refl _

    g : is-equiv φ
    g = invertibles-are-equivs φ (γ , η , ε)
```

The above constructions are short thanks to computations-under-the-hood performed by Agda, and may require some effort to unravel.

The above automatically gives a characterization of equality of preorders. But this characterization uses another equality, of hom types. The second step translates this equality into an equivalence:

```
lemma : (𝓧 𝓐 : Σ S) (F : Ob 𝓧 → Ob 𝓐)
      →
        (Σ p ꞉ hom 𝓧 ≡ (λ x y → hom 𝓐 (F x) (F y))
             , functorial 𝓧 𝓐 F (λ x y → transport (λ - → - x y) p))
      ≃
        (Σ 𝓕 ꞉ ((x y : Ob 𝓧) → hom 𝓧 x y → hom 𝓐 (F x) (F y))
             , (∀ x y → is-equiv (𝓕 x y))
             × functorial 𝓧 𝓐 F 𝓕)

lemma 𝓧 𝓐 F = γ
 where
  e = (hom 𝓧 ≡ λ x y → hom 𝓐 (F x) (F y))                          ≃⟨ i   ⟩
      (∀ x y → hom 𝓧 x y ≡ hom 𝓐 (F x) (F y))                       ≃⟨ ii  ⟩
      (∀ x y → hom 𝓧 x y ≃ hom 𝓐 (F x) (F y))                       ≃⟨ iii ⟩
      (∀ x → Σ φ ꞉ (∀ y → hom 𝓧 x y → hom 𝓐 (F x) (F y))
                 , ∀ y → is-equiv (φ y))                           ≃⟨ iv  ⟩
      (Σ 𝓕 ꞉ ((x y : Ob 𝓧) → hom 𝓧 x y → hom 𝓐 (F x) (F y))
           , (∀ x y → is-equiv (𝓕 x y)))                           ■
   where
    i   = hfunext₂-≃ hfe hfe (hom 𝓧 )  λ x y → hom 𝓐 (F x) (F y)
```

```
    ii   = Π-cong fe fe
            (λ x → Π-cong fe fe
            (λ y → univalence-≃ (ua 𝓥) (hom 𝓧 x y) (hom 𝓐 (F x) (F y))))
    iii = Π-cong fe fe (λ y → ΠΣ-distr-≃)
    iv   = ΠΣ-distr-≃
```

Here Agda silently performs a laborious computation to accept the definition of item `v`:

```
  v : (p : hom 𝓧 ≡ λ x y → hom 𝓐 (F x) (F y))
    → functorial 𝓧 𝓐 F (λ x y → transport (λ - → - x y) p)
    ≃ functorial 𝓧 𝓐 F (pr₁ (⌜ e ⌝ p))

  v (refl _) = Id→Eq _ _ (refl _)

  γ =

   (Σ p : hom 𝓧 ≡ (λ x y → hom 𝓐 (F x) (F y))
        , functorial 𝓧 𝓐 F (λ x y → transport (λ - → - x y) p)) ≃⟨ vi   ⟩

   (Σ p : hom 𝓧 ≡ (λ x y → hom 𝓐 (F x) (F y))
        , functorial 𝓧 𝓐 F (pr₁ (⌜ e ⌝ p)))                  ≃⟨ vii  ⟩

   (Σ σ : (Σ 𝓕 : ((x y : Ob 𝓧) → hom 𝓧 x y → hom 𝓐 (F x) (F y))
              , (∀ x y → is-equiv (𝓕 x y)))
        , functorial 𝓧 𝓐 F (pr₁ σ))                           ≃⟨ viii ⟩

   (Σ 𝓕 : ((x y : Ob 𝓧) → hom 𝓧 x y → hom 𝓐 (F x) (F y))
                , (∀ x y → is-equiv (𝓕 x y))
                × functorial 𝓧 𝓐 F 𝓕)                        ■
   where
    vi   = Σ-cong v
    vii  = ≃-sym (Σ-change-of-variable _ ⌜ e ⌝ (⌜⌝-is-equiv e))
    viii = Σ-assoc
```

Combining the two steps, we get the following characterization of equality of type-valued preorders in terms of equivalences:

```
 characterization-of-type-valued-preorder-≡ :

     (𝓧 𝓐 : Σ S)
   →
     (𝓧 ≡ 𝓐)
   ≃
     (Σ F : (Ob 𝓧 → Ob 𝓐)
          , is-equiv F
          × (Σ 𝓕 : ((x y : Ob 𝓧) → hom 𝓧 x y → hom 𝓐 (F x) (F y))
                 , (∀ x y → is-equiv (𝓕 x y))
                 × functorial 𝓧 𝓐 F 𝓕))

 characterization-of-type-valued-preorder-≡ 𝓧 𝓐 =

   (𝓧 ≡ 𝓐)                                                          ≃⟨ i  ⟩
   (Σ F : (Ob 𝓧 → Ob 𝓐)
        , is-equiv F
        × (Σ p : hom 𝓧 ≡ (λ x y → hom 𝓐 (F x) (F y))
               , functorial 𝓧 𝓐 F (λ x y → transport (λ - → - x y) p))) ≃⟨ ii ⟩
   (Σ F : (Ob 𝓧 → Ob 𝓐)
     , is-equiv F
     × (Σ 𝓕 : ((x y : Ob 𝓧) → hom 𝓧 x y → hom 𝓐 (F x) (F y))
            , (∀ x y → is-equiv (𝓕 x y))
            × functorial 𝓧 𝓐 F 𝓕))                                 ■

  where
   i  = characterization-of-≡ (ua 𝓤) sns-data 𝓧 𝓐
   ii = Σ-cong (λ F → Σ-cong (λ _ → lemma 𝓧 𝓐 F))
```

*Exercise*. The above equivalence is characterized by induction on identifications as the function that maps the reflexive identification to the identity functor.

Now we consider type-valued preorders subject to arbitrary axioms. The only reason we need to consider this explicitly is that again we need to combine two steps. The second step is the same, but the first step is modified to add axioms.

```
module type-valued-preorder-with-axioms
        (𝓤 𝓥 𝓦 : Universe)
        (ua : Univalence)
        (axioms  : (X : 𝓤 ̇ ) → type-valued-preorder-S {𝓤} {𝓥} X → 𝓦 ̇ )
        (axiomss : (X : 𝓤 ̇ ) (s : type-valued-preorder-S X) → is-subsingleton (axioms X s))
       where

 open sip
 open sip-with-axioms
 open type-valued-preorder 𝓤 𝓥 ua

 S' : 𝓤 ̇ → 𝓤 ⊔ (𝓥 ⁺) ⊔ 𝓦 ̇
 S' X = Σ s ꞉ S X , axioms X s

 sns-data' : SNS S' (𝓤 ⊔ (𝓥 ⁺))
 sns-data' = add-axioms axioms axiomss sns-data
```

Recall that `[_]` is the map that forgets the axioms.

```
 characterization-of-type-valued-preorder-≡-with-axioms :

      (𝓧' 𝓐' : Σ S')
    →
      (𝓧' ≡ 𝓐')
    ≃
      (Σ F ꞉ (Ob [ 𝓧' ] → Ob [ 𝓐' ])
           , is-equiv F
           × (Σ 𝓕 ꞉ ((x y : Ob [ 𝓧' ]) → hom [ 𝓧' ] x y → hom [ 𝓐' ] (F x) (F y))
                    , (∀ x y → is-equiv (𝓕 x y))
                    × functorial [ 𝓧' ] [ 𝓐' ] F 𝓕))

 characterization-of-type-valued-preorder-≡-with-axioms 𝓧' 𝓐' =

  (𝓧' ≡ 𝓐')                      ≃⟨ i  ⟩
  ([ 𝓧' ] ≃[ sns-data ] [ 𝓐' ])  ≃⟨ ii ⟩
  _                              ■

  where
   i  = characterization-of-≡-with-axioms (ua 𝓤) sns-data axioms axiomss 𝓧' 𝓐'
   ii = Σ-cong (λ F → Σ-cong (λ _ → lemma [ 𝓧' ] [ 𝓐' ] F))
```

### Categories

We now characterize equality of categories. By choosing suitable axioms for type-valued preorders, we get categories:

```
module category
        (𝓤 𝓥 : Universe)
        (ua : Univalence)
       where

 open type-valued-preorder-with-axioms 𝓤 𝓥 (𝓤 ⊔ 𝓥) ua

 fe : global-dfunext
 fe = univalence-gives-global-dfunext ua

 S : 𝓤 ̇ → 𝓤 ⊔ (𝓥 ⁺) ̇
 S = type-valued-preorder-S {𝓤} {𝓥}
```

The axioms say that

1. the homs form sets, rather than arbitrary types,
2. the identity is a left and right neutral element of composition,

3. composition is associative.

```
category-axioms : (X : 𝓤 ̇ ) → S X → 𝓤 ⊔ 𝓥 ̇
category-axioms X (homX , idX , compX) = hom-sets × identityl × identityr × associativity
 where
  _o_ : {x y z : X} → homX y z → homX x y → homX x z
  g o f = compX _ _ _ f g

  hom-sets      = ∀ x y → is-set (homX x y)

  identityl     = ∀ x y (f : homX x y) → f o (idX x) ≡ f

  identityr     = ∀ x y (f : homX x y) → (idX y) o f ≡ f

  associativity = ∀ x y z t (f : homX x y) (g : homX y z) (h : homX z t)
                → (h o g) o f ≡ h o (g o f)
```

The first axiom is subsingleton valued because the property of being a set is a subsingleton type. The second and the third axioms are subsingleton valued in the presence of the first axiom, because equations between elements of sets are subsingletons, by definition of set. And because subsingletons are closed under products, the category axioms form a subsingleton type:

```
category-axioms-subsingleton : (X : 𝓤 ̇ ) (s : S X) → is-subsingleton (category-axioms X s)
category-axioms-subsingleton X (homX , idX , compX) ca = γ ca
 where
  i : ∀ x y → is-set (homX x y)
  i = pr₁ ca

  γ : is-subsingleton (category-axioms X (homX , idX , compX))
  γ = ×-is-subsingleton ss (×-is-subsingleton ls (×-is-subsingleton rs as))
   where
    ss = Π-is-subsingleton fe
          (λ x → Π-is-subsingleton fe
          (λ y → being-set-is-subsingleton fe))

    ls = Π-is-subsingleton fe
          (λ x → Π-is-subsingleton fe
          (λ y → Π-is-subsingleton fe
          (λ f → i x y (compX x x y (idX x) f) f)))

    rs = Π-is-subsingleton fe
          (λ x → Π-is-subsingleton fe
          (λ y → Π-is-subsingleton fe
          (λ f → i x y (compX x y y f (idX y)) f)))

    as = Π-is-subsingleton fe
          (λ x → Π-is-subsingleton fe
          (λ y → Π-is-subsingleton fe
          (λ z → Π-is-subsingleton fe
          (λ t → Π-is-subsingleton fe
          (λ f → Π-is-subsingleton fe
          (λ g → Π-is-subsingleton fe
          (λ h → i x t (compX x y t f (compX y z t g h))
                       (compX x z t (compX x y z f g) h))))))))
```

We are now ready to define the type of categories, as a subtype of that of type-valued preorders:

```
Cat : (𝓤 ⊔ 𝓥)⁺ ̇
Cat = Σ X ꞉ 𝓤 ̇ , Σ s ꞉ S X , category-axioms X s
```

We reuse of above names in a slightly different way, taking into account that now we have axioms, which we simply ignore:

```
Ob : Cat → 𝓤 ̇
Ob (X , (homX , idX , compX) , _) = X

hom : (𝓧 : Cat) → Ob 𝓧 → Ob 𝓧 → 𝓥 ̇
hom (X , (homX , idX , compX) , _) = homX
```

```
id : (𝓧 : Cat) (x : Ob 𝓧) → hom 𝓧 x x
id (X , (homX , idX , compX) , _) = idX

comp : (𝓧 : Cat) (x y z : Ob 𝓧) (f : hom 𝓧 x y) (g : hom 𝓧 y z) → hom 𝓧 x z
comp (X , (homX , idX , compX) , _) = compX

is-functorial : (𝓧 𝓐 : Cat)
              → (F : Ob 𝓧 → Ob 𝓐)
              → ((x y : Ob 𝓧) → hom 𝓧 x y → hom 𝓐 (F x) (F y))
              → 𝓤 ⊔ 𝓥 ̇

is-functorial 𝓧 𝓐 F 𝓕' = pidentity × pcomposition
 where
  _o_ : {x y z : Ob 𝓧} → hom 𝓧 y z → hom 𝓧 x y → hom 𝓧 x z
  g o f = comp 𝓧 _ _ _ f g

  _□_ : {a b c : Ob 𝓐} → hom 𝓐 b c → hom 𝓐 a b → hom 𝓐 a c
  g □ f = comp 𝓐 _ _ _ f g

  𝓕 : {x y : Ob 𝓧} → hom 𝓧 x y → hom 𝓐 (F x) (F y)
  𝓕 f = 𝓕' _ _ f

  pidentity    = (λ x → 𝓕 (id 𝓧 x)) ≡ (λ x → id 𝓐 (F x))

  pcomposition = (λ x y z (f : hom 𝓧 x y) (g : hom 𝓧 y z) → 𝓕 (g o f))
               ≡ (λ x y z (f : hom 𝓧 x y) (g : hom 𝓧 y z) → 𝓕 g □ 𝓕 f)
```

*Exercise*. For type-valued preorders, `functorial 𝓧 𝓐 F 𝓕` is not in general a subsingleton, but for categories, `is-functorial 𝓧 𝓐 F 𝓕` is always a subsingleton.

We now apply the module `type-valued-preorder-with-axioms` to get the following characterization of identity of categories:

```
_≊_ : Cat → Cat → 𝓤 ⊔ 𝓥 ̇

𝓧 ≊ 𝓐 = Σ F ꞉ (Ob 𝓧 → Ob 𝓐)
             , is-equiv F
             × (Σ 𝓕 ꞉ ((x y : Ob 𝓧) → hom 𝓧 x y → hom 𝓐 (F x) (F y))
                    , (∀ x y → is-equiv (𝓕 x y))
                    × is-functorial 𝓧 𝓐 F 𝓕)

Id→EqCat : (𝓧 𝓐 : Cat) → 𝓧 ≡ 𝓐 → 𝓧 ≊ 𝓐
Id→EqCat 𝓧 𝓧 (refl 𝓧) = id (Ob 𝓧 ) ,
                         id-is-equiv (Ob 𝓧 ) ,
                         (λ x y → id (hom 𝓧 x y)) ,
                         (λ x y → id-is-equiv (hom 𝓧 x y)) ,
                         refl (id 𝓧) ,
                         refl (comp 𝓧)

characterization-of-category-≡ : (𝓧 𝓐 : Cat) → (𝓧 ≡ 𝓐) ≃ (𝓧 ≊ 𝓐)
characterization-of-category-≡ = characterization-of-type-valued-preorder-≡-with-axioms
                                  category-axioms category-axioms-subsingleton

Id→EqCat-is-equiv : (𝓧 𝓐 : Cat) → is-equiv (Id→EqCat 𝓧 𝓐)
Id→EqCat-is-equiv 𝓧 𝓐 = equivs-closed-under-∼
                            (⌜⌝-is-equiv (characterization-of-category-≡ 𝓧 𝓐))
                            (γ 𝓧 𝓐)
 where
  γ : (𝓧 𝓐 : Cat) → Id→EqCat 𝓧 𝓐 ∼ ⌜ characterization-of-category-≡ 𝓧 𝓐 ⌝
  γ 𝓧 𝓧 (refl 𝓧) = refl _
```

The HoTT book has a characterization of identity of categories as equivalence of categories in the traditional sense of category theory, assuming that the categories are univalent in a certain sense. We have chosen not to include the univalence requirement in our notion of category, although it may be argued that *univalent category* is the correct notion of category for univalent mathematics (because a univalent category may be equivalently defined as a category object in a 1-groupoid). In any case, the characterization of equality given here is not affected by the univalence requirement, or any subsingleton-valued property of categories.

## Subsingleton truncation

The subsingleton truncation of a type `X`, if it exists, is the universal solution of the problem of mapping `X` into a subsingleton type. The purpose of this section is to make this precise.

### Voevodsky's approach to subsingleton truncation

The following is Voevosky's approach to saying that a type is inhabited in such a way that the statement of inhabitation is a subsingleton, relying on function extensionality.

```
is-inhabited : 𝓤 ̇ → 𝓤 ⁺ ̇
is-inhabited {𝓤} X = (P : 𝓤 ̇ ) → is-subsingleton P → (X → P) → P
```

This says that if we have a function from `X` to a subsingleton `P`, then `P` must have a point. So this fails when `X=𝟘`. Considering `P=𝟘`, we conclude that `¬¬ X` if `X` is inhabited, which says that `X` is non-empty.

For simplicity in the formulation of the theorems, we assume *global* function extensionality. A type can be pointed in many ways, but inhabited in at most one way:

```
inhabitation-is-subsingleton : global-dfunext
                             → (X : 𝓤 ̇ )
                             → is-subsingleton (is-inhabited X)

inhabitation-is-subsingleton fe X =
 Π-is-subsingleton fe
   (λ P → Π-is-subsingleton fe
   (λ (s : is-subsingleton P) → Π-is-subsingleton fe
   (λ (f : X → P) → s)))
```

The following is the introduction rule for inhabitation, which says that pointed types are inhabited:

```
inhabited-intro : {X : 𝓤 ̇ } → X → is-inhabited X
inhabited-intro x = λ P s f → f x
```

And its recursion principle:

```
inhabited-recursion : {X P : 𝓤 ̇ } → is-subsingleton P → (X → P) → is-inhabited X → P
inhabited-recursion s f φ = φ (codomain f) s f
```

And its computation rule:

```
inhabited-recursion-computation : {X P : 𝓤 ̇ }
                                  (i : is-subsingleton P)
                                  (f : X → P)
                                  (x : X)
                                → inhabited-recursion i f (inhabited-intro x) ≡ f x

inhabited-recursion-computation i f x = refl (f x)
```

So the point `x` inside `inhabited x` is available for use by any function `f` into a subsingleton, via `inhabited-recursion`.

We can derive induction from recursion in this case, but its "computation rule" holds up to an identification, rather than judgmentally:

```
inhabited-induction : global-dfunext
                    → {X : 𝓤 ̇ } {P : is-inhabited X → 𝓤 ̇ }
                      (i : (s : is-inhabited X) → is-subsingleton (P s))
                      (f : (x : X) → P (inhabited-intro x))
                    → (s : is-inhabited X) → P s
```

```
inhabited-induction fe {X} {P} i f s = φ' s
 where
  φ : X → P s
  φ x = transport P (inhabitation-is-subsingleton fe X (inhabited-intro x) s) (f x)

  φ' : is-inhabited X → P s
  φ' = inhabited-recursion (i s) φ

inhabited-computation : (fe : global-dfunext) {X : 𝓤 ̇ } {P : is-inhabited X → 𝓤 ̇ }
                        (i : (s : is-inhabited X) → is-subsingleton (P s))
                        (f : (x : X) → P (inhabited-intro x))
                        (x : X)
                      → inhabited-induction fe i f (inhabited-intro x) ≡ f x

inhabited-computation fe i f x = i (inhabited-intro x)
                                   (inhabited-induction fe i f (inhabited-intro x))
                                   (f x)
```

The definition of inhabitation looks superficially like double negation. However, although we don't necessarily have that `¬¬ P → P`, we do have that `is-inhabited P → P` if `P` is a subsingleton.

```
inhabited-subsingletons-are-pointed : (P : 𝓤 ̇ )
                                    → is-subsingleton P → is-inhabited P → P

inhabited-subsingletons-are-pointed P s = inhabited-recursion s (id P)
```

*Exercise*. Show that `is-inhabited X ⇔ ¬¬X` if and only if excluded middle holds.

```
inhabited-functorial : global-dfunext → (X : 𝓤 ⁺ ̇ ) (Y : 𝓤 ̇ )
                     → (X → Y) → is-inhabited X → is-inhabited Y

inhabited-functorial fe X Y f = inhabited-recursion
                                  (inhabitation-is-subsingleton fe Y)
                                  (inhabited-intro ∘ f)
```

This universe assignment for functoriality is fairly restrictive, but is the only possible one.

With this notion, we can define the image of a function as follows:

```
image' : {X : 𝓤 ̇ } {Y : 𝓥 ̇ } → (X → Y) → (𝓤 ⊔ 𝓥)⁺ ̇
image' f = Σ y ꞉ codomain f , is-inhabited (Σ x ꞉ domain f , f x ≡ y)
```

An attempt to define the image of `f` without the inhabitation predicate would be to take it to be

```
    Σ y ꞉ codomain f , Σ x ꞉ domain f , f x ≡ y.
```

But we already know that this is equivalent to `X`. This is similar to what happens in set theory: the graph of any function is in bijection with its domain.

We can define the restriction and corestriction of a function to its image as follows:

```
restriction' : {X : 𝓤 ̇ } {Y : 𝓥 ̇ } (f : X → Y)
             → image' f → Y

restriction' f (y , _) = y

corestriction' : {X : 𝓤 ̇ } {Y : 𝓥 ̇ } (f : X → Y)
               → X → image' f

corestriction' f x = f x , inhabited-intro (x , refl (f x))
```

And we can define the notion of surjection as follows:

```
is-surjection' : {X : 𝓤 ̇ } {Y : 𝓥 ̇ } → (X → Y) → (𝓤 ⊔ 𝓥)⁺ ̇
is-surjection' f = (y : codomain f) → is-inhabited (Σ x ꞉ domain f , f x ≡ y)
```

*Exercise*. The type `(y : codomain f) → Σ x ꞉ domain f , f x ≡ y` is equivalent to the type `has-section f`, which is stronger than saying that `f` is a surjection.

There are two problems with this definition of inhabitation:

- Inhabitation has values in the next universe.

- We can eliminate into subsingletons of the same universe only.

In particular, it is not possible to show that the map `X → is-inhabited X` is a surjection, or that `X → Y` gives `is-inhabited X → is-inhabited Y` for `X` and `Y` in arbitrary universes.

There are two proposed ways to solve this kind of problem:

1. Voevodsky works with certain resizing rules for subsingletons. At the time of writing, the (relative) consistency of the system with such rules is an open question.

2. The HoTT book works with certain higher inductive types (HIT's), which are known to have models and hence to be (relatively) consistent. This is the same approach adopted by cubical type theory and cubical Agda.

### An axiomatic approach

A third possibility is to work with subsingleton truncations axiomatically, which is compatible with the above two proposals. We write this axiom as a record type rather than as an iterated `Σ` type for simplicity, where we use the HoTT-book notation `∥ X ∥` for the inhabitation of `X`, called the propositional, or subsingleton, or truth-value, truncation of `X`:

```
record subsingleton-truncations-exist : 𝓤ω where
 field
  ∥_∥                  : {𝓤 : Universe} → 𝓤 ̇ → 𝓤 ̇
  ∥∥-is-subsingleton   : {𝓤 : Universe} {X : 𝓤 ̇ } → is-subsingleton ∥ X ∥
  ∣_∣                  : {𝓤 : Universe} {X : 𝓤 ̇ } → X → ∥ X ∥
  ∥∥-recursion         : {𝓤 𝓥 : Universe} {X : 𝓤 ̇ } {P : 𝓥 ̇ }
                       → is-subsingleton P → (X → P) → ∥ X ∥ → P
 infix 0 ∥_∥
```

This is the approach we adopt in our personal Agda development.

We now assume that subsingleton truncations exist in the next few constructions, and we `open` the assumption to make the above fields visible.

```
module basic-truncation-development
        (pt  : subsingleton-truncations-exist)
        (hfe : global-hfunext)
       where

  open subsingleton-truncations-exist pt public

  hunapply : {X : 𝓤 ̇ } {A : X → 𝓥 ̇ } {f g : Π A} → f ∼ g → f ≡ g
  hunapply = hfunext-gives-dfunext hfe

  ∥∥-recursion-computation : {X : 𝓤 ̇ } {P :  𝓥 ̇ }
                           → (i : is-subsingleton P)
                           → (f : X → P)
                           → (x : X)
                           → ∥∥-recursion i f ∣ x ∣ ≡ f x

  ∥∥-recursion-computation i f x = i (∥∥-recursion i f ∣ x ∣) (f x)

  ∥∥-induction : {X : 𝓤 ̇ } {P : ∥ X ∥ → 𝓥 ̇ }
              → ((s : ∥ X ∥) → is-subsingleton (P s))
              → ((x : X) → P ∣ x ∣)
```

```
              → (s : ∥ X ∥) → P s

 ∥∥-induction {𝓤} {𝓥} {X} {P} i f s = φ' s
  where
   φ : X → P s
   φ x = transport P (∥∥-is-subsingleton ∣ x ∣ s) (f x)
   φ' : ∥ X ∥ → P s
   φ' = ∥∥-recursion (i s) φ

 ∥∥-computation : {X : 𝓤 ̇ } {P : ∥ X ∥ → 𝓥 ̇ }
                → (i : (s : ∥ X ∥) → is-subsingleton (P s))
                → (f : (x : X) → P ∣ x ∣)
                → (x : X)
                → ∥∥-induction i f ∣ x ∣ ≡ f x

 ∥∥-computation i f x = i ∣ x ∣ (∥∥-induction i f ∣ x ∣) (f x)

 ∥∥-functor : {X : 𝓤 ̇ } {Y : 𝓥 ̇ } → (X → Y) → ∥ X ∥ → ∥ Y ∥
 ∥∥-functor f = ∥∥-recursion ∥∥-is-subsingleton (λ x → ∣ f x ∣)
```

The subsingleton truncation of a type and its inhabitation are logically equivalent propositions:

```
 ∥∥-agrees-with-inhabitation : (X : 𝓤 ̇ ) → ∥ X ∥ ⇔ is-inhabited X
 ∥∥-agrees-with-inhabitation X = a , b
  where
   a : ∥ X ∥ → is-inhabited X
   a = ∥∥-recursion (inhabitation-is-subsingleton hunapply X) inhabited-intro

   b : is-inhabited X → ∥ X ∥
   b = inhabited-recursion ∥∥-is-subsingleton ∣_∣
```

Hence they differ only in size, and when size matters don't get on the way, we can use `is-inhabited` instead of ∥_∥ if we wish.

### Disjunction and existence

Disjunction and existence are defined as the truncation of `+` and `Σ`:

```
 _∨_ : 𝓤 ̇ → 𝓥 ̇ → 𝓤 ⊔ 𝓥 ̇
 A ∨ B = ∥ A + B ∥

 infixl 20 _∨_

 ∃ : {X : 𝓤 ̇ } → (X → 𝓥 ̇ ) → 𝓤 ⊔ 𝓥 ̇
 ∃ A = ∥ Σ A ∥

 -∃ : {𝓤 𝓥 : Universe} (X : 𝓤 ̇ ) (Y : X → 𝓥 ̇ ) → 𝓤 ⊔ 𝓥 ̇
 -∃ X Y = ∃ Y

 syntax -∃ A (λ x → b) = ∃ x ꞉ A , b

 infixr -1 -∃

 ∨-is-subsingleton : {A : 𝓤 ̇ } {B : 𝓥 ̇ } → is-subsingleton (A ∨ B)
 ∨-is-subsingleton = ∥∥-is-subsingleton

 ∃-is-subsingleton : {X : 𝓤 ̇ } {A : X → 𝓥 ̇ } → is-subsingleton (∃ A)
 ∃-is-subsingleton = ∥∥-is-subsingleton
```

The author's slides on univalent logic discuss further details about these notions of disjunction and existence.

### Images and surjections

The image of a function `f : X → Y` is the type of `y : Y` for which there exists `x : X` with `f x ＝ y`.

```
image : {X : 𝓤 ̇ } {Y : 𝓥 ̇ } → (X → Y) → 𝓤 ⊔ 𝓥 ̇
image f = Σ y ꞉ codomain f , ∃ x ꞉ domain f , f x ＝ y

restriction : {X : 𝓤 ̇ } {Y : 𝓥 ̇ } (f : X → Y)
            → image f → Y

restriction f (y , _) = y

corestriction : {X : 𝓤 ̇ } {Y : 𝓥 ̇ } (f : X → Y)
              → X → image f

corestriction f x = f x , ∣ (x , refl (f x)) ∣

is-surjection : {X : 𝓤 ̇ } {Y : 𝓥 ̇ } → (X → Y) → 𝓤 ⊔ 𝓥 ̇
is-surjection f = (y : codomain f) → ∃ x ꞉ domain f , f x ＝ y

being-surjection-is-subsingleton : {X : 𝓤 ̇ } {Y : 𝓥 ̇ } (f : X → Y)
                                 → is-subsingleton (is-surjection f)

being-surjection-is-subsingleton f = Π-is-subsingleton hunapply
                                      (λ y → ∃-is-subsingleton)

corestriction-surjection : {X : 𝓤 ̇ } {Y : 𝓥 ̇ } (f : X → Y)
                         → is-surjection (corestriction f)

corestriction-surjection f (y , s) = ∥∥-functor g s
 where
  g : (Σ x ꞉ domain f , f x ＝ y) → Σ x ꞉ domain f , corestriction f x ＝ (y , s)
  g (x , p) = x , to-subtype-＝ (λ _ → ∃-is-subsingleton) p

surjection-induction : {X : 𝓤 ̇ } {Y : 𝓥 ̇ } (f : X → Y)
                     → is-surjection f
                     → (P : Y → 𝓦 ̇ )
                     → ((y : Y) → is-subsingleton (P y))
                     → ((x : X) → P (f x))
                     → (y : Y) → P y

surjection-induction f i P j α y = ∥∥-recursion (j y) φ (i y)
 where
  φ : fiber f y → P y
  φ (x , r) = transport P r (α x)
```

*Exercise*. A map is an equivalence if and only if it is both an embedding and a surjection. (To be solved shortly.)

This time we can prove that the map `x ↦ ∣ x ∣` of `X` into `∥ X ∥` is a surjection without the universe levels getting in our way:

```
∣∣-is-surjection : (X : 𝓤 ̇ ) → is-surjection (λ (x : X) → ∣ x ∣)
∣∣-is-surjection X s = γ
 where
  f : X → ∃ x ꞉ X , ∣ x ∣ ＝ s
  f x = ∣ (x , ∥∥-is-subsingleton ∣ x ∣ s) ∣

  γ : ∃ x ꞉ X , ∣ x ∣ ＝ s
  γ = ∥∥-recursion ∥∥-is-subsingleton f s
```

Saying that this surjection `X → ∥ X ∥` has a section for all `X` (we can pick a point of every inhabited type) amounts to global choice, which contradicts univalence, and also gives classical logic.

The subsingleton truncation of a type is also known as its support, and a type `X` is said to have split support if there is a *choice function* `‖ X ‖ → X`, which is automatically a section of the surjection `X → ‖ X ‖`.

*Exercise.* Show that a type has split support if and only it is logically equivalent to a subsingleton. In particular, the type of invertibility data has split support, as it is logically equivalent to the equivalence property.

*Exercise* (hard). If `X` and `Y` are types obtained by summing `x`- and `y`-many copies of the type `𝟙`, respectively, as in `𝟙 + 𝟙 + ... + 𝟙`, where `x` and `y` are natural numbers, then `‖ X ＝ Y ‖ ≃ (x ＝ y)` and the type `X ＝ X` has `x!` elements.

### A characterization of equivalences

Singletons are inhabited, of course:

```
singletons-are-inhabited : (X : 𝓤 ̇ )
                         → is-singleton X
                         → ‖ X ‖

singletons-are-inhabited X s = ∣ center X s ∣
```

And inhabited subsingletons are singletons:

```
inhabited-subsingletons-are-singletons : (X : 𝓤 ̇ )
                                       → ‖ X ‖
                                       → is-subsingleton X
                                       → is-singleton X

inhabited-subsingletons-are-singletons X t i = c , φ
 where
  c : X
  c = ‖‖-recursion i (id X) t

  φ : (x : X) → c ＝ x
  φ = i c
```

Hence a type is a singleton if and only if it is inhabited and a subsingleton:

```
singleton-iff-inhabited-subsingleton : (X : 𝓤 ̇ )
                                     → is-singleton X
                                     ⇔ (‖ X ‖ × is-subsingleton X)

singleton-iff-inhabited-subsingleton X =

  (λ (s : is-singleton X) → singletons-are-inhabited     X s ,
                            singletons-are-subsingletons X s) ,
  Σ-induction (inhabited-subsingletons-are-singletons X)
```

By considering the unique map `X → 𝟙`, this can be regarded as a particular case of the fact that a map is an equivalence if and only if it is both an embedding and a surjection:

```
equiv-iff-embedding-and-surjection : {X : 𝓤 ̇ } {Y : 𝓥 ̇ } (f : X → Y)
                                   →  is-equiv f
                                   ⇔ (is-embedding f × is-surjection f)

equiv-iff-embedding-and-surjection f = (a , b)
 where
  a : is-equiv f → is-embedding f × is-surjection f
  a e = (λ y → singletons-are-subsingletons (fiber f y) (e y)) ,
        (λ y → singletons-are-inhabited     (fiber f y) (e y))

  b : is-embedding f × is-surjection f → is-equiv f
  b (e , s) y = inhabited-subsingletons-are-singletons (fiber f y) (s y) (e y)

equiv-＝-embedding-and-surjection : {X : 𝓤 ̇ } {Y : 𝓥 ̇ } (f : X → Y)
                                  → propext (𝓤 ⊔ 𝓥)
```

```
                                          → is-equiv f
                                          ＝ (is-embedding f × is-surjection f)

  equiv-＝-embedding-and-surjection f pe =
    pe (being-equiv-is-subsingleton hunapply hunapply f)
       (×-is-subsingleton
         (being-embedding-is-subsingleton hunapply f)
         (being-surjection-is-subsingleton f))
       (lr-implication (equiv-iff-embedding-and-surjection f))
       (rl-implication (equiv-iff-embedding-and-surjection f))
```

### Exiting subsingleton truncations

We will see that global choice

```
(X : 𝓤 ̇ ) → ∥ X ∥ → X
```

is inconsistent with univalence, and also implies excluded middle. However, for some types `X`, we can prove that `∥ X ∥ → X`. We characterize such types as those that have `wconstant` endomaps.

Because, as we have seen, we have a logical equivalence

```
∥ X ∥ ⇔ is-inhabited X,
```

it suffices to discuss

```
is-inhabited X → X,
```

which can be done in our spartan MLTT without any axioms for univalent mathematics (and hence also *with* axioms for univalent mathematics, including non-constructive ones such as excluded middle and choice).

For any type `X`, we have `is-inhabited X → X` iff `X` has a designated wconstant endomap. To prove this we first show that the type of fixed points of a `wconstant` endomap is a subsingleton.

We first define the type of fixed points:

```
fix : {X : 𝓤 ̇ } → (X → X) → 𝓤 ̇
fix f = Σ x ꞉ domain f , f x ＝ x
```

Of course any fixed point of `f` gives an element of `X`:

```
from-fix : {X : 𝓤 ̇ } (f : X → X)
         → fix f → X

from-fix f = pr₁
```

Conversely, if `f` is `wconstant` then for any `x : X` we have that `f x` is a fixed point of `f`, and hence from any element of `X` we get a fixed point of `f`:

```
to-fix : {X : 𝓤 ̇ } (f : X → X) → wconstant f
       → X → fix f

to-fix f κ x = f x , κ (f x) x
```

The following is trivial if the type `X` is a set. What may be surprising is that it holds for arbitrary types, because in this case the identity type `f x ＝ x` is in general not a subsingleton.

```
fix-is-subsingleton : {X : 𝓤 ̇ } (f : X → X)
                    → wconstant f → is-subsingleton (fix f)

fix-is-subsingleton {𝓤} {X} f κ = γ
 where
  a : (y x : X) → (f x ＝ x) ≃ (f y ＝ x)
```

```
  a y x = transport (_≡ x) (κ x y) , transport-is-equiv (_≡ x) (κ x y)

  b : (y : X) → fix f ≃ singleton-type' (f y)
  b y = Σ-cong (a y)

  c : X → is-singleton (fix f)
  c y = equiv-to-singleton (b y) (singleton-types'-are-singletons X (f y))

  d : fix f → is-singleton (fix f)
  d = c ∘ from-fix f

  γ : is-subsingleton (fix f)
  γ = subsingleton-criterion d
```

*Exercise.* Formulate and prove the fact that the type `fix f` has the universal property of the subsingleton truncation of `X` if `f` is `wconstant`. Moreover, argue that the computation rule for recursion holds definitionally in this case. This is an example of a situation where the truncation of a type just is available in MLTT without axioms or extensions.

We use `fix-is-subsingleton` to show that the type `is-inhabited X → X` is logically equivalent to the type `wconstant-endomap X`, where one direction uses function extensionality. We refer to a function `is-inhabited X → X` as a *choice function* for `X`. So the claim is that a type has a choice function if and only if it has a designated `wconstant` endomap.

```
choice-function : 𝓤 ̇ → 𝓤 ⁺ ̇
choice-function X = is-inhabited X → X
```

With a constant endomap of `X`, we can exit the truncation `is-inhabited X` in pure MLTT:

```
wconstant-endomap-gives-choice-function : {X : 𝓤 ̇ }
                                        → wconstant-endomap X → choice-function X

wconstant-endomap-gives-choice-function {𝓤} {X} (f , κ) = from-fix f ∘ γ
 where
  γ : is-inhabited X → fix f
  γ = inhabited-recursion (fix-is-subsingleton f κ) (to-fix f κ)
```

For the converse we use function extensionality (to know that the type `is-inhabited X` is a subsingleton in the construction of the `wconstant` endomap):

```
choice-function-gives-wconstant-endomap : global-dfunext
                                        → {X : 𝓤 ̇ }
                                        → choice-function X → wconstant-endomap X

choice-function-gives-wconstant-endomap fe {X} c = f , κ
 where
  f : X → X
  f = c ∘ inhabited-intro

  κ : wconstant f
  κ x y = ap c (inhabitation-is-subsingleton fe X (inhabited-intro x)
                                                 (inhabited-intro y))
```

As an application, we show that if the type of roots of a function `f : ℕ → ℕ` is inhabited, then it is pointed. In other words, with the information that there is some root, we can find an explicit root.

```
module find-hidden-root where

 open basic-arithmetic-and-order public
```

Given a root, we find a minimal root (below it, of course) by bounded search, and this gives a constant endomap of the type of roots:

```
 μρ : (f : ℕ → ℕ) → root f → root f
 μρ f r = minimal-root-is-root f (root-gives-minimal-root f r)

 μρ-root : (f : ℕ → ℕ) → root f → ℕ
 μρ-root f r = pr₁ (μρ f r)

 μρ-root-is-root : (f : ℕ → ℕ) (r : root f) → f (μρ-root f r) ≡ 0
```

```
μρ-root-is-root f r = pr₂ (μρ f r)

μρ-root-minimal : (f : ℕ → ℕ) (m : ℕ) (p : f m ≡ 0)
                → (n : ℕ) → f n ≡ 0 → μρ-root f (m , p) ≤ n

μρ-root-minimal f m p n q = not-<-gives-≥ (μρ-root f (m , p)) n γ
 where
  φ : ¬(f n ≢ 0) → ¬(n < μρ-root f (m , p))
  φ = contrapositive (pr₂(pr₂ (root-gives-minimal-root f (m , p))) n)

  γ : ¬ (n < μρ-root f (m , p))
  γ = φ (dni (f n ≡ 0) q)
```

The crucial property of the function `μρ f` is that it is `wconstant`:

```
μρ-wconstant : (f : ℕ → ℕ) → wconstant (μρ f)
μρ-wconstant f (n , p) (n' , p') = r
 where
  m m' : ℕ
  m  = μρ-root f (n , p)
  m' = μρ-root f (n' , p')

  l : m ≤ m'
  l = μρ-root-minimal f n p m' (μρ-root-is-root f (n' , p'))

  l' : m' ≤ m
  l' = μρ-root-minimal f n' p' m (μρ-root-is-root f (n , p))

  q : m ≡ m'
  q = ≤-anti _ _ l l'

  r : μρ f (n , p) ≡ μρ f (n' , p')
  r = to-subtype-≡ (λ _ → ℕ-is-set (f _) 0) q
```

Using the `wconstancy` of `μρ f`, if a root of `f` exists, then we can find one (which in fact will be the minimal one):

```
find-existing-root : (f : ℕ → ℕ) → is-inhabited (root f) → root f
find-existing-root f = h ∘ g
  where
   γ : root f → fix (μρ f)
   γ = to-fix (μρ f) (μρ-wconstant f)

   g : is-inhabited (root f) → fix (μρ f)
   g = inhabited-recursion (fix-is-subsingleton (μρ f) (μρ-wconstant f)) γ

   h : fix (μρ f) → root f
   h = from-fix (μρ f)
```

In the following example, we first hide a root with `inhabited-intro` and then find the minimal root with search bounded by this hidden root:

```
module find-existing-root-example where

 f : ℕ → ℕ
 f 0 = 1
 f 1 = 1
 f 2 = 0
 f 3 = 1
 f 4 = 0
 f 5 = 1
 f 6 = 1
 f 7 = 0
 f (succ (succ (succ (succ (succ (succ (succ (succ x)))))))) = x
```

We hide the root `8` of `f`:

```
 root-existence : is-inhabited (root f)
 root-existence = inhabited-intro (8 , refl 0)
```

```
  r : root f
  r = find-existing-root f root-existence

  x : ℕ
  x = pr₁ r

  x-is-root : f x ＝ 0
  x-is-root = pr₂ r
```

We have that `x` evaluates to `2`, which is clearly the minimal root of `f`:

```
  p : x ＝ 2
  p = refl _
```

Thus, even though the type `is-inhabited A` is a subsingleton for any type `A`, the function `inhabited-intro : A → is-inhabited A` doesn't erase information. We used the information contained in `root-existence` as a bound for searching for the minimal root.

Notice that this construction is in pure (spartan) MLTT without assumptions. Now we repeat part of the above using the existence of small truncations and functional extensionality as assumptions, replacing `is-inhabited` by ∥_∥:

```
module exit-∥∥
        (pt  : subsingleton-truncations-exist)
        (hfe : global-hfunext)
       where

 open basic-truncation-development pt hfe
 open find-hidden-root

 find-∥∥-existing-root : (f : ℕ → ℕ)
                       → (∃ n ꞉ ℕ , f n ＝ 0)
                       →  Σ n ꞉ ℕ , f n ＝ 0

 find-∥∥-existing-root f = k
  where
   γ : root f → fix (μρ f)
   γ = to-fix (μρ f) (μρ-wconstant f)

   g : ∥ root f ∥ → fix (μρ f)
   g = ∥∥-recursion (fix-is-subsingleton (μρ f) (μρ-wconstant f)) γ

   h : fix (μρ f) → root f
   h = from-fix (μρ f)

   k : ∥ root f ∥ → root f
   k = h ∘ g

 module find-∥∥-existing-root-example where

  f : ℕ → ℕ
  f 0 = 1
  f 1 = 1
  f 2 = 0
  f 3 = 1
  f 4 = 0
  f 5 = 1
  f 6 = 1
  f 7 = 0
  f (succ (succ (succ (succ (succ (succ (succ (succ x)))))))) = x

  root-∥∥-existence : ∥ root f ∥
  root-∥∥-existence = ∣ 8 , refl 0 ∣

  r : root f
  r = find-∥∥-existing-root f root-∥∥-existence

  x : ℕ
  x = pr₁ r
```

```
  x-is-root : f x ≡ 0
  x-is-root = pr₂ r
```

This time, because the existence of propositional truncations is an assumption for this submodule, we don't have that `x` evaluates to `2`, because the computation rule for truncation doesn't hold definitionally. But we do have that `x` is `2`, applying the computation rule manually.

```
  NB : find-∥∥-existing-root f
     ≡ from-fix (μρ f) ∘ ∥∥-recursion
                          (fix-is-subsingleton (μρ f) (μρ-wconstant f))
                          (to-fix (μρ f) (μρ-wconstant f))
  NB = refl _

  p : x ≡ 2
  p = ap (pr₁ ∘ from-fix (μρ f))
         (∥∥-recursion-computation
            (fix-is-subsingleton (μρ f) (μρ-wconstant f))
            (to-fix (μρ f) (μρ-wconstant f))
            (8 , refl _))
```

In cubical Agda, with the truncation defined as a higher inductive type, `x` would compute to `2` automatically, like in our previous example using Voevodsky's truncation `is-inhabited`. This concludes the example. We also have:

```
 wconstant-endomap-gives-∥∥-choice-function : {X : 𝓤 ̇ }
                                            → wconstant-endomap X
                                            → (∥ X ∥ → X)

 wconstant-endomap-gives-∥∥-choice-function {𝓤} {X} (f , κ) = from-fix f ∘ γ
  where
   γ : ∥ X ∥ → fix f
   γ = ∥∥-recursion (fix-is-subsingleton f κ) (to-fix f κ)

 ∥∥-choice-function-gives-wconstant-endomap : {X : 𝓤 ̇ }
                                            → (∥ X ∥ → X)
                                            → wconstant-endomap X

 ∥∥-choice-function-gives-wconstant-endomap {𝓤} {X} c = f , κ
  where
   f : X → X
   f = c ∘ ∣_∣

   κ : wconstant f
   κ x y = ap c (∥∥-is-subsingleton ∣ x ∣ ∣ y ∣)
```

There is another situation in which we can eliminate truncations that is often useful in practice. The universal property of subsingleton truncation says that we can get a function `∥ X ∥ → Y` provided the type `Y` is a subsingleton and we have a given function `X → Y`. Because `Y` is a subsingleton, the given function is automatically `wconstant`. Hence the following generalizes this to the situation in which `Y` is a set:

```
 ∥∥-recursion-set : (X : 𝓤 ̇ ) (Y : 𝓥 ̇ )
                  → is-set Y
                  → (f : X → Y)
                  → wconstant f
                  → ∥ X ∥ → Y

 ∥∥-recursion-set {𝓤} {𝓥} X Y s f κ = f'
  where
   ψ : (y y' : Y) →  (Σ x ꞉ X , f x ≡ y) → (Σ x' ꞉ X , f x' ≡ y') → y ≡ y'
   ψ y y' (x , r) (x' , r') = y    ≡⟨ r ⁻¹    ⟩
                              f x  ≡⟨ κ x x' ⟩
                              f x' ≡⟨ r'     ⟩
                              y'   ∎

   φ : (y y' : Y) → (∃ x ꞉ X , f x ≡ y) → (∃ x' ꞉ X , f x' ≡ y') → y ≡ y'
   φ y y' u u' = ∥∥-recursion (s y y') (λ - → ∥∥-recursion (s y y') (ψ y y' -) u') u
```

```
  P : 𝓤 ⊔ 𝓥 ̇
  P = image f

  i : is-subsingleton P
  i (y , u) (y' , u') = to-subtype-≡ (λ _ → ∃-is-subsingleton) (φ y y' u u')

  g : ∥ X ∥ → P
  g = ∥∥-recursion i (corestriction f)

  h : P → Y
  h = restriction f

  f' : ∥ X ∥ → Y
  f' = h ∘ g
```

If we try to do this with Voevodsky's truncation `is-inhabited`, we stumble into an insurmountable problem of size.

### Equality of Noetherian local rings

This section has that on rings as a pre-requisite. We now apply the notion of subsingleton truncation to give the promised examples of Noetherian rngs and commutative Noetherian local rings. Subsingleton truncation is needed to have the existential quantifier ∃ available, in order to be able to define the notion of Noetherian ring.

```
module noetherian-local-ring
         (pt : subsingleton-truncations-exist)
         {𝓤 : Universe}
         (ua : Univalence)
       where

 open ring {𝓤} ua
 open basic-truncation-development pt hfe
 open ℕ-order
```

We first consider left Noetherian rngs and then Noetherian local rings.

For this we need the notion of left ideal of a rng $\mathcal{R}$, which is an element of the powerset `𝓟 ⟨ 𝓡 ⟩` of the underlying set `⟨ 𝓡 ⟩` of $\mathcal{R}$:

```
 is-left-ideal : (𝓡 : Rng) → 𝓟 ⟨ 𝓡 ⟩ → 𝓤 ̇
 is-left-ideal (R , (_+_ , _·_) , (i , ii , iii , (O , _) , _)) I =

     (O ∈ I)
   × ((x y : R) → x ∈ I → y ∈ I → (x + y) ∈ I)
   × ((x y : R) → y ∈ I → (x · y) ∈ I)

 is-left-noetherian : (𝓡 : Rng) → 𝓤 ⁺ ̇
 is-left-noetherian 𝓡 = (I : ℕ → 𝓟 ⟨ 𝓡 ⟩)
                       → ((n : ℕ) → is-left-ideal 𝓡 (I n))
                       → ((n : ℕ) → I n ⊆ I (succ n))
                       → ∃ m ꞉ ℕ , ((n : ℕ) → m ≤ n → I m ≡ I n)

 LNRng : 𝓤 ⁺ ̇
 LNRng = Σ 𝓡 ꞉ Rng , is-left-noetherian 𝓡
```

In order to be able to characterize equality of left Noetherian rngs, we again need to show that `is-left-noetherian` is property rather than data:

```
 being-ln-is-subsingleton : (𝓡 : Rng) → is-subsingleton (is-left-noetherian 𝓡)
 being-ln-is-subsingleton 𝓡 = Π-is-subsingleton fe
                               (λ I → Π-is-subsingleton fe
                               (λ _ → Π-is-subsingleton fe
                               (λ _ → ∃-is-subsingleton)))

 forget-LN : LNRng → Rng
```

```
forget-LN (𝓡 , _) = 𝓡

forget-LN-is-embedding : is-embedding forget-LN
forget-LN-is-embedding = pr₁-is-embedding being-ln-is-subsingleton
```

Isomorphism of left Noetherian rngs:

```
_≅[LNRng]_ : LNRng → LNRng → 𝓤 ̇

((R , (_+_ , _·_) , _) , _) ≅[LNRng] ((R' , (_+'_ , _·'_) , _) , _) =

                              Σ f ꞉ (R → R')
                                  , is-equiv f
                                  × ((λ x y → f (x + y)) ＝ (λ x y → f x +' f y))
                                  × ((λ x y → f (x · y)) ＝ (λ x y → f x ·' f y))

NB : (𝓡 𝓡' : LNRng) → (𝓡 ≅[LNRng] 𝓡') ＝ (forget-LN 𝓡 ≅[Rng] forget-LN 𝓡')
NB 𝓡 𝓡' = refl _
```

Again the identity type of Noetherian rngs is in bijection with the type of Noetherian rng isomorphisms:

```
characterization-of-LNRng-＝ : (𝓡 𝓡' : LNRng) → (𝓡 ＝ 𝓡') ≃ (𝓡 ≅[LNRng] 𝓡')
characterization-of-LNRng-＝ 𝓡 𝓡' = (𝓡 ＝ 𝓡')                    ≃⟨ i  ⟩
                                    (forget-LN 𝓡 ＝ forget-LN 𝓡') ≃⟨ ii ⟩
                                    (𝓡 ≅[LNRng] 𝓡')              ■
  where
   i = ≃-sym (embedding-criterion-converse forget-LN
                forget-LN-is-embedding 𝓡 𝓡')
   ii = characterization-of-rng-＝ (forget-LN 𝓡) (forget-LN 𝓡')
```

Hence properties of left Noetherian rngs are invariant under isomorphism. More generally, we can transport along type-valued functions of left Noetherian rngs, with values in an arbitrary universe $\mathcal{V}$, rather than just truth-valued ones:

```
isomorphic-LNRng-transport : (A : LNRng → 𝓥 ̇ ) (𝓡 𝓡' : LNRng)
                           → 𝓡 ≅[LNRng] 𝓡' → A 𝓡 → A 𝓡'

isomorphic-LNRng-transport A 𝓡 𝓡' i a = a'
 where
  p : 𝓡 ＝ 𝓡'
  p = ⌜ ≃-sym (characterization-of-LNRng-＝ 𝓡 𝓡') ⌝ i

  a' : A 𝓡'
  a' = transport A p a
```

In particular, any theorem about a left Noetherian rng automatically applies to any left Noetherian rng isomorphic to it.

One can similarly define right Noetherian rng and Noetherian rng, and then obtain the expected characterizations of their identity types (exercise).

We now consider Noetherian local rings.

```
is-commutative : Rng → 𝓤 ̇
is-commutative (R , (_+_ , _·_) , _) = (x y : R) → x · y ＝ y · x

being-commutative-is-subsingleton : (𝓡 : Rng) → is-subsingleton (is-commutative 𝓡)
being-commutative-is-subsingleton (R , (_+_ , _·_) , i , ii-vii) =
  Π-is-subsingleton fe
   (λ x → Π-is-subsingleton fe
   (λ y → i (x · y) (y · x)))
```

In the presence of commutativity, there is no difference between left ideal, right ideal and two-sided ideal, and so one speaks of simply ideals. A commutative rng is said to be local if it has a unique maximal proper ideal.

```
is-Noetherian-Local : Ring → 𝓤 ⁺ ̇
is-Noetherian-Local (R , (𝟏 , _+_ , _·_) , i-vii , viii) = is-commutative 𝓡
```

```
                                                          × is-noetherian
                                                          × is-local
 where
  𝓡 : Rng
  𝓡 = (R , (_+_ , _·_) , i-vii)

  is-ideal      = is-left-ideal 𝓡
  is-noetherian = is-left-noetherian 𝓡

  is-proper-ideal : 𝓟 R → 𝓤 ̇
  is-proper-ideal I = is-ideal I × (∃ x : ⟨ 𝓡 ⟩ , x ∉ I)

  is-local = ∃! I : 𝓟 R , is-proper-ideal I
                        × ((J : 𝓟 R) → is-proper-ideal J → J ⊆ I)

being-NL-is-subsingleton : (𝓡 : Ring) → is-subsingleton (is-Noetherian-Local 𝓡)
being-NL-is-subsingleton (R , (𝟏 , _+_ , _·_) , i-vii , viii) =

   ×-is-subsingleton (being-commutative-is-subsingleton 𝓡)
  (×-is-subsingleton (being-ln-is-subsingleton 𝓡)
                     (∃!-is-subsingleton _ fe))
 where
  𝓡 : Rng
  𝓡 = (R , (_+_ , _·_) , i-vii)

NL-Ring : 𝓤 ⁺ ̇
NL-Ring = Σ 𝓡 : Ring , is-Noetherian-Local 𝓡

_≅[NL]_ : NL-Ring → NL-Ring → 𝓤 ̇

((R , (𝟏 , _+_ , _·_) , _) , _) ≅[NL] ((R' , (𝟏' , _+'_ , _·'_) , _) , _) =

                                 Σ f : (R → R')
                                     , is-equiv f
                                     × (f 𝟏 ≡ 𝟏')
                                     × ((λ x y → f (x + y)) ≡ (λ x y → f x +' f y))
                                     × ((λ x y → f (x · y)) ≡ (λ x y → f x ·' f y))

forget-NL : NL-Ring → Ring
forget-NL (𝓡 , _) = 𝓡

forget-NL-is-embedding : is-embedding forget-NL
forget-NL-is-embedding = pr₁-is-embedding being-NL-is-subsingleton

NB' : (𝓡 𝓡' : NL-Ring) → (𝓡 ≅[NL] 𝓡') ≡ (forget-NL 𝓡 ≅[Ring] forget-NL 𝓡')
NB' 𝓡 𝓡' = refl _

characterization-of-NL-ring-≡ : (𝓡 𝓡' : NL-Ring) → (𝓡 ≡ 𝓡') ≃ (𝓡 ≅[NL] 𝓡')
characterization-of-NL-ring-≡ 𝓡 𝓡' = (𝓡 ≡ 𝓡')                          ≃⟨ i  ⟩
                                      (forget-NL 𝓡 ≡ forget-NL 𝓡') ≃⟨ ii ⟩
                                      (𝓡 ≅[NL] 𝓡')                     ■
   where
    i  = ≃-sym (embedding-criterion-converse forget-NL
                 forget-NL-is-embedding 𝓡 𝓡')
    ii = characterization-of-ring-≡ (forget-NL 𝓡) (forget-NL 𝓡')

isomorphic-NL-Ring-transport : (A : NL-Ring → 𝓥 ̇ ) (𝓡 𝓡' : NL-Ring)
                             → 𝓡 ≅[NL] 𝓡' → A 𝓡 → A 𝓡'

isomorphic-NL-Ring-transport A 𝓡 𝓡' i a = a'
 where
  p : 𝓡 ≡ 𝓡'
```

```
  p = ⌜ ≃-sym (characterization-of-NL-ring-≡ 𝓡 𝓡') ⌝ i

  a' : A 𝓡'
  a' = transport A p a
```

We remark that alternative definitions of the above notions are adopted for the purposes of constructive algebra.

## Choice in univalent mathematics

We discuss unique choice, univalent choice and global choice.

1. A simple form of unique choice just holds in our spartan MLTT.

2. The full form of unique choice is logically equivalent to function extensionality.

3. Univalent choice implies excluded middle and is not provable or disprovable, but is consistent with univalence.

4. Global choice contradicts univalence, because it is not possible to choose an element of every inhabited type in way that is invariant under automorphisms.

### The principle of unique choice

The principle of *simple unique choice* says that

> if for every `x : X` there is a unique `a : A x` with `R x a`,

then

> there is a specified function `f : (x : X) → A x` such that `R x (f x)` for all `x : X`.

This just holds and is trivial, given by projection:

```
simple-unique-choice : (X : 𝓤 ̇ ) (A : X → 𝓥 ̇ ) (R : (x : X) → A x → 𝓦 ̇ )

                     → ((x : X) → ∃! a ꞉ A x , R x a)
                     → Σ f ꞉ Π A , ((x : X) → R x (f x))

simple-unique-choice X A R s = f , φ
 where
  f : (x : X) → A x
  f x = pr₁ (center (Σ a ꞉ A x , R x a) (s x))

  φ : (x : X) → R x (f x)
  φ x = pr₂ (center (Σ a ꞉ A x , R x a) (s x))
```

Below we also consider a variation of simple unique choice that works with `∃` (truncated `Σ`) rather than `∃!`.

A full form of unique choice is Voevodsky's formulation `vvfunext` of function extensionality, which says that products of singletons are singletons. We show that this is equivalent to our official formulation of unique choice:

```
Unique-Choice : (𝓤 𝓥 𝓦 : Universe) → (𝓤 ⊔ 𝓥 ⊔ 𝓦)⁺ ̇
Unique-Choice 𝓤 𝓥 𝓦 = (X : 𝓤 ̇ ) (A : X → 𝓥 ̇ ) (R : (x : X) → A x → 𝓦 ̇ )
                     → ((x : X) → ∃! a ꞉ A x , R x a)
                     → ∃! f ꞉ Π A , ((x : X) → R x (f x))
```

This can be read as saying that every single-valued relation is the graph of a unique function.

```
vvfunext-gives-unique-choice : vvfunext 𝓤 (𝓥 ⊔ 𝓦) → Unique-Choice 𝓤 𝓥 𝓦
vvfunext-gives-unique-choice vv X A R s = c
 where
```

```
  a : ((x : X) → Σ a ꞉ A x , R x a)
    ≃ (Σ f ꞉ ((x : X) → A x), ((x : X) → R x (f x)))

  a = ΠΣ-distr-≃

  b : is-singleton ((x : X) → Σ a ꞉ A x , R x a)
  b = vv s

  c : is-singleton (Σ f ꞉ ((x : X) → A x), ((x : X) → R x (f x)))
  c = equiv-to-singleton' a b

unique-choice-gives-vvfunext : Unique-Choice 𝓤 𝓥 𝓥 → vvfunext 𝓤 𝓥
unique-choice-gives-vvfunext {𝓤} {𝓥} uc {X} {A} φ = γ
 where
  R : (x : X) → A x → 𝓥 ̇
  R x a = A x

  s' : (x : X) → is-singleton (A x × A x)
  s' x = ×-is-singleton (φ x) (φ x)

  s : (x : X) → ∃! y ꞉ A x , R x y
  s = s'

  e : ∃! f ꞉ Π A , ((x : X) → R x (f x))
  e = uc X A R s

  e' : is-singleton (Π A × Π A)
  e' = e

  ρ : Π A ◁ Π A × Π A
  ρ = pr₁ , (λ y → y , y) , refl

  γ : is-singleton (Π A)
  γ = retract-of-singleton ρ e'
```

Here is an alternative proof that derives `hfunext` instead:

```
unique-choice-gives-hfunext : Unique-Choice 𝓤 𝓥 𝓥 → hfunext 𝓤 𝓥
unique-choice-gives-hfunext {𝓤} {𝓥} uc = ∀-hfunext γ
 where
  γ : (X : 𝓤 ̇ ) (A : X → 𝓥 ̇ ) (f : Π A) → ∃! g ꞉ Π A , f ∼ g
  γ X A f = uc X A R e
   where
    R : (x : X) → A x → 𝓥 ̇
    R x a = f x ≡ a
    e : (x : X) → ∃! a ꞉ A x , f x ≡ a
    e x = singleton-types'-are-singletons (A x) (f x)
```

The above is not quite the converse of the previous, as there is a universe mismatch, but we do get a logical equivalence by taking $\mathcal{W}$ to be $\mathcal{V}$:

```
unique-choice⇔vvfunext : Unique-Choice 𝓤 𝓥 𝓥 ⇔ vvfunext 𝓤 𝓥
unique-choice⇔vvfunext = unique-choice-gives-vvfunext ,
                         vvfunext-gives-unique-choice
```

We now give a different derivation of unique choice from function extensionality, in order to illustrate transport along the inverse of `happly`. For simplicity, we assume global function extensionality in the next few constructions.

```
module _ (hfe : global-hfunext) where

 private
   hunapply : {X : 𝓤 ̇ } {A : X → 𝓥 ̇ } {f g : Π A} → f ∼ g → f ≡ g
   hunapply = inverse (happly _ _) (hfe _ _)

 transport-hunapply : {X : 𝓤 ̇ } (A : X → 𝓥 ̇ ) (R : (x : X) → A x → 𝓦 ̇ )
                      (f g : Π A)
                      (φ : (x : X) → R x (f x))
```

```
                    (h : f ∼ g)
                    (x : X)
                  → transport (λ - → (x : X) → R x (- x)) (hunapply h) φ x
                  ≡ transport (R x) (h x) (φ x)

transport-hunapply A R f g φ h x =

  transport (λ - → ∀ x → R x (- x)) (hunapply h) φ x ≡⟨ i  ⟩
  transport (R x) (happly f g (hunapply h) x) (φ x)  ≡⟨ ii ⟩
  transport (R x) (h x) (φ x)                        ∎

 where
  a : {f g : Π A} {φ : ∀ x → R x (f x)} (p : f ≡ g) (x : domain A)
    → transport (λ - → ∀ x → R x (- x)) p φ x
    ≡ transport (R x) (happly f g p x) (φ x)

  a (refl _) x = refl _

  b : happly f g (hunapply h) ≡ h
  b = inverses-are-sections (happly f g) (hfe f g) h

  i  = a (hunapply h) x
  ii = ap (λ - → transport (R x) (- x) (φ x)) b

unique-choice : (X : 𝓤 ̇ ) (A : X → 𝓥 ̇ ) (R : (x : X) → A x → 𝓦 ̇ )

              → ((x : X) → ∃! a ꞉ A x , R x a)
              → ∃! f ꞉ ((x : X) → A x), ((x : X) → R x (f x))

unique-choice X A R s = C , Φ
 where
  f₀ : (x : X) → A x
  f₀ x = pr₁ (center (Σ a ꞉ A x , R x a) (s x))

  φ₀ : (x : X) → R x (f₀ x)
  φ₀ x = pr₂ (center (Σ a ꞉ A x , R x a) (s x))

  C : Σ f ꞉ ((x : X) → A x), ((x : X) → R x (f x))
  C = f₀ , φ₀

  c : (x : X) → (τ : Σ a ꞉ A x , R x a) → f₀ x , φ₀ x ≡ τ
  c x = centrality (Σ a ꞉ A x , R x a) (s x)

  c₁ : (x : X) (a : A x) (r : R x a) → f₀ x ≡ a
  c₁ x a r = ap pr₁ (c x (a , r))

  c₂ : (x : X) (a : A x) (r : R x a)
     → transport (λ - → R x (pr₁ -)) (c x (a , r)) (φ₀ x) ≡ r

  c₂ x a r = apd pr₂ (c x (a , r))

  Φ : (σ : Σ f ꞉ ((x : X) → A x), ((x : X) → R x (f x))) → C ≡ σ
  Φ (f , φ) = to-Σ-≡ (p , hunapply q)
   where
    p : f₀ ≡ f
    p = hunapply (λ x → c₁ x (f x) (φ x))

    q : transport (λ - → (x : X) → R x (- x)) p φ₀ ∼ φ
    q x = transport (λ - → (x : X) → R x (- x)) p φ₀ x           ≡⟨ i   ⟩
          transport (R x) (ap pr₁ (c x (f x , φ x))) (φ₀ x)      ≡⟨ ii  ⟩
          transport (λ σ → R x (pr₁ σ)) (c x (f x , φ x)) (φ₀ x) ≡⟨ iii ⟩
          φ x                                                    ∎
     where
      i   = transport-hunapply A R f₀ f φ₀ (λ x → c₁ x (f x) (φ x)) x
      ii  = (transport-ap (R x) pr₁ (c x (f x , φ x)) (φ₀ x))⁻¹
      iii = c₂ x (f x) (φ x)
```

Simple unique choice can be reformulated as follows using `∃` rather than `∃!`. The statement

```
is-subsingleton (Σ a ꞉ A x , R x a)
```

can be read as

> there is at most one `a ꞉ A x` with `R x a`.

So the hypothesis of the following is that there is at most one such `a` and at least one such `a`, which amounts to saying that there is a unique such `a`, and hence `simple-unique-choice'` amounts to the same thing as `simple-unique-choice`. However, `simple-unique-choice` can be formulated and proved in our spartan MLTT, whereas `simple-unique-choice'` requires the assumption of the existence of subsingleton truncations so that `∃` is available for its formulation.

```
module choice
        (pt  : subsingleton-truncations-exist)
        (hfe : global-hfunext)
       where

  open basic-truncation-development pt hfe public

  simple-unique-choice' : (X : 𝓤 ̇ ) (A : X → 𝓥 ̇ ) (R : (x : X) → A x → 𝓦 ̇ )

                        → ((x : X) → is-subsingleton (Σ a ꞉ A x , R x a))

                        → ((x : X) → ∃ a ꞉ A x , R x a)
                        → Σ f ꞉ Π A , ((x : X) → R x (f x))

  simple-unique-choice' X A R u φ = simple-unique-choice X A R s
   where
    s : (x : X) → ∃! a ꞉ A x , R x a
    s x = inhabited-subsingletons-are-singletons (Σ a ꞉ A x , R x a) (φ x) (u x)
```

In the next few subsections we continue working within the submodule `choice`, in order to have the existence of propositional truncations available, so that we can use the existential quantifier `∃`.

### The univalent axiom of choice

The axiom of choice in univalent mathematics says that

> if for every `x : X` there exists `a : A x` with `R x a`,

where `R` is some given relation,

> then there exists a choice function `f : (x : X) → A x` with `R x (f x)` for all `x : X`,

provided

1. `X` is a set,
2. `A` is a family of sets,
3. the relation `R` is subsingleton valued.

This is not provable or disprovable in spartan univalent type theory, but, with this proviso, it does hold in Voevodsky's simplicial model of our univalent type theory, and hence is consistent. It is important that we have the condition that `A` is a set-indexed family of sets and that the relation `R` is subsingleton valued. For arbitrary higher groupoids, it is not in general possible to perform the choice functorially.

```
  AC : ∀ 𝓣 (X : 𝓤 ̇ ) (A : X → 𝓥 ̇ )
     → is-set X → ((x : X) → is-set (A x)) → 𝓣 ⁺ ⊔ 𝓤 ⊔ 𝓥 ̇

  AC 𝓣 X A i j = (R : (x : X) → A x → 𝓣 ̇ )
               → ((x : X) (a : A x) → is-subsingleton (R x a))
```

```
       → ((x : X) → ∃ a ꞉ A x , R x a)
       → ∃ f ꞉ Π A , ((x : X) → R x (f x))
```

We define the axiom of choice in the universe $\mathcal{U}$ to be the above with $\mathcal{T} = \mathcal{U}$, for all possible `X` and `A` (and `R`).

```
Choice : ∀ 𝓤 → 𝓤 ⁺ ̇
Choice 𝓤 = (X : 𝓤 ̇ ) (A : X → 𝓤 ̇ ) (i : is-set X) (j : (x : X) → is-set (A x))
         → AC 𝓤 X A i j
```

### A second formulation of univalent choice

The above is equivalent to another familiar formulation of choice, namely that a set-indexed product of non-empty sets is non-empty, where in a constructive setting we strengthen `non-empty` to `inhabited` (but this strengthening is immaterial, because choice implies excluded middle, and excluded middle implies that non-emptiness and inhabitation are the same notion).

```
IAC : (X : 𝓤 ̇ ) (Y : X → 𝓥 ̇ ) → is-set X → ((x : X) → is-set (Y x)) → 𝓤 ⊔ 𝓥 ̇
IAC X Y i j = ((x : X) → ∥ Y x ∥) → ∥ Π Y ∥

IChoice : ∀ 𝓤 → 𝓤 ⁺ ̇
IChoice 𝓤 = (X : 𝓤 ̇ ) (Y : X → 𝓤 ̇ ) (i : is-set X) (j : (x : X) → is-set (Y x))
          → IAC X Y i j
```

These two forms of choice are logically equivalent (and hence equivalent, as both are subsingletons assuming function extensionality):

```
Choice-gives-IChoice : Choice 𝓤 → IChoice 𝓤
Choice-gives-IChoice {𝓤} ac X Y i j φ = γ
 where
  R : (x : X) → Y x → 𝓤 ̇
  R x y = x ≡ x -- Any singleton type in 𝓤 will do.

  k : (x : X) (y : Y x) → is-subsingleton (R x y)
  k x y = i x x

  h : (x : X) → Y x → Σ y ꞉ Y x , R x y
  h x y = (y , refl x)

  g : (x : X) → ∃ y ꞉ Y x , R x y
  g x = ∥∥-functor (h x) (φ x)

  c : ∃ f ꞉ Π Y , ((x : X) → R x (f x))
  c = ac X Y i j R k g

  γ : ∥ Π Y ∥
  γ = ∥∥-functor pr₁ c

IChoice-gives-Choice : IChoice 𝓤 → Choice 𝓤
IChoice-gives-Choice {𝓤} iac X A i j R k ψ = γ
 where
  Y : X → 𝓤 ̇
  Y x = Σ a ꞉ A x , R x a

  l : (x : X) → is-set (Y x)
  l x = subsets-of-sets-are-sets (A x) (R x) (j x) (k x)

  a : ∥ Π Y ∥
  a = iac X Y i l ψ

  h : Π Y → Σ f ꞉ Π A , ((x : X) → R x (f x))
  h g = (λ x → pr₁ (g x)) , (λ x → pr₂ (g x))

  γ : ∃ f ꞉ Π A , ((x : X) → R x (f x))
  γ = ∥∥-functor h a
```

### A third formulation of univalent choice

We will see that exiting trunctions in the sense of

```
(A : X → 𝓥 ̇ ) (x : X) → ∥ A x ∥ → A x
```

amounts to global choice and is inconsistent with univalence. However, exiting truncations in an unspecified way, in the sense of

```
(A : X → 𝓥 ̇ ) → ∥ (x : X) → ∥ A x ∥ → A x ∥
```

is consistent and equivalent to the above two versions of univalent choice, as we show now.

```
TAC : (X : 𝓤 ̇ ) (A : X → 𝓥 ̇ ) → is-set X → ((x : X) → is-set (A x)) → 𝓤 ⊔ 𝓥 ̇
TAC X A i j = ∥((x : X) → ∥ A x ∥ → A x)∥

TChoice : ∀ 𝓤 → 𝓤 ⁺ ̇
TChoice 𝓤 = (X : 𝓤 ̇ ) (A : X → 𝓤 ̇ ) (i : is-set X) (j : (x : X) → is-set (A x))
          → TAC X A i j
```

Notice that we use the hypothesis twice to get the conclusion in the following:

```
IChoice-gives-TChoice : IChoice 𝓤 → TChoice 𝓤
IChoice-gives-TChoice {𝓤} iac X A i j = γ
 where
  B : (x : X) → ∥ A x ∥ → 𝓤 ̇
  B x s = A x

  k : (x : X) (s : ∥ A x ∥) → is-set (B x s)
  k x s = j x

  l : (x : X) → is-set ∥ A x ∥
  l x = subsingletons-are-sets ∥ A x ∥ ∥∥-is-subsingleton

  m : (x : X) →  is-set (∥ A x ∥ → A x)
  m x = Π-is-set hfe (λ s → j x)

  φ : (x : X) → ∥ (∥ A x ∥ → A x) ∥
  φ x = iac ∥ A x ∥ (B x) (l x) (k x) (id ∥ A x ∥)

  γ : ∥((x : X) → ∥ A x ∥ → A x)∥
  γ = iac X (λ x → ∥ A x ∥ → A x) i m φ

TChoice-gives-IChoice : TChoice 𝓤 → IChoice 𝓤
TChoice-gives-IChoice tac X A i j = γ
 where
  γ : ((x : X) → ∥ A x ∥) → ∥ Π A ∥
  γ g = ∥∥-functor φ (tac X A i j)
   where
    φ : ((x : X) → ∥ A x ∥ → A x) → Π A
    φ f x = f x (g x)
```

*Exercise*. A fourth formulation of the axiom of choice is that every surjection of sets has an unspecified section.

### Univalent choice gives excluded middle

We apply the third formulation of univalent choice to show that choice implies excluded middle. We begin with the following lemma.

```
decidable-equality-criterion : {X : 𝓤 ̇ } (α : 𝟚 → X)
                             → ((x : X) → (∃ n ꞉ 𝟚 , α n ≡ x)
                                        → (Σ n ꞉ 𝟚 , α n ≡ x))
                             → decidable(α ₀ ≡ α ₁)

decidable-equality-criterion α c = γ d
 where
  r : 𝟚 → image α
  r = corestriction α

  σ : (y : image α) → Σ n ꞉ 𝟚 , r n ≡ y
  σ (x , t) = f u
   where
    u : Σ n ꞉ 𝟚 , α n ≡ x
    u = c x t

    f : (Σ n ꞉ 𝟚 , α n ≡ x) → Σ n ꞉ 𝟚 , r n ≡ (x , t)
    f (n , p) = n , to-subtype-≡ (λ _ → ∃-is-subsingleton) p

  s : image α → 𝟚
  s y = pr₁ (σ y)

  η : (y : image α) → r (s y) ≡ y
  η y = pr₂ (σ y)

  l : left-cancellable s
  l = sections-are-lc s (r , η)

  αr : {m n : 𝟚} → α m ≡ α n → r m ≡ r n
  αr p = to-subtype-≡ (λ _ → ∃-is-subsingleton) p

  rα : {m n : 𝟚} → r m ≡ r n → α m ≡ α n
  rα = ap pr₁

  αs : {m n : 𝟚} → α m ≡ α n → s (r m) ≡ s (r n)
  αs p = ap s (αr p)

  sα : {m n : 𝟚} → s (r m) ≡ s (r n) → α m ≡ α n
  sα p = rα (l p)

  γ : decidable (s (r ₀) ≡ s (r ₁)) → decidable(α ₀ ≡ α ₁)
  γ (inl p) = inl (sα p)
  γ (inr u) = inr (contrapositive αs u)

  d : decidable (s (r ₀) ≡ s (r ₁))
  d = 𝟚-has-decidable-equality (s (r ₀)) (s (r ₁))
```

The first consequence of this lemma is that choice implies that every set has decidable equality.

```
choice-gives-decidable-equality : TChoice 𝓤
                                → (X : 𝓤 ̇ ) → is-set X → has-decidable-equality X

choice-gives-decidable-equality {𝓤} tac X i x₀ x₁ = γ
 where
  α : 𝟚 → X
  α ₀ = x₀
  α ₁ = x₁

  A : X → 𝓤 ̇
  A x = Σ n ꞉ 𝟚 , α n ≡ x

  l : is-subsingleton (decidable (x₀ ≡ x₁))
  l = +-is-subsingleton' hunapply (i (α ₀) (α ₁))

  δ : ∥((x : X) → ∥ A x ∥ → A x)∥ → decidable(x₀ ≡ x₁)
  δ = ∥∥-recursion l (decidable-equality-criterion α)
```

```
  j : (x : X) → is-set (A x)
  j x = subsets-of-sets-are-sets 𝟚 (λ n → α n ≡ x) 𝟚-is-set (λ n → i (α n) x)

  h : ∥((x : X) → ∥ A x ∥ → A x)∥
  h = tac X A i j

  γ : decidable (x₀ ≡ x₁)
  γ = δ h
```

Applying the above to the object `Ω 𝓤` of truth-values in the universe `𝓤`, we get excluded middle:

```
choice-gives-EM : propext 𝓤 → TChoice (𝓤 ⁺) → EM 𝓤
choice-gives-EM {𝓤} pe tac = em
 where
  ⊤ : Ω 𝓤
  ⊤ = (Lift 𝓤 𝟙 , equiv-to-subsingleton (Lift-≃ 𝟙) 𝟙-is-subsingleton)

  δ : (ω : Ω 𝓤) → decidable (⊤ ≡ ω)
  δ = choice-gives-decidable-equality tac (Ω 𝓤) (Ω-is-a-set hunapply pe) ⊤

  em : (P : 𝓤 ̇ ) → is-subsingleton P → P + ¬ P
  em P i = γ (δ (P , i))
   where
    γ : decidable (⊤ ≡ (P , i)) → P + ¬ P

    γ (inl r) = inl (Id→fun s (lift ⋆))
     where
      s : Lift 𝓤 𝟙 ≡ P
      s = ap pr₁ r

    γ (inr n) = inr (contrapositive f n)
     where
      f : P → ⊤ ≡ P , i
      f p = Ω-ext hunapply pe (λ (_ : Lift 𝓤 𝟙) → p) (λ (_ : P) → lift ⋆)
```

For more information with Agda code, see this.

### Global choice

The following says that, for any given `X`, we can either choose a point of `X` or tell that `X` is empty:

```
global-choice : (𝓤 : Universe) → 𝓤 ⁺ ̇
global-choice 𝓤 = (X : 𝓤 ̇ ) → X + is-empty X
```

And the following says that we can pick a point of every inhabited type:

```
global-∥∥-choice : (𝓤 : Universe) → 𝓤 ⁺ ̇
global-∥∥-choice 𝓤 = (X : 𝓤 ̇ ) → ∥ X ∥ → X
```

We first show that these two forms of global choice are logically equivalent, where one direction requires propositional extensionality (in addition to function extensionality, which is an assumption for this local module).

```
open exit-∥∥ pt hfe

global-choice-gives-wconstant : global-choice 𝓤
                              → (X : 𝓤 ̇ ) → wconstant-endomap X

global-choice-gives-wconstant g X = decidable-has-wconstant-endomap (g X)

global-choice-gives-global-∥∥-choice : global-choice  𝓤
                                     → global-∥∥-choice 𝓤

global-choice-gives-global-∥∥-choice {𝓤} c X = γ (c X)
 where
```

```
  γ : X + is-empty X → ∥ X ∥ → X
  γ (inl x) s = x
  γ (inr n) s = !𝟘 X (∥∥-recursion 𝟘-is-subsingleton n s)

global-∥∥-choice-gives-all-types-are-sets : global-∥∥-choice 𝓤
                                          → (X : 𝓤 ̇ ) → is-set X

global-∥∥-choice-gives-all-types-are-sets {𝓤} c X =
  types-with-wconstant-＝-endomaps-are-sets X
      (λ x y → ∥∥-choice-function-gives-wconstant-endomap (c (x ＝ y)))

global-∥∥-choice-gives-universe-is-set : global-∥∥-choice (𝓤 ⁺)
                                       → is-set (𝓤 ̇ )

global-∥∥-choice-gives-universe-is-set {𝓤} c =
  global-∥∥-choice-gives-all-types-are-sets c (𝓤 ̇ )

global-∥∥-choice-gives-choice : global-∥∥-choice 𝓤
                              → TChoice 𝓤

global-∥∥-choice-gives-choice {𝓤} c X A i j = ∣(λ x → c (A x))∣

global-∥∥-choice-gives-EM : propext 𝓤
                          → global-∥∥-choice (𝓤 ⁺)
                          → EM  𝓤

global-∥∥-choice-gives-EM {𝓤} pe c =
  choice-gives-EM pe (global-∥∥-choice-gives-choice c)

global-∥∥-choice-gives-global-choice : propext 𝓤
                                     → global-∥∥-choice 𝓤
                                     → global-∥∥-choice (𝓤 ⁺)
                                     → global-choice 𝓤

global-∥∥-choice-gives-global-choice {𝓤} pe c c⁺ X = γ
 where
  d : decidable ∥ X ∥
  d = global-∥∥-choice-gives-EM pe c⁺ ∥ X ∥ ∥∥-is-subsingleton

  f : decidable ∥ X ∥ → X + is-empty X
  f (inl i) = inl (c X i)
  f (inr φ) = inr (contrapositive ∣_∣ φ)

  γ : X + is-empty X
  γ = f d
```

Two forms of globally global choice:

```
Global-Choice Global-∥∥-Choice : 𝓤ω
Global-Choice    = ∀ 𝓤 → global-choice  𝓤
Global-∥∥-Choice = ∀ 𝓤 → global-∥∥-choice 𝓤
```

Which are equivalent, where one direction uses propositional extensionality:

```
Global-Choice-gives-Global-∥∥-Choice : Global-Choice → Global-∥∥-Choice
Global-Choice-gives-Global-∥∥-Choice c 𝓤 =
  global-choice-gives-global-∥∥-choice (c 𝓤)

Global-∥∥-Choice-gives-Global-Choice : global-propext
                                     → Global-∥∥-Choice → Global-Choice

Global-∥∥-Choice-gives-Global-Choice pe c 𝓤 =
  global-∥∥-choice-gives-global-choice pe (c 𝓤) (c (𝓤 ⁺))
```

And which are inconsistent with univalence:

```
global-∥∥-choice-inconsistent-with-univalence : Global-∥∥-Choice
                                              → Univalence
                                              → 𝟘

global-∥∥-choice-inconsistent-with-univalence g ua = γ (g 𝓤₁) (ua 𝓤₀)
 where
  open example-of-a-nonset

  γ : global-∥∥-choice 𝓤₁ → is-univalent 𝓤₀ → 𝟘
  γ g ua = 𝓤₀-is-not-a-set ua (global-∥∥-choice-gives-universe-is-set g)

global-choice-inconsistent-with-univalence : Global-Choice
                                           → Univalence
                                           → 𝟘

global-choice-inconsistent-with-univalence g =
  global-∥∥-choice-inconsistent-with-univalence
    (Global-Choice-gives-Global-∥∥-Choice g)
```

See also Theorem 3.2.2 and Corollary 3.2.7 of the HoTT book for a different argument that works with a single, arbitrary universe.

```
global-choice-gives-all-types-are-sets : global-choice 𝓤
                                       → (X : 𝓤 ̇ ) → is-set X

global-choice-gives-all-types-are-sets {𝓤} c X = hedberg (λ x y → c (x ≡ y))
```

## Propositional resizing, truncation and the powerset

Voevodsky considered resizing rules for a type theory for univalent foundations. These rules govern the syntax of the formal system, and hence are of a meta-mathematical nature.

Here we instead formulate, in our type theory without such rules, mathematical resizing principles. These principles are provable in the system with Voevodsky's rules.

The consistency of the resizing *rules* is an open problem at the time of writing, but the resizing *principles* are consistent relative to ZFC with Grothendieck universes, because they follow from excluded middle, which is known to be validated by the simplicial-set model.

It is also an open problem whether the resizing principles discussed below have a computational interpretation.

### Propositional resizing

We say that a type `X` has size $\mathcal{V}$ if it is equivalent to a type in the universe $\mathcal{V}$:

```
_has-size_ : 𝓤 ̇ → (𝓥 : Universe) → 𝓥 ⁺ ⊔ 𝓤 ̇
X has-size 𝓥 = Σ Y ꞉ 𝓥 ̇ , X ≃ Y
```

The propositional resizing principle from a universe $\mathcal{U}$ to a universe $\mathcal{V}$ says that every subsingleton in $\mathcal{U}$ has size $\mathcal{V}$:

```
propositional-resizing : (𝓤 𝓥 : Universe) → (𝓤 ⊔ 𝓥)⁺ ̇
propositional-resizing 𝓤 𝓥 = (P : 𝓤 ̇ ) → is-subsingleton P → P has-size 𝓥
```

We also consider global propositional resizing:

```
Propositional-resizing : 𝓤ω
Propositional-resizing = {𝓤 𝓥 : Universe} → propositional-resizing 𝓤 𝓥
```

Resizing from a universe to a higher universe just holds, of course:

```
upper-resizing : ∀ {𝓤} 𝓥 (X : 𝓤 ̇ ) → X has-size (𝓤 ⊔ 𝓥)
upper-resizing 𝓥 X = (Lift 𝓥 X , ≃-Lift X)
```

Moreover, the notion of size is upper closed:

```
has-size-is-upper : (X : 𝓤 ̇ ) → X has-size 𝓥 → X has-size (𝓥 ⊔ 𝓦)
has-size-is-upper {𝓤} {𝓥} {𝓦} X (Y , e) =  Z , c
 where
  Z : 𝓥 ⊔ 𝓦 ̇
  Z = Lift 𝓦 Y

  d : Y ≃ Z
  d = ≃-Lift Y

  c : X ≃ Z
  c = e ● d

upper-propositional-resizing : propositional-resizing 𝓤 (𝓤 ⊔ 𝓥)
upper-propositional-resizing {𝓤} {𝓥} P i = upper-resizing 𝓥 P
```

We say that a type is small if it has a copy in the first universe:

```
is-small : 𝓤 ̇  → 𝓤 ⊔ 𝓤₁ ̇
is-small X = X has-size 𝓤₀
```

Then propositional resizing is equivalent to saying that all propositions of every universe are small:

```
all-propositions-are-small : ∀ 𝓤 → 𝓤 ⁺ ̇
all-propositions-are-small 𝓤 = (P : 𝓤 ̇ ) → is-prop P → is-small P

all-propositions-are-small-means-PR₀ : all-propositions-are-small 𝓤
                                      ≡ propositional-resizing 𝓤 𝓤₀

all-propositions-are-small-means-PR₀ = refl _

all-propositions-are-small-gives-PR : all-propositions-are-small 𝓤
                                    → propositional-resizing 𝓤 𝓥

all-propositions-are-small-gives-PR {𝓤} {𝓥} a P i = γ
 where
  δ : P has-size 𝓤₀
  δ = a P i

  γ : P has-size 𝓥
  γ = has-size-is-upper P δ
```

A global version:

```
All-propositions-are-small : 𝓤ω
All-propositions-are-small = ∀ 𝓤 → all-propositions-are-small 𝓤

PR-gives-All-propositions-are-small : Propositional-resizing
                                    → All-propositions-are-small

PR-gives-All-propositions-are-small PR 𝓤 = PR

All-propositions-are-small-gives-PR : All-propositions-are-small
                                    → Propositional-resizing

All-propositions-are-small-gives-PR a {𝓤} {𝓥} = all-propositions-are-small-gives-PR (a 𝓤)
```

We use the following to work with propositional resizing more abstractly:

```
resize : propositional-resizing 𝓤 𝓥
       → (P : 𝓤 ̇ ) (i : is-subsingleton P) → 𝓥 ̇

resize ρ P i = pr₁ (ρ P i)

resize-is-subsingleton : (ρ : propositional-resizing 𝓤 𝓥)
                         (P : 𝓤 ̇ ) (i : is-subsingleton P)
                       → is-subsingleton (resize ρ P i)

resize-is-subsingleton ρ P i = equiv-to-subsingleton (≃-sym (pr₂ (ρ P i))) i

to-resize : (ρ : propositional-resizing 𝓤 𝓥)
            (P : 𝓤 ̇ ) (i : is-subsingleton P)
          → P → resize ρ P i

to-resize ρ P i = ⌜ pr₂ (ρ P i) ⌝

from-resize : (ρ : propositional-resizing 𝓤 𝓥)
              (P : 𝓤 ̇ ) (i : is-subsingleton P)
            → resize ρ P i → P

from-resize ρ P i = ⌜ ≃-sym(pr₂ (ρ P i)) ⌝
```

### Excluded middle gives propositional resizing

Propositional resizing is consistent because it is implied by excluded middle, which is consistent (with or without univalence):

```
EM-gives-all-propositions-are-small : EM 𝓤 → all-propositions-are-small 𝓤
EM-gives-all-propositions-are-small em P i = γ
 where
   Q : P + ¬ P → 𝓤₀ ̇
   Q (inl _) = 𝟙
   Q (inr _) = 𝟘

   j : (d : P + ¬ P) → is-subsingleton (Q d)
   j (inl p) = 𝟙-is-subsingleton
   j (inr n) = 𝟘-is-subsingleton

   f : (d : P + ¬ P) → P → Q d
   f (inl _) _ = ⋆
   f (inr n) p   = !𝟘 𝟘 (n p)

   g : (d : P + ¬ P) → Q d → P
   g (inl p) _ = p
   g (inr _) q = !𝟘 P q

   e : P ≃ Q (em P i)
   e = logically-equivalent-subsingletons-are-equivalent
        P (Q (em P i)) i (j (em P i)) (f (em P i) , g (em P i))

   γ : is-small P
   γ = Q (em P i) , e
```

Hence excluded middle implies propositional resizing:

```
EM-gives-PR : EM 𝓤 → propositional-resizing 𝓤 𝓥
EM-gives-PR {𝓤} {𝓥} em = all-propositions-are-small-gives-PR
                            (EM-gives-all-propositions-are-small em)
```

### The propositional resizing axiom is a subsingleton

To show that the propositional resizing principle is a subsingleton, we use univalence here.

```
has-size-is-subsingleton : Univalence
                         → (X : 𝓤 ̇ ) (𝓥 :  Universe)
                         → is-subsingleton (X has-size 𝓥)

has-size-is-subsingleton {𝓤} ua X 𝓥 = univalence→' (ua 𝓥) (ua (𝓤 ⊔ 𝓥)) X

PR-is-subsingleton : Univalence → is-subsingleton (propositional-resizing 𝓤 𝓥)
PR-is-subsingleton {𝓤} {𝓥} ua =
 Π-is-subsingleton (univalence-gives-global-dfunext ua)
  (λ P → Π-is-subsingleton (univalence-gives-global-dfunext ua)
  (λ i → has-size-is-subsingleton ua P 𝓥))
```

*Exercise*. It is possible to show that the propositional resizing principle is a subsingleton using propositional and functional extensionality instead of univalence.

### Propositional impredicativity

We consider binary and unary notions of propositional impredicativity.

The binary notion of impredicativity, for two universes $\mathcal{U}$ and $\mathcal{V}$ says that the type of propositions in the universe $\mathcal{U}$ has a copy in the universe $\mathcal{V}$.

```
Impredicativity : (𝓤 𝓥 : Universe) → (𝓤 ⊔ 𝓥 )⁺ ̇
Impredicativity 𝓤 𝓥 = Ω 𝓤 has-size 𝓥
```

The unary notion of impredicativity, for one $\mathcal{U}$, says that the type `Ω 𝓤` of propositions, which lives in the universe $\mathcal{U}^+$, has a copy in $\mathcal{U}$ itself:

```
is-impredicative : (𝓤 : Universe) → 𝓤 ⁺ ̇
is-impredicative 𝓤 = Impredicativity 𝓤 𝓤
```

We can rephrase this in terms of a notion of relative smallness. We say that a type `X` in a successor universe $\mathcal{U}^+$ is *relatively small* if it has a copy in the universe $\mathcal{U}$:

```
is-relatively-small : 𝓤 ⁺ ̇  → 𝓤 ⁺ ̇
is-relatively-small {𝓤} X = X has-size 𝓤

impredicativity-is-Ω-smallness : ∀ {𝓤} → is-impredicative 𝓤 ≡ is-relatively-small (Ω 𝓤)
impredicativity-is-Ω-smallness {𝓤} = refl _
```

Propositional resizing gives impredicativity, in the presence of propositional and functional extensionality:

```
PR-gives-Impredicativity⁺ : global-propext
                          → global-dfunext
                          → propositional-resizing 𝓥 𝓤
                          → propositional-resizing 𝓤 𝓥
                          → Impredicativity 𝓤 (𝓥 ⁺)

PR-gives-Impredicativity⁺ {𝓥} {𝓤} pe fe ρ σ = γ
 where
  φ : Ω 𝓥 → Ω 𝓤
  φ (Q , j) = resize ρ Q j , resize-is-subsingleton ρ Q j

  ψ : Ω 𝓤 → Ω 𝓥
  ψ (P , i) = resize σ P i , resize-is-subsingleton σ P i

  η : (p : Ω 𝓤) → φ (ψ p) ≡ p
  η (P , i) = Ω-ext fe pe a b
   where
```

```
    Q : 𝓥 ̇
    Q = resize σ P i

    j : is-subsingleton Q
    j = resize-is-subsingleton σ P i

    a : resize ρ Q j → P
    a = from-resize σ P i ∘ from-resize ρ Q j

    b : P → resize ρ Q j
    b = to-resize ρ Q j ∘ to-resize σ P i

  ε : (q : Ω 𝓥) → ψ (φ q) ≡ q
  ε (Q , j) = Ω-ext fe pe a b
   where
    P : 𝓤 ̇
    P = resize ρ Q j

    i : is-subsingleton P
    i = resize-is-subsingleton ρ Q j

    a : resize σ P i → Q
    a = from-resize ρ Q j ∘ from-resize σ P i

    b : Q → resize σ P i
    b = to-resize σ P i ∘ to-resize ρ Q j

  γ : (Ω 𝓤) has-size (𝓥 ⁺)
  γ = Ω 𝓥 , invertibility-gives-≃ ψ (φ , η , ε)
```

Propositional resizing doesn't imply that the first universe $\mathcal{U}_0$ is impredicative, but it does imply that all other, successor, universes $\mathcal{U}^+$ are.

```
PR-gives-impredicativity⁺ : global-propext
                          → global-dfunext
                          → propositional-resizing (𝓤 ⁺) 𝓤
                          → is-impredicative (𝓤 ⁺)

PR-gives-impredicativity⁺ {𝓤} pe fe = PR-gives-Impredicativity⁺
                                        pe fe (λ P i → upper-resizing (𝓤 ⁺) P)
```

What we get with propositional resizing is that all types of propositions of any universe $\mathcal{U}$ are equivalent to $\Omega\ \mathcal{U}_0$, which lives in the second universe $\mathcal{U}_1$:

```
PR-gives-impredicativity₁ : global-propext
                          → global-dfunext
                          → propositional-resizing 𝓤 𝓤₀
                          → Impredicativity 𝓤 𝓤₁

PR-gives-impredicativity₁ {𝓤} pe fe = PR-gives-Impredicativity⁺
                                       pe fe (λ P i → upper-resizing 𝓤 P)
```

We can rephrase this as follows:

```
all-propositions-are-small-gives-impredicativity₁ :

     global-propext
   → global-dfunext
   → all-propositions-are-small 𝓤
   → Ω 𝓤 has-size 𝓤₁

all-propositions-are-small-gives-impredicativity₁ = PR-gives-impredicativity₁
```

*Exercise*. Excluded middle gives the impredicativity of the first universe, and of all other universes.

We also have that moving `Ω` around universes moves subsingletons around universes:

```
Impredicativity-gives-PR : propext 𝓤
                         → dfunext 𝓤 𝓤
```

```
                                  → Impredicativity 𝓤 𝓥
                                  → propositional-resizing 𝓤 𝓥

Impredicativity-gives-PR {𝓤} {𝓥} pe fe (O , e) P i = Q , ε
 where
  𝟙' : 𝓤 ̇
  𝟙' = Lift 𝓤 𝟙

  k : is-subsingleton 𝟙'
  k (lift ⋆) (lift ⋆) = refl (lift ⋆)

  down : Ω 𝓤 → O
  down = ⌜ e ⌝

  O-is-set : is-set O
  O-is-set = equiv-to-set (≃-sym e) (Ω-is-a-set fe pe)

  Q : 𝓥 ̇
  Q = down (𝟙' , k) ≡ down (P , i)

  j : is-subsingleton Q
  j = O-is-set (down (Lift 𝓤 𝟙 , k)) (down (P , i))

  φ : Q → P
  φ q = Id→fun
         (ap _holds (equivs-are-lc down (⌜⌝-is-equiv e) q))
         (lift ⋆)

  γ : P → Q
  γ p = ap down (to-subtype-≡
                    (λ _ → being-subsingleton-is-subsingleton fe)
                    (pe k i (λ _ → p) (λ _ → lift ⋆)))

  ε : P ≃ Q
  ε = logically-equivalent-subsingletons-are-equivalent P Q i j (γ , φ)
```

*Exercise*. `propext` and `funext` and excluded middle together imply that `Ω 𝓤` has size $\mathcal{U}_0$.

### Propositional resizing gives subsingleton truncation

Using Voevodsky's construction and propositional resizing, we get that function extensionality implies that subsingleton truncations exist:

```
PR-gives-existence-of-truncations : global-dfunext
                                  → Propositional-resizing
                                  → subsingleton-truncations-exist

PR-gives-existence-of-truncations fe R =
 record
 {
   ∥_∥ =

    λ {𝓤} X → resize R
               (is-inhabited X)
               (inhabitation-is-subsingleton fe X) ;

   ∥∥-is-subsingleton =

    λ {𝓤} {X} → resize-is-subsingleton R
                 (is-inhabited X)
                 (inhabitation-is-subsingleton fe X) ;

   ∣_∣ =

    λ {𝓤} {X} x → to-resize R
                   (is-inhabited X)
                   (inhabitation-is-subsingleton fe X)
```

```
                      (inhabited-intro x) ;

   ∥∥-recursion =

    λ {𝓤} {𝓥} {X} {P} i u s → from-resize R P i
                                 (inhabited-recursion
                                   (resize-is-subsingleton R P i)
                                   (to-resize R P i ∘ u)
                                   (from-resize R
                                     (is-inhabited X)
                                     (inhabitation-is-subsingleton fe X) s))
 }
```

### The powerset in the presence of propositional resizing

As a second, important, use of resizing, we revisit the powerset. First, given a set of subsets, that is, an element of the double powerset, we would like to consider its union. We investigate its existence in a submodule with assumptions.

```
module powerset-union-existence
        (pt  : subsingleton-truncations-exist)
        (hfe : global-hfunext)
       where

 open basic-truncation-development pt hfe
```

Unions of *families* of subsets exist under the above assumption of the existence of truncations, provided the index set `I` is in a universe equal or below the universe of the type `X` we take the powerset of:

```
 family-union : {X : 𝓤 ⊔ 𝓥 ̇ } {I : 𝓥 ̇ } → (I → 𝓟 X) → 𝓟 X
 family-union {𝓤} {𝓥} {X} {I} A = λ x → (∃ i ꞉ I , x ∈ A i) , ∃-is-subsingleton
```

Notice the increase of universe levels in the iteration of powersets:

```
 𝓟𝓟 : 𝓤 ̇ → 𝓤 ⁺⁺ ̇
 𝓟𝓟 X = 𝓟 (𝓟 X)
```

The following doesn't assert that unions of collections of subsets are available. It says what it means for them to be available:

```
 existence-of-unions : (𝓤 : Universe) → 𝓤 ⁺⁺ ̇
 existence-of-unions 𝓤 =
  (X : 𝓤 ̇ ) (𝓐 : 𝓟𝓟 X) → Σ B ꞉ 𝓟 X , ((x : X) → (x ∈ B) ⇔ (∃ A ꞉ 𝓟 X , (A ∈ 𝓐) × (x ∈ A)))
```

One may wonder whether there are different universe assignments for the above definition, for instance whether we can instead assume `𝓐 : (X → Ω 𝓥) → Ω 𝓦` for suitable universes 𝓥 and 𝓦, possibly depending on 𝓤. That this is not the case can be checked by writing the above definition in the following alternative form:

```
 existence-of-unions₁ : (𝓤 :  Universe) → _ ̇
 existence-of-unions₁ 𝓤 =
  (X : 𝓤 ̇ )
  (𝓐 : (X → Ω _) → Ω _)
     → Σ B ꞉ (X → Ω _) , ((x : X) → (x ∈ B) ⇔ (∃ A ꞉ (X → Ω _) , (A ∈ 𝓐) × (x ∈ A)))
```

The fact that Agda accepts the above definition without errors means that there is a unique way to fill each _, which has to be the following.

```
 existence-of-unions₂ : (𝓤 :  Universe) → 𝓤 ⁺⁺ ̇
 existence-of-unions₂ 𝓤 =
  (X : 𝓤 ̇ )
  (𝓐 : (X → Ω 𝓤) → Ω (𝓤 ⁺))
     → Σ B ꞉ (X → Ω 𝓤) , ((x : X) → (x ∈ B) ⇔ (∃ A ꞉ (X → Ω 𝓤) , (A ∈ 𝓐) × (x ∈ A)))

 existence-of-unions-agreement : existence-of-unions 𝓤 = existence-of-unions₂ 𝓤
 existence-of-unions-agreement = refl _
```

To get the universe assigments explicitly, using Agda's interactive mode, we can write holes `?` instead of _ and then ask Agda to fill them uniquely by solving constraints, which is what we did to construct `existence-of-unions₂`.

*Exercise*. Show that the existence of unions is a subsingleton type.

Without propositional resizing principles, it is not possible to establish the existence.

```
existence-of-unions-gives-PR : existence-of-unions 𝓤
                              → propositional-resizing (𝓤 ⁺) 𝓤

existence-of-unions-gives-PR {𝓤} α = γ
 where
  γ : (P : 𝓤 ⁺ ̇ ) → (i : is-subsingleton P) → P has-size 𝓤
  γ P i = Q , e
   where
   𝟙ᵤ : 𝓤 ̇
   𝟙ᵤ = Lift 𝓤 𝟙

   ⋆ᵤ : 𝟙ᵤ
   ⋆ᵤ = lift ⋆

   𝟙ᵤ-is-subsingleton : is-subsingleton 𝟙ᵤ
   𝟙ᵤ-is-subsingleton = equiv-to-subsingleton (Lift-≃ 𝟙) 𝟙-is-subsingleton

   𝓐 : 𝓟𝓟 𝟙ᵤ
   𝓐 = λ (A : 𝓟 𝟙ᵤ) → P , i

   B : 𝓟 𝟙ᵤ
   B = pr₁ (α 𝟙ᵤ 𝓐)

   φ : (x : 𝟙ᵤ) → (x ∈ B) ⇔ (∃ A ꞉ 𝓟 𝟙ᵤ , (A ∈ 𝓐) × (x ∈ A))
   φ = pr₂ (α 𝟙ᵤ 𝓐)

   Q : 𝓤 ̇
   Q = ⋆ᵤ ∈ B

   j : is-subsingleton Q
   j = ∈-is-subsingleton B ⋆ᵤ

   f : P → Q
   f p = b
    where
     a : Σ A ꞉ 𝓟 𝟙ᵤ , (A ∈ 𝓐) × (⋆ᵤ ∈ A)
     a = (λ (x : 𝟙ᵤ) → 𝟙ᵤ , 𝟙ᵤ-is-subsingleton) , p , ⋆ᵤ

     b : ⋆ᵤ ∈ B
     b = rl-implication (φ ⋆ᵤ) ∣ a ∣

   g : Q → P
   g q = ∥∥-recursion i b a
    where
     a : ∃ A ꞉ 𝓟 𝟙ᵤ , (A ∈ 𝓐) × (⋆ᵤ ∈ A)
     a = lr-implication (φ ⋆ᵤ) q

     b : (Σ A ꞉ 𝓟 𝟙ᵤ , (A ∈ 𝓐) × (⋆ᵤ ∈ A)) → P
     b (A , m , _) = m

   e : P ≃ Q
   e = logically-equivalent-subsingletons-are-equivalent P Q i j (f , g)
```

The converse also holds, with an easier construction:

```
PR-gives-existence-of-unions : propositional-resizing (𝓤 ⁺) 𝓤
                              → existence-of-unions 𝓤

PR-gives-existence-of-unions {𝓤} ρ X 𝓐 = B , (λ x → lr x , rl x)
 where
  β : X → 𝓤 ⁺ ̇
  β x = ∃ A ꞉ 𝓟 X , (A ∈ 𝓐) × (x ∈ A)
```

```
   i : (x : X) → is-subsingleton (β x)
   i x = ∃-is-subsingleton

   B : 𝓟 X
   B x = (resize ρ (β x) (i x) , resize-is-subsingleton ρ (β x) (i x))

   lr : (x : X) → x ∈ B → ∃ A ꞉ 𝓟 X , (A ∈ 𝓐) × (x ∈ A)
   lr x = from-resize ρ (β x) (i x)

   rl : (x : X) → (∃ A ꞉ 𝓟 X , (A ∈ 𝓐) × (x ∈ A)) → x ∈ B
   rl x = to-resize ρ (β x) (i x)
```

We now close the above submodule and start another one with different assumptions:

```
module basic-powerset-development
        (hfe : global-hfunext)
        (ρ   : Propositional-resizing)
       where

  pt : subsingleton-truncations-exist
  pt = PR-gives-existence-of-truncations (hfunext-gives-dfunext hfe) ρ

  open basic-truncation-development pt hfe
  open powerset-union-existence pt hfe

  ⋃ : {X : 𝓤 ̇ } → 𝓟𝓟 X → 𝓟 X
  ⋃ 𝓐 = pr₁ (PR-gives-existence-of-unions ρ _ 𝓐)

  ⋃-property : {X : 𝓤 ̇ } (𝓐 : 𝓟𝓟 X)
             → (x : X) → (x ∈ ⋃ 𝓐) ⇔ (∃ A ꞉ 𝓟 X , (A ∈ 𝓐) × (x ∈ A))

  ⋃-property 𝓐 = pr₂ (PR-gives-existence-of-unions ρ _ 𝓐)
```

The construction of intersections is as that of unions using propositional resizing:

```
  intersections-exist :
    (X : 𝓤 ̇ )
    (𝓐 : 𝓟𝓟 X)
       → Σ B ꞉ 𝓟 X , ((x : X) → (x ∈ B) ⇔ ((A : 𝓟 X) → A ∈ 𝓐 → x ∈ A))

  intersections-exist {𝓤} X 𝓐 = B , (λ x → lr x , rl x)
   where
    β : X → 𝓤 ⁺ ̇
    β x = (A : 𝓟 X) → A ∈ 𝓐 → x ∈ A

    i : (x : X) → is-subsingleton (β x)
    i x = Π-is-subsingleton hunapply
           (λ A → Π-is-subsingleton hunapply
           (λ _ → ∈-is-subsingleton A x))

    B : 𝓟 X
    B x = (resize ρ (β x) (i x) , resize-is-subsingleton ρ (β x) (i x))

    lr : (x : X) → x ∈ B → (A : 𝓟 X) → A ∈ 𝓐 → x ∈ A
    lr x = from-resize ρ (β x) (i x)

    rl : (x : X) → ((A : 𝓟 X) → A ∈ 𝓐 → x ∈ A) → x ∈ B
    rl x = to-resize ρ (β x) (i x)

  ⋂ : {X : 𝓤 ̇ } → 𝓟𝓟 X → 𝓟 X
  ⋂ {𝓤} {X} 𝓐 = pr₁ (intersections-exist X 𝓐)

  ⋂-property : {X : 𝓤 ̇ } (𝓐 : 𝓟𝓟 X)
             → (x : X) → (x ∈ ⋂ 𝓐) ⇔ ((A : 𝓟 X) → A ∈ 𝓐 → x ∈ A)

  ⋂-property {𝓤} {X} 𝓐 = pr₂ (intersections-exist X 𝓐)
```

```
∅ full : {X : 𝓤 ̇ } → 𝓟 X

∅    = λ x → (Lift _ 𝟘 , equiv-to-subsingleton (Lift-≃ 𝟘) 𝟘-is-subsingleton)

full = λ x → (Lift _ 𝟙 , equiv-to-subsingleton (Lift-≃ 𝟙) 𝟙-is-subsingleton)

∅-property : (X : 𝓤 ̇ ) → (x : X) → ¬(x ∈ ∅)
∅-property X x = lower

full-property : (X : 𝓤 ̇ ) → (x : X) → x ∈ full
full-property X x = lift ⋆

_∩_ _∪_ : {X : 𝓤 ̇ } → 𝓟 X → 𝓟 X → 𝓟 X

(A ∪ B) = λ x → ((x ∈ A) ∨ (x ∈ B)) , ∨-is-subsingleton

(A ∩ B) = λ x → ((x ∈ A) × (x ∈ B)) ,
                ×-is-subsingleton
                  (∈-is-subsingleton A x)
                  (∈-is-subsingleton B x)

∪-property : {X : 𝓤 ̇ } (A B : 𝓟 X)
           → (x : X) → x ∈ (A ∪ B) ⇔ (x ∈ A) ∨ (x ∈ B)

∪-property {𝓤} {X} A B x = id , id

∩-property : {X : 𝓤 ̇ } (A B : 𝓟 X)
           → (x : X) → x ∈ (A ∩ B) ⇔ (x ∈ A) × (x ∈ B)

∩-property {𝓤} {X} A B x = id , id

infix  20 _∩_
infix  20 _∪_
```

**Topological spaces in the presence of propositional resizing**

For example, with this we can define the type of topological spaces as follows, where 𝓞 consists of designated sets, conventionally called *open* and collectively referred to as the *topology* on `X`, which are stipulated to be closed under finite intersections and arbitrary unions. For finite intersections we consider the unary case `full` and the binary case `∩` . Because the empty set is the union of the empty collection (exercise), it is automatically included among the open sets.

```
Top : (𝓤 : Universe) → 𝓤 ⁺⁺ ̇
Top 𝓤 = Σ X ꞉ 𝓤 ̇ , is-set X
                 × (Σ 𝓞 ꞉ 𝓟𝓟 X , full ∈ 𝓞
                              × ((U V : 𝓟 X) → U ∈ 𝓞 → V ∈ 𝓞 → (U ∩ V) ∈ 𝓞)
                              × ((𝓖 : 𝓟𝓟 X) → 𝓖 ⊆ 𝓞 → ⋃ 𝓖 ∈ 𝓞))
```

Notice that this jumps two universes. It is also possible, with `Ω`-resizing, to construct the powerset in such a way that the powerset of any type lives in the same universe as the type (exercise), and hence so that the type of topological spaces in a base universe lives in the next universe (exercise), rather than two universes above the base universe.

*Exercise*. For a function `X → Y`, define its inverse image `𝓟 Y → 𝓟 X` and its direct image `𝓟 X → 𝓟 Y`. Define the notion of a continuous map of topological spaces, namely a function of the underlying sets whose inverse images of open sets are open. Show that the identity function is continuous and that continuous maps are closed under composition.

# Quotients

We now construct quotients using a technique proposed by Voevodsky, who assumed propositional resizing for that purpose, so that the quotient of a given type by a given equivalence relation would live in the same universe as the type. But the requirement that the quotient lives in the same universe is not needed to prove the universal property of the quotient.

We construct the quotient using propositional truncations, assuming functional and propositional extensionality, *without assuming resizing*.

A binary relation `_≈_` on a type `X : 𝓤` with values in a universe `𝓥` (which can of course be `𝓤`) is called an *equivalence relation* if it is subsingleton-valued, reflexive, symmetric and transitive. All these notions

```
is-subsingleton-valued
 reflexive
 symmetric
 transitive
 is-equivalence-relation :
```

have the same type

```
 {X : 𝓤 ̇ } → (X → X → 𝓥 ̇ ) → 𝓤 ⊔ 𝓥 ̇
```

and are defined by

```
is-subsingleton-valued  _≈_  = ∀ x y → is-subsingleton (x ≈ y)
reflexive               _≈_  = ∀ x → x ≈ x
symmetric               _≈_  = ∀ x y → x ≈ y → y ≈ x
transitive              _≈_  = ∀ x y z → x ≈ y → y ≈ z → x ≈ z

is-equivalence-relation _≈_  = is-subsingleton-valued _≈_
                             × reflexive _≈_
                             × symmetric _≈_
                             × transitive _≈_
```

We now work with a submodule with parameters to quotient a given type `X` by a given equivalence relation `_≈_`. We assume not only the existence of propositional truncations, as discussed above, but also functional and propositional extensionality.

```
module quotient
       {𝓤 𝓥 : Universe}
       (pt  : subsingleton-truncations-exist)
       (hfe : global-hfunext)
       (pe  : propext 𝓥)
       (X   : 𝓤 ̇ )
       (_≈_ : X → X → 𝓥 ̇ )
       (≈p  : is-subsingleton-valued _≈_)
       (≈r  : reflexive _≈_)
       (≈s  : symmetric _≈_)
       (≈t  : transitive _≈_)
      where

 open basic-truncation-development pt hfe
```

From the given relation

```
     _≈_ : X → X → 𝓥 ̇
```

we define a function

```
     X → (X → Ω 𝓥),
```

and we take the quotient `X/≈` to be the image of this function. It is for constructing the image that we need subsingleton truncations. Functional and propositional extensionality are then used to prove that the quotient is a set.

```
 equiv-rel : X → (X → Ω 𝓥)
 equiv-rel x y = (x ≈ y) , ≈p x y

 X/≈ : 𝓥 ⁺ ⊔ 𝓤 ̇
```

```
 X/≈ = image equiv-rel

 X/≈-is-set : is-set X/≈
 X/≈-is-set = subsets-of-sets-are-sets (X → Ω 𝓥) _
               (powersets-are-sets (dfunext-gives-hfunext hunapply) hunapply pe)
               (λ _ → ∃-is-subsingleton)

 η : X → X/≈
 η = corestriction equiv-rel
```

We show that η is the universal solution to the problem of transforming equivalence _≈_ into equality _=_.

By construction, η is a surjection, of course:

```
 η-surjection : is-surjection η
 η-surjection = corestriction-surjection equiv-rel
```

It is convenient to use the following induction principle for reasoning about the image X/≈.

```
 η-induction : (P : X/≈ → 𝓦 ̇ )
             → ((x' : X/≈) → is-subsingleton (P x'))
             → ((x : X) → P (η x))
             → (x' : X/≈) → P x'

 η-induction = surjection-induction η η-surjection
```

The first part of the universal property of η says that equivalent points are mapped to identified points:

```
 η-equiv-equal : {x y : X} → x ≈ y → η x = η y
 η-equiv-equal {x} {y} e =
  to-subtype-=
    (λ _ → ∃-is-subsingleton)
    (hunapply (λ z → to-subtype-=
                        (λ _ → being-subsingleton-is-subsingleton hunapply)
                        (pe (≈p x z) (≈p y z) (≈t y x z (≈s x y e)) (≈t x y z e))))
```

To prove the required universal property, we also need the fact that η reflects equality into equivalence:

```
 η-equal-equiv : {x y : X} → η x = η y → x ≈ y
 η-equal-equiv {x} {y} p = equiv-rel-reflect (ap pr₁ p)
  where
   equiv-rel-reflect : equiv-rel x = equiv-rel y → x ≈ y
   equiv-rel-reflect q = b (≈r y)
    where
     a : (y ≈ y) = (x ≈ y)
     a = ap (λ - → pr₁(- y)) (q ⁻¹)

     b : y ≈ y → x ≈ y
     b = Id→fun a
```

We are now ready to formulate and prove the required universal property of the quotient. What is noteworthy here, regarding universes, is that the universal property says that we can eliminate into any set A of any universe 𝓦.

```
 universal-property : (A : 𝓦 ̇ )
                    → is-set A
                    → (f : X → A)
                    → ({x x' : X} → x ≈ x' → f x = f x')
                    → ∃! f' ꞉ (X/≈ → A), f' ∘ η = f

 universal-property {𝓦} A i f τ = e
  where
   G : X/≈ → 𝓥 ⁺ ⊔ 𝓤 ⊔ 𝓦 ̇
   G x' = Σ a ꞉ A , ∃ x ꞉ X , (η x = x') × (f x = a)

   φ : (x' : X/≈) → is-subsingleton (G x')
   φ = η-induction _ γ induction-step
    where
     induction-step : (y : X) → is-subsingleton (G (η y))
```

```
    induction-step x (a , d) (b , e) = to-subtype-≡ (λ _ → ∃-is-subsingleton) p
     where
      h : (Σ x' ꞉ X , (η x' ≡ η x) × (f x' ≡ a))
        → (Σ y' ꞉ X , (η y' ≡ η x) × (f y' ≡ b))
        → a ≡ b
      h (x' , r , s) (y' , t , u) = a    ≡⟨ s ⁻¹                         ⟩
                                    f x' ≡⟨ τ (η-equal-equiv (r ∙ t ⁻¹)) ⟩
                                    f y' ≡⟨ u                            ⟩
                                    b    ∎

      p : a ≡ b
      p = ∥∥-recursion (i a b) (λ σ → ∥∥-recursion (i a b) (h σ) e) d

    γ : (x' : X/≈) → is-subsingleton (is-subsingleton (G x'))
    γ x' = being-subsingleton-is-subsingleton hunapply

  k : (x' : X/≈) → G x'
  k = η-induction _ φ induction-step
   where
    induction-step : (y : X) → G (η y)
    induction-step x = f x , ∣ x , refl (η x) , refl (f x) ∣

  f' : X/≈ → A
  f' x' = pr₁ (k x')

  r : f' ∘ η ≡ f
  r = hunapply h
   where
    g : (y : X) → ∃ x ꞉ X , (η x ≡ η y) × (f x ≡ f' (η y))
    g y = pr₂ (k (η y))

    j : (y : X) → (Σ x ꞉ X , (η x ≡ η y) × (f x ≡ f' (η y))) → f'(η y) ≡ f y
    j y (x , p , q) = f' (η y) ≡⟨ q ⁻¹                ⟩
                      f x      ≡⟨ τ (η-equal-equiv p) ⟩
                      f y      ∎

    h : (y : X) → f'(η y) ≡ f y
    h y = ∥∥-recursion (i (f' (η y)) (f y)) (j y) (g y)

  c : (σ : Σ f'' ꞉ (X/≈ → A), f'' ∘ η ≡ f) → (f' , r) ≡ σ
  c (f'' , s) = to-subtype-≡ (λ g → Π-is-set hfe (λ _ → i) (g ∘ η) f) t
   where
    w : ∀ x → f'(η x) ≡ f''(η x)
    w = happly (f' ∘ η) (f'' ∘ η) (r ∙ s ⁻¹)

    t : f' ≡ f''
    t = hunapply (η-induction _ (λ x' → i (f' x') (f'' x')) w)

  e : ∃! f' ꞉ (X/≈ → A), f' ∘ η ≡ f
  e = (f' , r) , c
```

As mentioned above, if one so wishes, it is possible to resize down the quotient `X/≈` to the same universe as the given type `X` lives by assuming propositional resizing. But we don't see any mathematical need or benefit to do so, as the constructed quotient, regardless of the universe it inhabits, has a universal property that eliminates into any desired universe, lower, equal or higher than the quotiented type.

## Summary of consistent axioms for univalent mathematics

The following axioms are together consistent by considering Voevodsky's simplicial-set model:

1. Function extensionality.
2. Propositional extensionality.

3. Univalence.
4. Existence of propositional truncations.
5. Excluded middle.
6. Choice.
7. Propositional resizing and impredicativity.

We have that:

- The first four admit a constructive interpretation via cubical type theory with an implementation in cubical Agda.

- Univalence implies function extensionality and propositional extensionality.

- Simple unique choice just holds.

- Full unique choice is equivalent to function extensionality.

- Choice implies excluded middle, as usual, and both are non-constructive.

- Excluded middle implies propositional resizing and impredicativity.

- The constructive status of propositional resizing and impredicativity is open.

- Function extensionality and propositional resizing imply the existence of propositional truncations, and hence so do function extensionality and excluded middle.

The avoidance of excluded middle and choice makes the theory not only constructive but also applicable to more models. However, one is free to assume excluded middle and choice for pieces of mathematics that require them, or just if one simply prefers classical reasoning.

Univalent foundations have enough room for the constructive, non-constructive, pluralistic and neutral approaches to mathematics, and in this sense they are no different from e.g. set theoretic foundations.

An example of a theorem of univalent mathematics that requires excluded middle is Cantor-Schröder-Bernstein for homotopy types, or ∞-groupoids, which is also discussed in this blog post and is implemented in Agda.

A major omission in these notes is a discussion of higher-inductive types. On the other hand, these notes completely cover the foundational principles officially supported by UniMath, namely (1)-(7) above.

# Appendix

### Solutions to some exercises

```
module ℕ-order-exercise-solution where

  _≤'_ : ℕ → ℕ → 𝓤₀ ̇
  _≤'_ = ℕ-iteration (ℕ → 𝓤₀ ̇ ) (λ y → 𝟙)
           (λ f → ℕ-recursion (𝓤₀ ̇ ) 𝟘 (λ y P → f y))

  open ℕ-order

  ≤-and-≤'-coincide : (x y : ℕ) → (x ≤ y) ≡ (x ≤' y)
  ≤-and-≤'-coincide 0 y = refl _
  ≤-and-≤'-coincide (succ x) 0 = refl _
  ≤-and-≤'-coincide (succ x) (succ y) = ≤-and-≤'-coincide x y

module ℕ-more where

  open Arithmetic renaming (_+_ to _∔_)
  open basic-arithmetic-and-order
```

```
  ≤-prop-valued : (x y : ℕ) → is-subsingleton (x ≤ y)
  ≤-prop-valued 0 y               = 𝟙-is-subsingleton
  ≤-prop-valued (succ x) zero     = 𝟘-is-subsingleton
  ≤-prop-valued (succ x) (succ y) = ≤-prop-valued x y

  ≼-prop-valued : (x y : ℕ) → is-subsingleton (x ≼ y)
  ≼-prop-valued x y (z , p) (z' , p') = γ
   where
    q : z ≡ z'
    q = +-lc x z z' (x ∔ z  ≡⟨ p     ⟩
                     y      ≡⟨ p' ⁻¹ ⟩
                     x ∔ z' ∎)

    γ : z , p ≡ z' , p'
    γ = to-subtype-≡ (λ z → ℕ-is-set (x ∔ z) y) q

  ≤-charac : propext 𝓤₀ → (x y : ℕ) → (x ≤ y) ≡ (x ≼ y)
  ≤-charac pe x y = pe (≤-prop-valued x y) (≼-prop-valued x y)
                       (≤-gives-≼ x y) (≼-gives-≤ x y)

the-subsingletons-are-the-subtypes-of-a-singleton : (X : 𝓤 ̇ )
                                                  → is-subsingleton X ⇔ (X ↪ 𝟙)
the-subsingletons-are-the-subtypes-of-a-singleton X = φ , ψ
 where
  i : is-subsingleton X → is-embedding (!𝟙' X)
  i s ⋆ (x , refl ⋆) (y , refl ⋆) = ap (λ - → - , refl ⋆) (s x y)

  φ : is-subsingleton X → X ↪ 𝟙
  φ s = !𝟙 , i s

  ψ : X ↪ 𝟙 → is-subsingleton X
  ψ (f , e) x y = d
   where
    a : x ≡ y → f x ≡ f y
    a = ap f {x} {y}

    b : is-equiv a
    b = embedding-gives-ap-is-equiv f e x y

    c : f x ≡ f y
    c = 𝟙-is-subsingleton (f x) (f y)

    d : x ≡ y
    d = inverse a b c

the-subsingletons-are-the-subtypes-of-a-singleton' : propext 𝓤 → global-dfunext
                                                   → (X : 𝓤 ̇ )
                                                   → is-subsingleton X ≡ (X ↪ 𝟙)
the-subsingletons-are-the-subtypes-of-a-singleton' pe fe X = γ
 where
  a : is-subsingleton X ⇔ (X ↪ 𝟙)
  a = the-subsingletons-are-the-subtypes-of-a-singleton X

  b : is-subsingleton (X ↪ 𝟙)
  b (f , e) (f' , e') = to-subtype-≡
                          (being-embedding-is-subsingleton fe)
                          (fe (λ x → 𝟙-is-subsingleton (f x) (f' x)))

  γ : is-subsingleton X ≡ (X ↪ 𝟙)
  γ = pe (being-subsingleton-is-subsingleton fe) b (pr₁ a) (pr₂ a)

G↑-≃-equation : (ua : is-univalent (𝓤 ⊔ 𝓥))
              → (X : 𝓤 ̇ )
              → (A : (Σ Y ꞉ 𝓤 ⊔ 𝓥 ̇ , X ≃ Y) → 𝓦 ̇ )
              → (a : A (Lift 𝓥 X , ≃-Lift X))
              → G↑-≃ ua X A a (Lift 𝓥 X) (≃-Lift X) ≡ a
```

```
G↑-≃-equation {𝓤} {𝓥} {𝓦} ua X A a =
  G↑-≃ ua X A a (Lift 𝓥 X) (≃-Lift X) ＝⟨ refl (transport A p a)       ⟩
  transport A p a                      ＝⟨ ap (λ - → transport A - a) q ⟩
  transport A (refl t) a               ＝⟨ refl a                       ⟩
  a                                    ∎
 where
  t : (Σ Y ꞉ 𝓤 ⊔ 𝓥 ̇ , X ≃ Y)
  t = (Lift 𝓥 X , ≃-Lift X)

  p : t ＝ t
  p = univalence→'' {𝓤} {𝓤 ⊔ 𝓥} ua X t t

  q : p ＝ refl t
  q = subsingletons-are-sets (Σ Y ꞉ 𝓤 ⊔ 𝓥 ̇ , X ≃ Y)
       (univalence→'' {𝓤} {𝓤 ⊔ 𝓥} ua X) t t p (refl t)

H↑-≃-equation : (ua : is-univalent (𝓤 ⊔ 𝓥))
              → (X : 𝓤 ̇ )
              → (A : (Y : 𝓤 ⊔ 𝓥 ̇ ) → X ≃ Y → 𝓦 ̇ )
              → (a : A (Lift 𝓥 X) (≃-Lift X))
              → H↑-≃ ua X A a (Lift 𝓥 X) (≃-Lift X) ＝ a
H↑-≃-equation ua X A = G↑-≃-equation ua X (Σ-induction A)

has-section-charac : {X : 𝓤 ̇ } {Y : 𝓥 ̇ } (f : X → Y)
                   → ((y : Y) → Σ x ꞉ X , f x ＝ y) ≃ has-section f
has-section-charac f = ΠΣ-distr-≃

retractions-into : 𝓤 ̇ → 𝓤 ⁺ ̇
retractions-into {𝓤} Y = Σ X ꞉ 𝓤 ̇ , Y ◁ X

pointed-types : (𝓤 : Universe) → 𝓤 ⁺ ̇
pointed-types 𝓤 = Σ X ꞉ 𝓤 ̇ , X

retraction-classifier : Univalence
                      → (Y : 𝓤 ̇ ) → retractions-into Y ≃ (Y → pointed-types 𝓤)
retraction-classifier {𝓤} ua Y =
 retractions-into Y                                               ≃⟨ i   ⟩
 (Σ X ꞉ 𝓤 ̇ , Σ f ꞉ (X → Y) , ((y : Y) → Σ x ꞉ X , f x ＝ y)) ≃⟨ ii  ⟩
 ((𝓤 /[ id ] Y))                                                  ≃⟨ iii ⟩
 (Y → pointed-types 𝓤)                                            ■
 where
  i   = ≃-sym (Σ-cong (λ X → Σ-cong (λ f → ΠΣ-distr-≃)))
  ii  = Id→Eq _ _ (refl _)
  iii = special-map-classifier (ua 𝓤)
         (univalence-gives-dfunext' (ua 𝓤) (ua (𝓤 ⁺)))
         id Y

module surjection-classifier
         (pt : subsingleton-truncations-exist)
         (ua : Univalence)
       where

  hfe : global-hfunext
  hfe = univalence-gives-global-hfunext ua

  open basic-truncation-development pt hfe public

  _↠_ : 𝓤 ̇ → 𝓥 ̇ → 𝓤 ⊔ 𝓥 ̇
  X ↠ Y = Σ f ꞉ (X → Y), is-surjection f

  surjections-into : 𝓤 ̇ → 𝓤 ⁺ ̇
  surjections-into {𝓤} Y = Σ X ꞉ 𝓤 ̇ , X ↠ Y

  inhabited-types : (𝓤 : Universe) → 𝓤 ⁺ ̇
  inhabited-types 𝓤 = Σ X ꞉ 𝓤 ̇ , ∥ X ∥

  surjection-classifier : Univalence
                        → (Y : 𝓤 ̇ )
```

```
                              → surjections-into Y ≃ (Y → inhabited-types 𝓤)
  surjection-classifier {𝓤} ua = special-map-classifier (ua 𝓤)
                                   (univalence-gives-dfunext' (ua 𝓤) (ua (𝓤 ⁺)))
                                   ∥_∥
```

## Additional exercises

Solutions are available at the end.

*Exercise*. A sequence of elements of a type `X` is just a function `ℕ → X`. Use Cantor's diagonal argument to show in Agda that the type of sequences of natural numbers is uncountable. Prove a positive version and then derive a negative version from it:

```
positive-cantors-diagonal : (e : ℕ → (ℕ → ℕ)) → Σ α ꞉ (ℕ → ℕ), ((n : ℕ) → α ≢ e n)
```

```
cantors-diagonal : ¬(Σ e ꞉ (ℕ → (ℕ → ℕ)) , ((α : ℕ → ℕ) → Σ n ꞉ ℕ , α ≡ e n))
```

*Hint*. It may be helpful to prove that the function `succ` has no fixed points, first.

*Exercise*.

```
𝟚-has-𝟚-automorphisms : dfunext 𝓤₀ 𝓤₀ → (𝟚 ≃ 𝟚) ≃ 𝟚
```

Now we would like to have `(𝟚 ≡ 𝟚) ≡ 𝟚` with univalence, but the problem is that the type `𝟚 ≡ 𝟚` lives in $\mathcal{U}_1$ whereas `𝟚` lives in $\mathcal{U}_0$ and so, having different types, can't be compared for equality. But we do have that

```
lifttwo : is-univalent 𝓤₀ → is-univalent 𝓤₁ → (𝟚 ≡ 𝟚) ≡ Lift 𝓤₁ 𝟚
```

*Exercise*. Having decidable equality is a subsingleton type.

*Exercise*. We now discuss alternative formulations of the principle of excluded middle.

```
DNE : ∀ 𝓤 → 𝓤 ⁺ ̇
DNE 𝓤 = (P : 𝓤 ̇ ) → is-subsingleton P → ¬¬ P → P

ne : (X : 𝓤 ̇ ) → ¬¬(X + ¬ X)

DNE-gives-EM : dfunext 𝓤 𝓤₀ → DNE 𝓤 → EM 𝓤

EM-gives-DNE : EM 𝓤 → DNE 𝓤
```

The following says that excluded middle holds if and only if every subsingleton is the negation of some type.

```
SN : ∀ 𝓤 → 𝓤 ⁺ ̇
SN 𝓤 = (P : 𝓤 ̇ ) → is-subsingleton P → Σ X ꞉ 𝓤 ̇ , P ⇔ ¬ X

SN-gives-DNE : SN 𝓤 → DNE 𝓤

DNE-gives-SN : DNE 𝓤 → SN 𝓤
```

## Solutions to additional exercises

```
succ-no-fixed-point : (n : ℕ) → succ n ≢ n
succ-no-fixed-point 0        = positive-not-zero 0
succ-no-fixed-point (succ n) = γ
 where
  IH : succ n ≢ n
  IH = succ-no-fixed-point n

  γ : succ (succ n) ≢ succ n
  γ p = IH (succ-lc p)
```

```agda
positive-cantors-diagonal = sol
 where
  sol : (e : ℕ → (ℕ → ℕ)) → Σ α ꞉ (ℕ → ℕ), ((n : ℕ) → α ≢ e n)
  sol e = (α , φ)
   where
    α : ℕ → ℕ
    α n = succ(e n n)

    φ : (n : ℕ) → α ≢ e n
    φ n p = succ-no-fixed-point (e n n) q
     where
      q = succ (e n n)  ≡⟨ refl (α n)       ⟩
          α n           ≡⟨ ap (λ - → - n) p ⟩
          e n n         ∎

cantors-diagonal = sol
 where
  sol : ¬(Σ e ꞉ (ℕ → (ℕ → ℕ)) , ((α : ℕ → ℕ) → Σ n ꞉ ℕ , α ≡ e n))
  sol (e , γ) = c
   where
    α : ℕ → ℕ
    α = pr₁ (positive-cantors-diagonal e)

    φ : (n : ℕ) → α ≢ e n
    φ = pr₂ (positive-cantors-diagonal e)

    b : Σ n ꞉ ℕ , α ≡ e n
    b = γ α

    c : 𝟘
    c = φ (pr₁ b) (pr₂ b)

𝟚-has-𝟚-automorphisms = sol
 where
  sol : dfunext 𝓤₀ 𝓤₀ → (𝟚 ≃ 𝟚) ≃ 𝟚
  sol fe = invertibility-gives-≃ f (g , η , ε)
   where
    f : (𝟚 ≃ 𝟚) → 𝟚
    f (h , e) = h ₀

    g : 𝟚 → (𝟚 ≃ 𝟚)
    g ₀ = id , id-is-equiv 𝟚
    g ₁ = swap₂ , swap₂-is-equiv

    η : (e : 𝟚 ≃ 𝟚) → g (f e) ≡ e
    η (h , e) = γ (h ₀) (h ₁) (refl (h ₀)) (refl (h ₁))
     where
      γ : (m n : 𝟚) → h ₀ ≡ m → h ₁ ≡ n → g (h ₀) ≡ (h , e)

      γ ₀ ₀ p q = !𝟘 (g (h ₀) ≡ (h , e))
                     (₁-is-not-₀ (equivs-are-lc h e (h ₁ ≡⟨ q    ⟩
                                                     ₀   ≡⟨ p ⁻¹ ⟩
                                                     h ₀ ∎)))

      γ ₀ ₁ p q = to-subtype-≡
                     (being-equiv-is-subsingleton fe fe)
                     (fe (𝟚-induction (λ n → pr₁ (g (h ₀)) n ≡ h n)
                               (pr₁ (g (h ₀)) ₀ ≡⟨ ap (λ - → pr₁ (g -) ₀) p ⟩
                                pr₁ (g ₀) ₀     ≡⟨ refl ₀                    ⟩
                                ₀                ≡⟨ p ⁻¹                      ⟩
                                h ₀              ∎)
                               (pr₁ (g (h ₀)) ₁ ≡⟨ ap (λ - → pr₁ (g -) ₁) p ⟩
                                pr₁ (g ₀) ₁     ≡⟨ refl ₁                    ⟩
                                ₁                ≡⟨ q ⁻¹                      ⟩
                                h ₁              ∎)))
```

```
      γ ₁ ₀ p q = to-subtype-≡
                   (being-equiv-is-subsingleton fe fe)
                   (fe (𝟚-induction (λ n → pr₁ (g (h ₀)) n ≡ h n)
                             (pr₁ (g (h ₀)) ₀ ≡⟨ ap (λ - → pr₁ (g -) ₀) p ⟩
                              pr₁ (g ₁) ₀     ≡⟨ refl ₁                    ⟩
                              ₁               ≡⟨ p ⁻¹                      ⟩
                              h ₀             ∎)
                             (pr₁ (g (h ₀)) ₁ ≡⟨ ap (λ - → pr₁ (g -) ₁) p ⟩
                              pr₁ (g ₁) ₁     ≡⟨ refl ₀                    ⟩
                              ₀               ≡⟨ q ⁻¹                      ⟩
                              h ₁             ∎)))

      γ ₁ ₁ p q = !𝟘 (g (h ₀) ≡ (h , e))
                   (₁-is-not-₀ (equivs-are-lc h e (h ₁ ≡⟨ q    ⟩
                                                  ₁   ≡⟨ p ⁻¹ ⟩
                                                  h ₀ ∎)))

    ε : (n : 𝟚) → f (g n) ≡ n
    ε ₀ = refl ₀
    ε ₁ = refl ₁

lifttwo = sol
 where
  sol : is-univalent 𝓤₀ → is-univalent 𝓤₁ → (𝟚 ≡ 𝟚) ≡ Lift 𝓤₁ 𝟚
  sol ua₀ ua₁ = Eq→Id ua₁ (𝟚 ≡ 𝟚) (Lift 𝓤₁ 𝟚) e
   where
    e = (𝟚 ≡ 𝟚)   ≃⟨ Id→Eq 𝟚 𝟚 , ua₀ 𝟚 𝟚                                  ⟩
        (𝟚 ≃ 𝟚)   ≃⟨ 𝟚-has-𝟚-automorphisms (univalence-gives-dfunext ua₀) ⟩
        𝟚         ≃⟨ ≃-sym (Lift-≃ 𝟚)                                     ⟩
        Lift 𝓤₁ 𝟚 ■

hde-is-subsingleton : dfunext 𝓤 𝓤₀
                    → dfunext 𝓤 𝓤
                    → (X : 𝓤 ̇ )
                    → is-subsingleton (has-decidable-equality X)
hde-is-subsingleton fe₀ fe X h h' = c h h'
 where
  a : (x y : X) → is-subsingleton (decidable (x ≡ y))
  a x y = +-is-subsingleton' fe₀ b
   where
    b : is-subsingleton (x ≡ y)
    b = hedberg h x y

  c : is-subsingleton (has-decidable-equality X)
  c = Π-is-subsingleton fe (λ x → Π-is-subsingleton fe (a x))

ne = sol
 where
  sol : (X : 𝓤 ̇ ) → ¬¬(X + ¬ X)
  sol X = λ (f : ¬(X + ¬ X)) → f (inr (λ (x : X) → f (inl x)))

DNE-gives-EM = sol
 where
  sol : dfunext 𝓤 𝓤₀ → DNE 𝓤 → EM 𝓤
  sol fe dne P i = dne (P + ¬ P) (+-is-subsingleton' fe i) (ne P)

EM-gives-DNE = sol
 where
  sol : EM 𝓤 → DNE 𝓤
  sol em P i = γ (em P i)
   where
    γ : P + ¬ P → ¬¬ P → P
    γ (inl p) φ = p
    γ (inr n) φ = !𝟘 P (φ n)

SN-gives-DNE = sol
 where
```

```
  sol : SN 𝓤 → DNE 𝓤
  sol {𝓤} sn P i = h
   where
    X : 𝓤 ̇
    X = pr₁ (sn P i)

    f : P → ¬ X
    f = pr₁ (pr₂ (sn P i))

    g : ¬ X → P
    g = pr₂ (pr₂ (sn P i))

    f' : ¬¬ P → ¬(¬¬ X)
    f' = contrapositive (contrapositive f)

    h : ¬¬ P → P
    h = g ∘ tno X ∘ f'

    h' : ¬¬ P → P
    h' φ = g (λ (x : X) → φ (λ (p : P) → f p x))

DNE-gives-SN = sol
 where
  sol : DNE 𝓤 → SN 𝓤
  sol dne P i = ¬ P , dni P , dne P i
```

## Operator fixities and precedences

Without the following list of operator precedences and associativities (left or right), this agda file doesn't parse and is rejected by Agda.

```
infix   0 _∼_
infixr 50 _,_
infixr 30 _×_
infixr 20 _+_
infixl 70 _∘_
infix   0 Id

infix   0 _≡_
infix  10 _⇔_
infixl 30 _·_

infixr  0 _≡⟨_⟩_
infix   1 _∎
infix  40 _⁻¹
infix  10 _◁_
infixr  0 _◁⟨_⟩_
infix   1 _◀
infix  10 _≃_
infixl 30 _●_
infixr  0 _≃⟨_⟩_
infix   1 _■
infix  40 _∈_
infix  30 _[_,_]
infixr -1 -Σ
infixr -1 -Π
infixr -1 -∃!
```